\documentclass{iopart}
\usepackage{graphicx}

\begin{document}
\title[Exact prefactors of correlation functions of 1D quantum integrable models]{Exact prefactors in static and dynamic correlation functions of 1D 
quantum integrable models:
applications to the Calogero-Sutherland, Lieb-Liniger and XXZ models \\  \ \ }

{\large \bf \today}
\author{Aditya Shashi}
\address{Department of Physics and Astronomy, Rice
University, Houston, Texas 77005, USA}
\author{Mi\l osz Panfil}
\address{Institute for Theoretical Physics, Universiteit van
Amsterdam, 1090 GL Amsterdam, The Netherlands}
\author{Jean-S{\'e}bastien Caux}
\address{Institute for Theoretical Physics, Universiteit van
Amsterdam, 1090 GL Amsterdam, The Netherlands}
\author{Adilet~Imambekov}
\address{Department of Physics and Astronomy, Rice
University, Houston, Texas 77005, USA}

\newcommand{\tpsi}{\tilde{\psi}}
\newcommand{\nn}{\nonumber}
\newcommand{\tV}{\tilde{V}}
\newcommand{\hpsi}{\hat{\psi}}
\renewcommand{\d}{\hat{d}}
\newcommand{\trho}{\tilde{\rho}}
\newcommand{\hrho}{\hat{\rho}}
\newcommand{\alphabar}{{\bar{\alpha}}}
\newcommand{\expct}[1]{\left\langle #1 \right\rangle}
\newcommand{\expcts}[1]{\langle #1 \rangle}
\renewcommand{\th}{{\rm th}}
\newcommand{\bw}{\begin{widetext}}
\newcommand{\ew}{\end{widetext}}
\newcommand{\be}{\bea}
\newcommand{\ee}{\eea}
\newcommand{\bea}{\begin{eqnarray}}
\newcommand{\eea}{\end{eqnarray}}
\newcommand{\p}{{\rm ^{'}}}

\newcommand{\ceq}[1]{Eq.~(\ref{#1})}
\newcommand{\dg}{^{\dagger}}

\newcommand{\V}{\hat{V}}
\newcommand{\al}{|\alpha\rangle}
\newcommand{\pd}{\hat{\psi}\dg}
\newcommand{\bab}{\bigg |}

\begin{abstract}
In this article we demonstrate a recently developed technique which addresses the problem of obtaining non-universal prefactors of the correlation functions of 1D systems at zero temperature. Our approach combines the effective field theory description of generic 1D quantum liquids with the finite size scaling of form factors (matrix elements) which are obtained using microscopic techniques developed in the context of integrable models. We thus establish exact analytic forms for the prefactors of the long-distance behavior of equal time correlation functions as well as prefactors of singularities of dynamic response functions. In this article our focus is on three specific integrable models: the Calogero-Sutherland, Lieb-Liniger, and XXZ models. 

\end{abstract}
\maketitle

\section{Introduction}

One finds for one-dimensional (1D) quantum mechanical systems of interacting bosons, fermions and spins, that the highly restrictive kinematics leads to many examples of exactly solvable models.  Consequently there often exist explicit expressions for the wavefunctions of such models, and for some of their bulk properties like energy and the excitation spectrum. However in the context of condensed matter physics one is interested in the correlation functions of these systems, which characterize their response to external probes. Unfortunately the complexity of the wavefunction makes the problem of calculating correlation functions notoriously difficult. 

On the other hand, the effective low energy physics of a broad set of 1D quantum liquids can be described using a hydrodynamic approach known as the Luttinger liquid theory~\cite{EL,Haldane,GNT,Caza04,Gbook}. This theory predicts the universal long-range behavior of the correlation functions of such systems. One obtains, for example, asymptotic expressions for equal-time correlation functions at zero temperature as series expansions with power laws controlled by a dimensionless Luttinger liquid parameter $K>0,$ see Eqs.~(\ref{Amdef})-(\ref{Cmdef}), (\ref{Dmdef})-(\ref{Emdef}). While the ``universal'' parameter $K$ is related to thermodynamic properties and can be easily extracted from numerical or exact solutions, the ``non-universal'' (model-specific) prefactors in the series expansion, e.g. $A_m, B_m, C_m, D_m$, see Eqs.~(\ref{Amdef})-(\ref{Cmdef}), (\ref{Dmdef})-(\ref{Emdef}), are usually not known analytically except for a few cases~\cite{Popov_prefactor,K1,correlation_results1,correlation_results2,correlation_numerical, Kozlowski, lukyanov03,CS_results,Lyon_group}. These prefactors set the actual scale of observable correlations, and are an impediment to establishing the complete asymptotic behavior of the static correlation functions of a given 1D quantum model.

In a recent publication \cite{PRB}, by combining the analysis of the Luttinger liquid Hamiltonian with the finite-size properties of certain matrix elements (form factors), a general technique for calculating these non-universal prefactors was developed for a generic 1D quantum liquid. Moreover, it has been shown recently~\cite{Pustilnik2006Fermions}-\cite{Affleck_preprint} that dynamic response functions generically have singularities which can be described by effective Hamiltonians of impurities moving in Luttinger liquids.
Analysis of the finite-size properties of these effective Hamiltonians can also be used to obtain prefactors of various dynamic response functions. 

In what follows, we apply this technique to the calculation of prefactors of the correlation functions of integrable models, focusing specifically on three integrable models: the Calogero-Sutherland model (CSM)~\cite{Calogero, Sutherland, CSM} of fermions interacting via a long range inverse-squared-distance potential, the Lieb-Liniger model~\cite{LL} of bosons with pairwise contact interactions and the XXZ model~\cite{XXZ_Heisenberg,1958_Orbach_PR_112} of a Heisenberg spin chain. These models have the advantage of being solvable~\cite{CSM, LL,XXX_soln, XXZ_soln} and additionally have readily available expressions for finite size form factors~\cite{Korepin, CSMexact, Slavnov, XXZ_formfactors1,XXZ_formfactors}, which we will investigate to obtain analytic expressions for prefactors of their correlation functions valid in the thermodynamic limit. Furthermore, the Lieb-Liniger model has been realized with ultracold atomic gases~\cite{dweiss}, and its correlation functions can be measured using
interference~\cite{interference, interference2, exp2}, analysis of particle losses~\cite{3decay}, photoassociation~\cite{3decay}, or Bragg and photoemission spectroscopy~\cite{BraggPE}, and density fluctuation statistics 
~\cite{density_fluctuations}. The XXZ model has enjoyed a long history as a well studied model in statistical physics \cite{Korepin,LMbook,Gaudin}.

This article is organized as follows.
Sec. 2 we outline the analysis of the effective field theory description of 1D quantum systems. In Sec. 2.1, we use the linear Luttinger liquid theory to work out the connection 
between prefactors of equal-time correlation functions and lowest energy form factors. In Sec. 2.2, we show that analysis of the effective field theory of impurities moving in Luttinger liquids gives the relations between form factors and prefactors of dynamic response functions.  In Sec. 3 we present details of the calculation of the prefactor in the density structure factor (DSF) in the vicinity of the threshold singularity for the CSM, by working out the thermodynamic limit of form factors. This is to serve as an introduction to the more technically involved calculations presented in Sec. 4 where we obtain several prefactors of static and dynamic correlation functions of the Lieb-Liniger model, with details on the calculation of thermodynamic limits of form factors of this model. In Sec. 5 we present the derivation of the prefactors of the equal time spin-spin correlator of the XXZ spin chain with a finite magnetization. Additional technical details are contained in the Appendixes.

\section{Results from effective field theory}

The Luttinger liquid theory~\cite{EL,Haldane,Caza04,Gbook, GNT}  predicts the behavior of the correlation functions for spinless bosons and fermions of density $\rho_0$, when $\rho_0
x \gg 1$ as (here $k_F=\pi \rho_0$) 

\bea
 \frac{\langle \hat \rho(x) \hat \rho(0)\rangle }{\rho_0^2} \approx&&\!\!\!\!\!\!\!\!\!\! 1\!-\! \frac{K}{2(\pi \rho_0 x)^2}+\!\!\sum_{m\geq 1}
\frac{A_m \cos(2 m k_F x)}{\left(\rho_0x\right)^{2m^2 K}}, \label{Amdef} \\
\frac{\langle \hat \psi_B^{\dagger}(x) \hat \psi_B(0) \rangle}{\rho_0} \!\!&\approx&\!\! \sum_{m\geq 0} \frac{B_m \cos(2m k_F x)}{\left(\rho_0x\right)^{2m^2
K+1/(2K)}},\label{Bmdef}\\
\frac{\langle \hat \psi_F^{\dagger}(x) \hat \psi_F(0) \rangle}{\rho_0} \!\!&\approx&\!\!\sum_{m\geq 0} \frac{C_m \sin{[(2m+1)k_F x]}}{(\rho_0 x)^{(2m+1)^2 K/2+1/(2K)}},
\label{Cmdef}
\eea
Here $\hat \rho$ is the density operator, $\hat \psi_B (\hat \psi_F)$ is the bosonic (fermionic) annihilation operator. The Hamiltonian describing these correlations is written as (we follow notations of Ref.~\cite{Gbook})
\bea
H_0= \frac{v}{2\pi}\int dx\;\left( K (\nabla \theta)^2 +\frac1{K}(\nabla \phi)^2\right ), \label{hbosonic}
\eea
where $v$ is the sound velocity, the canonically conjugate fields $\phi(x),\theta(x)$ have the
commutation relation $ [\phi(x),\nabla \theta(x')]=i \pi \delta(x -x'),$ and the components of the fermionic (bosonic) fields with momenta $(2m+1/2\pm 1/2)k_F$ are written as
\bea
\label{fieldops}
 \psi_{F(B)}(x,t)\sim  e^{i(2m+1/2\pm 1/2)[k_Fx -\phi(x,t)]+i\theta(x,t)},
 \eea
 where each component is defined up to a non-universal prefactor. For repulsive bosons, one has $K>1,$ while for repulsive (attractive) fermions $K<1(>1).$
  In Eqs. (\ref{Amdef})-(\ref{Cmdef}), we included only slowest decaying power laws for each oscillating component. In principle, irrelevant corrections ( see e.g. Ref.~\cite{Pereira_PRL_06}) to Hamiltonian (\ref{hbosonic})  and operators (\ref{fieldops}) generate various faster decaying  power-law terms for each oscillating component in addition to the one presented above \cite{Didier_PRA_2009}.
    
In the case of the spin chain, we consider specifically the XXZ model given by the Hamiltonian
\bea
\label{xxz_hamiltonian}
H_{XXZ} = \sum_{j=1}^{N} J(S_j^{x}S_{j+1}^x + S_j^yS_{j+1}^y + \Delta S_{j}^zS_{j+1}^z - hS_j^z),
\eea
where $S_j^{\alpha=x,y,z} = \sigma_j^\alpha/2$ is a spin half operator acting at site $j$, satisfying the algebra $[S^\alpha,S^\beta] = i \epsilon_{\alpha \beta \gamma}S^\gamma$, and the spins at different sites commute. $\Delta$ is a measure of the anisotropy of the coupling in the $z-$direction and $h$ is an external magnetic field aligned in the $z-$direction. 
One typically performs a Jordan-Wigner transformation to map the Hamiltonian (\ref{xxz_hamiltonian}) to that of a 1D lattice of fermions with nearest neighbor interactions. We can then calculate the asymptotic behavior of the correlators of the XXZ model, by first representing the spins in terms of fermions and then carry out the usual bosonization procedure \cite{Gbook, GNT, spin_bosonization} for the fermionic fields.  A finite magnetization, $s^z= \langle S^z \rangle$  is easily accommodated in this framework and enters as a chemical potential for the fermions, see e.g. Ref.~\cite{Gbook}. Following this prescription one obtains the following bosonic representations for the spin operators:
 \bea
 S^z(x,t) \sim s_z - \frac{\nabla \phi}{\pi} +  e^{i2m[(s_z+1/2)\pi x- \phi(x,t)]},\nn\\
 S^+(x,t) \sim  e^{-i2m[(s_z+1/2)x - \phi(x,t)]-i\theta(x,t)}.
  \eea
where again each oscillating component is defined up to a non-universal prefactor.

Similarly to Eqs. (\ref{Amdef})-(\ref{Cmdef}), Hamiltonian (\ref{hbosonic}) then results in the following long distance behavior

\bea 
 \label{Dmdef}
 \langle S^z(x)S^z(0)\rangle = s_z^2 -  \frac{K}{2(\pi x)^2} + \sum_{m\geq 1} \frac{D_m cos(2m(s_z+1/2)\pi x)}{x^{2m^2 K}},\\
 \label{Emdef}
 \langle S^+(x)S^-(0)\rangle = \sum_{m\geq 0} \frac{E_m cos(2m(s_z+1/2)\pi x )}{x^{2m^2 K+1/(2K)}}.
\eea

We note however, that in the presence of the lattice, the definition of the prefactors for higher $m$ requires more care. Indeed, for rational fillings $1/2+s_z=p/q$, the momenta of oscillating terms  for $m_1 \equiv m_2  ({\rm mod}~q)$ are not distinguishable, since momentum is no longer a good quantum number, but quasimomentum is. Thus in principle ``subleading" terms from smaller $m_1$ need to be distinguished from the leading term for $m_2.$ While this might not be possible for rational fillings (in particular, half-filling), and some rational values of the Luttinger liquid parameters $K$, it should be possible for generic filling and interaction strength. In what follows, we will focus on the case of  generic interaction strength and finite magnetic field  to avoid the complications 
related to half-filling (the case of half-filling for the XXZ model was considered in Ref.~\cite{lukyanov03} by different techniques).

\subsection{Prefactors of equal-time correlators 
from the Luttinger liquid theory}

Let us consider a system of interacting bosons. Using the resolution of the identity in the expectation value $\langle \hat \psi_B^{\dagger}(x,t)\hat \psi_B(0)\rangle,$
we get
\bea
\langle \hat \psi_B^{\dagger}(x,t)\hat \psi_B(0)\rangle=\sum_{k,\omega}e^{i (kx-\omega t)}\left|\langle k ,\omega|\hat \psi_B|N\rangle\right|^2, \label{Lehmann}
\eea
where $\langle k ,\omega  |\hat \psi_B|N\rangle$ is a form factor of the annihilation operator, $|k ,\omega \rangle$ denotes  an eigenstate of $N-1$ particles with momentum $k$ and energy $\omega$, and $|N\rangle$ is the ground state of $N$ particles. For a finite system, $k$  and $\omega$ are not continuous, but will be quantized and consequently the spectral function is a collection of delta functions in $(k,\omega).$ We will now obtain a similar representation from the Luttinger liquid theory and compare it with Eq.~(\ref{Lehmann}) to obtain the non-universal prefactors $B_m.$ 
Hamiltonian (\ref{hbosonic}) can be written using left- and
right-moving components $\varphi_{L(R)}=\theta \sqrt{K} \pm
\varphi/\sqrt{K},$ ~\cite{Caza04} which dictates the time dependence of the
$\cos{(2mk_Fx)}$ component of $\langle \hat \psi_B(x,t) \psi_B(0,0)\rangle/\rho_0,$ at $\rho_0|x\pm vt| \gg 1$ as
\bea
\frac{(-1)^m B_m \rho_0^{-2m^2K-1/2K}\cos{(2mk_Fx)}}{\left(i
(vt+x)+0\right)^{\mu_L}\left(i
(vt-x)+0\right)^{\mu_R}},\label{tdep}
\eea
where $\mu_{L(R)}=m^2K\pm m+1/4K>0.$ The coefficients $B_m$ appeared in \ceq{tdep} because we relate the $t=0$ limit of $\langle \hat{\psi}_B^\dagger(x,t)\hat{\psi}_B(0,0)\rangle/\rho_0$ to the right hand side of \ceq{Bmdef}.  The two factors in the denominator
describe contributions from left (right)-going excitations which
propagate with velocities $\mp v,$ and signs of the infinitesimal
shifts in the denominators ensure that only excitations with
negative (positive) momenta can be created at the respective
branches. For a finite system with periodic boundary conditions on a circle of length $L$, conformal invariance dictates (see e.g. Ref.~\cite{Caza04}) , the right-going component $\left(i
(vt-x)+0\right)^{-\mu_R}$ is replaced by
\bea
\left(\frac{\pi
e^{i\pi(vt-x)/L}}{iL\sin{\frac{\pi(vt-x)}{L}}+0}\right)^{\mu_R}=\sum_{n_r\geq
0}  C(n_r, \mu_R)\frac{e^{2i\pi
n_r(x-vt)/L}}{(L/2\pi)^{\mu_R}},\label{fourier} 
\eea
\bea
\label{finite_size}
C(n_r,
\mu_R)=\frac{\Gamma(\mu_R+n_r)}{\Gamma(\mu_R)\Gamma(n_r+1)},
\eea
and similarly for left-going components with $\mu_R,n_r$
substituted by $\mu_L, n_l.$ These equations lead to nontrivial
predictions for the exact scaling of the form factors of a model describing interacting 1D bosons, e.g. the
Lieb-Liniger model. By comparing the finite
size excitation spectrum of $H_0$ with the exact solution of the Lieb-Liniger model, up to $\propto 1/L$ terms, we identify $\mp 2\pi n_{l(r)}/L$ as the total
momenta of excitations created near the left (right) quasi-Fermi
points, i.e. the edges of the distribution of quasimomenta characterizing an eigenstate of the model. 
Considering $n_r=n_l=0$ then leads
to the scaling law
\be
\left|\langle m,N-1|\hat
\psi_B|0,N\rangle\right|^2=\frac{(-1)^m B_m \rho_0}{2 - \delta_{0,m}} \left(\frac{2\pi}{\rho_0L}\right)^{\frac{4m^2
K^2+1}{2K}}, \label{boson_scaling}
\ee
where $|m,N\rangle$ denotes an eigenstate of $N$ bosons having
center of mass momentum $2mk_F,$. We see that as a consequence of the criticality of the Luttinger liquid,
form factors of the annihilation operator have nontrivial scaling with the system size, and the prefactors of these nontrivial powers of $L$ are directly related to the prefactors
of the correlation functions. 

For density correlations, field correlation functions for fermions and correlators of spins, similar relations can be worked out as long as we represent the relevant operators in the bosonized language, and are given by 
\bea
\left|\langle m, N-1|\hat \psi_F|N\rangle\right|^2 &\approx& \frac{C_m \rho_0}{2(-1)^{m}}\left(\frac{2\pi}{\rho_0 L}\right)^{\frac{(2m+1)^2 K^2+1}{2K}}, \ \ \ \ \ \ \label{fermion_scaling}\\
\left|\langle m, N|\hat \rho |N\rangle\right|^2 &\approx& \frac{A_m \rho_0}{2}  \left(\frac{2\pi}{\rho_0 L}\right)^{2m^2 K}\label{density_scaling},\\
\left|\langle m,N|\hat S^z|N\rangle\right|^2 &\approx& \frac{D_m}{2}\left(\frac{2\pi}{L}\right)^{2m^2K}\label{spin_scaling},\\
\left|\langle m,N-1|\hat S^-|N\rangle\right|^2 &\approx& \frac{(-1)^mE_m}{2}\left(\frac{2\pi}{L}\right)^{2m^2K+1/(2K)}\label{spin_scaling2}.
\eea

Eqs.~(\ref{boson_scaling})-(\ref{spin_scaling2}) allow one to evaluate the prefactors in Eqs.~(\ref{Amdef})-(\ref{Cmdef}), (\ref{Dmdef})-(\ref{Emdef}), by identifying a single, simplest ``parent" form factor for each of the operators $\hat{\rho},\hat{\psi}_B$, $\hat{\psi}_F$, $\hat S^z$ and $\hat{S}^-$.

We note here that it recently came to our attention that a method to obtain prefactors of equal-time correlators, similar in spirit to ours is presented in Ref.~\cite{Bogoliubov_J_Phys}, however recent theoretical developments allow us to push the technique further and obtain e.g. the precise splitting of spectral weight among low energy form factors, and also to obtain prefactors of dynamic response functions. 


Field theoretical considerations allow to fix not only the
form factor with  $n_r=n_l=0,$ but also the form factors for all
low-energy states. However, since for $n_{r(l)}>1$ states can be
degenerate, one needs to
understand how the spectral weight $C(n_r,\mu_R)$ is split between
different form factors. This question can be answered by matching
contributions from each form factor with the free fermionic
quasiparticle representation of the Luttinger Liquid~\cite{universal}.
Such representation has the same degeneracies as the exact
solution, and we calculated~\cite{PRB} its form factors using the
results of Ref.~\cite{BAW06}. For a state with $k$
particle-hole excitations near the right quasi-Fermi point specified by integers $p_1>...>p_k\geq0$ (particles) and $q_1<...<q_k<0$ (holes), with total momentum $(2\pi/L) \sum_i(p_i-q_i)=2\pi n_r/L,$  we obtain that the ratio of its form factor to the one with $n_r = n_l =0$ equals
\be
f(\{p_i,q_i\})={\rm Det}_{i,j\leq
k}\left(\frac{1}{p_i-q_j}\right)\prod_{i\leq k}
f^+(p_i)f^-(q_i),\label{overlap}
\ee
where
\bea f^+(p)=\frac{\Gamma(p + 1 -
\sqrt{\mu_R} )}{\Gamma(-\sqrt{\mu_R})
\Gamma(p+1)},f^-(q)=\frac{\Gamma(-q +\sqrt{\mu_R}
)}{\Gamma(1+\sqrt{\mu_R}) \Gamma(-q)}.\nonumber
\eea
Normalization of the spectral weight leads to the following ``multiplet summation rule" (see Ref.~\cite{PRB} and appendix of the same for details):
\bea
\sum_{\sum p_i-q_i=n_r}\left|f(\{p_i,q_i\})\right|^2= C(n_r,\mu_R).
\label{mrule}
\eea
When $ n_l \neq 0,$ contributions from the left quasi-Fermi point are accounted for similarly, and the total form factor is a product of these two terms.

\subsection{Prefactors of singularities in DRFs using three subband model}

We now apply the techniques described above to the prefactors of singularities in dynamic response functions \cite{PRB}. For more comprehensive discussions of singularities in the response functions of 1D quantum liquids as well as the field theoretical description which captures these phenomena, see Refs.~\cite{Pustilnik2006Fermions}-\cite{Affleck_preprint}. For concreteness we focus here on the Lieb-Liniger model of bosons. We will consider the density structure factor
\bea
S(k,\omega) = \int dx dt^{i(\omega t - k x)} \langle \hat{\rho}(x,t)\hat{\rho}(0,0)\rangle,\label{DSF}
\eea
and the spectral function $A(k,\omega) = -\frac{1}{\pi}{\rm Im} G(k,\omega){\rm sign} \omega$ where the Green's function $G(k,\omega)$ is defined as~\cite{AGDbook}
\bea
G(k,\omega) = -i\int dx dt e^{i(\omega t - kx)} \langle T[\hat{\psi}(x,t)\hat{\psi}^{\dagger}(0,0)]\rangle.\label{Spectral}
\eea

We consider the density structure factor $S(k,\omega)$ in more detail below, and present only final results for the spectral function $A(k,\omega).$ The exponents of $S(k,\omega)$,
$\mu_{1,2}$ at Lieb's collective modes~\cite{LL, Korepin}
$\varepsilon_{1,2}(k)$ can be written as
 $\mu_{1,2}=1-\tilde\mu_R- \tilde \mu_L,$ where $ \tilde
\mu_{R(L)}$ denote contributions from right (left) branches and are given by~\cite{PRL_09}.
\bea
\tilde\mu_{R(L)} = \left(\frac{\sqrt{K}}{2} \pm \frac{1}{2\sqrt{K}} + \frac{\delta_{\pm}(k)}{2\pi}\right)^2.
\label{tildemu} 
\eea
The phases $\delta_\pm$ can be obtained explicitly by knowing the analytic form of the dispersion curve, the Luttinger parameter $K$ and momentum $k$ \cite{PRL_09}, or by directly extracting them from the microscopic model (see Ref.~\cite{PRL_08} for the calculation of the phase shifts for bosons using the Bethe ansatz for the Lieb-Liniger model).

 \begin{figure}
\includegraphics[width=11 cm,height=7 cm]{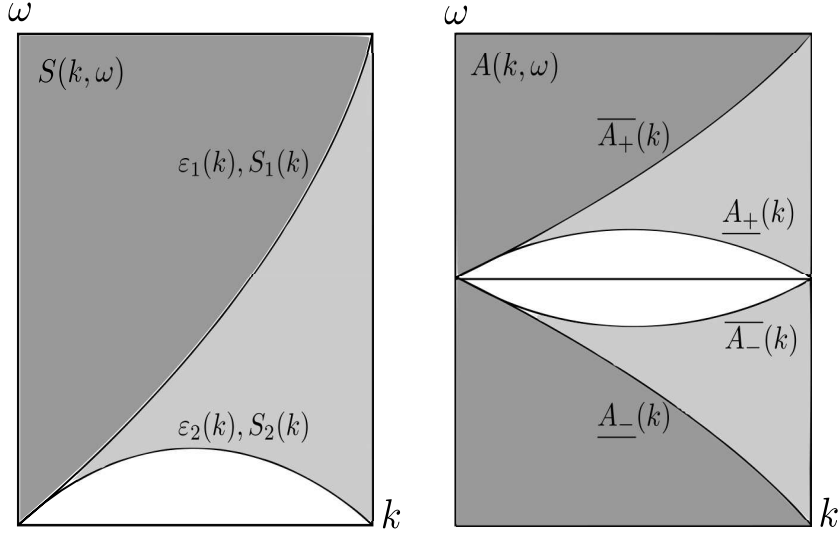}
\caption{\label{drf_notation} (a) Dynamic structure factor $S(k,\omega)$ and
(b) spectral function $A(k,\omega).$ Shaded areas indicate the regions
where they are non-vanishing. LiebÕs particle mode $\varepsilon_1(k)$ and 
hole excitation mode $\varepsilon_2(k)$ are indicated. For spectral function $A(k,\omega)$ region with $\omega > 0(\omega < 0)$ corresponds to the particle
(hole) part of the spectrum. The figure above indicates the notation for the prefactor of the singularity in the response functions at a given edge of support.}
\end{figure}

 In addition to the quasi-Fermi
points, the field theoretical description of the singularities ~\cite{PRL_09}
involves an impurity moving with velocity
$v_d=\partial\varepsilon_{1,2}(k)/\partial k,$ and 
$S(k,\omega)$ in the vicinities of collective modes are
written as
\bea
S(k,\omega)=S_{1,2}(k)\int dx dt
e^{i\delta \omega t}D(x,t)L(x,t)R(x,t),\label{Sint}
\eea
where $\delta\omega=\omega-\varepsilon_{1,2}(k),
D(x,t)=\delta(x-v_d t)$ is the impurity correlator,
$L(R)(x,t)=(i(vt \pm x)+0)^{-\tilde \mu_{L(R)}},$ and we
introduced prefactors $S_{1,2}(k)$ which will be fixed from the
comparison with form factors. In the vicinity of
$\varepsilon_2(k),$ Eq.~(\ref{Sint}) results in
\bea
S(k,\omega)= \theta(\delta\omega ) \frac{2\pi S_{2}(k)
\delta\omega^{\tilde \mu_R+\tilde \mu_L-1} }{\Gamma(\tilde
\mu_R+\tilde \mu_L)(v+v_d)^{\tilde \mu_L}|v-v_d|^{\tilde
\mu_R}},\label{S2def}
\eea
while in the vicinity of $\varepsilon_1(k)$ one has a two-sided
singularity, 
\bea
\fl
S(k,\omega)= \frac{\sin{\pi \tilde \mu_L}\theta(\delta\omega)+\sin{\pi
  \tilde \mu_R}\theta(-\delta\omega)}{\sin{\pi (\tilde \mu_L+\tilde \mu_R)}} \frac{2\pi S_{1}(k)
\delta\omega^{\tilde \mu_R+\tilde \mu_L-1} }{\Gamma(\tilde
\mu_R+\tilde \mu_L)(v+v_d)^{\tilde \mu_L}|v-v_d|^{\tilde
\mu_R}}, \label{S1def}
\eea
 In finite
size systems, $L(x,t)$ and $R(x,t)$ get modified, see Eq.~(\ref{fourier}). The change of $D(x,t)$ to  $\sum_{n_D} e^{2i\pi
n_D(x-v_dt)/L}$ corresponds to the quantization of the impurity momentum. 
Analysis of the scaling of the form factor with
$n_r=n_l=0$ then leads to
\bea
\left|\langle k;N|\hat
\rho|N\rangle\right|^2\approx \frac{S_{1(2)}(k)}{L}\left(\frac{2\pi}{L}\right)^{\tilde \mu_R +\tilde
\mu_L}, \label{DSFscaling}\eea
where the exponents are specified in \ceq{tildemu}, and $|k;N\rangle$ denotes a state of $N$ bosons 
with
a single
particle (hole) with high momentum, and a hole (particle) at the
right quasi-Fermi point, such that the total momentum is
$k.$ The state on the right is the full $N$ particle ground state (in this case that of the bosons). We note that in Eq.~(\ref{DSFscaling}) $k$ has to be
fixed before taking the limit $L\rightarrow \infty,$ since 
the $k\rightarrow 0$ and $L\rightarrow \infty$ limits do not commute 
in the nonlinear Luttinger Liquid
theory~\cite{PRL_08,universal,PRL_09}. It should also be noted that while our discussion focused on bosons in the Lieb-Liniger model, the relation in \ceq{DSFscaling} is general and will apply equally well to the DSF of the CSM, albeit with different exponents $\tilde{\mu}_{R,L}$. 

Similar to the density structure factor, the spectral function also displays singular behavior near the Lieb modes \cite{PRL_08,PRL_09} and we find relations for the prefactors of the spectral function in terms of form factors of the creation/annihilation operators. We set the following notation before presenting equations similar to \ceq{DSFscaling} for the prefactors of the spectral function. In the vicinities of $\varepsilon_1(k)$ and
$-\varepsilon_2(k),$ respectively, the spectral function behaves like
\bea
A(k,\omega) = \theta(\omega \mp \varepsilon_{1(2)}(k)) \frac{2\pi \overline{A_{\pm}(k)}
(\omega \mp \varepsilon_{1(2)}(k))^{ \overline{\mu_R}+\overline{ \mu_L}-1} }{\Gamma(\overline{
\mu_R}+\overline{ \mu_L})(v+v_d)^{\overline{ \mu_L}}|v-v_d|^{\overline{
\mu_R}}},\label{overline_behavior}
\eea
with exponents \cite{PRL_09}
\bea
\overline{\mu_{\pm}}=1-\overline{\mu_{R}}-\overline{\mu_{L}},\nn\\
\overline{\mu_{R(L)}} = \left(\frac{\sqrt{K}}{2} - \frac{\delta_{\pm}(k)}{2\pi}\right)^2.
\label{overlinemu}
\eea
Note that in the above the phase shifts $\delta_{\pm}$ can be calculated by knowing the dispersion curve and $k$~\cite{PRL_09} or from microscopics~\cite{PRL_08}. In Ref.~\cite{PRL_08}, the phase shifts are bosonic and calculated from the Bethe ansatz for the Lieb-Liniger model. In \ceq{overlinemu} above, the expressions for the exponents refers to fermionic $\delta_{\pm}$, consistent with the notations of Refs.~\cite{PRL_08,PRB}. The final answer for the exponents in Refs.~\cite{PRL_08,PRL_09} are all in agreement provided the correct $\delta_\pm$ are used. 

By analyzing  the scaling of the form factor of the creation/annihilation operator with
$n_r=n_l=0$ we obtain
\bea
|\langle k,N\mp1|\psi^{(\dagger)}|N\rangle|^2 \approx \frac{2\pi \overline{A_{\mp}(k)}}{L}\left(\frac{2\pi}{L}\right)^{\overline{\mu_R} + \overline{\mu_L}},\label{spectralscaling1}
\eea
with $\overline{\mu_{R(L)}}$ specified in \ceq{overlinemu}. The state $|k,N\pm 1\rangle$ refers to a state with an additional particle or hole such that the total momentum is $k$, while $|N\rangle$ is the ground state of $N$ bosons.

 Similarly, near  $\varepsilon_2(k)$ and $-\varepsilon_1(k)$ we have
\bea
A(k,\omega) = \theta(\omega \mp \varepsilon_{2(1)}(k)) \frac{2\pi \underline{A_{\pm}(k)}
(\omega \mp \varepsilon_{2(1)}(k))^{ \underline{\mu_R}+\underline{ \mu_L}-1} }{\Gamma(\underline{
\mu_R}+\underline{ \mu_L})(v+v_d)^{\underline{ \mu_L}}|v-v_d|^{\underline{
\mu_R}}},\label{underline_behavior}
\eea
with exponents \cite{PRL_09}
\bea
\underline{\mu_{\pm}}=1-\underline{\mu_{R}}-\underline{\mu_{L}},\nn\\
\underline{\mu_{R(L)}} = \left(\frac{3\sqrt{K}}{2} \mp \frac{1}{\sqrt{K}} - \frac{\delta_{\pm}(k)}{2\pi}\right)^2.
\label{underlinemu}\eea
We note again that the phase shifts $\delta_{\pm}$ are meant to be fermionic in line with notations of Refs.~\cite{PRL_09,PRB}. Alternate expressions for the bosonic exponents with bosonic phase shifts calculated in terms of microsocopics of the Lieb-Liniger model can be found in Ref.~\cite{PRL_08}. Moreover for the prefactors we obtain
\bea
|\langle k,N\mp1|\psi^{(\dagger)}|N\rangle|^2 \approx \frac{2\pi \underline{A_{\pm}(k)}}{L}\left(\frac{2\pi}{L}\right)^{\underline{\mu_R} + \underline{\mu_L}},\label{spectralscaling2}
\eea
with $\underline{\mu_{R(L)}}$ specified in \ceq{overlinemu}. The state $|k,N\pm 1\rangle$ refers to a state with two particles (holes) near the right quasi-Fermi point and an additional hole (particle) such that the total momentum is $k$, while $|N\rangle$ is the ground state of $N$ bosons. Let us also mention that for certain parameters
$A(k,\omega)$ can have a non-analyticity at
$-\varepsilon_1(k)$ instead of a divergence~\cite{PRL_09}. In this case $\underline{A_{-}}(k)$ refers to the prefactor of the
non-analytic part.  The results for prefactors of the singularities in the spectral function for fermionic models are presented in Ref.~\cite{PRB} along with expressions for the relevant exponents.

Finally let us note that the methods described here can be also used to fix the prefactors of dynamic response functions of  integrable lattice models, such as  the model of spinless fermions with nearest neighbor interactions~\cite{spinless_fermions} or XXZ model. For that,  one  needs to combine their form factors~\cite{XXZ_formfactors,lattice_formfactors}
with the corresponding field theoretical description of the singularities in their DRFs~\cite{XXZ,spinless_fermions, Affleck_preprint}. We postpone the discussion of DRFs of these lattice models for future  due to more complicated nature of their exact eigenstates.

\section{Density structure factor of the Calogero-Sutherland Model }

We demonstrate here how the relationship between non-universal prefactors of correlation functions and form factors of excited states calculated for the corresponding operators, established in Eqs.~(\ref{density_scaling}) and (\ref{DSFscaling}) can be used to calculate prefactors in the correlation functions of the Calogero-Sutherland Model (CSM)\cite{Calogero,Sutherland,CSM} described by the Hamiltonian
\bea
\hat{H}_{\rm CSM} = -\sum_{j=1}^{N}\frac{\partial^2}{\partial z_j^2} + \sum_{1\leq j<k\leq N} \frac{\lambda(\lambda-1)\pi^2}{L^2{\rm sin}^2\left[\pi(z_j-z_k)/L\right]}.
\eea
Here $N$ is the total number of particles of mass $1/2$, contained in a system of length $L$ held at density $\rho_0=N/L$ and $\lambda=p/q>1/2$ with $p,q$ coprime controls the strength of the long range interactions between the constituent particles. 

 The CSM admits an exact solution and consequently one obtains the energies of the ground state and the spectrum of excitations. Furthermore, the CSM ground state wavefunction has a product form \cite{Calogero}
 \bea
 \psi_{\rm GS} = \prod_{i<j}(z_i - z_j)^\lambda \prod_{k}z_k^{-\lambda(N-1)/2}, {\rm where} \hspace{.3 cm} z_j = e^{\frac{2\pi i x_j}{L}},
 \eea
 and can be interpreted qualitatively as the wavefunction of a gas of non-interacting particles with fractional exchange statistics. The excited states can moreover be constructed from the ground state by multiplying by Jack symmetric polynomials \cite{Sutherland} and are completely characterized by a set of $N$ quantum numbers $n_1,...,n_N$. These quantum numbers allow us to solve for the ``asymptotic'' quasimomenta which allow us to calculate physical observables like energy and excitation spectra, using \cite{CSM}:
 \bea 
 \label{CSM_BA}
 p_j = \frac{2\pi n_j}{L} + \frac{\pi(\lambda -1)(2j-N-1)}{L}, \ \ \ \ \ j = 1,...,N.
 \eea
 One may use the intuitive appeal of asymptotic plane wave states to interpret the quantum numbers $n_j$ as the wave numbers of these asymptotic states. The algebraic structure of these polynomials is efficiently captured in terms of operations performed over Young tableaux \cite{Jack, Young}.
 
 The CSM exhibits an additional property that leads to the existence of closed form expressions for form factors of physically relevant operators. The action of, e.g. the density operator, on the ground state can only result in particular types of states allowed under ``selection rules'', corresponding to states with a finite number of excitations \cite{CSMexact}. Such a structure is a consequence of the fact that the ground state wavefunction and the operators acting on it admit symmetrized representations in terms of Jack polynomials which have to satisfy orthogonality relations. Qualitatively, creation of states with finite numbers of excitations can be thought of as a generalization of the action of the operator on the gas of free fermions where it results in a finite number of (real) particle-hole processes. Thus, the space of eigenstates of the CSM possesses a structure reminiscent of Fock space, which can equivalently be seen from the form of the ground state wavefunction. This property greatly constrains the complexity of the form factors of the CSM and allows one to obtain analytic expressions. 
 
An expression for the density-density correlator is available for a finite sized system for the CSM due to Ref.~\cite{CSMexact}. One begins with a representation of the density fluctuation operator as
\bea
\label{density_jack}
\fl \rho(x) &=& \frac{1}{L}\sum_{j=1}^N \delta(x-x_j) - \frac{N}{L}=\frac{1}{L}\left(\sum_{m=1}^\infty e^{\frac{2\pi i m}{L}x}\sum_{j=1}^{N} e^{-\frac{2\pi i m}{L}x_j} + {\rm c. c.}\right),
\eea
which may in turn be expanded in terms of Jack polynomials. Using the orthogonality relation for these polynomials the density-density correlator may be written as a sum over Young Tableaux\cite{CSMexact}:
\bea
\label{ddcf}
\langle \rho(x,t)\rho(0,0)\rangle = \frac{1}{L^2\lambda^2}\sum_\kappa \frac{|\kappa|^2 ([0']^\lambda_\kappa)^2[N]^{\lambda}_{\kappa}}{j^\lambda_\kappa [N-1+1/\lambda]^\lambda_\kappa}e^{i(2\pi|\kappa|/L)x -i(E_\kappa - \mu)t}.  \hspace{1 cm}
\eea
where in the above expression the following notation is employed:
\bea
\label{CSM_notation}
E_{\kappa} &=& \left(\frac{2\pi}{L}\right)^2\sum_{j=1}^{N}(\kappa_j^2+\lambda(N+1-2j)\kappa_j),\nn\\
\left[a\right]^\lambda_\kappa &=& \prod_{(i,j) \in \kappa} (a+(j-1)/\lambda - (i-1)),\nonumber \\
j^\lambda_\kappa &=& \prod_{(i,j)\in \kappa}\left(\kappa'_j - i + \frac{1+\kappa_i - j}{\lambda}\right)\left(\kappa'_j - i + 1 +\frac{\kappa_i - j}{\lambda}\right),\nonumber \\
\eea
where $\kappa_i$ and $\kappa_j'$ refer to the total length of the $i^{th}$ row and the $j^{th}$ column respectively. Moreover $|\kappa|$, called the ``weight'' of a partition is the total number of blocks appearing in a given Young diagram. The Young diagrams above are indexed by $\kappa$ and particular ``cells'' or blocks contained in the diagram are indexed by $(i,j)$, with $i,j \geq 1$. Note that the prime in $[0']$ is an instruction to skip over the cell $i=1,j=1$ which will cause all terms to evaluate to 0, see \ceq{CSM_notation} in the product.

The expression in Eq.~(\ref{ddcf}) is in direct correspondence to the form factor expansion of the correlation functions, see Eq.~(\ref{Lehmann}), which involves a sum over all excited eigenstates of the system; every term in the sum Eq.~(\ref{ddcf}) explicitly contains the momentum $k$ and energy $\omega$ dependence of the state in the exponential term. Thus we may uniquely identify the  energy and momentum of the state associated with a Young diagram in the sum. This provides a recipe for constructing Young diagrams associated with excited states created by the action of the density operator on the ground state; we start by ordering the quantum numbers of the ground state and record the shift in the $i^{th}$ quantum number of the excited state as the length of the $i^{th}$ row of the Young diagram ($\kappa_i$ in the above notation). It should be noted that the selection rule mentioned earlier is encoded in the term $[0']^\lambda_\kappa$ which appears as a coefficient when expanding the density operator in terms of Jack polynomials in \ceq{density_jack}. This term is zero unless the Young diagram $\kappa$ contains a $p\times q$ block. We may interpret this as an excitation of $p$ quasiparticles and $q$ quasiholes, and the mathematical structure reflects the fundamental way in which the density operator may act on the ground state.

 We may directly obtain the form factor contribution from an excited state with a definite energy and momentum in terms of algebraic operations on a Young diagram. However to extract the prefactor of the correlation function in the thermodynamic limit from the form factor, one needs to carefully separate the non-trivial power-law of the system size $L$, see Eq.~(\ref{DSFscaling}), from the remaining contribution. We carry out this analysis below for the prefactor of the density structure factor and obtain analytic results.

\begin{center}
 \begin{figure}
\includegraphics[width=12 cm,height=10 cm]{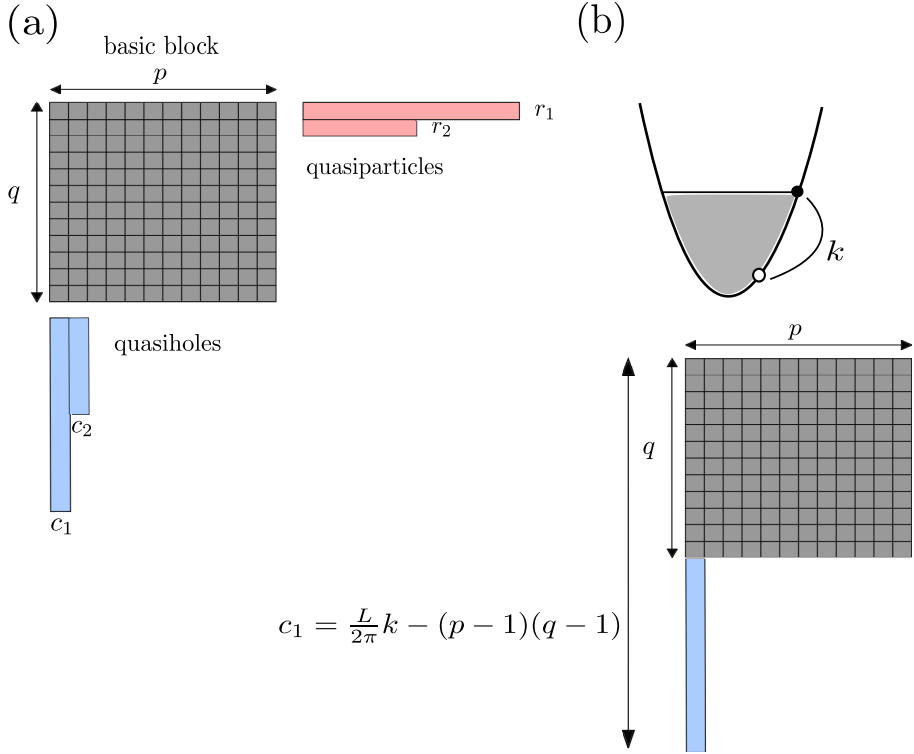}
\caption{\label{Fig1} (a) Typical Young diagram which appears in the sum in the expression for the density-density correlator see \ceq{ddcf}.  The basic block appears in all non-vanishing contributions to the density-density correlation function and is a $q \times p$ block. Excited states correspond to states with quasiparticle (quasihole) excitations, rows (columns) of total length $r_1, r_2,...(c_1, c_2,...)$ above the basic configuration, where $\lim_{N\to \infty} c,r/N \neq 0$; (b) The Young diagram corresponding to an excited state of momentum $k$, with energy $\varepsilon(k) = \lambda(k^2-k_F^2)$ which defines the edge of support. For free fermions this state would correspond to a state with a particle-hole pair with momentum transfer $k$ on top of the $N$ particle ground state}
\end{figure}
\end{center}

We wish to identify a single diagram that captures the contribution to the density structure factor when  $\omega \approx |\varepsilon(k)|$, i.e. in the vicinity of the edge of support with $\varepsilon(k) = \lambda (k^2 -k_F^2)$, and evaluate the corresponding form factor in the thermodynamic limit. We know from the finite size analysis of the effective three subband model, see Refs.~\cite{PRB,PRL_09}, that the contribution to the prefactor $S(k)$ comes from the form factor of the lowest energy state of momentum  $0 < k < 2k_F$. Such a state is the CSM analog of free fermions with a particle near the right branch, and a hole of momentum $k_F - k$. Following our prescription for constructing Young diagrams for states with particle-hole excitations over the ground state, this corresponds to a Young diagram with a single column of length $c_1 = \frac{L}{2\pi}k - (p-1)(q-1)$. We note here that there are also singularities for $k$ outside the range (0,$2k_F)$ and the prefactors of these singularities can be obtained from form factors of the state with additional umklapp excitations (a hole on the left branch, a particle on the right) on top of the configuration described above. For simplicity we present only the result for the first prefactor.
We will choose groups of terms appearing in the form factor expansion in \ceq{ddcf} to evaluate together. This is necessitated by the fact that some terms may individually diverge factorially ($\sim N^N$) in the thermodynamic limit, but their divergence is suppressed by other terms in the expression in such a way that the full expression will contain a non-trivial power law in $L$ and a prefactor. 

We first consider the following group of terms:
\bea
\mathcal{T}_1 &=& \frac{[N]^\lambda_{\kappa}}{[N+\frac{1}{\lambda}-1]^\lambda_\kappa} .
\eea

Let us first evaluate the contribution from the basic block (see Fig.~(\ref{Fig1})) which should be present in all terms:
\bea
\mathcal{T}_1^{\rm b.b} &=& \prod_{i=1}^{q}\prod_{j=1}^{p} \frac{\left(N+\frac{(j-1)}{\lambda} - (i-1)\right)}{\left(N+\frac{j}{\lambda} -i\right)}\nonumber\\
&=& \left(1+\frac{1}{N}\right)\prod_{j=1}^p\left(\frac{\lambda N + j-1}{\lambda N - p + \lambda + j}\right)\prod_{i=1}^{q}\left(\frac{N+1-i}{N+q-i}\right)\nonumber\\
&=& 1+ O(1/N).
\label{group1,a}
\eea

Additionally, the presence of particle-hole like excitations i.e. columns with length large in the thermodynamic sense, will lead to the following contribution:
\bea
\mathcal{T}_1^{\rm ex} &=&\prod_{i=q+1}^{c_1}\frac{N - i + 1}{N+ \frac{1}{\lambda}-i}\nn\\
&=& \frac{\Gamma(N-q+1)\Gamma(N-c_1+1/\lambda)}{\Gamma(N-q+1/\lambda)\Gamma(N-c_1+1 )} = \left(1- \frac{k}{2k_F}\right)^{\frac{1}{\lambda} - 1}.
\label{group1,b}
\eea

Next we consider the term $\mathcal{T}_2 =\frac{([0']^\lambda_\kappa)^2}{j^\lambda_\kappa}.$

We again calculate separately the contribution from the first column corresponding to the particle-hole like excitation:
\bea
\eqalign
\fl
\mathcal{T}_2^{\rm ex}=\prod_{i=2}^{q}\frac{(i-1)^2}{(c_1-i + q)(c_1-i+1+q-1/\lambda)}\times\prod_{i=q+1}^{c_1}\frac{ (i-1)^2}{(c_1-i + 1/\lambda)(c_1-i+1)}\nn\\
= \Gamma\left(\frac{1}{\lambda}\right)\frac{\Gamma(c_1)^3 \Gamma(c_1 +1 - 1/\lambda)}{\Gamma(c_1 + q)\Gamma(c_1 + q + 1- 1/\lambda)\Gamma(c_1 -q + 1/\lambda)\Gamma(c_1 - q + 1)}\nn\\
= \left(\frac{L}{2\pi}\right)^{-1-\frac{1}{\lambda}}\Gamma\left(\frac{1}{\lambda}\right)k^{-1-\frac{1}{\lambda}}. 
\eea

The remaining terms evaluate to unity when properly normalized by requiring the basic block form factor evaluates to $\rho_0$. We may now combine the above expressions to obtain a prefactor of the DSF using \ceq{DSFscaling}, the correspondence between the Luttinger parameter $K$ and the interaction parameter $\lambda$, $\frac{1}{\lambda} = K$, and the exponents $\tilde{\mu}_R, \tilde{\mu}_L$ given by
\bea
\tilde{\mu}_{R(L)} = \left[\frac{\sqrt{K}}{2} +(-) \frac{1}{2\sqrt{K}} + \frac{\delta_{+(-)}(k)}{2\pi}\right]^2,\nn\\
\delta_{\pm} = \mp\left( \frac{\sqrt{K}}{2} - \frac{1}{2\sqrt{K}}\right). 
\eea

Thus we have
\bea
\eqalign
\fl
S_{\rm CSM}(k) &=& \lim_{L\to \infty}2\pi \left(\frac{L}{2\pi}\right)^{1+\frac{1}{\lambda} }|\langle k;N|\hat{\rho}|0,N\rangle|^2\nn\\
&=& \frac{2\pi}{\lambda}\Gamma\left(1+\frac{1}{\lambda}\right)\left(\frac{2k_Fk }{2k_F-k}\right)^{1-\frac{1}{\lambda}}.
\eea 

We were also able to evaluate the equal-time density-density correlation prefactor $A_1$, see \ceq{Amdef}, by considering the form factor for a state with a single column of length $N$ over the basic configuration. We obtain for the prefactor
\bea
A_1 = 2\frac{\Gamma\left(1+\frac{1}{\lambda}\right)^2}{(2\pi)^{\frac{2}{\lambda}}},
\eea
in agreement with the result of Ref.~\cite{CS_results}, where general expressions for the prefactors of oscillating components of the density-density correlators were obtained using the Replica Method. 

\section{ Prefactors of the Lieb-Liniger Bose Gas}

The one dimensional Bose gas with contact interactions of strength {\bf $c>0$} is described by the Hamiltonian
\bea
H = \int dx \left[ \partial_x \psi^{\dagger}(x) \partial_x\psi(x) + c\psi^{\dagger}(x)\psi^{\dagger}(x)\psi(x)\psi(x)\right],\eea
where $\psi^{\dagger}, \psi$ are the boson creation and annihilation operators, and where we assume the particles to have mass $1/2$. The problem has an exact solution given by the Bethe Ansatz~\cite{LL}. The solution of the $N$-particle interacting Bose gas is obtained in terms of well defined ground and excited states characterized by sets of $N$ quasimomenta $\{\lambda\}$ and $\{\mu\}$, respectively. 

The quasimomenta of the ground state for the problem are given as the solutions of the Bethe equations \cite{Korepin}:
\bea
\fl
L \lambda_j + \sum_{k=1}^{N} \theta(\lambda_j - \lambda_k) =2\pi n_j =  2\pi (j - (N+1)/2), \;\;j =1, 2,..., N,\eea
where $\theta(\lambda) = i\log\left(\frac{ic + \lambda}{ic - \lambda}\right) = 2\arctan\left({\frac{\lambda}{c}}\right)$, and where the equation carries the physical meaning of a particle traversing an entire turn of a ring and returning to its origin having picked up a total phase shift resulting from pairwise two particle scattering from the other particles in the way.

An arbitrary excited state may be created from the ground state by simply erasing some finite $m < N$ of the $n_j$, $\{n^-_{1},n^-_{2},...,n^-_{m}\}$, and replacing them with new quantum numbers $\{n^+_{1},n^+_{2},..., n^+_{m}\}$, that are not from the original set of $n_j$. If we call the new set of quantum numbers $\{\tilde{n}_j\}$, then $\{\tilde{n}_j\}$ is obtained as $\{n_j\}/ \{n^-\} \cup \{n^+\}$. The quasimomenta $\mu_j$ characterizing this excited state are given by the solutions of the equations:
\bea
\label{baex}
L \mu_j + \sum_{k=1}^{N} \theta(\mu_j - \mu_k) =2\pi \tilde{n}_j, \;\; j =1, 2,..., N.\eea

\subsection{Form Factors}

As demonstrated in the previous sections, full knowledge of matrix elements calculated from the microscopic theory is sufficient to obtain the associated prefactor of a correlation function. Thus, 
we wish to obtain the thermodynamic limits of various form factors between ground state and  excited states specified by the field theory. The final expression for the limit of the form factors is expected to have a power law behavior as a function of the system length $L$, see Eqs.~(\ref{fermion_scaling})-(\ref{density_scaling}), and Eq.~(\ref{DSFscaling}). Again our task is to separate the power law from the prefactor in the form factor when faced with terms that are badly divergent ($O(N^N)$), so we proceed in a similar spirit to the calculation of the DSF prefactor of the CSM. 

 We begin by considering the exact expressions for the form factors of the finite size Bose gas that arise from the machinery of the Algebraic Bethe Ansatz \cite{Korepin}.

The form factor of the density operator between normalized eigenstates  $| \{\mu\}_N \rangle$ and $| \{\lambda\}_N \rangle$ is given by the following expression \cite{Slavnov}: 
\bea
\label{ffdsf}
\mathcal{F} &=& \langle \{\mu\}_N| \rho(0)|\{\lambda\}_N \rangle = \left|\frac{iP_{ex}}{{\rm exp}[-i x P_{ex}]-1}\langle \{\mu\}_{N}\mid\int^{x}_0{\rm dy}\rho(y)\mid\{\lambda\}_{N}\rangle\right| \nn\\
&=& \left|\frac{i P_{ex} \Omega}{||\{\mu\}_N||^{1/2} ||\{\lambda\}_N||^{1/2}}\right|,
\eea
where
\bea
\label{omega}
 \Omega = \prod_j (V_j^+ - V_j^-)\prod_{j,k} \left(\frac{\lambda_{jk} + i c}{\mu_j - \lambda_k}\right) \frac{{\rm Det}(\delta_{jk} + U_{jk})}{V^+_p - V^-_p},\\
 \label{vjdef}
V_j^{\pm} = \prod_{k} \frac{\mu_k - \lambda_j \pm ic}{\lambda_k - \lambda_j \pm ic},\\
U_{jk} = \frac{i(\mu_j - \lambda_j)}{V_j^+ - V_j^-}\prod_{m\neq j}\left(\frac{\mu_m - \lambda_j}{\lambda_m - \lambda_j}\right)(K(\lambda_j -  \lambda_k) - K(\lambda_p - \lambda_k)),\\
K(\lambda_j -  \lambda_k) = K(\lambda_j,\lambda_k) = \frac{2c}{(\lambda_j-\lambda_k)^2 + c^2}.
\eea
Note that in the above expression $p$ is any integer and this freedom of choice is due to the rank deficient structure of $U_{jk}$ \cite{Slavnov}, \cite{Caux}.

Similarly, the form factor for the boson annihilation operator is given by ~\cite{Slavnov},~\cite{Caux}
\bea
\label{ffspec}
\mathcal{G}^- = \left| \langle \{ \mu \}_{N-1} |\psi(0)| \{ \lambda \}_{N}\rangle\right|= \left|\frac{\Xi}{||\{\mu\}_{N-1}||^{1/2} ||\{\lambda\}_N||^{1/2}}\right|,\eea
where
\bea
\Xi= \frac{1}{c^{1/2}}\frac{\displaystyle{\prod_{j,k}( \lambda_{jk}^2 + c^2)^{1/2}\prod_{j}^{N-1}(\tilde{V}_j^+ - \tilde{V}_j^-)}}{\displaystyle{\prod_{j,k,k\neq N}} (\lambda_j - \mu_k)}{\rm Det}_{N-1}(\delta_{jk} + S^-_{jk}),\\
\label{vjdef2}
\tilde{V}_j^{\pm} = \frac{\displaystyle{\prod_{k=1}^{N-1} \mu_k - \lambda_j \pm i c}}{\displaystyle{\prod_{k=1}^{N} \lambda_k - \lambda_j \pm ic}},\\
S^-_{jk} =  \frac{i(\mu_j -\lambda_j)}{\tilde{V}_j^+-\tilde{V}_j^-} \frac{\prod_{m \neq j}^{N-1}(\mu_m - \lambda_j)}{\prod_{m \neq j}^{N}(\lambda_m - \lambda_j)}(K(\lambda_j - \lambda_k) - K(\lambda_N - \lambda_k)). 
\eea

Moreover, the form factor for the creation operator $\mathcal{G}^+$ is obtained from the one for the annihilation operator $\mathcal{G}^-$ by Hermitian conjugation.

In the above expressions the norms are given by ~\cite{Gaudin},~\cite{Korepin_proof},~\cite{Slavnov},~\cite{Korepin}
\bea
\fl
\label{normdisc}
||\{\lambda\}_N|| = c^N\left\{\prod_{j\neq k}\frac{\lambda_j - \lambda_k + ic}{\lambda_j - \lambda_k}\right\}{\rm Det}_N\left(\delta_{jk}\left[L + \sum_{m=1}^N K(\lambda_j - \lambda_m)\right] - K(\lambda_j - \lambda_k)\right).\nn\\\eea
Using the density operator and boson creation and annihilation operators, we may obtain expressions for the density structure factor (DSF) and spectral function (SF) using the Lehmann representation \cite{AGDbook}.

\subsection{Thermodynamic Limits and Finte Size corrections}\label{sec:TL_FS}

 The various terms in the expressions for the form factors, Eqs. (\ref{ffdsf}) - (\ref{ffspec}), depend directly on the quasimomenta of the ground and excited states. A thermodynamic description should not make reference to these discrete quasimomenta. We wish to define ``thermodynamic quantities'' that provide an equivalent, but sensibly defined description in the limit of large $N, L$. In the thermodynamic limit, the ground state of the Bose gas can be described using a continuous distribution of quasimomenta $\lambda$ belonging to the set of real numbers $(-q,q)$, while some excited state with $n$ holes and $m$ particles consists of quasimomenta in the set $(-q,q)/(\lambda^-_1,...\lambda^-_n) \cup (\lambda^+_1,...,\lambda^+_m)$. For this purpose we will use the density function $\rho(\lambda)$ and the shift function $F(\lambda|\nu)$, that are conventionally used in thermodynamic descriptions of the Bose gas. These functions are given by \cite{Korepin}
 \bea
 \label{rhoint}
 \rho(\lambda) -\frac{1}{2\pi}\int_{-q}^{q}d\mu K(\lambda - \mu)\rho(\mu)  = \frac{1}{2\pi},\\
 \label{fint}
 F(\lambda |\nu) - \frac{1}{2\pi}\int_{-q}^{q}d\mu K(\lambda - \mu)F(\mu|\nu) = \frac{\theta(\lambda - \nu)}{2\pi},
 \eea
 where the shift function picks up an overall positive or negative sign depending on whether the excitation $\nu$ in \ceq{fint} is a particle or hole. Thus, the shift function for a particle-hole pair excitation can be described by $F(\lambda| \mu^+, \lambda^-) = F(\lambda|\mu^+) - F(\lambda|\lambda^-)$. Moreover using the linearity of \ceq{fint}, a general excited state with multiple excitations of both types can be written as a sum of shift functions taken with appropriate signs. 
 
 Furthermore, we may express the number density, $\rho_0$, of the Bose gas using the ground state density function
\bea \int_{-q}^{q} d\lambda \rho(\lambda) = \rho_0.
 \eea
 
 To successfully account for all constant factors in the thermodynamic limit, we will need to consider finite size corrections (in orders of $1/L$) to these functions and collect any corrections that sum to finite values in the thermodynamic limit.  
  We begin by writing the system of $N$ Bethe Ansatz equations that give the ground state quasimomenta of the $N$ particle interacting Bose gas
\bea
\label{rhodisc}\fl
L \lambda_j + \sum_{k=1}^{N} \theta(\lambda_j - \lambda_k) =2\pi n_j =  2\pi (j - (N+1)/2), \;\; j =1, 2,..., N.\eea
The quantum numbers $n_j$ on the right hand side give us a natural way to index the quasimomenta as we take the above system over to the thermodynamic limit - we will want to keep the ratio of $n_j/L$ a constant as we take $L \rightarrow \infty$. It is convenient to construct a generalized continuum version of the $n_i's$ that smoothly interpolates between them in the thermodynamic limit, as described in Ref.~\cite{Korepin}.

Let us define a variable $x$ that satisfies
\bea
\label{xdef}
L\lambda(x)  + \sum_{k=1}^{N} \theta(\lambda(x) - \lambda_k) = 2L\pi x.\eea
We now define a function $\rho_L$ as
\bea
\rho_L(\lambda(x)) = \frac{dx}{d\lambda}.\eea
This function carries the usual meaning of the density of quasiparticles in a given quasimomentum interval, but we will require it to retain finite size corrections. 

 To obtain finite size corrections to our thermodynamic quantities we may use the Euler-McLaurin formula~\cite{hbook} which quantifies the difference between a sum and an integral in terms of a series in powers of the discretization (in our case $1/L$). We obtain for $\rho_L$ \cite{Korepin}:
\bea
\fl
\rho_L(\lambda(x)) = \frac{1}{2\pi} + \frac{1}{2\pi L} \sum_{k=1}^{N} K(\lambda(x)-\lambda_k)\nn\\
\fl
=\frac{1}{2\pi} + \frac{1}{2\pi L} \int_{-N/2L}^{N/2L} K(\lambda(x)-\mu(y)) Ldy  -  \frac{1}{2\pi L}\frac{B_2}{2}\left(1-\frac{1}{2}\right)\left(\frac{dK(\lambda(x)-\mu(y))}{Ldy}\right)\bigg|_{y=-N/2L}^{y=N/2L}\nn\\
\fl
=\frac{1}{2\pi} + \frac{1}{2\pi} \int_{-q}^{q} K(\lambda-\mu) \rho_L(\mu)d\mu - \frac{1}{2\pi L^2}\frac{1}{24}\left(\frac{K'(\lambda-\mu)}{\rho_L(\mu)}\right)\bigg|_{\mu=-q}^{\mu=q}\nn\\ \fl
=\frac{1}{2\pi} + \frac{1}{2\pi} \int_{-q}^{q} K(\lambda-\mu) \rho_L(\mu)d\mu - \frac{1}{48\pi L^2 \rho_L(q)}\left(K'(\lambda-q)-K'(\lambda+q)\right) + O(1/L^3).
\eea

Let us note two things about the above calculation. Firstly, in the above expression the discrepancy between the boundary value of $x_{\rm b.c.} = N/2L$ and  $n_N/L$, the quantum number associated with the edge of the distribution $q$ can be resolved by modifying the definition of $q$ to absorb the linear term from the Euler-McLaurin expansion~\cite{Korepin}, i.e. we redefine $q$ as, $q \rightarrow \lambda_N + \frac{1}{2L\rho(\lambda_N)}$. Secondly, we will only need to keep track of corrections to $O(1/L)$ in the subsequent calculations since this captures all finite results in the thermodynamic limit.

We would like to characterize the excited state with the same precision as the ground state. Excited states are obtained from the ground state by replacing a finite number of ground state quantum numbers $n_j$ with new quantum numbers $\tilde{n}_j$ that are not from the set of ground state $n_j$. For a choice of finite $N, L$ these excitations are well defined in that we will have two finite sets of quasimomenta $\{\mu^+\}, \{\lambda^-\}$ corresponding to the particles and holes respectively. The notation is intended to be suggestive of the fact that $\mu^+$ are ``actual'' quasimomenta in the excited state, while $\lambda^-$ will occur only as a subset of the set of ground state quasimomenta and are absent in the excited state. 

It will be more convenient for our purposes to also consider ``artificial'' hole quasimomenta $\{\mu^-\}$ and attribute them to the excited state. This is meant in the following sense - all the quasimomenta of the excited state are shifted with respect to the ground state quasimomenta. We expect this shift to be of $O(1/L)$. Had the hole quasimomenta been present in the excited state, they too would have been shifted with respect to their ground state counterparts, $\{\lambda^-\}$, on the order of $1/L$. Consequently we may realize  $\mu^-_i$ as $\lambda^-_i + O(1/L)$. 

We would now like to obtain thermodynamic versions of $\{\mu^\pm\}, \{\lambda^-\}$ and retain $O(1/L)$ corrections to them.

Let us start by considering the simplest case of a single particle-hole pair and some finite $N, L$. Furthermore let us denote the quantum number corresponding to the particle, $n^+$, and the quasimomentum, $\mu^+$, with the corresponding hole and its counterpart in the excited state defined as $\lambda^-, \mu^-$ respectively.

 For convenience, let us define the following notation. A single prime accompanying a sum, $\sum '$, means we are leaving out terms corresponding to $\mu^+, \lambda^-$ in the sum. A double prime accompanying the sum, $\sum ''$, means we are adding in $\mu^-$ in place of $\mu^+$. To go from $\sum$ to $\sum '$, we explicitly treat the finite number of excitation terms. To go from $\sum '$ to $\sum ''$ we compensate the extra terms with terms outside the sum of the opposite sign. 

Starting from the Bethe equation for the particle excitation we proceed as follows:
\be
\label{primes}
\eqalign
L \mu^+ = 2\pi n^+ - \sum_{k=1}^N \theta(\mu^+ - \mu_k)\nn\\
L \mu^+ = 2\pi n^+ - \sum \p \theta(\mu^+ - \mu_k)  - \theta(\mu^+ - \mu^-) + \theta(\mu^+ - \mu^-)  \nn\\
L \mu^+ = 2\pi n^+ - \sum_{k=1}^N (\theta(\mu^+ - \lambda_k) - K(\mu^+ - \lambda_k)(\mu_k - \lambda_k)) + \theta(\mu^+ - \lambda^-)\nn\\
\ \ \  \ \ \ \ \ \  - K(\mu^+ - \lambda^-)(\mu^- - \lambda^-) + O(1/L^2)\nn\\
\mu^+ = 2\pi \frac{n^+}{L} - \sum_{k=1}^N \left(\frac{\theta(\mu^+ - \lambda_k) + K(\mu^+ - \lambda_k)\frac{F(\lambda_k)}{L\rho_L(\lambda)}}{L}\right) + \frac{\theta(\mu^+ - \lambda^-)}{L} + O(1/L^2)\nn\\
\mu^+ = 2\pi \frac{n^+}{L} - \int_{-q}^q d\lambda \theta(\mu^+ - \lambda)\rho_L(\lambda) + \frac{1}{L}\left(-\int_{-q}^{q}d\lambda K(\mu^+ - \lambda) F(\lambda) + \theta(\mu^+ - \lambda^-)\right)\nn\\
 + O(1/L^2).
\ee
Note that in the above derivation we have used the fact that we expect $\mu_k - \lambda_k, \mu^- - \lambda^-$ to be $O(1/L)$, and the relation, $(\lambda_j - \mu_j) = \frac{F(\lambda_j)}{L\rho(\lambda_j)}+O(1/L^2)$. Using the last step of the above as our starting point, we may now take the thermodynamic limit. The prescription we use is to keep $n^+/L$ constant as we send $N,L \rightarrow \infty$. Furthermore let us separate the $O(1)$ and $O(1/L)$ contributions to $\mu^+$ for the sake of clarity, i.e. $\mu^+ = \mu^+_0 + \mu^+_{1/L} + O(1/L^2)$. We obtain
\bea 
\mu^+_0 + \mu^+_{1/L} &=& 2\pi\frac{n^+}{L} - \int_{-q}^{q} d\lambda \theta(\mu^+_0 - \lambda)\rho(\lambda) - \mu^+_{1/L}\int_{-q}^q d\lambda K(\mu^+_0 - \lambda)\rho(\lambda) \nn\\
&+&  \frac{1}{L}\left(-\int_{-q}^{q}d\lambda K(\mu^+_0 - \lambda) F(\lambda) + \theta(\mu^+_0 - \lambda^-)\right) +O(1/L^2),\nn\\
\eea
where on the right hand side we have substituted $\rho$ for $\rho_L$, etc. because the lowest order difference between such terms is higher order in $1/L$ than we are keeping. 

Thus we obtain as the thermodynamic limit and the first order correction
\bea
\mu^+_0 =  2\pi\frac{n^+}{L} - \int_{-q}^{q} d\lambda \theta(\mu^+_0 - \lambda)\rho(\lambda),\nn\\
\mu^+_{1/L} = \frac{1}{L}\left(-\frac{\int_{-q}^{q}d\lambda K(\mu^+ - \lambda) F(\lambda) -\theta(\mu^+ - \lambda^-)}{1 +\int_{-q}^q d\lambda K(\mu^+_0 - \lambda)\rho(\lambda)} \right) = - \frac{F(\mu^+_0)}{L\rho(\mu^+_0)}.\nn\\
\eea
In the last step above, we have used Eqs.~(\ref{rhoint}),~(\ref{fint}).

Similarly, let us consider the Bethe equation for the ground state quasimomentum, $\lambda^-$,
\bea
L\lambda^- = 2\pi n^- - \sum_{k=1}^{N} \theta(\lambda^- - \lambda_k)\nn\\
\lambda^- = 2\pi\frac{n^-}{L} - \int_{-q}^q d\lambda \theta(\lambda^- - \lambda)\rho(\lambda) + O(1/L^2).\eea

We wish to now obtain $\mu^-$. Since we expect $\mu^- = \lambda^- + O(1/L)$, and the lowest order correction comes from the usual shift in the excited state quasimomenta, we can quantify this exactly as
\bea
\mu^-_0 = 2\pi\frac{n^-}{L} - \int_{-q}^q d\lambda \theta(\mu^-_0 - \lambda)\rho(\lambda),\nn\\
\mu^-_{1/L} = -\frac{F(\mu^-_0)}{L \rho(\mu^-_0)}.
\eea

We are now ready to obtain the lowest order finite size corrections to the shift function. Let us start with a definition for the finite sized shift function. For an excited state with some set of particles and holes, $\{\mu^+_1,\mu^+_2,...,\mu^+_n\}, \{\mu^-_1,\mu^-_2,...,\mu^-_n\}$, where $\mu^-$ is used in the sense described above, we have
\bea
\label{fl}
F_L(\lambda_j|\{\mu^+_1,\mu^+_2,...,\mu^+_n\}\{\mu^-_1,\mu^-_2,...,\mu^-_n\}) = (\lambda_j - \mu_j)L\rho_L(\lambda_j),\eea
where $\lambda_j, \mu_j$ are the ground and excited state quasimomenta respectively.

We may obtain an equation for $F_L$ by subtracting the Bethe equations for $\mu_j$ and $\lambda_j$ for $\mu_j$ not in $\{\mu^+,\mu^-\}$:
\bea
\fl
L(\lambda_j - \mu_j) + \sum_{k=1}^N(\theta(\lambda_j - \lambda_k) - \theta(\mu_j - \mu_k)) = 0\nn\nn\\ \fl
L(\lambda_j - \mu_j) + \sum \p (\theta(\lambda_j - \lambda_k) - \theta(\mu_j - \mu_k))+\theta(\mu_j - \mu^-) - \theta(\mu_j - \mu^-) - \theta(\mu_j - \mu^+) = 0\nn\nn\\ \fl
L(\lambda_j - \mu_j) + \sum \p\p(\theta(\lambda_j - \lambda_k) - \theta(\lambda_j - \lambda_k) - K(\lambda_j, \lambda_k)(\mu_j - \mu_k -\lambda_j + \lambda_k) \nn\nn\\ \fl
-\frac{1}{2}K'(\lambda_j-\lambda_k)(\mu_j - \mu_k -\lambda_j + \lambda_k)^2) =  \theta(\mu_j - \mu^+) - \theta(\mu_j - \mu^-)\nn\nn\\ \fl
L(\lambda_j - \mu_j)\left(1 + \frac{1}{L} \sum\p\p K(\lambda_j, \lambda_k) \right) = \sum \p\p K(\lambda_j,\lambda_k)(\lambda_k - \mu_k)+ \theta(\lambda_j - \mu^+_0) - \theta(\lambda_j - \mu^-_0)\nn\nn\\ \fl
+\sum \p\p \frac{1}{2}K'(\lambda_j-\lambda_k)(\mu_j - \mu_k -\lambda_j + \lambda_k)^2 +K(\lambda_j,\mu^+_0)(\mu_j - \lambda_j - \mu^+_{1/L})\nn\nn\\ \fl
 - K(\lambda_j, \mu^-_0)(\mu_j - \lambda_j - \mu^-_{1/L}),\nn\nn\\ \fl
2\pi L (\lambda_j - \mu_j)\rho_L(\lambda_j) = \sum \p\p K(\lambda_j,\lambda_k)\frac{\lambda_k - \mu_k}{L\rho_L(\lambda_k)}L\rho_L(\lambda_k) + \theta(\lambda_j - \mu^+_0) - \theta(\lambda_j - \mu^-_0)\nn\nn\\ \fl
+\sum \p\p \frac{1}{2}K'(\lambda_j-\lambda_k)(\mu_j - \mu_k -\lambda_j + \lambda_k)^2 +K(\lambda_j,\mu^+_0)(\mu_j - \lambda_j - \mu^+_{1/L})\nn\nn\\ \fl
 - K(\lambda_j, \mu^-_0)(\mu_j - \lambda_j - \mu^-_{1/L})\nn\nn\\ \fl
2\pi F_L(\lambda) = \theta(\lambda - \mu^+_0) - \theta(\lambda - \mu^-_0) + \int_{-N/2L}^{N/2L} L dy K(\lambda,\mu(y))\frac{F_L(\mu(y))}{L\rho_L(\mu(y))}\nn\nn\\ \fl
-\frac{1}{L}\left[K(\lambda,\mu^+_0)\left(\frac{F(\lambda)}{\rho_L(\lambda)} - \frac{F(\mu^+_0)}{\rho_L(\mu^+_0)}\right) - K(\lambda, \mu^-_0)\left(\frac{F(\lambda)}{\rho_L(\lambda)} - \frac{F(\mu^-_0)}{\rho_L(\mu^-_0)}\right)\right] \nn\\ \fl
+\frac{1}{2L}\int_{-q}^{q}d\mu \rho_L(\mu)K'(\lambda-\mu)\left(\frac{F_L(\lambda)}{\rho_L(\lambda)} - \frac{F_L(\mu)}{\rho_L(\mu)}\right)^2+ O(1/L^2).
\eea
This makes the final equation for $F_L(\lambda)$ up to $O(1/L^2)$
\bea
\label{fdis}
\eqalign
\fl
F_L(\lambda) = \frac{\theta(\lambda - \mu^+_0) - \theta(\lambda - \mu^-_0)}{2\pi}+ \frac{1}{2\pi}\int_{-q}^{q} d\mu  K(\lambda,\mu)F_L(\mu) \nn\\
-\frac{1}{2\pi L}\left[K(\lambda,\mu^+_0)\left(\frac{F_L(\lambda)}{\rho_L(\lambda)} - \frac{F_L(\mu^+_0)}{\rho_L(\mu^+_0)}\right) - K(\lambda, \mu^-_0)\left(\frac{F_L(\lambda)}{\rho_L(\lambda)} - \frac{F_L(\mu^-_0)}{\rho_L(\mu^-_0)}\right)\right]\nn\\
+\frac{1}{4\pi L}\int_{-q}^{q}d\mu \rho_L(\mu)K'(\lambda-\mu)\left(\frac{F_L(\lambda)}{\rho_L(\lambda)} - \frac{F_L(\mu)}{\rho_L(\mu)}\right)^2+ O(1/L^2).
\eea
Note that in the above derivation, $F_L(\lambda)$ is a condensed notation for $F_L(\lambda|\mu^+,\mu^-)$, and moreover this $F_L$ is related to the more general one defined in \ceq{fl} as
\bea
\label{fcomp}\fl
 F_L(\lambda|\{\mu^+_1,\mu^+_2,...,\mu^+_n\};\{\mu^-_1,\mu^-_2,...,\mu^-_n\}) = F_L(\lambda|\mu^+_1, \mu^-_1) + ... + F_L(\lambda|\mu^+_n,\mu^-_n).
 \eea
 
 Furthermore, we expect that in the thermodynamic limit $F_L(\lambda| \mu^+_i,\mu^-_i) = F(\lambda|\mu^+_i) - F(\lambda|\mu^-_i) + O(1/L)$, where $F(\lambda|\mu^{\pm}_i)$ is given by \ceq{fint}.

It is also convenient to define an analog of the ground state momentum density function for the excited state. This will help in calculating the normalization of the excited state that appears in \ceq{ffdsf} and \ceq{ffspec}.
Before proceeding, let us write down the integral equations for the derivatives of $\rho_L, F_L$ as these will be used in the simplifications to follow.
\bea
\label{deriv1}
2\pi F_L'(\lambda) = \int_{-q}^{q}d\mu K'(\lambda - \mu)F_L(\mu) + K(\lambda_j-\mu^+_0)- K(\lambda_j-\mu^-_0) + O(1/L),\nn\\
\label{deriv2}
2\pi \rho_L'(\lambda) = \int_{-q}^{q}d\mu K'(\lambda - \mu)\rho_L(\mu) + O(1/L^2).
\eea

 To obtain the density function for the excited state, we start with the Bethe equation for a generic quasimomentum from the excited state, \ceq{baex}, and differentiate once with respect to $\mu_j$:
\bea
\label{derirhoex}
\fl
L = 2\pi L\rho_{ex,L}(\mu_j) - \sum_k\p K(\mu_j - \mu_k) - K(\mu_j - \mu^-) + K(\mu_j - \mu^-) - K(\mu_j - \mu^+)\nn\nn\\ \fl
L = 2\pi L\rho_{ex,L}(\mu_j) - \sum_k \p\p K(\mu_j - \mu_k) - K(\mu_j - \mu^+) + K(\mu_j - \mu^-)\nn\nn\\ \fl
L = 2\pi L\rho_{ex,L} (\lambda_j)  - \sum_{k=1}^N \left[ K(\lambda_j - \lambda_k)- K'(\lambda_j - \lambda_k)\left(\frac{F_L(\lambda_j)}{L \rho_L(\lambda_j)} - \frac{ F_L(\lambda_k)}{L\rho_L(\lambda_k)} \right)\right]\nn\nn\\ \fl
- 2\pi\rho_L'(\lambda_j)\frac{F_L(\lambda_j)}{\rho_L(\lambda_j)}  - K(\lambda_j - \mu^+_0) + K(\lambda_j - \mu^-_0) + O(1/L)\nn\nn\\ \fl
2\pi \rho_{ex}(\lambda)  - 2\pi\rho_L'(\lambda)\frac{F_L(\lambda)}{L\rho_L(\lambda)} = 1 - \frac{K(\lambda - \mu^+_0) - K(\lambda- \mu^-_0)}{L} \nn\nn\\ \fl
 + \int_{-q}^{q} d\nu \left[K(\lambda - \nu) \rho_{L}(\nu)  - K'(\lambda - \nu)\left(\rho_L(\nu)\frac{F_L(\lambda)}{L \rho_L(\lambda)} - \frac{F_L(\nu)}{L} \right)\right]+ O(1/L^2)\nn\nn\\ \fl
2\pi \rho_{L}(\lambda) + 2\pi \delta \rho(\lambda)  - 2\pi\rho_L'(\lambda)\frac{F_L(\lambda)}{L\rho_L(\lambda)} = 1 +   \int_{-q}^{q} d\nu K(\lambda - \nu) \rho_{L}(\nu) \nn\nn\\ \fl
 -2\pi \rho_L'(\lambda)\frac{F_L(\lambda)}{L \rho_L(\lambda)} + 2\pi \frac{F_L'(\lambda)}{L}.
\eea

The first two steps of the above set of equations is exact. From the third step on we retain only terms up to $O(1/L)$. In going to the third step we have expanded all terms dependent on $\mu_j$ in terms of $\lambda_j$ as this is what is useful for us in our final expression (see the following sections). In going to the last step we have used the relations given by Eq.~(\ref{deriv2}).

From \ceq{derirhoex} we have
\bea
\label{rhodiff}
\rho_{ex,L}(\mu_j) - \rho_L(\lambda_j) =\frac{1}{L}\left(F_L'(\lambda_j) - \frac{F_L(\lambda_j)\rho_L'(\lambda_j)}{\rho_L(\lambda_j)}\right) + O(1/L^2).\eea

It is useful to consider the notion of a ``partial thermodynamic limit''. For instance the algebraic expression for the norm of the eigenstate appearing in \ceq{normdisc} can be expressed in terms of a partial limit as established in Refs.~\cite{Korepin},~\cite{Gaudin},~\cite{Slavnov}
\bea
\label{normg}
\lim_{{\rm partial}}||\{\lambda\}||^2 = \prod_{j>k}\frac{\lambda_{jk}^2 + c^2}{\lambda_{jk}^2}\prod_j (2\pi L c \rho(\lambda_j)){\rm Det}\left(1-\frac{1}{2\pi} K\right),\eea
where the determinant is intended to be a Fredholm Determinant (see Ref.~\cite{Smirnov}, Appendix D for more details). The above expression is a ``partial'' limit in the sense that it contains both the quasimomenta of the ground state which should be absent when the ``full'' thermodynamic limit  is taken, as well as the Fredholm Determinant which is well defined in the TDL. 

Similarly, we may write down the norm for the excited state as
\bea
\label{normx}
\lim_{{\rm partial}}||\{\mu\}||^2 = \prod_{j>k}\frac{\mu_{jk}^2 + c^2}{\mu_{jk}^2}\prod_j (2\pi L c \rho_{ex}(\mu_j)){\rm Det}\left(1-\frac{1}{2\pi} K\right).
\eea

We will frequently use the notion of such partial limits because individual terms in the algebraic expressions for the form factors may not have good thermodynamic limits - they may often diverge in a non polynomial way. However, they can be regrouped to obtain expressions that have good (finite or power law divergent) limits. Thus we want the final expression for the form factor to be well defined in the limit, on the other hand the intermediate steps leading up to this will be a mixture of thermodynamically well defined quantities and the aforementioned partial thermodynamic limits. We will take partial limits of simple groupings occurring in the form factors in the following sections and later collect them and express the final answer for the form factors. 

 Let us consider the term $V_j^+/V_j^-$ with $V_j^{\pm}$ as defined in \ceq{vjdef}
\bea
\label{vj1}
\eqalign
\log\left(\frac{V_j^+}{V_j^-}\right) = \sum_{i=1}^{n}\log\left(\frac{\mu^+_i - \lambda_j + i c}{\mu^+_i - \lambda_j - ic}\right)-\sum_{i=1}^{n}\log\left(\frac{\lambda^-_i - \lambda_j + i c}{\lambda^-_i - \lambda_j - ic}\right)\nn\\
+ \sum_{m=1}^N \log\left(1 - \frac{F_L(\lambda_m)}{L\rho_L(\lambda_m)(\lambda_m - \lambda_j + ic)}\right)\nn\\
 - \sum_{m=1}^N \log\left(1 - \frac{F_L(\lambda_m)}{L\rho_L(\lambda_m)(\lambda_m - \lambda_j - ic)}\right)\nn\\
-\sum_{i=1}^n \log\left(\frac{\mu^-_i - \lambda_j + ic}{\lambda^-_i- \lambda_j + ic}\right) + \sum_{i=1}^n \log\left(\frac{\mu^-_i - \lambda_j - ic}{\lambda_i^- - \lambda_j - ic}\right)\nn\\
= \sum_{i=1}^{n}\log\left(\frac{\mu^+_i - \lambda_j + i c}{\mu^+_i - \lambda_j - ic}\right)-\sum_{i=1}^{n}\log\left(\frac{\mu^-_i - \lambda_j + i c}{\mu^-_i - \lambda_j - ic}\right)\nn\\
+ \sum_{m=1}^N i\frac{2c F_L(\lambda_m)}{L\rho_L(\lambda_m)((\lambda_m - \lambda_j)^2 + c^2)} \nn\\
+ \sum_{m=1}^N\frac{i}{2}\frac{4c(\lambda_m - \lambda_j) F_L^2(\lambda_m)}{L^2\rho_L^2(\lambda_m)((\lambda_m - \lambda_j)^2 + c^2)^2}+O(1/L^2)\nn\\
= \sum_{i=1}^n i \left[\theta(\lambda_j - \mu^+_{i,0}) - K(\lambda_j-\mu^+_{i,0})\mu_{i,1/L}^+ - \theta(\lambda_j - \mu^-_{i,0}) +K(\lambda_j - \mu^-_{i,0})\mu_{i,1/L}^-\right] \nn\\
+ \sum_{m=1}^N i\frac{ F_L(\lambda_m) K(\lambda_m-\lambda_j)}{L\rho_L(\lambda_m)} - \sum_{m=1}^N \frac{i}{2}\frac{K'(\lambda_m-\lambda_j) F_L^2(\lambda_m)}{L^2\rho_L^2(\lambda_m)} + O(1/L^2)\nn\\
= 2\pi i F_L(\lambda_j)+ \frac{i}{L}\left(K(\lambda_j-\mu^+_0)\frac{F_L(\lambda_j)}{\rho_L(\lambda_j)} - K(\lambda_j-\mu^-_0)\frac{F_L(\lambda_j)}{\rho_L(\lambda_j)} \right) \nn\\
-\frac{i}{2L}\int_{-q}^{q}d\mu K'(\lambda_j -\mu) \rho_L(\mu) \left(\frac{F_L(\lambda)}{\rho_L(\lambda)} - \frac{F_L(\mu)}{\rho_L(\mu)}\right)^2 \nn\\
+ \frac{i}{2L} \int_{-q}^{q}d\mu \frac{F_L^2(\mu)}{\rho_L(\mu)} K'(\lambda_j - \mu)+O(1/L^2)\nn\\
=2\pi i F_L(\lambda_j)+ \frac{i}{L}\left(K(\lambda_j-\mu^+_0)\frac{F_L(\lambda_j)}{\rho_L(\lambda_j)} - K(\lambda_j-\mu^-_0)\frac{F_L(\lambda_j)}{\rho_L(\lambda_j)} \right)\nn\\
+\frac{i}{L}\int_{-q}^{q}d\mu K'(\lambda_j -\mu) \frac{F_L(\lambda_j)}{\rho_L(\lambda_j)} \left(F_L(\mu) - \frac{F_L(\lambda_j)\rho_L(\mu)}{2\rho_L(\lambda_j)} \right) +O(1/L^2).
\eea

Now using the relations given by \ceq{deriv1} we have
\bea
\left(K(\lambda_j-\mu^+_0) - K(\lambda_j-\mu^-_0) + \int_{-q}^{q}d\mu K'(\lambda_j -\mu) \left(F_L(\mu) - \frac{F_L(\lambda_j)\rho_L(\mu)}{2\rho_L(\lambda_j)} \right) \right)\nn\\
= 2\pi\left( F_L'(\lambda_j) - \frac{F_L(\lambda_j)\rho_L'(\lambda_j)}{2\rho_L(\lambda_j)}\right).
\eea

Substituting in \ceq{vj1} we obtain
\bea
\fl\log\left(\frac{V_j^+}{V_j^-}\right)=2\pi i F_L(\lambda_j)\left[1 +  \frac{1}{L\rho_L(\lambda_j)}\left( F_L'(\lambda_j) - \frac{F_L(\lambda_j)\rho_L'(\lambda_j)}{2\rho_L(\lambda_j)}\right)\right] +O(1/L^2).\eea
Thus we may handle the following term appearing in $\Omega$, \ceq{omega}, as follows:
\bea
\fl
\prod_j^N(V_j^+ - V_j^-)\nn\\ \fl
= \prod_j^N V_j^+ \left(1 - {\rm exp}\left[-2\pi i F_L(\lambda_j)\left(1 +  \frac{1}{L\rho_L(\lambda_j)}\left( F_L'(\lambda_j) - \frac{F_L(\lambda_j)\rho_L'(\lambda_j)}{2\rho_L(\lambda_j)}\right)\right)\right]\right) + O\left(\frac{1}{L}\right)\nn\\ \fl
=\prod_j^N V_j^+ e^{i\phi(j)} 2{\rm sin}\left[\pi F_L(\lambda_j)\left(1 +  \frac{1}{L\rho_L(\lambda_j)}\left( F_L'(\lambda_j) - \frac{F_L(\lambda_j)\rho_L'(\lambda_j)}{2\rho_L(\lambda_j)}\right)\right)\right] + O(1/L),
\eea
where in the above expression the term $\phi(j)$ is purely a phase. Such terms will drop out in the final answer because we only need the absolute square of the form factor. The product of sine terms can be expanded in the following way
\bea
\fl V = \prod_j^N {\rm sin}\left[\pi F_L(\lambda_j)\left(1 +  \frac{1}{L\rho_L(\lambda_j)}\left( F_L'(\lambda_j) - \frac{F_L(\lambda_j)\rho_L'(\lambda_j)}{2\rho_L(\lambda_j)}\right)\right)\right]\nn\\ \fl
=\left(\prod_j^N {\rm sin}[\pi F_L(\lambda_j)]\right){\rm exp}\left[\sum_j^N \frac{\pi F_L(\lambda_j)\cos(\pi F_L(\lambda_j))}{L \rho_L(\lambda_j) \sin(\pi F_L(\lambda_j))}\left( F_L'(\lambda_j) - \frac{F_L(\lambda_j)\rho_L'(\lambda_j)}{2\rho_L(\lambda_j)}\right)\right]\nn\\ \fl
= \left(\prod_j^N {\rm sin}[\pi F_L(\lambda_j)]\right){\rm exp}\left[\pi \int_{-q}^q \frac{{\rm cos}(\pi F_L(\lambda))}{ sin(\pi F_L(\lambda))}F_L(\lambda)\left( F_L'(\lambda) - \frac{F_L(\lambda)\rho_L'(\lambda)}{2\rho_L(\lambda)}\right)\right].
\eea

The other terms in $\Omega$ may also be regrouped 
\bea
\label{det}
 \fl
\Omega=  \Theta \times V\prod_j 2 i e^{i\phi(j)}V_j^+\prod_{j,k} \left(\frac{\lambda_{jk} + i c}{\mu_j - \lambda_k}\right)\nn\\ \fl
=\Theta \times V\prod_j e^{i\phi(j) +3i\pi/2 }V_j^+ \frac{2}{\lambda_j - \mu_j} \prod_{j,k} \left(\lambda_{jk} + i c \right)\prod_{j\neq k}\frac{1}{\mu_j-\lambda_k}\nn\\ \fl
=\Theta \times V\prod_j e^{i\phi'(j)}V_j^+ \frac{2L\rho_L(\lambda_j)}{F_L(\lambda_j)} \prod_{j,k} \left(\lambda_{jk} + i c \right)\prod_{j\neq k}\frac{1}{\mu_j-\lambda_k}\prod_{i=1}^{n}\left(\frac{F_L(\lambda_i^-)}{L\rho_L(\lambda^-_i)(\lambda^-_i - \mu^+_i)}\right).\nn\\ \fl
\eea
where we use $\Theta$ to denote $\frac{{\rm Det}(\delta_{jk} + U_{jk})}{V^+_p - V^-_p}$

With this form for $\Omega$ we may combine it with the norm terms :
\bea
\fl
|\mathcal{F}|=\bigg|iP_{ex} \Theta \times V\prod_j e^{i\phi'(j)}V_j^+ \frac{2L\rho_L(\lambda_j)}{F_L(\lambda_j)} \prod_{j,k} \left(\lambda_{jk} + i c \right)\prod_{j\neq k}\frac{1}{\mu_j-\lambda_k}\prod_{i=1}^{n}\left(\frac{F_L(\lambda_i^-)}{L\rho_L(\lambda^-_i)(\lambda^-_i - \mu^+_i)}\right)\nn\nn\\ \fl
\times \prod_{j\neq k}\frac{\lambda_{jk}^{1/2}\mu_{jk}^{1/2}}{(\lambda_{jk} +i c)^{1/4}(\lambda_{jk} - i c)^{1/4}(\mu_{jk} +i c)^{1/4}(\mu_{jk} - i c)^{1/4}} \times {\rm Det} \left(1-\frac{1}{2\pi} K\right)^{-1}\nn\nn\\ \fl
\times \prod_j \frac{1}{(2\pi L c \rho_L(\lambda_j))}\left(1+\frac{\rho_{ex}(\mu_j) - \rho_L(\lambda_j)}{\rho_L(\lambda_j)}\right)^{-1/2}\bigg|\nn\nn\\ \fl
=\bigg|iP_{ex} \Theta \times V\prod_j V_j^+ \frac{1}{\pi F_L(\lambda_j)} \prod_{j,k} \left(\lambda_{jk} + i c \right)\prod_{j\neq k}\frac{1}{\mu_j-\lambda_k}\nn\nn\\ \fl
\times \prod_{j\neq k}\frac{\lambda_{jk}^{1/2}\mu_{jk}^{1/2}}{(\lambda_{jk} +i c)^{1/4}(\lambda_{jk} - i c)^{1/4}(\mu_{jk} +i c)^{1/4}(\mu_{jk} - i c)^{1/4}}{\rm Det}\left(1-\frac{1}{2\pi} K\right)^{-1}\nn\nn\\ \fl
\times\prod_{i=1}^{n}\left(\frac{F_L(\lambda_i^-)}{L(\rho_L(\lambda^-_i)\rho_{ex}(\mu^+_i))^{1/2}(\lambda^-_i - \mu^+_i)}\right)\times \prod_j \left(1+\frac{F_L'(\lambda_j) - \frac{F_L(\lambda_j)\rho_L'(\lambda_j)}{\rho_L(\lambda_j)}}{L\rho_L(\lambda_j)}\right)^{-1/2} \bigg |\nn\nn\\ \fl
=\bigg| iP_{ex} \Theta \times V\prod_j \frac{1}{\pi F_L(\lambda_j)}\prod_{j\neq k}\frac{\lambda_{jk}^{1/2}\mu_{jk}^{1/2}}{\mu_j-\lambda_k}\prod_{i=1}^{n}\left(\frac{F_L(\lambda_i^-)}{L(\rho_L(\lambda^-_i)\rho_{ex}(\mu^+_i))^{1/2}(\lambda^-_i - \mu^+_i)}\right)\nn\nn\\ \fl
\times \prod_{j,k}\frac{(\mu_k - \lambda_j + ic)}{(\lambda_{jk} +i c)^{1/2}(\mu_{jk} +i c)^{1/2}}{\rm Det}\left(1-\frac{1}{2\pi} K\right)^{-1}\times\prod_{j,k}\frac{(\lambda_{jk} +i c)^{1/4}(\mu_{jk} +i c)^{1/4}}{(\lambda_{jk} -i c)^{1/4}(\mu_{jk} -i c)^{1/4}}\nn\nn\\ \fl
{\rm exp}\left[-\frac{1}{2}\int_{-q}^{q}d\lambda \left(F_{L}'(\lambda) - \frac{F_{L}(\lambda)\rho_L'(\lambda)}{\rho_L(\lambda)}\right)\right] \bigg |.\nn\nn\\ \fl
\eea

In going to the last line we have collected a purely phase term in the last product.

After substituting the expression for $V$ and regrouping terms we obtain the final form
\bea
\label{startf}
 \fl
|\mathcal{F}|=\bigg |\left[\prod_{j,k} \left(\frac{(\lambda_j - \mu_k +ic)(\lambda_j - \mu_k + ic)}{(\lambda_{jk} + ic)(\mu_{jk} + ic)}\right)^{1/2}\right] \times \left\{\prod_j\frac{{\rm sin}(\pi F_L(\lambda_j))}{\pi F_L(\lambda_j)} \prod_{j\neq k}\left(\frac{\lambda_{jk} \mu_{jk}}{(\mu_j - \lambda_k)^2}\right)^{1/2}\right\}\nn\\  \fl
\times \prod_{i=1}^n\left( \frac{F_L(\lambda^-_i)}{L(\rho_L(\lambda^-_i)\rho_{ex}(\mu^+_i))^{1/2}(\lambda_i^- - \mu_i^+)}\right)\frac{i P_{ex}{\rm Det}(\delta_{jk} + U_{jk})}{(V_p^+ - V_p^-){\rm Det}\left(1-\frac{\hat{K}}{2\pi} \right)}\nn\\ \fl
{\rm exp}\left[\int_{-q}^q \left\{\frac{\pi F_L(\lambda) {\rm cos}(\pi F_L(\lambda))}{ {\rm sin}(\pi F_L(\lambda))}\left( F_L'(\lambda) - \frac{F_L(\lambda)\rho_L'(\lambda)}{2\rho_L(\lambda)}\right) - \frac{1}{2}\left(F_L'(\lambda) - \frac{F_L(\lambda)\rho_L'(\lambda)}{\rho_L(\lambda)}\right)\right\}\right] \bigg|.
\eea

 Similar manipulations can be performed to obtain the partial thermodynamic limit of the form factors associated with the boson creation/annihilation operators.  For the creation/annihilation operators we consider a matrix element between states with different numbers of particles (see \ceq{ffspec}). For the creation operator form factor, the ground state has $N$ particles while the excited state has $N+1$ particles, while for the annihilation operator form factor the ground state has $N$ particles while the excited state has $N-1$ particles. This means that in the construction of the excited state, there is either a missing or extra quantum number, relative to the ground state. This difference should manifest as an extra hole (for annihilation operator) or particle (for creation operator) in the excited state. To account for this key difference from the calculation of the density operator matrix element, we will introduce modified shift functions to express terms in the form factors of the creation and annihilation operators. We define these modified shift functions as
  \bea
 \label{fplus}
 F_{+}(\lambda) = F(\lambda| \{\mu^+_1,...,\mu^+_{n+1}\};\{\mu^-_1,..., \mu^-_n\}) + \pi \rho(\lambda),\\
 \label{fminus}
 F_{-}(\lambda) =  F(\lambda| \{\mu^+_1,...,\mu^+_n\};\{\mu^-_1,..., \mu^-_{n+1}\}) - \pi \rho(\lambda),
 \eea
 for $n \geq 0$ and with $F_{+}$ appearing in the creation operator form factor and $F_-$ appearing in the annihilation operator. The terms in Eqs. (\ref{fplus}), (\ref{fminus}) can be justified as follows. 

 The first term in Eqs. (\ref{fplus}), (\ref{fminus})  corresponds to the particle/hole excitations of the state $|\mu \rangle$, including the extra particle or hole excitation due to the difference in number of particles in  $|\mu \rangle$ versus $|\lambda \rangle$. This term is analogous to the shift function used for the density operator in \ceq{fcomp}, $F(\lambda|\{\mu^+_1,...,\mu^+_n\};\{\mu^-_1,..., \mu^-_n\}) = F(\lambda|\mu^+_1) - F(\lambda|\mu^-_1) +...+ F(\lambda|\mu^+_n) - F(\lambda|\mu^-_n)$ where $\mu^+_i$ and $\mu^-_i$ are particle and hole quasimomenta with respect to the $N$ particle ground state.  
 The second term, an additional factor of $\pm \pi \rho(\lambda)$ has a more subtle origin - the derivation of the equation for the shift function, \ceq{fint} from~\cite{Korepin}, assumes a change of periodic boundary conditions to antiperiodic ones with the change of particle number. Phase shifts calculated from this shift function correspond to the fermionic Cheon-Shigehara model~\cite{PhysRevLett.82.2536} dual to the bosonic Lieb-Liniger model. To obtain the desired phase shifts, one may ``correct'' the shift function by adding $\pi \rho(\lambda)$ to the unpaired shift function in \ceq{fplus} and \ceq{fminus}~\cite{PRL_08}. The sign before this correction is picked up according to whether the unpaired shift comes from an extra particle or hole. The reason these two terms are simply summed to give the correct total shift functions comes from the fact that we can sum the integral equation for the shift function for two different particle-hole configurations and obtain the integral equation for the composite shift function, i.e. the shift when all the individual particle hole configurations occur together.  
 
 Furthermore, there are a few modified terms in the form factor of the creation/annihilation operators that need to be carefully accounted for. To do so we will use the following procedure, focusing on the annihilation operator for concreteness. Let us  choose one of the $n$ holes in the excited state and consider its quantum number. There should be a ground state quasimomentum corresponding to this quantum number, with some index which we will call $n^-$. We are assured of at least one such hole since there is always one less quasiparticle in the excited state used for obtaining the creation operator form factor. We will seemingly treat this hole specially, but this is just a convenience as we could have picked any of the potentially $n$ available holes - but we will commit to one to simplify calculations. The final answers for the prefactors will be independent of the particular choice of $n^-$.
 
 Now we reindex the excited state quasimomenta as follows. For $i < n^-, \mu_i \to \tilde{\mu}_i$, for $i \geq n^-, \mu_i \to \tilde{\mu}_{i+1}$. Thus, in all subsequent steps there is a ``gap'' in the indices for $\mu$ at $\mu_{n^-}$. We will artificially define $\mu^-_{n+1} = \lambda_{n^-} - \frac{F_-(\lambda_{n^-})}{L\rho(\lambda_{n^-})}$. Thus we have $n+1$ holes and $n$ particles in the excited state with quasimomenta, $\mu^{\pm}_i$ for $n \geq 0$ and we will ``remember'' the index of one of the holes, $n^-$. Note that $\tilde{\mu_i}$ is simply a reindexing of the original set of $\mu_{i}$ with no change in numerical values implied. Under this reordering it becomes possible to define $F_-(\lambda_i) = (\lambda_i - \tilde{\mu_i})L\rho(\lambda_i), i\neq n^-$ and $F_-(\lambda_{n^-}) = (\lambda_{n^-} - \mu^-_{n+1})L\rho(\lambda_{n^-})$, as in \ceq{fl}, where the $F_-(\lambda)$ is the one defined in \ceq{fminus}.
 
 Let us consider the combination $\tilde{V}_j^+/\tilde{V}_j^-$ appearing in the form factors in \ceq{vjdef2}
\bea
\fl \frac{\tilde{V}_j^+}{\tilde{V}_j^-} 
=\left(\prod_{i =1}^n \frac{\mu^+_i - \lambda_j + ic}{\mu^+_i - \lambda_j - ic} \right)\left(\prod_{i =1}^{n+1} \frac{\mu^-_i - \lambda_j + ic}{\mu^-_i - \lambda_j - ic} \right)^{-1} \left(\prod_{k=1}^{N} \p\p \frac{\tilde{\mu}_k - \lambda_j + ic}{\lambda_k - \lambda_j + ic}\right)\left(\prod_{k=1}^{N} \p\p\frac{\tilde{\mu}_k - \lambda_j - ic}{\lambda_k - \lambda_j - ic}\right)^{-1}\nn\\
\fl= \left(\prod_{i=1}^n \frac{\mu^+_i - \lambda_j + ic}{\mu^+_i - \lambda_j - ic} \right)\left(\prod_{i=1}^{n+1} \frac{\mu^-_i - \lambda_j + ic}{\mu^-_i - \lambda_j - ic} \right)^{-1} \left(\prod_{k=1}^{N} \p\p \frac{\tilde{\mu}_k - \lambda_j + ic}{\lambda_k - \lambda_j + ic}\right)\left(\prod_{k=1}^{N} \p\p\frac{\tilde{\mu}_k - \lambda_j - ic}{\lambda_k - \lambda_j - ic}\right)^{-1}\nn\\
\fl ={\rm exp}\left\{2\pi i F_{-,L}(\lambda_j)\left[1 +  \frac{1}{L\rho_L(\lambda_j)}\left( F_{-,L}'(\lambda_j) - \frac{F_{-,L}(\lambda_j)\rho_L'(\lambda_j)}{2\rho_L(\lambda_j)}\right)\right] \right\}+O(1/L^2).
\eea
 
 We use the double prime notation as described in earlier sections (see the text prior to \ceq{primes}) and collect the hole created by having one less particle in $|\mu\rangle$ with the other $n$ holes of the excited state and raise the index for such terms to $n+1$. To obtain the last line we have used the same steps as in \ceq{vj1}.

We will now consider the following grouping of terms appearing in \ceq{ffspec}
\bea
\fl
\left |\frac{\displaystyle{\prod_{j,k}( \lambda_{jk}^2 + c^2)^{1/2}\prod_{j \neq n^-}^{N} \tilde{V}_j^+}}{\displaystyle{\prod_{j,k}^{N}(\lambda_{jk}^2 + c^2)^{1/4}\prod_{j,k \neq n^-}^{N}(\tilde{\mu}_{jk}^2 + c^2)^{1/4}}}\right |\nn\\ \fl
=\left|\frac{\displaystyle{\prod_{j,k}^N( \lambda_{jk}^2 + c^2)^{1/2}\prod_{j,k \neq n^-}^{N}(\tilde{\mu}_k - \lambda_j + ic)}}{\displaystyle{\prod_{j \neq n^-}^{N}\prod_k^N(\lambda_{kj} + ic)\prod_{j,k}^{N}(\lambda_{jk}^2 + c^2)^{1/4}\prod_{j,k \neq n^-}^{N}(\tilde{\mu}_{jk}^2 + c^2)^{1/4}}}\right |\nn\\ \fl
=\left |\frac{\displaystyle{\prod_{j,k}^N( \lambda_{jk}^2 + c^2)^{1/4}\prod_{j,k \neq n^-}^{N}(\tilde{\mu}_k - \lambda_j + ic)}}{\displaystyle{\prod_{j \neq n^-}^{N}\prod_k^N(\lambda_{kj} + ic)\prod_{j,k \neq n^-}^{N}(\tilde{\mu}_{jk}^2 + c^2)^{1/4}}}\right |\nn\\ \fl
=\left |\frac{\displaystyle{\prod_{j,k \neq n^-}^{N}( \lambda_{jk}^2 + c^2)^{1/4}\prod_{j,k \neq n^-}^{N}(\tilde{\mu}_k - \lambda_j + ic)}}{\displaystyle{\prod_{j,k \neq n^-}^{N}(\lambda_{kj} + ic)\prod_{j,k \neq n^-}^{N}(\tilde{\mu}_{jk}^2 + c^2)^{1/4}}}
\times \frac{\displaystyle{c^{1/2} \prod_{j\neq n^-}^N( \lambda_{jn^-}^2 + c^2)^{1/2}}}{\displaystyle{\prod_{j\neq n^-}^N( \lambda_{jn^-} - ic)}}\right |\nn\\ \fl
=\left |\frac{c^{1/2} \displaystyle{\prod_{j,k \neq n^-}^{N}(\tilde{\mu}_k - \lambda_j + ic)}}{\displaystyle{\prod_{j,k \neq n^-}^{N}(\lambda_{kj} + ic)^{1/2}\prod_{j,k \neq n^-}^{N}(\tilde{\mu}_{jk} + ic)^{1/2}}} \right |.
\eea

Moreover, the following terms from \ceq{ffspec} give
 \bea
 \label{m2der}\fl
\left(\frac{ \prod_{j\neq k}^N \lambda_{jk}\prod_{j \neq k \neq n^-}^{N} \tilde{\mu}_{jk}}{\prod_{j=1}^N\prod_{k \neq n^-}^{N}( \lambda_j - \tilde{\mu}_k)^2}\right)^{1/2} =\prod_{j\neq n^-}^{N}\frac{L \rho_L(\lambda_j)}{F_{-,L}(\lambda_j)}\left( \prod_{j\neq k \neq n^-}^{N}\frac{\lambda_{jk} \tilde{\mu}_{jk}}{( \lambda_j - \tilde{\mu}_k)^2}\right)^{1/2} \prod_{j\neq n^-}^{N}\frac{\lambda_{n^-}- \lambda_j}{\lambda_{n^-} - \tilde{\mu}_j}.\nn\\
\eea
 
On combining these results with the rest of the annihilation form factor we obtain as the partial limit of \ceq{ffspec}

\bea
\label{startg}
\fl
\mathcal{G}^- = \bigg |\prod_{j,k \neq n^-}^{N} \left(\frac{(\lambda_j - \tilde{\mu}_k +ic)(\lambda_j - \tilde{\mu}_k-ic)}{(\lambda_{jk} + ic)(\tilde{\mu}_{jk} + ic)}\right)^{1/2}\times \left \{ \prod_{j\neq n^-}^{N}\frac{{\rm sin}(F_{-,L}(\lambda_j))}{\pi F_{-,L}(\lambda_j)}\left( \prod_{j\neq k \neq n^-}^{N}\frac{\lambda_{jk} \tilde{\mu}_{jk}}{( \lambda_j - \tilde{\mu}_k)^2}\right)^{1/2}\right\} \nn\\ \fl
 \left(\prod_{j\neq n^-}^{N}\frac{\lambda_{n^-} -  \lambda_j}{\lambda_{n^-} - \tilde{\mu}_j}\right)\prod_{i=1}^n\left( \frac{F_{-,L}(\mu^-_i)}{L(\rho_L(\mu^-_i)\rho_{L}(\mu^+_i))^{1/2}(\mu_i^- - \mu_i^+)}\right) \frac{{\rm Det}(\delta_{jk} + S^-_{jk})}{(2\pi L \rho_L(\mu_{n+1}^-))^{1/2}{\rm Det}\left(1-\frac{\hat{K}}{2\pi} \right)}\nn\\ \fl
{\rm exp}\left[\int_{-q}^q d\lambda \left\{\frac{\pi F_{-,L}(\lambda) {\rm cos}(\pi F_{-,L}(\lambda))}{ {\rm sin}(\pi F_{-,L}(\lambda))}\left( F_{-,L}'(\lambda) - \frac{F_{-,L}(\lambda)\rho_L'(\lambda)}{2\rho_L(\lambda)}\right)\right\}\right]\nn\\ \fl
{\rm exp}\left[  - \frac{1}{2}\int_{-q}^{q} d\lambda\left(F_{-,L}'(\lambda) - \frac{F_{-,L}(\lambda)\rho_L'(\lambda)}{\rho_L(\lambda)}\right)\right] \bigg |.
\eea
The derivation of the result for $\mathcal{G}^+$ is almost the same as the one for $\mathcal{G}^-$, hence we present only the answer:
\bea
\label{startgp}
\fl
\mathcal{G}^+ = \bigg |\left[\prod_{j,k}^{N} \left(\frac{(\lambda_j - \tilde{\mu}_k +ic)(\lambda_j - \tilde{\mu}_k + ic)}{(\lambda_{jk} + ic)(\tilde{\mu}_{jk} + ic)}\right)^{1/2}\right]\times\left \{ \prod_{j=1}^{N}\frac{{\rm sin}(F_{+,L}(\lambda_j))}{\pi F_{+,L}(\lambda_j)}\left( \prod_{j\neq k}^{N}\frac{\lambda_{jk} \tilde{\mu}_{jk}}{( \lambda_j - \tilde{\mu}_k)^2}\right)^{1/2}\right\}\nn\\ \fl
\prod_{j=1}^{N} \left(\frac{\tilde{\mu}_j - \mu^+_{n+1}}{\lambda_j - \mu^+_{n+1}}\right)\times\prod_{i=1}^n\left( \frac{F_{+,L}(\mu^-_i)}{L(\rho_L(\mu^-_i)\rho_{L}(\mu^+_i))^{1/2}(\mu_i^- - \mu_i^+)}\right)\frac{{\rm Det}(\delta_{jk} + S^+_{jk})}{(2\pi L \rho_{ex}(\mu_{N+1}))^{1/2}{\rm Det}\left(1-\frac{\hat{K}}{2\pi} \right)}\nn\\ \fl
 {\rm exp}\left[\int_{-q}^q d\lambda \left\{\frac{\pi F_{+,L}(\lambda) {\rm cos}(\pi F_{+,L}(\lambda))}{ {\rm sin}(\pi F_{+,L}(\lambda))}\left( F_{+,L}'(\lambda) - \frac{F_{+,L}(\lambda)\rho_L'(\lambda)}{2\rho_L(\lambda)}\right)\right\}\right]\nn\\ \fl
 {\rm exp}\left[-\frac{1}{2} \int_{-q}^{q}d\lambda\left(F_{+,L}'(\lambda) - \frac{F_{+,L}(\lambda)\rho_L'(\lambda)}{\rho_L(\lambda)}\right)\right] \bigg |.
\eea

Note that in \ceq{startg} and \ceq{startgp} the notation $S^{\pm}_{jk}$ is used to differentiate between the matrices appearing in \ceq{ffspec} when the excited state has $N\pm1$ quasimomenta. The matrices are given by
\bea\fl
 S^-_{jk} = \frac{i(\tilde{\mu}_j -\lambda_j)}{\tilde{V}_j^+-\tilde{V}_j^-} \frac{\prod_{m \neq j}^{N-1}(\tilde{\mu}_m - \lambda_j)}{\prod_{m \neq j}^{N}(\lambda_m - \lambda_j)}(K(\lambda_j - \lambda_k) - K(\mu^-_{n+1} - \lambda_k)),\nn\\ \fl 
 S^+_{jk} = \frac{i(\tilde{\mu}_j -\lambda_j)}{\tilde{V}_j^+-\tilde{V}_j^-} \frac{\prod_{m \neq j}^{N+1}(\tilde{\mu}_m - \lambda_j)}{\prod_{m \neq j}^{N}(\lambda_m - \lambda_j)}(K(\lambda_j - \lambda_k) - K(\lambda_j - \mu^+_{n+1})).
 \eea

It is more convenient to separately evaluate the thermodynamic limits of specific groups of terms appearing in the expressions Eqs.~(\ref{startf}),~(\ref{startg}),~(\ref{startgp}). We begin with the terms in the square brackets.
  

\subsection{Evaluation of M$_1$}
We will denote by $M_1$ the first group of terms appearing in square brackets in Eqs.~(\ref{startf}),~(\ref{startg}),~(\ref{startgp}). These terms have common origins and the evaluation of their limits is essentially the same. The only difference lies in the fact that the shift function, $F$, for the creation/annihilation operators, accounts for an extra particle or hole, as described earlier.

Let us start with the density form factor case
\bea
M_1 = \bab\prod_{j,k=1}^N \left(\frac{(\lambda_j - \mu_k +ic)(\lambda_j - \mu_k + ic)}{(\lambda_{jk} + ic)(\mu_{jk} + ic)}\right)^{1/2}\bab.\eea

We expect the first group of terms to be finite in the TDL because there are no obvious singularities in the individual terms. We start by breaking the term into four parts as follows
\begin{eqnarray}
M_1 = p_1\times p_2\times p_3\times p_4,\nn\\ 
p_1^2 = \bab\prod_{j,k} \p\p \frac{\lambda_j - \mu_k + i c}{\lambda_{jk} + ic}\bab,\nn\\ 
p_2^2 = \bab\prod_{j,k} \p\p \frac{\lambda_j - \mu_k - ic}{\mu_{jk} - ic}\bab,
\end{eqnarray}
\begin{eqnarray}
p_3 = \bab\prod_{h=1}^{n}\prod_{j}\left(\frac{(\lambda_j - \mu^+_h +ic)(\lambda_j - \mu^+_h + ic)}{(\lambda_j-\lambda^-_h + ic)(\mu_j - \mu^+_h + ic)}\right)\bab,\nn\\ 
p_4 = \bab\prod_{h_1,h_2}^{n} \left(\frac{(\lambda^-_{h_1}- \mu^+_{h_2}+ic)(\lambda^-_{h_1} - \mu^+_{h_2}+ ic)}{(\lambda^-_{h_1}-\lambda^-_{h_2} + ic)(\mu^+_{h_1}-\mu^+_{h_2} + ic)}\right)^{1/2}\bab, 
\end{eqnarray}
where the prime notation is the same as used in the earlier sections, see definitions before Eq. (21). Note that because we are only interested in the absolute value when evaluating the term $M_1$, we can freely replace terms by their complex conjugate in order to avoid generating unbounded imaginary parts in the various terms, see e.g. $p_2^2$. The $n$ quasimomenta corresponding to excitations are treated separately and are indexed by $h$. 

We start with the first two pairs of numerator and denominator terms
\bea
p_1^2 &=& \bab\prod_{j,k} \p\p \frac{\lambda_j - \mu_k + i c}{\lambda_{jk} + ic}\bab\nn\\ 
&=& \bab\prod_{j,k} \p\p \frac{\lambda_{jk} + \frac{F_L(\lambda_k)}{L\rho_L(\lambda_k)}+ic}{\lambda_{jk} + ic}\bab\nn\\ 
&=&  \bab\prod_{j,k} \p\p\left(1 + \frac{F_L(\lambda_k)}{L\rho_L(\lambda_k)(\lambda_{jk} + ic)}\right)\bab.
\eea

Similarly,
\bea
p_2^2&=& \bab \prod_{j,k} \p\p \frac{\lambda_j - \mu_k - ic}{\mu_{jk} - ic}\bab\nn\\ 
&=& \bab\prod_{j,k} \p\p\left( 1 - \frac{F_L(\lambda_j)}{L\rho(\lambda_j)(\lambda_{jk} - ic +\frac{ F_L(\lambda_k)}{L\rho(\lambda_k)})}\right)^{-1}\bab\nn\\
&=& \bab\prod_{j,k} \p\p\left( 1 + \frac{F_L(\lambda_k)}{L\rho(\lambda_k)(\lambda_{jk} + ic -\frac{ F_L(\lambda_j)}{L\rho(\lambda_j)})}\right)^{-1}\bab\nn\\
&=& \bab\prod_{j,k} \p\p\left( 1 + \frac{F_L(\lambda_k)}{L\rho(\lambda_k)(\lambda_{jk} + ic)}+ \frac{ F_L(\lambda_j)F_L(\lambda_k)}{L^2\rho(\lambda_j)\rho(\lambda_k)(\lambda_{jk} + ic)^2}\right)^{-1}\bab.\nn\\
\eea

To evaluate the products above we do a logarithmic expansion:
\bea \fl
\log(p_1 p_2)& &\nn\\ \fl
= \sum_{j,k} \log(p_1) + \log(p_2)\nn\\ \fl
= -\frac{1}{2}\int_{-q}^{q} d\lambda \int_{-q}^{q} d\mu \frac{F(\lambda)F(\mu)}{(\lambda - \mu + ic)^2}+ O(1/L).
\eea
Similarly, the term $p_3$ can be reduced to the following form
\bea
p_3 &=& \bab\prod_{j}\prod_{h=1}^{n}\left(\frac{\mu_j - \lambda^-_h + i c}{\lambda_j - \lambda_h^- + ic}\right)\left(\frac{\mu_j - \mu^+_h + ic}{\lambda_j-\mu^+_h + ic}\right)^{-1}\bab\nn\\ 
	&=&\bab\prod_{j}\prod_{h=1}^{n}\left(1+\frac{F_L(\lambda_j)}{L\rho_L(\lambda_j)}\frac{\mu^+_h - \lambda^-_h}{(\lambda_j -\lambda^-_h + ic)(\lambda_j - \mu^+_h + ic)}+ O\left(\frac{1}{L^2}\right)\right)\bab\nn\\
	&=&{\rm exp}\left[\sum_{h=1}^n \int_{-q}^{q}d\lambda F(\lambda) \frac{\mu^+_h - \lambda^-_h}{(\lambda -\lambda^-_h + ic)(\lambda - \mu^+_h + ic)}\right]  + O\left(\frac{1}{L}\right).
\eea
Thus, the overall factor $M_1$ takes the following form
\bea
\label{m1fin}
M_1&=&{\rm exp}\left[-\frac{1}{2}\int_{-q}^{q}d\mu \int_{-q}^{q}d\lambda \frac{F(\lambda)F(\mu)}{(\lambda - \mu + ic)^2}\right]\prod_{h_1,h_2}^{n} \left[\frac{(\mu^-_{h_1}- \mu^+_{h_2}+ic)^2}{(\mu^-_{h_1,h_2} + ic)(\mu^+_{h_1,h_2} + ic)}\right]^{1/2}\nn\\ 
&\times& {\rm exp}\left[\sum_{h=1}^n \int_{-q}^{q}d\lambda \frac{F(\lambda) (\mu^+_h - \mu^-_h)}{(\lambda -\mu^-_h + ic)(\lambda -\mu^+_h + ic)}\right]. 
\eea
To obtain the analogous term for boson creation/annihilation form factor, we use the same expression as above in \ceq{m1fin}, but substitute the shift function $F(\lambda)$, with $F_{\pm}(\lambda)$.

\subsection{Obtaining term $M_2$}\label{sec:M_2}
Here we will focus on the term
\bea
\label{m2start}
M_2 = \left\{ \prod_{j} \frac{{\rm sin}(\pi F_L(\lambda_j))}{\pi F_L(\lambda_j)}\prod_{j\neq k}\left(\frac{\lambda_{jk}\mu_{kj}}{(\mu_k-\lambda_j)^2}\right)^{1/2}\right\}.\eea
 This term is expected to contain a power law divergence and a prefactor, and appears in all form factors.  We require a consistent and physically clear method of extracting the divergence and all contributions to the prefactor which are finite in the thermodynamic limit. The intuition we will use to do this comes from the fact that we know the final answer is expected to scale as a power law of $L$. In the thermodynamic limit, the information about $L$ contained in the quantization of $\lambda, \mu$. Consequently it is reasonable to expect that terms which collect to give this power law divergence must appear as a difference in $\lambda_j - \lambda_k$ when $\lambda_j$ and $\lambda_k$ are near one another. Therefore it is important to isolate such terms on the basis of the nearness of the $\lambda's$. Thus we propose to use a cutoff in order to control the nearness of these parameters and perform controlled expansions to extract the divergence and the prefactor. The cutoff will drop out of the final answer. This is made more precise below. 
 
  The term $M_2$ appears in the form factors associated with the density operator and with the creation/annihilation operators, with a slight modification accounting for the difference. We will explicitly calculate the density form factor and point out how to modify the result for the creation/annihilation form factors. Furthermore we will consider first a single particle-hole pair, with quasimomenta, $\mu^-, \mu^+$ and indices, $i^-, i^+$, and lastly generalize the calculation to include multiple excitations. 
  
  It is easier to calculate the thermodynamic limit of $M_2$ after regrouping terms in a convenient way. Our procedure will be the following.
   
   First we rewrite the terms in round brackets in \ceq{m2start} as $T'$ as shown below
 \bea
 \label{mrt}\fl
 T^{\prime} &=& \prod_{j\neq k}\p  \left(\frac{\lambda_{jk}\mu_{kj}}{(\mu_k-\lambda_j)^2}\right)^{1/2}\nn\\ \fl
 &=&\left[ \prod_{j\neq k}\p \left(\frac{\lambda_j -\mu_k}{\lambda_j - \lambda_k}\right)^{-2}\left(\frac{\mu_j - \mu_k}{\lambda_j - \lambda_k}\right)\right]^{1/2}\nn\\ \fl
 &=& \prod_{j\neq k}\p \left(1+\frac{F_L(\lambda_k)}{L\rho_L(\lambda_k)(\lambda_j - \lambda_k)}\right)^{-1/2}\left(1+\frac{F_L(\lambda_k)}{L\rho_L(\lambda_k)(\lambda_j - \lambda_k)}-\frac{F_L(\lambda_j)}{L\rho_L(\lambda_j)(\lambda_j - \lambda_k)}\right)^{1/2}\nn\\ \fl
 &\times&\left(1-\frac{F_L(\lambda_j)}{L\rho_L(\lambda_j)(\lambda_j - \lambda_k)}\right)^{-1/2}.
\eea

The prime notation is the same as used earlier - i.e. we leave off the quasimomenta corresponding to particles and holes. 
  
   It is more convenient to evaluate a ``temporarily incorrect" term, which we will call  $T''$, where we include $\mu^-$ instead of $\mu^+$ and then correct the mistake with a separate term. Thus to obtain the correct $T'$, we will calculate a correction term $T_{hole}$ that must be multiplied to $T''$ in order to remove the extra terms:
 \bea
 \label{tp}
 T' = T'' \times T_{hole}
 \eea
   The terms that we need to remove are exactly those in which the difference, $\mu_k^- - \lambda_k^-$ appears. Consequently, we obtain $T_{hole}$ in the following way:
\bea
\label{Thole}\fl
T_{hole} &=&\prod_{j \neq i^-} \left(1 + \frac{F_L(\lambda^-)}{L\rho_L(\lambda^-)(\lambda_j - \lambda^-)} \right)\prod_{j\neq i^-} \left(1+ \frac{F_L(\mu^-)}{L\rho_L(\lambda^-)(\lambda_j - \lambda^-)} -  \frac{F_L(\lambda_j)}{L\rho_L(\lambda_j)(\lambda_j - \lambda^-)}\right)^{-1/2} \nn\\ \fl
& &\prod_{k\neq i^-} \left(1+ \frac{F_L(\lambda_k)}{L\rho(\lambda_k)(\lambda^- - \lambda_k)} -  \frac{F_L(\lambda^-)}{L\rho_L(\lambda^-)(\lambda^- - \lambda_k)}\right)^{-1/2}\nn\\ \fl
&=&\prod_{j \neq i^-} \left(1 + \frac{F_L(\lambda^-)}{L\rho_L(\lambda^-)(\lambda_j - \lambda^-)} \right)\prod_{j\neq i^-} \left(1+ \frac{F_L(\lambda^-)}{L\rho_L(\lambda^-)(\lambda_j - \lambda^-)} -  \frac{F_L(\lambda_j)}{L\rho_L(\lambda_j)(\lambda_j - \lambda^-)}\right)^{-1}\nn\\ \fl
&=& \prod_{j \neq i^-} \left(1-  \frac{F_L(\lambda_j)}{L\rho_L(\lambda_j)(\lambda_j - \lambda^-)\left(1 + \frac{F_L(\lambda^-)}{L\rho_L(\lambda^-)(\lambda_j - \lambda^-)} \right)}\right)^{-1}\nn\\ \fl
&=&\prod_{j \neq i^-} \left(1-  \frac{F_L(\lambda_j)}{L\rho_L(\lambda_j)(\lambda_j - \lambda^-) + \frac{F_L(\lambda^-)\rho_L(\lambda_j)}{\rho_L(\lambda^-)}}\right)^{-1}.
\eea
In the second line, we have relabeled index $k$ as $j$ to obtain line 3 in \ceq{Thole}.

  Finally, the terms that were removed must be replaced by the correct terms. Let us call the product of these correct terms $T_{particle}$. The way to obtain this term is similar to the process used to remove the incorrect terms as in \ceq{Thole}. We get
 \bea  
  \label{Tpart}\fl
  T_{particle} &=& \prod_{j \neq i^-} \left(1 + \frac{\lambda^- - \mu^+}{\lambda_j - \lambda^-} \right)^{-1}\prod_{j\neq i^-} \left(1+ \frac{\lambda^- -\mu^+}{\lambda_j - \lambda^-} -  \frac{F_L(\lambda_j)}{L\rho_L(\lambda_j)(\lambda_j - \lambda^-)}\right)^{1/2}\nn\\ \fl
& &\prod_{k\neq i^-} \left(1+ \frac{F_L(\lambda_k)}{L\rho(\lambda_k)(\lambda^- - \lambda_k)} -  \frac{\lambda^- - \mu^+}{\lambda^- - \lambda_k}\right)^{1/2}\nn \\ \fl
&=& \prod_{j \neq i^-}\left(1 - \frac{F_L(\lambda_j)}{\rho_L(\lambda_j)(\lambda_j - \lambda^-)\left(1 + \frac{\lambda^- - \mu^+}{\lambda_j - \lambda^-} \right)}\right)\nn \\ \fl
&=&  \prod_{j \neq i^-}\left(1 - \frac{F_L(\lambda_j)}{L\rho_L(\lambda_j)(\lambda_j - \mu^+)}\right).
\eea
  Thus, the final answer will be given by
 \bea
 M_2 = T'' \times T_{hole} \times T_{particle}.
 \eea
  
 We start with the expression
 \bea \fl
 \label{tstart}
 T'' &=& \prod_{j\neq k}\p\p  \left(\frac{\lambda_{jk}\mu_{kj}}{(\mu_k-\lambda_j)^2}\right)^{1/2}\nn\\ \fl
 &=&\left[ \prod_{j\neq k}\p\p \left(\frac{\lambda_j -\mu_k}{\lambda_j - \lambda_k}\right)^{-2}\left(\frac{\mu_j - \mu_k}{\lambda_j - \lambda_k}\right)\right]^{1/2}\nn\\ \fl
 &=& \prod_{j\neq k}\p\p  \left(1+\frac{F_L(\lambda_k)}{L\rho_L(\lambda_k)(\lambda_j - \lambda_k)}\right)^{-1/2}\left(1+\frac{F_L(\lambda_k)}{L\rho_L(\lambda_k)(\lambda_j - \lambda_k)}-\frac{F_L(\lambda_j)}{L\rho_L(\lambda_j)(\lambda_j - \lambda_k)}\right)^{1/2}\nn \\ \fl
 &\times&\left(1-\frac{F_L(\lambda_j)}{L\rho_L(\lambda_j)(\lambda_j - \lambda_k)}\right)^{-1/2}.
 \eea
 
At this stage there are two levels of approximations that can be made. One, we can express differences in $\lambda$ using
 \bea
 \label{app1}\fl
  j - k = L\left(x(\lambda_j) - x(\lambda_k)\right),\nn\\ \fl
   j - k = L\left(x(\lambda_j) - x(\lambda_j) - x'(\lambda_j)(\lambda_k - \lambda_j) - x''(\lambda_j)(\lambda_k - \lambda_j)^2/2\right) + O(\lambda_{jk}^3),\nn\\ \fl
   j - k = L\rho_L(\lambda_j)(\lambda_j - \lambda_k) - L\rho_L'(\lambda_j)(\lambda_j - \lambda_k)^2/2 + O(\lambda_{jk}^3),\nn\\ \fl
   L\rho_L(\lambda_j - \lambda_k) = (j-k) + \frac{\rho_L'(\lambda_j)}{2L\rho_L(\lambda_j)^2}(j-k)^2 + O(1/L^2). 
 \eea

Two, the products in \ceq{tstart} above can be exponentiated, and the resulting logarithms expanded order by order in $1/L$. 
  
  Both these procedures have a region of validity. For instance, it is incorrect to perform the logarithmic expansion when $j$ and $k$ are near one another. On the other hand, when $j$ and $k$ are far away, the approximation in \ceq{app1} breaks down.  
 We calculate $T''$ by combining these two approximations. The idea behind this is to distinguish two regions of the range of $\lambda_j, \lambda_k$ using a cutoff - a region when they are near one another, and a region when they are far away. Then the two approximations outlined above can be used in the region in which they are valid. 
 
\begin{figure}[htpb]
\includegraphics[scale=0.4]{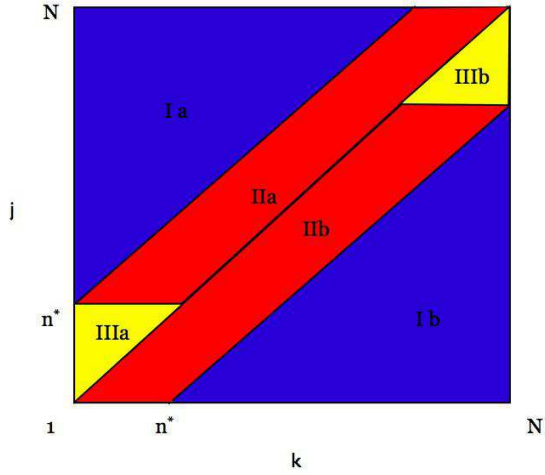}
 \caption{Schematic of range of $\lambda$'s. Cutoff parameter $n^*$ is used to separate regions I, II and III. In region I the indices $j,k$ are far enough to perform logarithmic expansions as in, for example, \ceq{t1start}. In region II and III, the approximation given by \ceq{app1} is valid, but the logarithmic expansion fails. Consequently products must be carefully and explicitly evaluated discretely. In region III proximity to the quasi-Fermi point demands greater care in evaluating products as demonstrated in \ceq{t3start}, for example. }
 \label{fig1}
 \end{figure}

 This is illustrated in Fig. \ref{fig1}. The blue region I, corresponds to the case when $j$ and $k$ are far enough from each other for the logarithmic expansion to become valid. The red region II corresponds to when $j$ and $k$ are near enough for \ceq{app1} to become valid. The third, yellow region III will be treated more carefully since \ceq{app1} is valid here, but proximity to the quasi-Fermi points plays a role in how terms are treated in this region. Moreover, the denotations $a$ and $b$ are used to differentiate the region where $j > k$ and $j < k$, respectively, to make the calculation more convenient.
 
 The separation of the regions described above is not arbitrary. To do this we introduce a cutoff in the following way - in region I, we only allow the index $k$ to get within $ n^* \gg 1$ of the index $j$ for all products/sums. In the figure this corresponds to the lines at the interface of the red and blue regions. Since we will be working in the continuum limit in region I, we would like to know what this condition about indices means for the quasi momenta. Since $n^*$ is at the crossover where the approximation (\ref{app1}) \emph{just starts} to break down, we will use this approximation to define the $\nu^*$ corresponding to $n^*$
\bea
\label{lamapp}
\nu^*(\lambda)=\left |\lambda_j - \lambda_k \right | \approx \left |\frac{j-j\pm n^*}{L\rho_L(\lambda_j = \lambda)}\right | =\frac{n^*}{L\rho_L(\lambda)}.
\eea
Note that we have a choice of either keeping $n^*$ or $\nu^*$ constant - fixing one parameter endows the other with a dependence on its position in the range [-q,q] or [1,N] respectively. Our particular choice keeps $n^*$ fixed, and makes $\nu^*(\lambda)$ dependent on $\lambda$.

 Using the two approximations in the relevant regions should produce, first, a (principal value) integral which captures the contribution to the prefactor from the ``far-away'' region (I a,b in Fig.\ref{fig1}). Second, a constant from the entire region where $\lambda_j$ is near $\lambda_k$, but when they are not both at the same quasi-Fermi point  (II a,b in Fig. \ref{fig1}). Lastly, a constant and a power law divergence from the region where both $\lambda$ are near the same quasi-Fermi point  (III a,b in Fig. \ref{fig1}). While the intermediate steps may produce cutoff dependent terms, these terms should mutually cancel. The final answer will be shown to be independent of the cutoff parameter, $n^* \gg 1$.

 Let us denote the value of $T''$ in region I as $T_I$. To treat the product in region I we may expand the products in $T ''$ logarithmically in orders of $1/L$. Only terms of order $1/L^2$ and higher survive after the double products are evaluated. Without any additional approximation, we obtain the following term
 \bea
 \label{t1start}\fl
 T_I&=& \prod_{j\neq k} \left(1+\frac{F_L(\lambda_k)}{L\rho_L(\lambda_k)(\lambda_j - \lambda_k)}\right)^{-1/2}\left(1+\frac{F_L(\lambda_k)}{L\rho_L(\lambda_k)(\lambda_j - \lambda_k)}-\frac{F_L(\lambda_j)}{L\rho_L(\lambda_j)(\lambda_j - \lambda_k)}\right)^{1/2}\nn\\ \fl
 &\times&\left(1-\frac{F_L(\lambda_j)}{L\rho_L(\lambda_j)(\lambda_j - \lambda_k)}\right)^{-1/2} \approx {\rm exp}\left[ \frac{1}{2} \sum_{j\neq k} \frac{F_L(\lambda_j)F_L(\lambda_k)}{L^2\rho_L(\lambda_j)\rho_L(\lambda_k)(\lambda_j - \lambda_k)^2}\right].
 \eea

With the definition of the cutoff in $\lambda$, \ceq{lamapp}, we may write
\bea
\label{isol}\fl
{\rm log}(T_I) &=& \frac{1}{2} \sum_{j\neq k} \frac{F_L(\lambda_j)F_L(\lambda_k)}{L^2\rho_L(\lambda_j)\rho_L(\lambda_k)(\lambda_j - \lambda_k)^2}\nn\\ \fl
&=&\frac{1}{2}\left(\int_{-q+\nu^*(-q)}^{q}d\lambda \int_{-q}^{\lambda-\nu^*(\lambda)}d\mu \frac{F_L(\lambda)F_L(\mu)}{(\lambda-\mu)^2} + \int_{-q}^{q-\nu^*(q)}d\lambda \int_{\lambda + \nu^*(\lambda) }^{q}d\mu \frac{F_L(\lambda)F_L(\mu)}{(\lambda-\mu)^2}\right)\nn\\ \fl
&=&\frac{1}{2}\int_{-q+\nu^*(-q)}^{q}F_L^2(\lambda) d\lambda \int_{-q}^{\lambda-\nu^*(\lambda)}d\mu \left(\frac{1}{\lambda-\mu}\right)^2\nn\\ \fl
& &+ \frac{1}{2}\int_{-q}^{q-\nu^*(q)}F_L^2(\lambda)d\lambda \int_{\lambda + \nu^*(\lambda) }^{q}d\mu \left(\frac{1}{\lambda-\mu}\right)^2\nn\\ \fl
& & -\frac{1}{4}\int_{-q+\nu^*(-q)}^{q}d\lambda \int_{-q}^{\lambda-\nu^*(\lambda)}d\mu \left(\frac{F_L(\lambda)-F_L(\mu)}{\lambda-\mu}\right)^2\nn\\ \fl
& &-\frac{1}{4} \int_{-q}^{q-\nu^*(q)}d\lambda \int_{\lambda + \nu^*(\lambda) }^{q}d\mu \left(\frac{F_L(\lambda)-F_L(\mu)}{\lambda-\mu}\right)^2.
\eea
In the last step we have isolated a straightforwardly integrable (lines~3 and 4) and a cutoff dependent part (lines~1 and 2). 

Let us first analyze the term
\bea
\fl
\frac{1}{2}\int_{-q+\nu^*(-q)}^{q}F_L^2(\lambda) d\lambda \int_{-q}^{\lambda-\nu^*(\lambda)}d\mu \frac{1}{(\lambda-\mu)^2}=\frac{1}{2}\int_{-1+\nu^*(-q)/q}^{1}dx F_L^2(q x) \left(\frac{1}{\nu^*(q x)/q}-\frac{1}{x+1}\right)\nn\\ \fl
=-\frac{1}{2}F_L^2(-q){\rm log}(qL) - \frac{1}{2}\left[F_L^2(-q){\rm log}\left(\frac{\rho_L(q)}{n^*}\right)-\int_{-q+\nu^*(q)}^{q}d\lambda \frac{F_L^2(\lambda)}{\nu^*(\lambda)}\right] -\frac{1}{2} P_-\int_{-1}^{1}dx \frac{F_L^2(qx)}{x+1},\nn\\
\eea where we introduced the generalized principal value integrals denoted by $P_-,P_+, P_{\pm}$, defined in Appendix B. Note that such integrals can only be well defined when the integrand is made dimensionless. This is achieved by mapping the range of integration from $(-q,q) \to (-1,1)$. 
 
 Similarly we may obtain the second cutoff dependent term of \ceq{isol}. The combined answer in region I becomes 
\bea
\label{tcut}\fl
T_I = \left(\frac{\nu^*(q)}{q}\right)^{\frac{1}{2}(F_L^2(q)+F_L^2(-q))}{\rm exp}\left\{P_{\pm}\int_{-1}^{1}dx\frac{ F_L^2(q x)}{x^2-1^2} -\frac{1}{4}\int_{-q}^{q}d\lambda \int_{-q}^{q}d\mu\left(\frac{F_L(\lambda)-F_L(\mu)}{\lambda - \mu}\right)^2 \right\}\nn\\ \fl
\times {\rm exp}\left\{\frac{ L}{n^*}\int_{-q}^{q-\nu^*(q)} d\lambda \rho_L(\lambda)F_L^2(\lambda)+\frac{ L}{n^*}\int_{-q+\nu^*(-q)}^{q} d\lambda \rho_L(\lambda)F_L^2(\lambda)\right\}.
\eea

In Region II we know that we can no longer expand the log of the products like in the previous section because here all terms are $O(1)$ or bigger. We will consider the two sets of products in this region, products in IIa, and, IIb, separately. In both these products we can approximate $\lambda_j - \lambda_k$ as in \ceq{app1}. The second thing is to treat terms like $F_L(\lambda_j) - F_L(\lambda_k)$ with the  same precision with which we treat differences in $\lambda$ in this region. Thus we need to evaluate the difference in $F_L$ to $O(1/L)$ as follows
\bea\frac{F_L(\lambda_j)}{\rho_L(\lambda_j)} - \frac{F_L(\lambda_k)}{\rho_L(\lambda_k)} = \frac{d}{d\lambda}\left(\frac{F_L(\lambda)}{\rho_L(\lambda)}\right)|_{\lambda = \lambda_j}(\lambda_j - \lambda_k).\eea

Let us define the following notation, $F_j = F_L(\lambda_j)$, $C^{a}_j = \left(\frac{F_L(\lambda_j)}{\rho_L(\lambda_j)}\right)'$, where the prime denotes that the derivative has been taken with respect to $\lambda$ and evaluated at $\lambda = \lambda_j$. Similarly, define $C^b_j = -\frac{\rho_L'(\lambda_j)}{2\rho_L(\lambda_j)^2}$.

Thus we have from \ceq{tstart}
\bea\fl\label{t2a}
T_{IIa} &=& \prod_{j=n^*+1}^{N}\prod_{k=j-n^*}^{j-1}\left(1+\frac{\frac{F_L(\lambda_k)}{L\rho_L(\lambda_k)}}{\lambda_{jk}}\right)^{-1/2}\left(1+\frac{\frac{F_L(\lambda_k)}{L\rho_L(\lambda_k)}}{\lambda_{jk}}-\frac{\frac{F_L(\lambda_j)}{L\rho_L(\lambda_j)}}{\lambda_{jk}}\right)^{1/2}\left(1-\frac{\frac{F_L(\lambda_j)}{L\rho_L(\lambda_j)}}{\lambda_{jk}}\right)^{-1/2}\nn\\ \fl
&=& \prod_{j=n^*+1}^{N}\prod_{k=j-n^*}^{j-1} \frac{(L\rho_L(\lambda_j)\lambda_{jk})^{1/2}(L\rho_L(\lambda_j)\lambda_{jk}(1-C^a_j/L))^{1/2}}{(L\rho_L(\lambda_j)\lambda_{jk} - F_j)^{1/2}(L\rho_L(\lambda_j)\lambda_{jk}(1-C^a_j/L) + F_j)^{1/2}}\nn\\ \fl
&=& \prod_{j=n^*+1}^{N}\prod_{k=j-n^*}^{j-1}\frac{(j-k)(1+\frac{1}{L}C_j^b(j-k))(1+\frac{1}{L}C_j^b(j-k))^{-1}(1-C^a_j/L)^{1/2}}{(j-k - \frac{F_j}{1+\frac{1}{L}C^b_j(j-k)})^{1/2}(j-k + \frac{F_j}{(1+\frac{1}{L}C^b_j(j-k))(1-C^a_j/L)})^{1/2}(1-C^a_j/L)^{1/2}}\nn\\ \fl
&=& \prod_{j=n^*+1}^{N}\prod_{k=j-n^*}^{j-1}\frac{(j-k)(1+\frac{1}{L}C_j^b(j-k))(1+\frac{1}{L}C_j^b(j-k))^{-1}(1-C^a_j/L)^{1/2}}{(j-k - \frac{F_j}{1+\frac{1}{L}C^b_j(j-k)})^{1/2}(j-k + \frac{F_j}{(1+\frac{1}{L}C^b_j(j-k))(1-C^a_j/L)})^{1/2}(1-C^a_j/L)^{1/2}}\nn\\ \fl
&=&\prod_{j=n^*+1}^{N}\prod_{k=j-n^*}^{j-1}\frac{(j-k)}{(j-k - \frac{F_j}{1+\frac{1}{L}C^b_j(j-k)})^{1/2}(j-k + \frac{F_j}{(1+\frac{1}{L}C^b_j(j-k))(1-C^a_j/L)})^{1/2}}\nn\\ \fl
&=&\prod_{j=n^*+1}^{N}\prod_{k=j-n^*}^{j-1}\frac{(j-k)(1+\frac{1}{L}F_jC^b_j)^{-1/2}(1-\frac{1}{L}F_j C_j^b)^{-1/2}}{(j-k - \frac{F_j}{1+\frac{1}{L}F_jC^b_j})^{1/2}(j-k + \frac{F_j}{1-\frac{1}{L}C_j^a-\frac{1}{L}F_jC^b_j})^{1/2}}\nn\\ \fl
&=&\prod_{j=n^*+1}^{N}\frac{ \Gamma(n^*+1)\Gamma(1- F_j +\frac{1}{L}F_j^2 C_j^b)^{1/2}\Gamma(1+F_j + \frac{1}{L}(F_j^2C_j^b + F_jC_j^a))^{1/2} }{\Gamma(n^*+1 - F_j + \frac{1}{L}F_j^2C_j^b)^{1/2}\Gamma(n^*+1+F_j + \frac{1}{L}(F_j^2C_j^b + F_jC_j^a))^{1/2}}.
 \eea

In the above expressions, only $O(1/L)$ terms will survive and consequently terms of higher order in $1/L$ have been dropped in the intermediate steps. Similarly for the region IIb
\bea \fl
T_{IIb}&=&\prod_{j=1}^{N-n^*}\prod_{k=j+1}^{j+n^*}\frac{(j-k)(1+\frac{1}{L}F_jC^b_j)^{-1/2}(1-\frac{1}{L}F_j C_j^b)^{-1/2}}{(j-k - \frac{F_j}{1+\frac{1}{L}F_jC^b_j})^{1/2}(j-k + \frac{F_j}{1-\frac{1}{L}C_j^a-\frac{1}{L}F_jC^b_j})^{1/2}}\nn\\ \fl
&=&\prod_{j=1}^{N-n^*}\frac{ \Gamma(n^*+1)\Gamma(1+ F_j - \frac{1}{L}F_j^2 C_j^b)^{1/2}\Gamma(1-F_j - \frac{1}{L}(F_j^2C_j^b + F_jC_j^a))^{1/2} }{\Gamma(n^*+1+ F_j - \frac{1}{L}F_j^2C_j^b)^{1/2}\Gamma(n^*+1-F_j -\frac{1}{L}(F_j^2C_j^b + F_jC_j^a))^{1/2}}.
 \eea

We may expand the log of the cutoff dependent terms in the $n^* \gg 1$ limit. This leaves the following terms
\bea \fl
log(T_{cutoff}) &=& -\sum_{j=1}^{N-n^*} \frac{F_j^2}{2 n^*} - \sum_{j=n^*}^{N} \frac{F_j^2}{2 n^*}+ O\left(\frac{1}{n^*}\right)\nn\\ \fl
&=&-\frac{L\int_{-q}^{q-\nu^*(q)}d\lambda F_L^2(\lambda)\rho_L(\lambda)}{n^*}-\frac{L\int_{-q+\nu^*(-q)}^{q}d\lambda F_L^2(\lambda)\rho_L(\lambda)}{n^*}+  O\left(\frac{1}{n^*}\right).\nn\\
\eea

The first two terms exactly cancel the cutoff dependent term from region I in \ceq{tcut}. The last term can be ignored with the precision to which we are considering expansions in the parameter $n^*$. 

Now we focus on the terms that are not cutoff dependent. First we note that all the $C_j^b$ terms drop out to O(1/L). We would like to treat the $C_j^a$ terms. Using the notation, $\psi(x) = \frac{d}{dx}\log(\Gamma(x))$, we expand the terms to precision $O(1/L)$ as follows
\bea
\label{deriv} \fl
\Gamma(1\pm \frac{F_j}{1-C_j/L}) = {\rm exp}\left[\log(\Gamma(1\pm \frac{F_j}{1-C_j/L}))\right]\nn\\ \fl
\approx {\rm exp}\left[\log(\Gamma(1\pm F_j \pm F_j C_j/L))\right]\nn\\ \fl
\approx {\rm exp}\left[\log(\Gamma(1\pm F_j)) \pm \psi(1\pm F_j)F_jC_j/L\right]\nn\\ \fl
= \Gamma(1 \pm F_L(\lambda_j)){\rm exp}\left[\pm\psi(1\pm F_L(\lambda_j))\frac{1}{L}F_L(\lambda_j)\left(\frac{F_L(\lambda_j)}{\rho_L(\lambda_j)}\right)'\right].
\eea
We will need to evaluate the product over $j$ for such terms. This can be done by first noting as a consequence of the chain rule
\bea
\frac{\partial}{\partial x} \log(\Gamma(x)) = \frac{\partial}{\partial \lambda}\log(\Gamma(x(\lambda)))\times\frac{1}{\frac{\partial x(\lambda)}{\partial \lambda}}.\eea

We will evaluate the product of such terms as follows

\bea \fl
\prod_{j=n^*}^{N-n^*}{\rm exp}\left[\frac{C_j F_j}{L}\left(\frac{\partial}{\partial x}\log(\Gamma(x))|_{x=1+F_L(\lambda_j)} - \frac{\partial}{\partial x}\log(\Gamma(x))|_{x=1-F_L(\lambda_j)}\right)\right]\nn\\ \fl
={\rm exp}\left[\sum_j \frac{C_j F_j}{L F_L'(\lambda_j)}\frac{\partial}{\partial \lambda}\left(\log(\Gamma(1-F_L(\lambda_j))\Gamma(1+ F_L(\lambda_j))\right)\right]\nn\\ \fl
={\rm exp}\left[\sum_j \frac{C_j F_j}{L F_L'(\lambda_j)}\frac{\partial}{\partial \lambda}\left(\log\left(\frac{\pi F_L(\lambda_j)}{{\rm sin}(\pi F_L(\lambda_j))}\right)\right)\right]\nn\\ \fl
={\rm exp}\left[\int_{-q + \nu^*}^{q-\nu^*} d\lambda \rho_L(\lambda)\frac{F_L(\lambda)}{F_L'(\lambda)}\frac{\partial}{\partial \lambda}\left(\frac{F_L(\lambda)}{\rho_L(\lambda)}\right)\frac{\partial}{\partial \lambda}\left(\log\left(\frac{\pi F_L(\lambda)}{{\rm sin}(\pi F_L(\lambda))}\right)\right)\right]\nn\\ \fl
={\rm exp}\left[\int_{-q }^{q} d\lambda \rho_L(\lambda)\frac{F_L(\lambda)}{F_L'(\lambda)}\frac{\partial}{\partial \lambda}\left(\frac{F_L(\lambda)}{\rho_L(\lambda)}\right)\frac{\partial}{\partial \lambda}\left(\log\left(\frac{\pi F_L(\lambda)}{{\rm sin}(\pi F_L(\lambda))}\right)\right)\right].\nn\\ \fl
\eea
 Note that we have let $\nu^*$ go to zero. In so doing, we have added in some terms that occur at the edges, i.e. form 1 to $n^*$ and from $N-n^*$ to $N$. However, these terms only make an order $1/L $ contribution and can be safely treated. Moreover,  we treat the  $\Gamma(1+F_j)\Gamma(1-F_j)$ terms by extending the product from $1$ to $N $ and dividing by the ``extra terms''
\bea\fl
\prod_{j=1}^{N-n^*}(\Gamma(1+F_j)\Gamma(1-F_j))^{1/2}\prod_{j=n^*+1}^{N}(\Gamma(1+F_j)\Gamma(1-F_j))^{1/2}  \nn\\ \fl
=\prod_{j=1}^N \left(\frac{\pi F_L(\lambda_j)}{{\rm sin}(\pi F_L(\lambda_j))}\right) \left(\frac{{\rm sin}(\pi F_L(q)){\rm sin}(\pi F_L(-q))}{\pi^2 F_L(-q) F_L(q)}\right)^{n^*/2},
\eea 
where, in the last step we have set all the $F_j$ within $n^*$ of the edges to be $F_L(\pm q)$ since this only generates a $1/L$ error that can be ignored. We have to carry out the above procedure since terms involving $\Gamma(1\pm F_j)$ are $O(1)$ and any terms added or removed may result in a finite constant in the final answer. 
We combine all the terms obtained in region II as follows
\bea
\label{t2region}\fl
T_{II} = \prod_j\left(\frac{\pi F_L(\lambda_j)}{{\rm sin}(\pi F_L(\lambda_j))}\right) {\rm exp}\left[\frac{1}{2}\int_{-q }^{q} d\lambda \rho_L(\lambda)\frac{F_L(\lambda)}{F_L'(\lambda)}\frac{\partial}{\partial \lambda}\left(\frac{F_L(\lambda)}{\rho_L(\lambda)}\right)\frac{\partial}{\partial \lambda}\left(\log\left(\frac{\pi F_L(\lambda)}{{\rm sin}(\pi F_L(\lambda))}\right)\right)\right]\nn\\ \fl
\times{\rm exp}\left[-\frac{L}{n^*}\int_{-q}^{q-\nu^*(q)} d\lambda \rho_L(\lambda)F_L^2(\lambda)-\frac{L}{n^*}\int_{-q+\nu^*(-q)}^{q} d\lambda \rho_L(\lambda)F_L^2(\lambda) \right] \nn\\ \fl
\times \left(\frac{1}{\Gamma(1-F_L(q))\Gamma(1+F_L(q))\Gamma(1+F_L(-q))\Gamma(1-F_L(-q))}\right)^{n^*/2}.
\eea

In this region both $\lambda_j, \lambda_k$ are near the same quasi-Fermi point. Moreover, because there are only at most $n^*$ terms to consider in this region, we do not require the precision we used in the rest of region II. This is because $n^*/L$ can be neglected in the thermodynamic limit. Thus, by ignoring $1/L$ terms in the near edge region, and, simply sending $\nu^*$ to zero in the integral in \ceq{t2region}, we will only be making $O(1/L)$ errors in the final answer. 

This simplifies the analysis considerably, since we will be setting $F_j = F_k = F_L(\pm q)$ in this region. We obtain two sets of products
\bea \fl
T_{IIIa} = \prod_{j=1}^{n^*} \prod_{k=1}^{j-1} \frac{(j-k)}{(j-k + F_L(-q))^{1/2}(j-k - F_L(-q))^{1/2}},\nn\\ \fl
T_{IIIb} = \prod_{j=N-n^*+1}^{N} \prod_{k=j+1}^{N} \frac{(j-k)}{(j-k + F_L(q))^{1/2}(j-k - F_L(q))^{1/2}}.
\eea

Starting with the first expression and expanding the answer for $n^*\gg1$, we obtain
\bea
\label{t3start}\fl\label{t3a}
T_{IIIa} = \prod_{j=1}^{n^*} \frac{\Gamma(j)\Gamma(1+F_L(-q))^{1/2}\Gamma(1-F_L(-q))^{1/2}}{\Gamma(j+F_L(-q))^{1/2}\Gamma(j-F_L(-q))^{1/2}}\nn\\ \fl
= \frac{G(n^*+1)G(1+F_L(-q))^{1/2}G(1-F_L(-q))^{1/2}\Gamma(1+F_L(-q))^{n^*/2}\Gamma(1-F_L(-q))^{n^*/2}}{G(n^*+F_L(-q)+1)^{1/2}G(n^*-F_L(-q)+1)^{1/2}}\nn\\ \fl
\approx \Gamma(1+F_L(-q))^{n^*/2}\Gamma(1-F_L(-q))^{n^*/2}G(1-F_L(-q))^{1/2}G(1+F_L(-q))^{1/2}(n^*)^{-\frac{F_L^2(-q)}{2}},\nn\\
\eea 
where $G(x)$ denotes the Barnes G function, (see Eqs.(A1) - (A4)).

Similarly, for the other product
\bea\fl\label{t3b}
T_{IIIb} = \Gamma(1+F_L(q))^{n^*/2}\Gamma(1-F_L(q))^{n^*/2}G(1-F_L(q))^{1/2}G(1+F_L(q))^{1/2}(n^*)^{-\frac{F_L^2(q)}{2}}.\nn\\
\eea

The final result for $T''$ is given by
\bea
\fl
T'' = \left(qL\rho_L(q) \right)^{-\frac{1}{2}(F_L^2(q)+F_L^2(-q))}(G(1+F_L(q))G(1+F_L(-q))G(1-F_L(q))G(1-F_L(-q)))^{1/2}\nn\\ \fl
\times \left(\prod_{j=1}^N \frac{\pi F_L(\lambda_j)}{\sin(\pi F_L(\lambda_j))}\right){\rm exp}\left\{P_{\pm} \int_{-1}^{1}dx \frac{ F_L^2(qx)}{x^2-1} -\frac{1}{4}\int_{-q}^{q}d\lambda \int_{-q}^{q}d\mu\left(\frac{F_L(\lambda)-F_L(\mu)}{\lambda - \mu}\right)^2 \right\}\nn\\ \fl
\times{\rm exp}\left[\frac{1}{2}\int_{-q}^{q} d\lambda \{1-\pi F_L(\lambda) {\rm cot}(\pi F_L(\lambda))\}\left(F_L'(\lambda) - \frac{F_L(\lambda)\rho_L'(\lambda)}{\rho_L(\lambda)}\right)\right].
\eea


 Having obtained the expression for $T''$, we need to remove the ``incorrect'' hole terms, \ceq{Thole}, to obtain $T'$ as in \ceq{tp}. The treatment of this term depends on the distance of the hole from the quasi-Fermi points. To quantify the notion of the distance from the nearest quasi-Fermi point, we define a quantity $q_k$ in the following way
\bea \label{qr}
 q_k^+=n_k^- -n_N - 1 .
 \ee
 Thus if the $k^{th}$ hole is at the right quasi-Fermi point, $q_k = -1$, if it is in the middle of distribution then $q_k \approx -N/2$ etc. Moreover, we say that a hole is near the right quasi-Fermi point, if $q_k/L$ is 0 in the thermodynamic limit, and conversely we say it is ``deep", or far from the right quasi-Fermi point, if $q_k/L$ is a constant in the thermodynamic limit.
 
 Similarly for holes near the left quasi-Fermi point, we define $q_k^-$ to be
\bea \label{ql}
 q_k^- =n^-_k - n_1+1.
 \ee
 
 Now we wish to evaluate
\be 
T_{hole} =\prod_{j \neq i^-} \left(1-  \frac{F_L(\lambda_j)}{L\rho_L(\lambda_j)(\lambda_j - \lambda^-) + \frac{F_L(\lambda^-)\rho_L(\lambda_j)}{\rho_L(\lambda^-)}}\right)^{-1}.
 \ee
 
 We would again like to use the cutoff procedure used to obtain $T''$. In the region where $|j-i^-| \geq n^*$, we expand $T_{hole}$ logarithmically as follows
\bea 
 \label{thole1}
T_{hole}^{I} &=& {\rm exp}\left[-\sum_{|j-i^-| \geq n^*} \log \left(1-  \frac{F_L(\lambda_j)}{L\rho_L(\lambda_j)(\lambda_j - \lambda^-) + \frac{F_L(\lambda^-)\rho_L(\lambda_j)}{\rho_L(\lambda^-)}}\right)\right]\nn\\ \fl
&=&  {\rm exp}\left[\sum_{|j-i^-| \geq n^*} \frac{F_L(\lambda_j)}{L\rho_L(\lambda_j)(\lambda_j - \lambda^-)}\right] + O\left(\frac{1}{L}\right).
 \eea
 
 Meanwhile, in the region where $|j-i^-| < n^*$, we use the approximation in \ceq{app1}, to evalute discrete products as follows
\bea\label{thole2}
T_{hole}^{II} = \prod_{|j-i^-| < n^*} \frac{j - i^- + F(\lambda^-)}{j - i^-}.
\eea 

{\bf Case I : Hole is deep inside the distribution}
 
When the hole is deep inside the distribution, we obtain from Eqs.~(\ref{thole1}), (\ref{thole2})
\bea
\label{tholedeep}
\log(T_{hole}^{k, deep}) =& \int_{-q}^{\lambda_k^- - \nu^*(\lambda_k^-)} d\lambda \frac{F_L(\lambda)}{\lambda - \lambda^-_k} + \int_{\lambda_k^- + \nu^*(\lambda_k^-)}^{q} d\lambda \frac{F_L(\lambda)}{\lambda - \lambda^-_k}\nn\\ \fl
& - \log\left(\frac{\Gamma(1+F_L(\lambda_k^-))\Gamma(1-F_L(\lambda_k^-))\Gamma(n^*)^2)}{\Gamma(n^* + F_L(\lambda_k^-))\Gamma(n^* - F_L(\lambda_k^-))}\right).
\eea We get cancellation between divergent and cutoff dependent terms from the integrals due to the symmetric cutoff.
Thus in the thermodynamic limit we have
\bea 
T_{hole}^{k, deep} = \frac{{\rm sin}[\pi F(\mu^-_k)]}{\pi F(\mu^-_k)} {\rm exp}\left(P \int_{-1}^{1}dx \frac{F(qx)}{x - \frac{\mu^-_k}{q}}\right).\eea

{\bf Case II: Hole near the quasi-Fermi point}

Meanwhile when the hole is near either $\pm q$, we obtain from Eqs.~(\ref{thole1}), (\ref{thole2}), the following expression
\bea\label{hfer}
\fl T_{hole}^{k, edge} = \left(q L\rho(q)\right)^{\mp F(\pm q)} \frac{{\rm sin}[\pi F(\pm q)]}{\pi F(\pm q)}\frac{\Gamma(\mp q_k^{\pm} \pm F(\pm q))}{\Gamma(\mp q_k^{\pm})}{\rm exp}\left(P_{\pm} \int_{-1}^{1}dx \frac{F(q x)}{x \mp 1}\right),\eea where we use $q_k^{\pm}$ to refer to the location of holes near a quasi-Fermi point as defined in Eqs.~(\ref{qr}),~(\ref{ql}).

 
In order to replace the ``correct'' term in the evaluation of $M_2$, we require the following term:
\bea\label{tpartfar}
T_{particle} = \prod_{j \neq i^-}\left(1 - \frac{F_L(\lambda_j)}{L\rho_L(\lambda_j)(\lambda_j - \mu^+_0) + \frac{F(\mu^+_0)\rho(\lambda_j)}{\rho(\mu^+_0)}}\right).\eea
Again, we distinguish the case where the particle excitation is far from either edge and the case where it is near the edge. For this purpose we define a means of quantifying nearness to a quasi-Fermi point. Let us define, $p_k^+ = n^+_k - n_N - 1$, for a particle near the right quasi-Fermi point, and $p_k^- = n^+_k - n_1+1$, for a particle near the left quasi-Fermi point.  We treat two distinct cases, 

{\bf Case I: Particle far from edge}

If we find $p_k/L$ is finite in the thermodynamic limit, then we have the following convergent term
\bea\label{tpart1}
T_{particle}^{k,far} ={\rm exp}\left(-\int_{-1}^{1}dx \frac{F(qx)}{x -\frac{\mu^+_k}{q}}\right).\eea

{\bf Case II: Particle near quasi-Fermi Point}

On the other hand if $p_k^{\pm}/L$ vanishes in the thermodynamic limit, we need to use our cutoff procedure to separately evaluate the contribution to $T_{particle}$ from the region where $\lambda_j$ is near $\mu^+$, and where they are far. We let the $j$ product extend as near as $\nu^*(\pm q)$ to the relevant quasi-Fermi point, and treat the terms within $\nu^*$ of the quasi-Fermi point separately
\bea
\label{tpart2}
\fl T_{particle}^{k, near} &=& \left(\frac{q L\rho(q)}{n^* }\right)^{\pm F(\pm q)} \frac{\Gamma(n^*)\Gamma(\pm p_k^{\pm} + 1 \mp F(\pm q))}{\Gamma(\pm p_k^{\pm} + 1)\Gamma(n^* \mp F(\pm q))}\nn\\ \fl
&=&\left(q L\rho(q)\right)^{\pm F(\pm q)}\frac{\Gamma(\pm p_k^{\pm} + 1\mp F(\pm q))}{\Gamma(\pm p_k^{\pm} + 1)} {\rm exp}\left(-P_{\pm} \int_{-1}^{1}dx \frac{F(qx)}{x \mp 1}\right).
\eea

 Now we consider how the term $M_2$ needs to be modified when there are $n$ particle-hole pairs. Most obviously, the shift function used, $F(\lambda)$, becomes the sum of the shift functions with the different particle-hole contributions. The term $T''$ is unaltered since it is evaluated with $\mu^-_i$ substituted for $\mu^+_i$, making the definition of $F(\lambda)$ in \ceq{fl} valid for all $\lambda_j$. In order to ``correct'' for this convenient substitution, we include the terms $T_{hole}$ associated with each of the holes, and terms $T_{particle}$ associated with each of the particles as described in the previous subsections. However, when there are multiple particle-hole pairs, this is no longer sufficient to correctly evaluate $M_2$. There will be another group of terms which we shall call $T_{cross}$ whose origin can be understood in the following way. Let us start by writing down an expression for $M_2$ and show how to arrive at the correct final answer
\bea
\label{tcrossd}
\fl
M_2 &=& T'' \times \prod_{k=1}^{n} T^{(k)}_{particle} \times T^{(k)}_{hole}\nn\\ \fl
&=&  \prod_{j\neq k}^N \p\p  \left(\frac{\lambda_{jk}\mu_{kj}}{(\mu_k-\lambda_j)^2}\right)^{1/2} \times \prod_{k=1}^{n} \prod_{j < n^-_k}\left(\frac{\mu_j - \mu^-_k}{\lambda_j - \mu^-_k}\right)^{-1}\left(\frac{\mu_j - \mu^+_k}{\lambda_j - \mu^+_k}\right)\nn\\ \fl
&=& \prod_{j\neq k}^N \p\p  \left(\frac{\lambda_{jk}\mu_{kj}}{(\mu_k-\lambda_j)^2}\right)^{1/2} \times \prod_{k=1}^{n} \prod_j^N \p\p \left(\frac{\mu_j - \mu^-_k}{\lambda_j - \mu^-_k}\right)^{-1}\left(\frac{\mu_j - \mu^+_k}{\lambda_j - \mu^+_k}\right)\nn\\ \fl
&\times& \prod_{k_1 \neq k_2}^{n}\left(\frac{(\mu_{k_1}^- - \mu_{k_2}^-)(\mu_{k_1}^{+} - \mu_{k_2}^{+})}{(\mu_{k_1}^- - \mu^+_{k_2})^2}\right)^{1/2}\nn\\ \fl
&=&T'' \times \left(\prod_{k=1}^{n} T^{(k)}_{particle} \times T^{(k)}_{hole}\right) \times T_{cross},
\eea
where the notation $T_{hole}^{(k)} (T_{particle}^{(k)})$ means the contribution from the hole (particle) term specific to the details of the excitation, $\mu_k^- (\mu^+_k)$, and $n_k^-$ refers to the index corresponding to the excitation pair $\mu^+_k, \mu^-_k$.

Thus we obtain the expression
\bea\label{tcross}
T_{cross} = \prod_{k_1 \neq k_2}^{n}\left(\frac{(\mu_{k_1}^- - \mu_{k_2}^-)(\mu_{k_1}^{+} - \mu_{k_2}^{+})}{(\mu_{k_1}^- - \mu^+_{k_2})^2}\right)^{1/2}.\eea
We expect that the contributions to $T_{particle}, T_{hole}$ from the excitations $\{\mu^+_1,..., \mu^+_n | \mu^-_1,..., \mu^-_n\}$ fall into one of the cases discussed in previous subsections. Thus, the form of these terms is always known in principle.

The term $T_{cross}$ is sensitive to the details of the excitations - for instance it will contain divergences when some number of the particles, or holes, are clustered near each other. For our purposes, this is only relevant to form factors of states containing high order Umklapp excitations, i.e. several particles and holes near the left (right) quasi-Fermi points. We will discuss the term $T_{cross}$ in more detail in the section devoted to Umklapp form factors.

The discussion above can be made applicable to $M_2$ in the creation/annihilation operators with two modifications.  First we need to replace $F_L(\lambda)$ by $F_{\pm,L}(\lambda)$ defined in Eqs.~(\ref{fplus}),~(\ref{fminus}), as discussed in the section following these definitions. 

Secondly, if we wish to retain the calculation performed in the previous subsection then we need to extend/reduce the upper bounds of the indices of the products in $M_2$ (see \ceq{m2start}) from $N -1$ to $N$ by using the relation, $\mu_{n+1}^- = \lambda_{n^-} - \frac{F_{-}(\lambda_{n^-})}{L\rho_L(\lambda_{n^-})} + O(1/L^2)$ to define the ``missing'' ground state or excited state quasi momentum (see Section IV.3), and for concreteness we again concentrate on the case of the annihilation operator. This will introduce some error to be corrected to obtain the final answer for $M_2^{\psi}$ and $M_2^{\psi^{\dagger}}$ - for the creation and annihilation operator respectively.

We will follow the consequences of such a replacement and correction for the annihilation operator and just present the final answer for the creation operator since the derivation is essentially the same. 

Consider the product
\bea
M_2^{\psi} = \left( \prod_{j\neq k \neq n^-}^{N}\frac{\lambda_{jk} \tilde{\mu}_{jk}}{( \lambda_j - \tilde{\mu}_k)^2}\right)^{1/2}.\eea
Now extending the upper limit of indices to $N$ introduces the following extra terms
\bea\prod_{j\neq n^-}^{N} \frac{(\lambda_{n^-} - \lambda_j)(\mu_{n+1}^- -\tilde{\mu}_j)}{(\lambda_{n^-} - \tilde{\mu}_j)(\lambda_j - \mu_{n+1}^-)}.\eea 
Notice, however that one group of terms in the above expression is present in the final answer for the annihilation operator, $\mathcal{G}^-$, which can be seen explicitly in \ceq{m2der}. Thus we may write
\bea\label{m2psid}\fl
M_2^{\psi} \times \prod_{j\neq n^-}^{N} \frac{\lambda_{n^-} - \lambda_j}{\lambda_{n^-} - \mu_j} =  \left( \prod_{j\neq k}^{N}\frac{\lambda_{jk} \tilde{\mu}_{jk}}{( \lambda_j - \tilde{\mu}_k)^2}\right)^{1/2} \times \prod_{j\neq n^-}^{N} \frac{\mu_{n+1}^- - \lambda_j}{\mu_{n+1}^- - \mu_j},\eea
where the last product serves to cancel the unwanted terms in the parentheses. For notational convenience let us call the product multiplying $M_2^{\psi}$ on the left hand side of the above equation, $T_{extra}^{\psi}$. However, as before it is more convenient to evaluate our double primed products where we substitute $\mu^-_i$ in place of $\mu^+_i$ and correct this mistake separately as in \ceq{tp}.
Consequently we obtain from \ceq{m2psid} and \ceq{tcrossd}
\bea\fl
M_2^{\psi} \times \prod_{j\neq n^-}^{N} \frac{\lambda_{n^-} - \lambda_j}{\lambda_{n^-} - \tilde{\mu}_j} &=& \left[T'' \times \left(\prod_{k=1}^{n} T^{(k)}_{particle} \times\prod_{k=1}^{n+1} T^{(k)}_{hole}\right) \times T^{\psi}_{cross}\right],
\eea
with
\bea 
T^{\psi}_{cross} = \left[\frac{\displaystyle \prod_{k_1\neq k_2}^{n+1}(\mu_{k_1}^- - \mu_{k_2}^-)\prod_{k_1\neq k_2}^{n}(\mu_{k_1}^+ - \mu_{k_2}^+)}{\displaystyle\prod_{k_1\neq k_2}^{n+1}\prod_{k_2=1}^{n}(\mu^-_{k_1}-\mu^+_{k_2})}\right]^{1/2}.
\eea
The corresponding term for the creation operator form factor can be obtained by switching particle and hole quasimomenta in the above expression. 

We obtain a similar expression for the creation operator and the answers are summarized in the next section.

The most general result for $M_2$ in the case of the density form factor is given by
\bea
\label{m2rhofin}\fl
M_2 = \left(q L\rho_L(q) \right)^{-\frac{1}{2}(F_L^2(q)+F_L^2(-q))}(G(1+F_L(q))G(1+F_L(-q))G(1-F_L(q))G(1-F_L(-q)))^{1/2}\nn\\ \fl
\times \left(\prod_{j=1}^N \frac{\pi F_L(\lambda_j)}{\sin(\pi F_L(\lambda_j))}\right){\rm exp}\left\{P_{\pm} \int_{-1}^{1}dx\frac{ F_L^2(qx)}{x^2-1} -\frac{1}{4}\int_{-q}^{q}d\lambda \int_{-q}^{q}d\mu\left(\frac{F_L(\lambda)-F_L(\mu)}{\lambda - \mu}\right)^2 \right\}\nn\\ \fl
\times{\rm exp}\left[\frac{1}{2}\int_{-q}^{q} d\lambda \{1-\pi F_L(\lambda) {\rm cot}(\pi F_L(\lambda))\}\left(F_L'(\lambda) - \frac{F_L(\lambda)\rho_L'(\lambda)}{\rho_L(\lambda)}\right)\right] \nn\\ \fl
\times \left[ \prod_{k=1}^{n} T_{hole}^{(k)} \times T_{particle}^{(k)}\right] \times T_{cross}.
\eea
Similarly, for the creation/annihilation operators it is given by
\bea
\label{m2psifin} \fl
M_2^{\psi/\psi^{\dagger}}=\nn\\ \fl
 \left(qL\rho_L(q) \right)^{-\frac{1}{2}(F_{\pm, L}^2(q)+F_{\pm, L}^2(-q))}(G(1+F_{\pm, L}(q))G(1+F_{\pm, L}(-q))G(1-F_{\pm, L}(q))G(1-F_{\pm, L}(-q)))^{1/2}\nn\\ \fl
\times \left(\prod_{j=1}^N \frac{\pi F_{\pm, L}(\lambda_j)}{\sin(\pi F_{\pm, L}(\lambda_j))}\right){\rm exp}\left\{P_{\pm} \int_{-1}^{1}dx \frac{ F_{\pm, L}^2(qx)}{x^2-1} -\frac{1}{4}\int_{-q}^{q}d\lambda \int_{-q}^{q}d\mu\left(\frac{F_{\pm, L}(\lambda)-F_{\pm, L}(\mu)}{\lambda - \mu}\right)^2 \right\}\nn\\ \fl
\times{\rm exp}\left[\frac{1}{2}\int_{-q}^{q} d\lambda \{1-\pi F_{\pm, L}(\lambda) {\rm cot}(\pi F_{\pm, L}(\lambda))\}\left(F_{\pm, L}'(\lambda) - \frac{F_{\pm, L}(\lambda)\rho_L'(\lambda)}{\rho_L(\lambda)}\right)\right]\nn\\ \fl
\times \left[ \prod_{k=1}^{n(+1)} T_{hole}^{(k)} \times \prod_{k=1}^{n+1(-1)}  T_{particle}^{(k)} \right] \times T^{\psi/\psi^{\dagger}}_{cross}.
\eea

\subsection{Fredholm Determinants}

The last terms left to express in the thermodynamic limit are the determinant terms appearing in the form factors, such as $\Theta$ in \ceq{det}. We would like to write such terms as  Fredholm determinants (see Appendix D and Ref.~\cite{Smirnov} for details on Fredholm determinants). The determinants are slightly different for the density and creation/annihilation operators. We will first do the calculation for the determinant in the density form factor, and then use some of these results to express the determinant for the creation/annihilation form factors. 

We need to evaluate
\bea
\Theta = \frac{{\rm Det} \left(\delta_{j k} + \frac{i(\mu_j -\lambda_j)}{V_j^+-V_j^-}\prod_{m\neq j} \frac{\mu_m - \lambda_j}{\lambda_m - \lambda_j}(K(\lambda_j,\lambda_k) - K(\lambda_p, \lambda_k)) \right)}{V^+_p - V^-_p}.\nn\\
\eea
Due to properties of the matrices involved in the determinant, the choice of $\lambda_p$ is entirely arbitrary \cite{Slavnov} and it need not be from  the set $\{\lambda_i\}$. We can use this fact by taking $\lambda_p$ to be much larger than all other parameters of the problem. Under these circumstances, $K(\lambda_p, \lambda_k) \rightarrow \frac{2 c}{\lambda_p^2}$. Meanwhile the denominator lends itself to the following expansion
\bea
V^{\pm}_p =& \prod_{h=1}^n \left(1 - \frac{\mu^+_h-\lambda^-_h}{\lambda_p} - \frac{(\lambda^-_h - \mu^+_h)(\lambda^-_h \pm ic)}{\lambda_p^2}\right) \nn\\ \fl
&\times \prod_m \p\p \left(1 + \frac{F(\lambda_m)}{L\rho(\lambda_m)\lambda_p} +  \frac{F(\lambda_m)(\lambda_m \pm ic)}{L\rho(\lambda_m)
\lambda_p}\right) + O\left(\frac{1}{\lambda_p^3}\right).
\eea
Using this, one can evaluate the denominator in this limit as
\bea
V_p^+ - V_p^- = \frac{2 i c}{\lambda_p^2}\left(\sum_h\lambda^-_h - \mu_h^+ + \int_{-q}^{q} F(\lambda) d\lambda \right) = \frac{-2ic P_{ex}}{\lambda_p^2},\eea
where we have used the relation between momentum and quasimomenta as in Ref.~\cite{Korepin}.

The numerator also lends itself to a convenient re-expression. Essentially the numerator has the form $Det(A + \alpha \frac{B}{c})$, where
\bea
A = \delta_{j k} + \frac{i(\mu_j -\lambda_j)}{V_j^+-V_j^-}\prod_{m\neq j} \frac{\mu_m - \lambda_j}{\lambda_m - \lambda_j}K(\lambda_j,\lambda_k),\nn\\ 
B =  -\frac{i(\mu_j -\lambda_j)}{c(V_j^+-V_j^-)}\prod_{m\neq j} \frac{\mu_m - \lambda_j}{\lambda_m - \lambda_j},\nn\\ 
\alpha = \frac{2c^2}{\lambda_p^2}.
\eea
 Using the fact that $B$ is a matrix of rank one, we can expand the determinant using rows and minors alternately from $A$ and $B$. Using the expression for the determinant of a sum of matrices \cite{Korepin}, we find that because of $B$'s rank, only terms with one row from $B$ will survive. Moreover the determinant of $A$ also vanishes permitting further simplification. Thus the determinant can be written
\bea
Det(A + \alpha B) &=& Det(A) + \alpha \sum B_{i j} a_{i j} + \alpha \sum A_{i j} a_{i j} - \alpha  \sum A_{i j} a_{i j} \nn\\ \fl
			    &=& Det(A) -\alpha Det(A) + \alpha Det(A+B)\nn\\ \fl
			    &=& \alpha Det(A+B).
\eea

Taken together with the previous statements we get
\bea
\label{detstart} \fl
\Theta = \frac{ic}{P_{ex}}{\rm Det} \left(\delta_{j k} + \displaystyle{ \frac{i(\mu_j -\lambda_j)}{V_j^+-V_j^-}}\displaystyle{\prod_{m\neq j}} \frac{\mu_m - \lambda_j}{\lambda_m - \lambda_j}\left(\frac{2c}{(\lambda_j - \lambda_k)^2 + c^2} -\frac{1}{c}\right) \right).\eea

To express the above as a Fredholm determinant, we consider the following term
\bea
a_j=  \frac{i(\mu_j -\lambda_j)}{V_j^+-V_j^-}\prod_{m\neq j} \frac{\mu_m - \lambda_j}{\lambda_m - \lambda_j}.\eea
The terms $V_j^{\pm}$ admit the following partial limit
\bea
V_j^{\pm} = \prod_{h=1}^n \frac{\mu^+_h - \lambda_j \pm ic}{\lambda^-_h - \lambda_j \pm ic} {\rm exp}\left[\int_{-q}^{q} \frac{F_L(\lambda)}{\lambda - \lambda_j \pm ic} + O(1/L) \right].
\eea
The infinite product in \ceq{detstart} is not as easy to treat. We cannot make the same kind of continuum approximation as above because of the bad behavior of such an integral around $\lambda_j$. This can be treated by temporarily introducing a cutoff into the continuum version of the product (an integral) and extracting any constant (cutoff independent) error committed in so doing to finally obtain an answer that is cutoff independent. We obtain
\bea \fl
\prod_{m\neq j}^{N} \left(\frac{\mu_m - \lambda_j}{\lambda_m - \lambda_j}\right) &=&\prod_{m\neq j}^{N-1} \left(1- \frac{F_L(\lambda_m)}{L \rho_L(\lambda_m)(\lambda_m - \lambda_j)}\right)\times\left(	\frac{\mu^+ - \lambda_j}{\lambda^- - \lambda_j}\right)\nn\\ \fl
&=&{\rm exp}\left\{- \int_{-q}^{\lambda_L} \frac{F_L(\lambda)}{\lambda - \lambda_j} - \int_{\lambda_R}^{q} \frac{F_L(\lambda)}{\lambda - \lambda_j}\right\}\frac{\Gamma(n^*+ 1+F_j)\Gamma(n^* + 1 - F_j)}{\Gamma(1+F_j)\Gamma(n^* + 1)\Gamma(n^*+1)\Gamma(1-F_j)}\nn\\ \fl
& \times&\prod_{h=1}^n\left(	\frac{\mu^+_h - \lambda_j}{\lambda^-_h - \lambda_j}\right).\nn\\ \fl
\eea
Here $\lambda_L$ and $\lambda_R$ correspond to the spectral parameters corresponding to the quantum number $j \pm n^*$. Under the assumption that $n^* \gg1$, we can use the Stirling approximation on the gamma functions:
\bea\fl
& &\frac{\Gamma(n^*+ 1+F_j)\Gamma(n^* + 1 - F_j)}{\Gamma(1+F_j)\Gamma(n^* + 1)\Gamma(n^*+1)\Gamma(1-F_j)}\nn\\ \fl
&=&{\rm exp}\left\{ (n^*+1+F_j)log(n^*+1+F_j) -n-1-F_j + (n^*+1-F_j)log(n^*+1-F_j)  \right\} \nn\\ \fl
&\times& {\rm exp}\left\{ -n -1+F_j - 2(n^*+1)log(n^*+1)+2(n*+1) \right\}\frac{1}{\Gamma(1+F_L(\lambda_j))\Gamma(1-F_L(\lambda_j))}\nn\\ \fl
&=&\frac{1}{\Gamma(1+F(\lambda_j))\Gamma(1-F(\lambda_j))} + O(1/L).
\eea
Where the exponential terms exactly cancel out. Furthermore the integrals contribute terms at the boundaries that also mutually cancel, leaving no constant terms upto $O(1/L)$
\bea
I_{boundaries} &=& -F(\lambda_j) log(\lambda_L - \lambda_j) + F(\lambda_j)log(\lambda_R - \lambda_j)\nn\\ \fl
&=& F(\lambda_j)\left[log\left(\frac{n^*}{L\rho(\lambda_j)}\right) - log\left(\frac{n^*}{L\rho(\lambda_j)}\right)\right]\nn\\ \fl
&=&0.
\eea

This allows us to express $a_j$ as
\bea \fl
a_j &=& \frac{i(\mu_j - \lambda_j)}{\prod_{h=1}^n \frac{\mu^+_h - \lambda_j + ic}{\lambda^-_h - \lambda_j + ic} {\rm exp}\left[\int_{-q}^{q}d\lambda \frac{F(\lambda)}{\lambda - \lambda_j + ic}\right] - c.c}\prod_{h=1}^n\frac{\mu^+_h-\lambda_j}{\lambda^-_h - \lambda_j}\frac{{\rm exp}\left[P\int_{-1}^{1} dx \frac{F(qx)}{x - \lambda_j/q}\right]}{\Gamma(1+F(\lambda_j))\Gamma(1-F(\lambda_j))}\nn\\ \fl
&=& \frac{1}{2\pi L \rho(\lambda_j)}\prod_{h=1}^n\frac{\lambda^-_h - \lambda_j + i c}{\mu^+_h - \lambda_j + ic} \times\frac{\mu^+_h-\lambda_j}{\lambda^-_h - \lambda_j}\nn\\ \fl
&\times&{\rm exp} \left [- P\int^{1}_{-1} dx \frac{F(qx)}{x - \lambda_j/q} +i\pi F(\lambda_j) + \int^{q}_{-q}d\lambda \frac{F(\lambda)}{\lambda - \lambda_j + ic}\right ].
\eea

Let us define an operator, $\hat{G}$ that acts on $[-q,q]\times[-q,q]$ and is given by
\bea
\label{Gdef}
G(\mu, \nu) = a(\mu)\left(K(\mu - \nu) - \frac{1}{c}\right),\eea
where the function $a(\mu$) appearing above is
\bea \fl
a(\mu) = \frac{1}{2\pi}\prod_{h=1}^{n}\frac{\lambda^-_h - \mu + i c}{\mu^+_h - \mu + ic} \times\frac{\mu^+_h-\mu}{\lambda^-_h - \mu}{\rm exp} \left [- P \int^{1}_{-1}dx \frac{F(qx)}{x - \mu/q} +i\pi F(\mu) + \int^{q}_{-q}d\lambda \frac{F(\lambda)}{\lambda - \mu + ic}\right ].\nn\\
\eea

Using Eqs.~(\ref{detstart}),~(\ref{Gdef}), we obtain
\bea\label{gfin}
\Theta = \frac{ic}{P_{ex}}{\rm Det} \left(I + \hat{G}\right).\nn\\
\eea

The determinant terms appearing in the creation/annihilation form factors are a slightly modified version of the one for the density form factor, and are given by
 \bea\fl
 \zeta^- = {\rm Det}\left(\delta_{jk} +  \frac{i(\tilde{\mu}_j -\lambda_j)}{\tilde{V}_j^+-\tilde{V}_j^-} \frac{\prod_{m \neq j \neq n^-}^{N}(\tilde{\mu}_m - \lambda_j)}{\prod_{m \neq j}^{N}(\lambda_m - \lambda_j)}(K(\lambda_j - \lambda_k) - K(\mu_{n+1}^- - \lambda_k))\right),\nn\\ \fl
 \zeta^+ = {\rm Det}\left(\delta_{jk} +  \frac{i(\tilde{\mu}_j -\lambda_j)}{\tilde{\tilde{V}}_j^+-\tilde{\tilde{V}}_j^-} \frac{\prod_{m \neq j}^{N+1}(\tilde{\mu}_m - \lambda_j)}{\prod_{m \neq j}^{N}(\lambda_m - \lambda_j)}(K(\lambda_j - \lambda_k) - K(\lambda_j - \mu^+_{n+1}))\right),
 \eea
 where the $+/-$ in the subscript correspond to the determinant for the creation/annihilation operators respectively. Note that by $\tilde{\tilde{V_j}}^{\pm}$ we mean the term occurring in the creation operator form factor defined analogously to \ceq{vjdef2}.
 
 The procedure we use to express $\zeta^{\pm}$ as a Fredholm determinant is similar to what we used for the case of the density form factor.   Let us define functions $b_j^{\pm}$ similar to $a_j$:
 \bea 
 b_j^-= \frac{i(\tilde{\mu}_j -\lambda_j)}{\tilde{V}_j^+-\tilde{V}_j^-} \frac{\prod_{m \neq j}^{N-1}(\tilde{\mu}_m - \lambda_j)}{\prod_{m \neq j}^{N}(\lambda_m - \lambda_j)},\nn\\ 
 b_j^+ = \frac{i(\tilde{\mu}_j -\lambda_j)}{\tilde{\tilde{V}}_j^+-\tilde{\tilde{V}}_j^-} \frac{\prod_{m \neq j}^{N-1}(\tilde{\mu}_m - \lambda_j)}{\prod_{m \neq j}^{N}(\lambda_m - \lambda_j)}.
 \eea

 In the thermodynamic limit, and to the same accuracy with which we calculated $a(\mu)$ we find
\bea 
 b^-(\mu) =& \frac{1}{2\pi}\prod_{i=1}^{n+1}\frac{\lambda^-_i - \mu + i c}{\lambda^-_i - \mu} \prod_{j=1}^n \frac{\mu^+_j-\mu}{\mu^+_j - \mu + ic}\nn\\ 
 &\times {\rm exp} \left [-P \int^{1}_{-1}dx\frac{F_-(qx)}{x - \mu/q} +i\pi F_-(\mu) + \int^{q}_{-q}d\lambda \frac{F_-(\lambda)}{\lambda - \mu + ic}\right ],\nn\\ 
  b^+(\mu) =& \frac{1}{2\pi}\prod_{i=1}^{n}\frac{\lambda^-_i - \mu + i c}{\lambda^-_i - \mu} \prod_{j=1}^{n+1} \frac{\mu^+_j-\mu}{\mu^+_j - \mu + ic}\nn\\
  &\times{\rm exp} \left [- P \int^{1}_{-1}dx \frac{F_+(qx)}{x- \mu/q} +i\pi F_+(\mu) + \int^{q}_{-q}d\lambda \frac{F_+(\lambda)}{\lambda - \mu + ic}\right ].\nn\\
 \ee
 
Now we may evaluate $\zeta^{\pm}$ as Fredholm determinants of $I+\hat{H}^{\pm}(\mu, \nu)$, where $\hat{H}^{\pm}$ act on $(-q,q)$ and are given by
\bea
\label{Hdef}
H^{-}(\mu, \nu) = b^-(\mu)(K(\mu - \nu) - K(\mu^-_{n+1}- \nu)),\nn\\ 
H^{+}(\mu, \nu) = b^+(\mu)(K(\mu - \nu) - K(\mu - \mu^+_{n+1})).
\eea

 
\subsection{Thermodynamic Limit of Form Factors}

We now present form factors for a few excited states that we would like to relate to the correlation prefactors of the Lieb-Liniger model. We first present a general expression for the density form factor and then obtain specific form factors from it, and then repeat the procedure with the creation/annihilation form factors.

By combining the results from Eqs.~(\ref{startf}),~(\ref{m1fin}),~(\ref{m2rhofin}),~(\ref{gfin}), and the simplification, (\ref{lemfin}), described in Appendix C, we obtain a general expression for the density form factor:
\bea
\label{tgen}\fl
\mathcal{F}= c\left(qL\rho(q)\right)^{-\frac{1}{2}(F^2(q)+F^2(-q))}(2\pi)^{\frac{1}{2}(F(q)-F(-q))}G(1-F(q))G(1+F(-q))\nn\\ \fl
{\rm exp}\left\{P_{\pm} \int_{-1}^{1}dx\frac{F^2(qx)}{x^2-1} -\frac{1}{4}\int_{-q}^{q}d\lambda \int_{-q}^{q}d\mu\left(\frac{F(\lambda)-F(\mu)}{\lambda - \mu}\right)^2 \right\}\nn\\ \fl
 {\rm exp}\left[\sum_{i=1}^n \int_{-q}^{q}d\lambda \frac{F(\lambda) (\mu^+_i - \mu^-_i)}{(\lambda -\mu^-_i + ic)(\lambda -\mu^+_i + ic)}\right]\prod_{i,j}^{n} \left[\frac{(\mu^-_{i}- \mu^+_{j}+ic)^2}{(\mu^-_{i,j} + ic)(\mu^+_{i,j} + ic)}\right]^{1/2}\nn \\ \fl
{\rm exp}\left[-\frac{1}{2}\int_{-q}^{q}d\mu \int_{-q}^{q}d\lambda \left(\frac{F(\lambda)F(\mu)}{(\lambda - \mu + ic)^2}\right)\right] \frac{{\rm Det}(1 + \hat{G})}{{\rm Det}\left(1-\frac{\hat{K}}{2\pi} \right)}\nn \\ \fl
\times\left[\prod_{k=1}^{n}T^{(k)}_{particle} \times T^{(k)}_{hole}\right]\times T_{cross} \prod_{i=1}^n\left( \frac{F(\mu^-_i)}{L(\rho(\mu^-_i)\rho(\mu^+_i))^{1/2}(\mu_i^- - \mu_i^+)}\right).
\eea
In the above expressions, $L$ is the length of the system, $q$ the quasi momentum at the edge of the distribution, $\mu^-,\mu^+$ are the quasi momenta corresponding to excitations, and, $\rho(\lambda)$ is the ground state distribution function defined in \ceq{rhoint}. Moreover, $F(\lambda)$ is the composite shift function defined in Eqs.~(\ref{fint}),~(\ref{fcomp}), $G(x)$ is the Barnes function (Appendix A), while $\hat{G}^{\pm}$ is obtained from Sec. 4.5, $K(\lambda) = \frac{2c}{c^2 + \lambda^2}$ and the terms $T_{particle}, T_{hole}, T_{cross}$ are described in Sec. 4.4 and cannot be written down in a general form.

Let us first calculate the density form factor for the Umklapp state. This state contains $m$ adjacent holes starting at the left quasi-Fermi point, and $m$ particles starting at the first available spot after the right quasi-Fermi point. Consequently we obtain $m$ contributions from $T_{particle}$, as in \ceq{tpart2}, of the following form:
\bea\label{firsteq}\fl
\prod_{k=1}^{m} T^{(k)}_{particle} = \left(qL\rho(q)\right)^{mF(q)}\prod_{k=1}^{m}\frac{\Gamma(k-F(q))}{\Gamma(k)}{\rm exp}\left(-P_{+} \int_{-1}^{1}dx \frac{m F(qx)}{x-1}\right),\eea
and similar contribution from the product of the $T_{hole}$ terms, given by \ceq{hfer}, 
\bea\fl
\prod_{k=1}^{m} T_{hole}^{(k)} = \left(qL\rho(q)\right)^{ mF(-q)}\prod_{k=1}^{m}\frac{\Gamma(k-F(-q))}{F(-q)\Gamma(k)\Gamma(F(-q))\Gamma(1-F(-q))}{\rm exp}\left(P_- \int_{-1}^{1}\frac{m F(qx)}{x+1}\right). \nn\nn\\ \fl
\eea
Next from the cross terms given by \ceq{tcross}, $T_{cross}$, and using the approximation \ceq{app1} combined with the fact that the holes and particles occupy $m$ adjacent spots near the two quasi-Fermi points, we obtain up to $O(1/L)$ terms of the following form:
\bea
\prod_{j \neq k}^m (\mu^-_{jk})^{1/2} = (L\rho(q))^{-1/2(m^2-m)} \prod_{i=1}^{m} \Gamma(i),\nn\\ 
\prod_{j \neq k}^m (\mu^+_{jk})^{1/2} = (L\rho(q))^{-1/2(m^2-m)} \prod_{i=1}^m \Gamma(i),\nn\\ \prod_{j \neq k}^m (\mu^+_j - \mu^-_k) = (2q)^{m^2 - m},\nn\\
\label{lasteq}
T_{cross} = (2Lq\rho(q))^{-m^2 + m}G^2(m+1).
\eea

Lastly, there is a term coming from the product (for instance appearing in \ceq{tgen}),
\bea\prod_{i=1}^m \left( \frac{F(\mu^-)}{(\mu^+-\mu^-)L(\rho(\mu^-)\rho(\mu^+))^{1/2}}\right) = \left(\frac{L \rho(q)2q}{F(-q)}\right)^{-m},\eea
where, we have only retained terms bigger than $O(1/L)$ by assuming all the holes and particles occur at approximately the quasi-Fermi points. Furthermore we have used the symmetry of $\rho(\lambda)$.

Combining Eqs.~(\ref{firsteq}-\ref{lasteq}) and using the symmetry of the shift function for the Umklapp, and, the relation (A1) we obtain
\bea
\label{clap}\fl
T_{\rm Umklapp}&=& \left(qL\rho(q)\right)^{2mF(q) - m^2}\frac{G(m+1-F(q))^2}{ (2)^{m^2}G(1-F(q))^2\Gamma(F(q))^m\Gamma(1-F(q))^m}\nn\\ \fl
&\times& {\rm exp}\left(P_{-} \int_{-1}^{1}dx \frac{m F(qx)}{x+1}-P_{+} \int_{-1}^{1}dx \frac{mF(qx)}{x-1}\right).
\eea
Furthermore we may also relate the shift function $F(\lambda)$ to the ground state density function $\rho(\lambda)$ for the Umklapp \cite{Slavnov}. Let us start from the equation for the shift due to a first order Umklapp, $F(\lambda|-q) - F(\lambda|q)$, i.e. the shift for a particle hole pair at $\{q;-q\}$
\bea
&F(\lambda) - \frac{1}{2\pi} \int_{-q}^{q}d\nu K(\lambda - \nu) F(\nu) = \frac{\theta(\lambda - q)}{2\pi} - \frac{\theta(\lambda + q)}{2\pi},\nn\\ \fl
&F(\lambda) - \frac{1}{2\pi} \int_{-q}^{q}d\nu K(\lambda - \nu) F(\nu) = -\frac{1}{2\pi}\int_{-q}^{q}d\nu K(\lambda - \nu),\nn\\ \fl
&F(\lambda) - \frac{1}{2\pi} \int_{-q}^{q}d\nu K(\lambda - \nu)(F(\nu) - 1) = 0.\nn\\ \fl
\eea

But this last equation is solved by $F(\lambda) = -(2\pi \rho(\lambda) - 1)$, since
\bea 2\pi \rho(\lambda) - \frac{1}{2\pi}\int_{-q}^{q}d\nu K(\lambda - \nu)2\pi \rho(\nu)  - 1 = 0.
 \eea
Moreover, since an order $m$ Umklapp can be constructed by repeatedly performing $m$ order one Umklapps, we may use the linearity of $F(\lambda)$ to conclude
\bea\label{fklap}
F_m(\lambda|{\rm Umklapp}) =m(F(\lambda|-q) - F(\lambda|q))= -m(2\pi \rho(\lambda)-1).\eea
This makes the exponent of $L$
\bea
-F^2(q)+2mF(q)-m^2 &=& -(m\sqrt{K} -m)^2 -2m(m\sqrt{K}-m) -m^2\nn\\ \fl
&=&-m^2 K,
\eea
where we have used the relation
\bea\label{rhoK}
 2\pi \rho(q) = \sqrt{K}, 
 \ee
  from Ref.~\cite{Korepin}, where $K$ is the Luttinger parameter \cite{Gbook}.

We may now obtain $A_m$ the coefficients in the equal time density-density correlator Eq.~(\ref{Amdef}) using Eq.~(\ref{density_scaling}). We substitute the various expressions obtained above into \ceq{tgen} to obtain
\bea
\label{klap}\fl
A_{m} =& 2\gamma^2\left(\frac{q\sqrt{K}}{\rho_0}\right)^{-2m^2K}\left(\frac{4q^2+c^2}{4 c^2}\right)^{m^2}\left(\frac{G(1+m\sqrt{K})^2G(m+1-m\sqrt{K})}{\Gamma(m-m\sqrt{K})^m\Gamma(1-m+m\sqrt{K})^mG(1-m+m\sqrt{K})} \right)^2 \nn\\ \fl
& {\rm exp}\left[4mq \int_{-q}^{q}d\lambda \frac{F(\lambda) }{((\lambda + ic)^2 -q^2)}-\int_{-q}^{q}d\mu \int_{-q}^{q}d\lambda\frac{F(\lambda)F(\mu)}{(\lambda - \mu + ic)^2}\right]\nn\\ \fl
& {\rm exp}\left[2 P_{\pm} \int_{-1}^{1}dx\frac{ F^2(qx)-2mF(qx)}{x^2-1} -\frac{1}{2}\int_{-q}^{q} \int_{-q}^{q}d\lambda d\mu\left(\frac{F(\lambda)-F(\mu)}{\lambda - \mu}\right)^2\right]\frac{{\rm Det}^2(1 + \hat{G})}{{\rm Det}^2\left(1-\frac{\hat{K}}{2\pi} \right)}.
\eea

To determine the prefactors of the singularities of the density structure factor (DSF) $S(k,\omega)$ at Lieb's colective modes $\varepsilon_{1,2}(k)$~\cite{LL, Korepin} using \ceq{DSFscaling} we need to find corresponding form factors of the density operator from \ceq{tgen}. 

By considering the form factor for a Bose gas with one hole at the right edge and a high momentum particle, we may determine the prefactors of the DSF singularity at $\varepsilon_1(k)$. To do this we need to relate the quasi momentum of the particle $\mu^+$ to the momentum $k$ of the excited bosonic state as follows \cite{Korepin, PRL_08}:
\bea
k = \mu^+ - \pi \rho_0 - \int_{-q}^{q}d\lambda \theta(\mu^+ - \lambda) \rho(\lambda),\eea
where, $\theta(\lambda)$ = i ln$\left(\frac{ic+\lambda}{ic-\lambda}\right)$, and $q$ is the quasimomentum at the edge of the distribution. 

For the excited state with the particle hole pair ($\mu^+, q$), we substitute $F(\lambda|\mu^+, q)$ from \ceq{fint} and substitute this function in the relevant places in \ceq{tgen}. For such a particle-hole pair, we obtain a form for $T_{hole}$ as in \ceq{hfer}, and for $T_{particle}$ as in \ceq{tpartfar}. Since there is only a single particle-hole pair there is no contribution like $T_{cross}$ we obtain
\bea \fl
S_1(k) & =c^2\sqrt{K}\left(q\sqrt{K}\right)^{-(F^2(q)+F^2(-q) - 2F(q)+2)}(2\pi)^{F(q)-F(-q)}\left(\frac{G(1+F(-q))G(1-F(q))}{\Gamma(1-F(q))}\right)^2\nn\\ \fl
&{\rm exp}\left\{P_{\pm} \int_{-1}^{1}dx\frac{2 F^2(qx)}{x^2-1} + P_+ \int_{-1}^{1}dx \frac{2F(qx)}{x - 1} -\frac{1}{2}\int_{-q}^{q}d\lambda \int_{-q}^{q}d\mu\left(\frac{F(\lambda)-F(\mu)}{\lambda - \mu}\right)^2 \right\}\nn\\ \fl
& {\rm exp}\left[ \int_{-q}^{q}d\lambda \frac{2F(\lambda) (\mu^+ - q)}{(\lambda -q+ ic)(\lambda -\mu^+ + ic)}\right]\frac{(q- \mu^+)^2 +c^2}{ c^2}\left(\frac{F^2(q)} {\rho(\mu^+)(\mu^+/q-1)^2}\right)\nn\\ \fl
&{\rm exp}\left[-\int_{-q}^{q}d\mu \int_{-q}^{q}d\lambda\frac{F(\lambda)F(\mu)}{(\lambda - \mu + ic)^2}  -\int_{-q}^{q}d\lambda \frac{2F(\lambda)}{\lambda - \mu^+}\right] \frac{{\rm Det}^2(1 + \hat{G})}{{\rm Det}^2\left(1-\frac{\hat{K}}{2\pi} \right)}.
\eea
Note that the momentum dependence in the above expression is contained in the shift function, $F(\lambda)=F(\lambda|\mu^+) - F(\lambda|q)$, as well as the terms directly involving $\mu^+$.  Furthermore we have used the relation in \ceq{rhoK} to express the answer in terms of the Luttinger parameter, $K$ \cite{Gbook}.

Similarly, we may use \ceq{tgen}, to obtain the DSF at the Lieb mode $\varepsilon_2(k)$, by considering the density form factor of a system with a single high momentum hole and a low momentum particle (at the right quasi-Fermi point). This time we relate the quasi momentum of the hole, $\mu^-$, to the momentum, $k$ using
\bea-k = \mu^-  - \pi \rho_0 - \int_{-q}^{q}d\lambda \theta(\mu^- - \lambda) \rho(\lambda),\eea
and, use the unique solution of the above to determine a shift function, $F(\lambda|q,\mu^-)$, using \ceq{fint}. We may then use this $F(\lambda)$ in \ceq{tgen} in conjunction with the expression for $T_{particle}$ given by \ceq{tpart2}, and $T_{hole}$ given by \ceq{tholedeep} to get the following expression for $S_2(k)$:
\bea \fl
S_2(k) & = c^2\sqrt{K}\left(q\sqrt{K}\right)^{-(F^2(q)+F^2(-q)+2F(q)+2)}(2\pi)^{F(q)-F(-q)}\left(\frac{G(1+F(-q))G(1-F(q))\Gamma(1-F(q))}{\Gamma(1+F(\mu^-))\Gamma(1-F(\mu^-))}\right)^2\nn\\ \fl
& {\rm exp}\left[ \int_{-q}^{q}d\lambda \frac{2F(\lambda) (q-\mu^-)}{(\lambda -\mu^-+ ic)(\lambda -q + ic)}\right]\frac{(q- \mu^-)^2 +c^2}{ c^2}\left(\frac{F^2(\mu^-)} {\rho(\mu^-)(\mu^-/q -1)^2}\right)\nn\\ \fl
&{\rm exp}\left\{-\int_{-q}^{q}d\mu \int_{-q}^{q}d\lambda \frac{F(\lambda)F(\mu)}{(\lambda - \mu + ic)^2}+ P_{\pm} \int_{-1}^{1}dx\frac{2 F^2(qx)}{x^2-1} - P_+ \int_{-1}^{1}dx\frac{2F(qx)}{x - 1}\right\}\nn\\ \fl
&{\rm exp}\left[ P \int_{-1}^{1}dx \frac{2F(qx)}{x-\mu^-/q} -\frac{1}{2}\int_{-q}^{q}d\lambda \int_{-q}^{q}d\mu\left(\frac{F(\lambda)-F(\mu)}{\lambda - \mu}\right)^2\right]\times \frac{{\rm Det}^2(1 + \hat{G})}{{\rm Det}^2\left(1-\frac{\hat{K}}{2\pi} \right)}.
\eea
Here too the dependence on momentum, $k$, is carried by $F(\lambda) = F(\lambda|q) - F(\lambda|\mu^-)$ and $\mu^-$. Furthermore we have used the relation in \ceq{rhoK} to express the answer in terms of the Luttinger parameter, $K$ \cite{Gbook}.

Using the results from Eqs.~(\ref{startg}),~(\ref{startgp}),~(\ref{m1fin}),~(\ref{m2psifin}) as well as the simplification (\ref{lemfin}) outlined in Appendix C, we obtain the following analytic expressions for the form factors of the creation and annihilation operators. 

The most general expression for the annihilation operator form factor is given by
\bea
\label{tgmin}\fl
\mathcal{G}^-&=\frac{\left(qL\rho(q)\right)^{-\frac{1}{2}(F_{-}^2(q)+F_{-}^2(-q))}}{L^{1/2}\rho(\mu^-)^{1/2}} (2\pi)^{\frac{1}{2}(F_{-}(q)-F_{-}(-q)-1)}\frac{\pi F_{-}(\mu^-)}{{\rm sin}(\pi F_{-}(\mu^-))}G(1+F_{-}(-q))G(1-F_{-}(q))\nn\\ \fl
&\prod_{i=1}^n\left( \frac{F_{-}(\mu^-_i)}{L(\rho(\mu^-_i)\rho(\mu^+_i))^{1/2}(\mu_i^- - \mu_i^+)}\right)\left[\prod_{k=1}^{n}T^{(k)}_{particle} \prod_{k=1}^{n+1} \times T^{(k)}_{hole}\right] \times T_{cross}\nn\\ \fl
&{\rm exp}\left\{P_{\pm} \int_{-1}^{1}dx\frac{F_{-}^2(qx)}{x^2-1} -\frac{1}{4}\int_{-q}^{q}d\lambda \int_{-q}^{q}d\mu\left(\frac{F_{-}(\lambda)-F_{-}(\mu)}{\lambda - \mu}\right)^2 \right\}\nn\\ \fl
& {\rm exp}\left[\sum_{i=1}^n \int_{-q}^{q}d\lambda \frac{F_{-}(\lambda) (\mu^+_i - \mu^-_i)}{(\lambda -\mu^-_i + ic)(\lambda -\mu^+_i + ic)}\right]\prod_{i,j}^{n} \left[\frac{(\mu^-_{i}- \mu^+_{j}+ic)^2}{(\mu^-_{i,j} + ic)(\mu^+_{i,j} + ic)}\right]^{1/2}\nn\\ \fl
&{\rm exp}\left[-\int_{-q}^{q}d\mu \int_{-q}^{q}d\lambda \left(\frac{F_{-}(\lambda)F_{-}(\mu)}{2(\lambda - \mu + ic)^2}\right)\right] \frac{{\rm Det}(1 + \hat{H}^-)}{{\rm Det}\left(1-\frac{\hat{K}}{2\pi} \right)}.
\eea
The creation operator form factor is given by
\bea
\label{tgplus} \fl
\mathcal{G}^+ &=  \frac{\left(qL\rho(q)\right)^{-\frac{1}{2}(F_{+}^2(q)+F_{+}^2(-q))}}{L^{1/2}\rho(\mu^+)^{1/2}}(2\pi)^{\frac{1}{2}(F_{+}(q)-F_{+}(-q)-1)}G(1+F_{+}(-q))G(1-F_{+}(q))\nn\\ \fl
&\prod_{i=1}^n\left( \frac{F_{+}(\mu^-_i)}{L(\rho(\mu^-_h)\rho(\mu^+_h))^{1/2}(\mu_i^- - \mu_i^+)}\right)\left[\prod_{k=1}^{n+1}T^{(k)}_{particle} \times \prod_{k=1}^{n}T^{(k)}_{hole}\right] \times T_{cross}\nn\\ \fl
&{\rm exp}\left\{P_{\pm} \int_{-1}^{1}dx\frac{ F_{+}^2(qx)}{x^2-1} -\frac{1}{4}\int_{-q}^{q}d\lambda \int_{-q}^{q}d\mu\left(\frac{F_{+}(\lambda)-F_{+}(\mu)}{\lambda - \mu}\right)^2 \right\}\nn\\ \fl
& {\rm exp}\left[\sum_{i=1}^n \int_{-q}^{q}d\lambda \frac{F_{+}(\lambda) (\mu^+_i - \mu^-_i)}{(\lambda -\mu^-_i + ic)(\lambda -\mu^+_i + ic)}\right]\prod_{i,j}^{n} \left[\frac{(\mu^-_{i}- \mu^+_{j}+ic)^2}{(\mu^-_{i,j} + ic)(\mu^+_{i,j} + ic)}\right]^{1/2}\nn\\ \fl
&{\rm exp}\left[-\int_{-q}^{q}d\mu \int_{-q}^{q}d\lambda \left(\frac{F_{+}(\lambda)F_{+}(\mu)}{2(\lambda - \mu + ic)^2} \right)\right]\frac{{\rm Det}(1 + \hat{H}^+)}{{\rm Det}\left(1-\frac{\hat{K}}{2\pi} \right)}.
\eea
In the above expressions, $L$ is the length of the system, $q$ the quasi momentum at the edge of the distribution, $\mu^-,\mu^+$ are the quasi momenta corresponding to excitations, and $\rho(\lambda)$ is the ground state distribution function defined in \ceq{rhoint}. Moreover, $F_{\pm}(\lambda)$ are the modified shift functions defined in Eqs.~(\ref{fplus}),~(\ref{fminus}), $G(x)$ is the Barnes function (Appendix A),  $\hat{H}^{\pm}$ is obtained from Sec.~4.6, $K(\lambda) = \frac{2c}{c^2 + \lambda^2}$ and the terms $T_{particle}, T_{hole}, T_{cross}$ are described in Sec. 4.4 and cannot be written down in a general form.

In order to obtain from \ceq{tgmin} the annihilation operator form factor for an order $m$ Umklapp state, we first collect terms from Eqs. (\ref{firsteq}) - (\ref{lasteq}) with the appropriate modified shift function and the extra hole to obtain
\bea \fl
&T_{Umklapp}\nn\\ \fl
&= \left(L\rho(q)\right)^{(m-1)F_-(q) + m F_-(-q) - m(m-1)}\times \frac{1}{(2q)^{m^2-m}}\nn\\ \fl
&\times \frac{G(m+1-F_-(-q))G(m+1-F_-(q))}{G(1-F_-(q))G(1-F_-(-q))\Gamma(F_-(-q))^{m-1}\Gamma(1-F_-(-q))^{m-1}\Gamma(1-F_-(q))\Gamma(1+F_-(q))}\nn\\ \fl
&\times{\rm exp}\left[-(m-1)P_{+} \int_{-1}^{1}dx \frac{F_-(qx)}{x - 1}  + m P_{-} \int_{-1}^{1} dx \frac{F_-(qx)}{x+1} \right].
\eea
For this excitation we may determine the shift function at the quasi-Fermi points, $F_{-}(\pm q)$, using \ceq{fminus}:
\bea\label{fminusfull}
F_-(\pm q) = F_m(\pm q| {\rm Umklapp}) - F(\pm q| q).\eea
We can substitute $F_m(\pm q|{\rm Umklapp}) = -m(\sqrt{K}-1)$, from \ceq{fklap} and \ceq{rhoK}, and substitue $F(\pm q| q) + \pi \rho(\pm q) =  1/2 \pm (1/2 - 1/(2\sqrt{K})$ from Ref.~\cite{PRL_08}, to determine the exponent of $L$ in the Umklapp annihilation form factor,
\bea\fl-\frac{1}{2}(F_{-}^2(q)+F_{-}^2(-q)-2(m-1)F_{-}(q)-2mF_{-}(-q)+2m^2-2m+1) = -m^2 K - \frac{1}{4K}.\nn\\ 
\eea
where, $K$ is the Luttinger parameter \cite{Gbook}.

We may then use \ceq{boson_scaling} to relate the absolute square of the annihilation operator form factor $\mathcal{G}^-$ for the Umklapp state to determine the coefficients $B_m$ of the Green's function of the Bose gas in \ceq{Bmdef} to obtain
\bea \fl
B_{m>0} &=q\frac{(q^2K)^{-m^2 K - 1/4K}}{2^{2m^2-2m-1}}(2\pi)^{-2+1/\sqrt{K}}\left(\frac{4q^2+c^2}{c^2}\right)^{(m-1)^2}\frac{{\rm Det}^2(1 + \hat{H}^-)}{{\rm Det}^2\left(1-\frac{\hat{K}}{2\pi} \right)}\nn\\ \fl
&\times \left(\frac{G(m-F_{-}(q))G(m+1-F_{-}(-q)))G(1+F_-(-q))}{\Gamma(F_-(-q))^{m-1}\Gamma(1-F_{-}(-q))^{m-1}G(1-F_-(-q))}\right)^2\nn\\ \fl
& {\rm exp}\left[4(m-1)q \int_{-q}^{q}d\lambda \frac{F_{-}(\lambda) }{((\lambda + ic)^2 -q^2)}-\int_{-q}^{q}d\mu \int_{-q}^{q}d\lambda\frac{F_{-}(\lambda)F_{-}(\mu)}{(\lambda - \mu + ic)^2} \right]\nn\\ \fl
& {\rm exp}\left[2 P_{\pm} \int_{-1}^{1}dx\frac{ F_{-}^2(qx)-2(m-1/2)F_{-}(qx) +x F_-(qx)}{x^2-1} -\frac{1}{2}\int_{-q}^{q} \int_{-q}^{q}d\lambda d\mu\left(\frac{F_{-}(\lambda)-F_{-}(\mu)}{\lambda - \mu}\right)^2\right].\nn\\ \fl \label{klap2}
\eea

 Note that we have used the relation in \ceq{rhoK} to express the answer in terms of the Luttinger parameter, $K$ \cite{Gbook}.

To determine the prefactors of the singularities of the spectral function $A(k,\omega)$ near Lieb's colective modes $\pm \epsilon_{1,2}(k)$~\cite{LL, Korepin} we need to find corresponding form factors of the creation/annihilation operators from Eqs. (\ref{tgmin}), (\ref{tgplus}). 

 We obtain $\overline{A_+(k)}$ from the creation operator form factor for a Bose gas with a high energy particle, and obtain $\overline{A_-(k)}$ from the annihilation operator form factor  for a Bose gas with one high energy hole. These functions give the prefactors for the spectral function singularities at $\epsilon_1(k)$ and  $-\epsilon_2(k)$ respectively. We must first relate the quasi momentum of the particle to the momentum $k$ of the excited bosonic state as follows \cite{Korepin, PRL_08}:
\bea\label{qpart}
k = \mu^+ - \pi \rho_0 - \int_{-q}^{q}d\lambda \theta(\mu^+ - \lambda) \rho(\lambda),\eea
and for the quasi momentum of the hole
\bea\label{qhole}
-k = \mu^-  - \pi \rho_0 - \int_{-q}^{q}d\lambda \theta(\mu^-  - \lambda) \rho(\lambda),\eea
where $\theta(\lambda)$ = i ln$\left(\frac{ic+\lambda}{ic-\lambda}\right)$, and $q$ is the quasimomentum at the edge of the distribution. 

Particle-hole pairs in Eqs.~(\ref{fplus}),(\ref{fminus}) are defined with respect to ground states of $N\pm1$ particles, so to obtain a state with a single high energy particle (hole), a particle-hole pair needs to contain a hole (particle) at $q$. Such a procedure results in the following expressions for shift functions $\overline{F_{\pm}(\lambda)}$
\bea
\overline{F_+(\lambda)} = F(\lambda|\mu^+) +\pi \rho(\lambda),\nn\\ 
\overline{F_-(\lambda)} = -F(\lambda|\mu^-) - \pi \rho(\lambda).
\eea

We may then substitute these functions in the relevant places in \ceq{tgplus} and \ceq{tgmin}, combine terms from $T_{hole}$ given by \ceq{hfer}, and $T_{particle}$ given by \ceq{tpartfar}, to obtain

\bea
\fl
\overline{A_{\pm}(k)} & = \frac{1}{\rho(\mu^-)}\left(q\sqrt{K}\right)^{-(\overline{F_{\pm}^2(q)}+\overline{F_{\pm}^2(-q)})}(2\pi)^{\overline{F_{\pm}(q)}-\overline{F_{\pm}(-q)}-1}(G(1+\overline{F_{\pm}(-q)})G(1-\overline{F_{\pm}(q)}))^2\nn\\ \fl
&{\rm exp}\left\{P_{\pm} \int_{-1}^{1}dx\frac{2 \overline{F_{\pm}^2(qx)}}{x^2-1} \mp (P) \int_{-1}^{1}dx \frac{2\overline{F_{\pm}(qx)}}{x-\mu^{\pm}/q} -\frac{1}{2}\int_{-q}^{q}d\lambda \int_{-q}^{q}d\mu\left(\frac{\overline{F_{\pm}(\lambda)}-\overline{F_{\pm}(\mu)}}{\lambda - \mu}\right)^2 \right\}\nn\\ \fl
&{\rm exp}\left[-\int_{-q}^{q}d\mu \int_{-q}^{q}d\lambda \frac{\overline{F_{\pm}(\lambda)}\ \overline{F_{\pm}(\mu)}}{(\lambda - \mu + ic)^2} \right]\times \frac{{\rm Det}^2(1 + \hat{H}^-)}{{\rm Det}^2\left(1-\frac{\hat{K}}{2\pi} \right)}\frac{((q- \mu^{\pm})^2 +c^2)}{c^2}\nn\\ \fl
& {\rm exp}\left[\pm 2 \int_{-q}^{q}d\lambda \frac{\overline{F_{\pm}(\lambda)} (\mu^{\pm} - q)}{(\lambda -\mu^{\pm}+ ic)(\lambda -q + ic)}\right].\label{oapm}
\eea
Note that the momentum dependence in the above expressions is contained in the shift function, $\overline{F_{\pm}(\lambda)}$ defined above, as well as the terms directly involving $\mu^{\pm}$. Furthermore we have used the relation in \ceq{rhoK} to express the answer in terms of the Luttinger parameter, $K$ \cite{Gbook}.

 We obtain the prefactors of the spectral function singularities, $\underline{A_+(k)}$ and $\underline{A_-(k)}$, near the Lieb modes $\epsilon_2(k)$ and $-\epsilon_1(k)$ from the form factors of the creation and annihilation operator respectively. We can obtain the prefactor $\underline{A_+(k)}$ from the creation operator form factor of a system with a high momentum hole and two particles at the right quasi-Fermi point. From the annihilation operator form factor of a system with a high momentum particle and two holes at the right quasi-Fermi point, we obtain the prefactor $\underline{A_-(k)}$. We again relate the quasimomenta of the particle and hole to the momentum $k$ using, \ceq{qpart} and \ceq{qhole} and determine $F_{\pm}(\lambda)$ using Eqs. (\ref{fplus}), (\ref{fminus}). 

Particle-hole pairs in Eqs.~(\ref{fplus}),(\ref{fminus}) are defined with respect to ground states of $N\pm1$ particles, so to obtain a state with a single high energy particle (hole), a particle-hole pair needs to contain a hole (particle) at $q$. Such a procedure results in the following expressions for shift functions  $\underline{F_{\pm}(\lambda)}$:
\bea
\underline{F_+(\lambda)} = 2F(\lambda|q) - F(\lambda|\mu^-) + \pi \rho(\lambda),\nn\\
\underline{F_-(\lambda)} = F(\lambda|\mu^+) - 2F(\lambda|q) - \pi \rho(\lambda).
\eea

 Substituting these results in \ceq{tgplus} and \ceq{tgmin} in conjunction with the expression for $T_{particle}$ given by \ceq{tpart2}, and $T_{hole}$ given by \ceq{tholedeep} we get the following expressions:
\bea
\fl
\underline{A_+(k)} & =\frac{1}{\rho(\mu^-)(\mu^-/q -1)^4}\left(q\sqrt{K}\right)^{-(\underline{F_{+}^2(q)}+\underline{F_{+}^2(-q)}-4\underline{F_{+}(q)}+4)}(2\pi)^{\underline{F_{+}(q)}-\underline{F_{+}(-q)}-2} \frac{((q- \mu^-)^2 +c^2)}{c^2} \nn\\ \fl
&\times \left(\frac{G(1+\underline{F_{+}(-q)})G(1-\underline{F_{+}(q)})\Gamma(1-\underline{F_{+}(q)})\Gamma(2-\underline{F_{+}(q)})}{\Gamma(1-\underline{F_{+}(\mu^-)})\Gamma(\underline{F_{+}(\mu^-)})}\right)^2\times \frac{{\rm Det}^2(1 + \hat{H}^+)}{{\rm Det}^2\left(1-\frac{\hat{K}}{2\pi} \right)}\nn\\ \fl
&{\rm exp}\left\{P_{\pm} \int_{-1}^{1}dx\frac{2 \underline{F_{+}^2(qx)}}{x^2-1} - P_+ \int_{-1}^{1}dx\frac{4\underline{F_{+}(qx)}}{x-1}+2 \int_{-q}^{q}d\lambda \frac{\underline{F_{+}(\lambda)} (q-\mu^-)}{(\lambda -\mu^-+ ic)(\lambda -q + ic)}\right\}\nn\\ \fl
&{\rm exp}\left[ P \int_{-1}^{1}d\lambda \frac{2\underline{F_{+}(qx)}}{x-\mu^-/q} -\frac{1}{2}\int_{-q}^{q}d\lambda \int_{-q}^{q}d\mu\left(\frac{\underline{F_{+}(\lambda)}-\underline{F_{+}(\mu)}}{\lambda - \mu}\right)^2 \right]\nn\\ \fl
&{\rm exp}\left[-\int_{-q}^{q}d\mu \int_{-q}^{q}d\lambda \frac{\underline{F_{+}(\lambda)}\ \underline{F_{+}(\mu)}}{(\lambda - \mu + ic)^2} \right].\label{uap}
\eea\bea
\fl
\underline{A_-(k)} & = \frac{1}{\rho(\mu^+)(\mu^+/q-1)^4}\left(q\sqrt{K}\right)^{-(\underline{F_{-}^2(q)}+\underline{F_{-}^2(-q)} + 4\underline{F_{-}(q)}+4)}(2\pi)^{\underline{F_{-}(q)}-\underline{F_{-}(-q)}-2}\nn\\ \fl
&\frac{{\rm Det}^2(1 + \hat{H}^-)}{{\rm Det}^2\left(1-\frac{\hat{K}}{2\pi} \right)}\frac{((q- \mu^+)^2 +c^2)}{c^2}\left(\frac{G(1+\underline{F_{-}(-q)})G(1-\underline{F_{-}(q)})\Gamma(1+\underline{F_{-}(q)})\Gamma(2+\underline{F_-(q)})}{\Gamma(\underline{F_-(q)})\Gamma(1-\underline{F_{-}(q)})}\right)^2\nn\\ \fl
&{\rm exp}\left\{P_{\pm} \int_{-1}^{1}dx \frac{2 \underline{F_{-}^2(qx)}}{x^2-1} + P_+ \int_{-1}^{1}dx\frac{4\underline{F_{-}(qx)}}{x - 1} -\frac{1}{2}\int_{-q}^{q}d\lambda \int_{-q}^{q}d\mu\left(\frac{\underline{F_{-}(\lambda)}-\underline{F_{-}(\mu)}}{\lambda - \mu}\right)^2 \right\}\nn\\ \fl
&{\rm exp}\left[-\int_{-q}^{q}d\mu \int_{-q}^{q}d\lambda \frac{\underline{F_{-}(\lambda)}\ \underline{F_{-}(\mu)}}{(\lambda - \mu + ic)^2}  -\int_{-q}^{q}d\lambda \frac{2\underline{F_{-}(\lambda)}}{\lambda - \mu^+}\right]\nn\\ \fl
&{\rm exp}\left[2 \int_{-q}^{q}d\lambda \frac{\underline{F_{-}(\lambda)} (\mu^+ - q)}{(\lambda -q+ ic)(\lambda -\mu^+ + ic)}\right].\label{uam}
\eea

Here too the dependence on momentum, $k$, is carried by $\underline{F_{\pm}(\lambda)}$ defined above and $\mu^{\pm}$.  Furthermore we have used the relation in \ceq{rhoK} to express the answer in terms of the Luttinger parameter, $K$ \cite{Gbook}.

\subsection{Numerical Results}
 Let us plot here some prefactors which are obtained using above analytical results.
 
\begin{figure}[htpb]
\includegraphics[scale=.4]{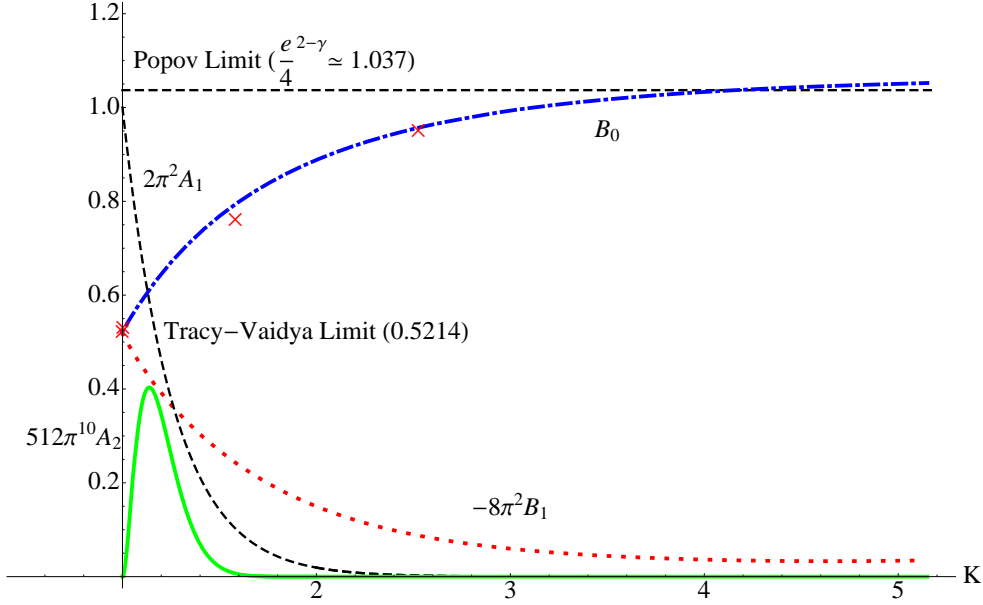}
\caption{\label{FigLL}(Color online) Results for the Lieb-Liniger model of 1D bosons:  $ 2\pi^{2}A_1$ (dashed black), $512\pi^{10} A_2$ (solid green) and $B_0$ (dot-dashed blue), $-8\pi^2B_1$ (dotted red) as functions of the Luttinger liquid parameter $K$. In the limit of strong interaction ($K \to 1$) our expressions for $B_0$ and $B_1$ agree with the known values~\cite{K1}, while $A_1 \to 1/2\pi^2, A_2 \to 0$ are in accordance with the density correlator of the free Fermi gas. We also match $B_0$ in the weakly interacting regime ($K \gg 1$) to Popov's result (dashed line) \cite{Popov_prefactor}, and  show some numerical results (crosses)~\cite{correlation_numerical}.
 }
 \end{figure}

 A few prefactors of equal time correlators $A_m, B_m$, see Eqs.~(\ref{Amdef}) and (\ref{Bmdef}) are plotted in Fig.~\ref{FigLL}

\begin{figure}[htpb]
 \begin{center}
 \includegraphics[scale=.5]{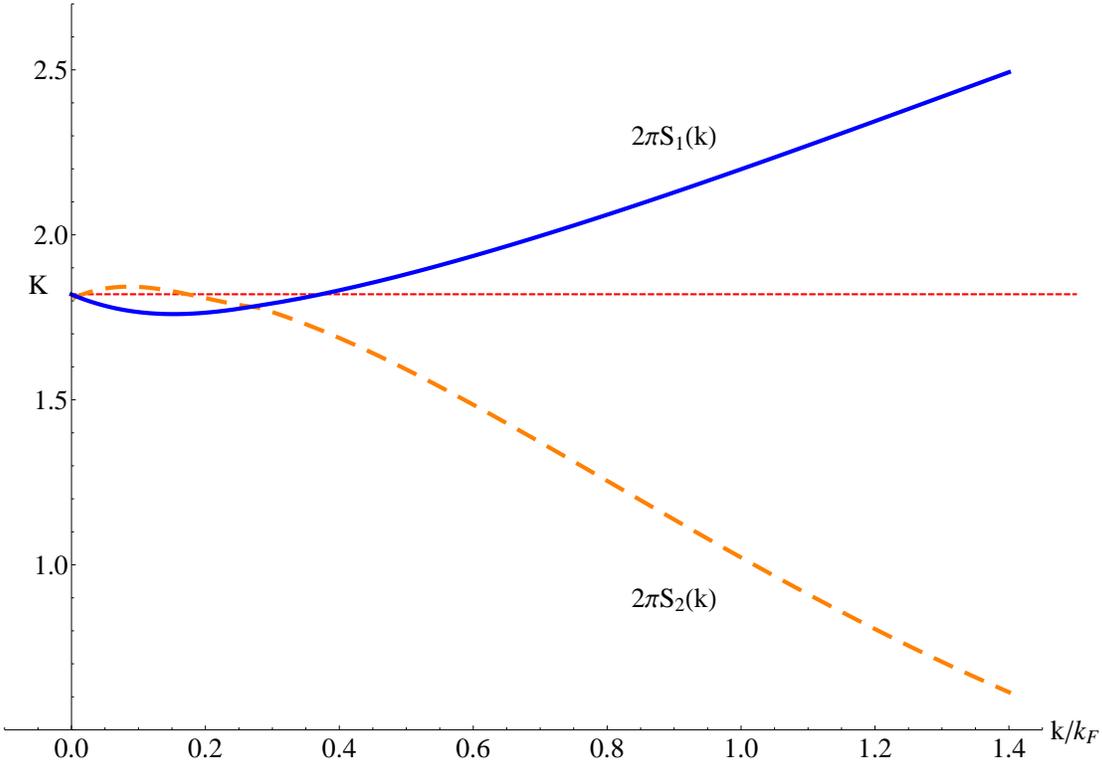}
 \caption{Plot of $2\pi S_1$ (blue), $2\pi S_2$ (orange, dashed)  as functions of $k/k_F$ for $K=1.81$.  As $k/k_F \to 0$ they approach $K$ indicated by a dashed red line.}
 \label{fig2}
 \end{center}
 \end{figure}

 $S_1(k)$(blue), $S_2(k)$ (orange, dashed) are plotted in Fig.~\ref{fig2} for $\gamma = 4.52, \rho_0 = 1, K = 1.81$ as functions of $k/k_F$. We can analyze the limiting behavior as $k \to 0$ of these prefactors using their forms
 \bea \fl
 S(k,\omega) = \frac{\sin{\pi \tilde \mu_L}\theta(\delta\omega)+\sin{\pi
  \tilde \mu_R}\theta(-\delta\omega)}{\sin{\pi (\tilde \mu_L+\tilde \mu_R)}}  \frac{2\pi S_{1}(k)
\delta\omega^{\tilde \mu_R+\tilde \mu_L-1} }{\Gamma(\tilde
\mu_R+\tilde \mu_L)(v+v_d)^{\tilde \mu_L}|v-v_d|^{\tilde
\mu_R}},\nn\\  |\delta \omega| = |\omega - \varepsilon_1(k)| \ll \varepsilon_1(k),\nonumber \nn\\ \fl
 S(k,\omega) = \theta(\delta \omega)  \frac{2\pi S_{2}(k)
\delta\omega^{\tilde \mu_R+\tilde \mu_L-1} }{\Gamma(\tilde
\mu_R+\tilde \mu_L)(v+v_d)^{\tilde \mu_L}|v-v_d|^{\tilde
\mu_R}}, \ \ \ \ \  \delta \omega = \omega - \varepsilon_2(k) \ll \varepsilon_2(k), \nonumber 
\eea
 and the universal relation for $S(k,\omega)$ given in Ref.~\cite{universal}, 
\bea S(k,\omega) = \frac{m_* K}{k} \theta(\frac{k^2}{2m_*} - |\omega - v|k||).
 \eea
 The exponents $\tilde{\mu}_{R(L)} = (F(\pm q|\lambda) + \sqrt{K}/2 + 1/2\sqrt{K})$ and $m_*$ is the effective mass. In the limit that $k \to 0$ we have $\mu_R \to 1, \mu_L \to 0$. Then from the above relations we obtain 
\bea \lim_{k\to 0} S_2(k) = S_1(k) = \frac{K}{2\pi}.
 \eea
This asymptote is indicated by the red dotted line. 

\begin{figure}[p]
 \begin{center}
 \includegraphics[scale=.4]{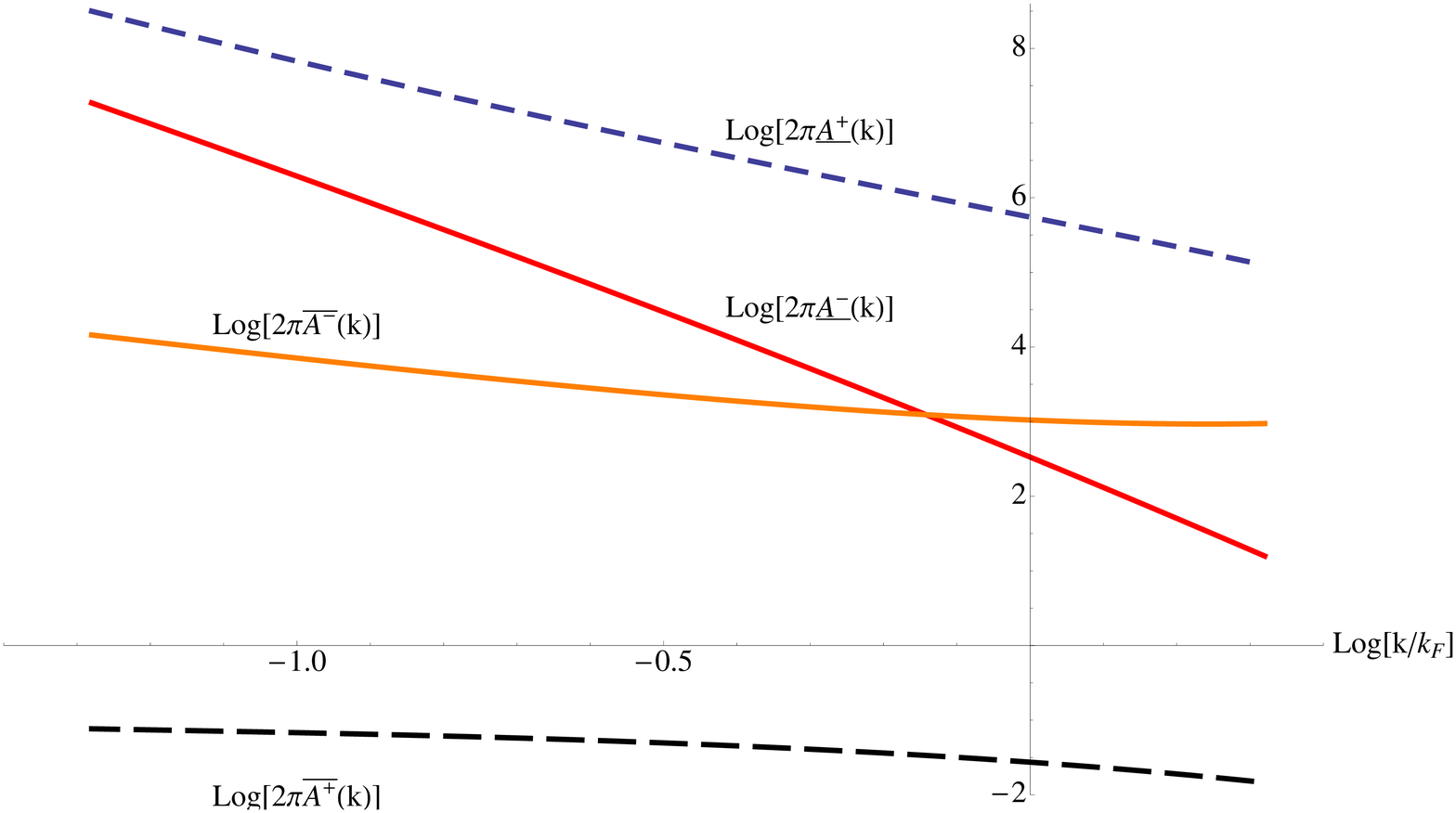}
 \caption{Log-log plot of $\overline{A^{+}(k)}$ (black, dashed), $\overline{A^-(k)}$ (orange), $\underline{A^{+}(k)}$ (blue, dashed) and $\underline{A^-(k)}$ (red) as functions of $k/k_F$ for $K=1.81$.}
 \label{fig3}
 \end{center}
 \end{figure}

Numerical results for $\overline{A^{\pm} (k)}, \underline{A^{\pm}(k)}$ as functions of $k$ are obtained from Eqs.~(\ref{oapm}),~(\ref{uap}) and (\ref{uam}), and are plotted in Fig. \ref{fig3} for $K = 1.81$.
 
 \newpage

\section{Prefactors of equal time spin correlators of XXZ model with finite magnetization}
\subsection{Introduction}
Let us now turn to the $XXZ$ magnet, whose Hamiltonian is given by equation (\ref{xxz_hamiltonian}).
For the sake of brevity, our discussion here will follow that of the Lieb-Liniger model, and omit
extensive discussion of the various steps, which are similar. 

All eigenstates of the $XXZ$ magnet can be obtained from the Bethe Ansatz \cite{1958_Orbach_PR_112}. 
Assuming periodic boundary conditions, the quasimomenta $k$ appearing in the wavefunction are 
parametrized in terms of rapidities according to the definition
\begin{equation}
e^{ik} = \frac{\sinh (\lambda + i\zeta/2)}{\sinh (\lambda - i\zeta/2)}, 
\hspace{1cm} k = \pi - 2~\mbox{atan}~ \frac{\tanh\lambda}{\tan \zeta/2}
\end{equation}
in which the anisotropy $\Delta$ (we restrict ourselves here to $0 < \Delta < 1$) is parametrized by 
$\zeta = \mbox{acos} \Delta$ with $\zeta \in ]0, \pi/2[$. 
The Bethe equations for an eigenstate with $M$ down spins are
\begin{equation}
\left[ \frac{\sinh (\lambda_j + i\zeta/2)}{\sinh(\lambda_j - i\zeta/2)} \right]^N = \prod_{k \neq j}^M 
\frac{\sinh(\lambda_j - \lambda_k + i\zeta)}{\sinh(\lambda_j - \lambda_k - i\zeta)}, \hspace{1cm} j = 1, ..., M.
\label{eq:BE_XXZ}
\end{equation}
or more conveniently in logarithmic form,
\begin{eqnarray}
\theta_1(\lambda_a) - \frac{1}{N} \sum_{b = 1}^M \theta_2(\lambda_a - \lambda_b) = 2\pi \frac{I_a}{N}
\label{eq:logBE_XXZ}
\end{eqnarray}
in which $I_a$ is half-odd integer for $N-M$ even, and integer for $N-M$ odd, and the kernels are defined as
\begin{equation}
\theta_n (\lambda) = 2~\mbox{atan}~ \frac{\tanh \lambda}{\tan (n\zeta/2)}.
\label{eq:theta_XXZ}
\end{equation}
Such a state has energy given by
\begin{eqnarray}
E = J \sum_{a = 1}^M \frac{-\sin^2 \zeta}{\cosh 2\lambda_a - \cos \zeta},
\label{eq:E_XXZ}
\end{eqnarray}
with momentum
\begin{eqnarray}
P = \sum_{a = 1}^M \frac{1}{i} \ln \left[\frac{\sinh(\lambda_a + i\zeta/2)}{\sinh(\lambda_a - i\zeta/2)}\right]
= \pi M - \frac{2\pi}{N}\sum_{a = 1}^M I_a \hspace{0.5cm} \mbox{mod} \hspace{0.2cm}2\pi.
\label{eq:P_XXZ}
\end{eqnarray}

In contrast to the Lieb-Liniger model, there exist both real and complex-valued solutions to the Bethe 
equations for the $XXZ$ model (we refer the reader for example to \cite{Takahashi, Gaudin} for
details). For simplicity here, we will confine our discussion to purely real solutions
only, similarly to the discussion contained in \cite{Pereira_PRL_06} (our equations are however directly
applicable to eigenstates with negative-parity one-strings; higher strings need to be treated following
the logic of \cite{2005_Caux_JSTAT_P09003}). In essence, we consider a finite field leading to a
generic incommensurate filling, for which the ground state is given by a filled Fermi interval of 
$M < N/2$ adjacent quantum numbers/rapidities centered on the origin. Starting from such a state, 
excitations can be constructed
using a similar logic as for the Lieb-Liniger case, namely by removing/adding quantum numbers and
applying particle-hole excitations. 

To obtain correlation prefactors, our starting point is once again the finite-size representation
of matrix elements of the considered spin operators. 
For the Heisenberg chain, these matrix elements
were obtained within the Algebraic Bethe Ansatz framework in \cite{XXZ_formfactors}.
We thus start here by writing down in our conventions all the formulas that we are going 
to need in further calculations. 
The form factors for the operators $S^{z}$, $S^{\pm}$ can be written as
\bea\label{eq:Sz_start}\fl\nonumber
|\langle\{\mu\}_M| S_q^z |\{\lambda\}_M\rangle| = \frac{\sqrt{N}}{2} \delta_{q, P_{\lambda}-P_{\mu}}\prod_{j=1}^M \left|\frac{\sinh(\mu_j+i\zeta/2)}{\sinh(\lambda_j+i\zeta/2)}\right| \\\nonumber
\times \prod_{j>k} |\sinh^{-1}(\mu_j-\mu_k-i\zeta)\sinh^{-1}(\lambda_j-\lambda_k-i\zeta)|\\
\times\frac{|\det_M \left(H_{ab}^z\left(\{\mu\},\{\lambda\}\right)-2P_{ab}(\{\mu\},\{\lambda\})\right)|}
{\det_M ^{1/2}\Phi\left(\{\mu\}\right)\det_M ^{1/2}\Phi\left(\{\lambda\}\right)},\\
\label{eq:S-_start}\fl\nonumber
|\langle\{\mu\}_{M+1}| S_q^- |\{\lambda\}_M\rangle| = \sqrt{N} \delta_{q, P_{\lambda}-P_{\mu}} \sin^{1/2}(\zeta)\frac{\prod_{j=1}^{M+1} |\sinh(\mu_j-i\zeta/2)|}{\prod_{j=1}^M |\sinh(\lambda_j-i\zeta/2)|}\\\nonumber
\times \prod_{j>k}^{M+1} |\sinh^{-1}(\mu_j-\mu_k-i\zeta)|\prod_{j>k}^{M}|\sinh^{-1}(\lambda_j-\lambda_k-i\zeta)|\\
\times\frac{|\det_{M+1} \left(H_{ab}^-\left(\{\mu\},\{\lambda\}\right)\right)|}{\det_{M+1} ^{1/2}\Phi\left(\{\mu\}\right)\det_{M} ^{1/2}\Phi\left(\{\lambda\}\right)}.
\eea
In the above expression $\{\lambda\}$ and $\{\mu\}$ fulfil the Bethe equations (\ref{eq:BE_XXZ}).

The norm of the eigenstates is given by the determinant of the Gaudin matrix \cite{1981_Gaudin_PRD_23, Gaudin, Korepin_proof}, 
\begin{eqnarray}
\fl {\bf \Phi}_{ab} (\{\lambda\}) = \delta_{ab} \left[ N\frac{\sin\zeta}{\sinh^2 \lambda_a + \sin^2 \zeta/2} - \sum_{k \neq a} 
\frac{\sin 2\zeta}{\sinh^2 (\lambda_a - \lambda_k) + \sin^2 \zeta} \right] \nonumber \\
+ (1 - \delta_{ab}) \frac{\sin 2\zeta}{\sinh^2 (\lambda_a - \lambda_b) + \sin^2 \zeta}.
\end{eqnarray}

The matrices $H_{ab}^z\left(\{\mu\}_M,\{\lambda\}_M\right)$, $H_{ab}^-\left(\{\mu\}_{M+1},\{\lambda\}_M\right)$, $P_{ab}(\{\mu\}_M,\{\lambda\}_M)$ are given by the following equations
\bea \fl\nonumber
H_{ab}^z\left(\{\mu\}_M,\{\lambda\}_M\right) = \frac{ \prod_{j\neq a}^M \sinh(\mu_j-\lambda_b-i\zeta)-\frac{\sinh^N(\lambda_b+i\zeta/2)}{\sinh^N(\lambda_b-i\zeta/2)}\prod_{j\neq a}^M \sinh(\mu_j-\lambda_b+i\zeta)}{\sinh(\mu_a-\lambda_b)}, \\\fl\nonumber
H_{ab}^-\left(\{\mu\}_{M+1},\{\lambda\}_M\right) = H_{ab}^z\left(\{\mu\}_M,\{\lambda\}_M\right),\,\,\, b=1,\dots, M,\\\fl\nonumber
H_{a,M+1}^-\left(\{\mu\}_{M+1},\{\lambda\}_M\right) = \frac{1}{\sinh^2 \mu_a + \sin^2\zeta/2},\\\fl\nonumber
P_{ab}(\{\mu\},\{\lambda\}) = \frac{\prod_{m=1}^M \sinh(\lambda_m-\lambda_b-i\zeta)}{\sinh^2\mu_a+\sin^2 \zeta/2}.
\eea
This form of the matrices is not really appropriate for taking the thermodynamic limit. As it is shown in \ref{sec:App_xxzDet} their determinants can be recast into the following form given by \ceq{eq:detHz_final},  
\bea\fl\nonumber
\det_M \left(H_{ab}^z -2P_{ab}\right)= (-1)^M \sinh^{-M}(i\zeta) \;\prod_{i,j}^M \sinh(\mu_i-\lambda_j-i\zeta)\det_M(\delta_{ab}+G_{ab}^z)\\\nonumber
\times\prod_{i=1}^M \left(1+\frac{\sinh^N(\lambda_i+i\zeta/2)}{\sinh^N(\lambda_i-i\zeta/2)}\frac{\prod_{m=1}^M \sinh(\mu_m-\lambda_i + i\zeta)}{\prod_{m=1}^M \sinh(\mu_m-\lambda_i- i\zeta)}\right)\\
\times\prod_{m=1}^M\frac{\sinh(\lambda_m+i\zeta/2)}{\sinh(\mu_m+i\zeta/2)}\frac{\prod_{j>i}^M\sinh(\mu_j-\mu_i)\sinh(\lambda_j-\lambda_i)}{\prod_{i,j}^M \sinh(\mu_i - \lambda_j)},
\eea
where the $M\times M$ matrix $G_{ab}^z$ is given by \ceq{eq:Gz_final}
\bea\nonumber\fl
G_{ab}^z = -\frac{\sinh(i\zeta)}{\sinh(\lambda_b+i\zeta/2)} \frac{\prod_{m=1}^M\sinh(\lambda_b-\mu_m)}{\prod_{m=1,\neq b}^M\sinh(\lambda_b-\lambda_m)} \prod_{m=1}^M \frac{\sinh(\lambda_m-\lambda_b-i\zeta)}{\sinh(\mu_m-\lambda_b- i\zeta)} \\
\times\left(1+\frac{\sinh^N(\lambda_b+i\zeta/2)}{\sinh^N(\lambda_b-i\zeta/2)}\prod_{m=1}^M\frac{ \sinh(\mu_m-\lambda_b + i\zeta)}{\sinh(\mu_m-\lambda_b- i\zeta)}\right)^{-1}\nonumber\\\nn
\times\left(\frac{\sinh(\lambda_b+3/2i\zeta)}{\sinh(\lambda_a-\lambda_b-i\zeta)}-\frac{\sinh(\lambda_b-i\zeta/2)}{\sinh(\lambda_a-\lambda_b+i\zeta)}-\right. \\
\left. -2\prod_{m=1}^M\frac{\sinh(\lambda_m-i\zeta/2)}{\sinh(\mu_m-i\zeta/2)}\frac{\sinh(i\zeta)}{\sinh(\lambda_a-i\zeta/2)}\right).
\eea
Similar calculations are also possible for the second determinant and lead to \ceq{eq:detH-_final}, 
\bea\fl\nonumber
\det_{M+1} H_{ab}^-(\{\mu\},\{\lambda\}) = \frac{1}{\sinh^{M}(i\zeta)}\prod_{i=1}^{M+1} \prod_{j=1}^{M}\sinh(\mu_i-\lambda_j-i\zeta)\det_{M}\left(\delta_{ab}+G_{ab}^-\right)\\\nn
 \times\prod_{i=1}^{M}\left(1+\frac{\sinh^N(\lambda_i+i\zeta/2)}{\sinh^N(\lambda_i-i\zeta/2)}\prod_{m=1}^{M+1}\frac{ \sinh(\mu_m-\lambda_i + i\zeta)}{\sinh(\mu_m-\lambda_i- i\zeta)} \right)\\\nn
\times \frac{\prod_{i=1}^{M}|\sinh(\lambda_i+i\zeta/2)|^2}{\prod_{i=1}^{M+1} |\sinh(\mu_i+i\zeta/2)|^2}\frac{\prod_{j>i}^{M+1} \sinh(\mu_j-\mu_i)\prod_{j>i}^{M}\sinh(\lambda_j-\lambda_i)}{\prod_{i=1}^{M+1}\prod_{j=1}^{M}\sinh(\mu_i-\lambda_j)},
\eea
where the $M\times M$ matrix $G_{ab}^-$ is given by \ceq{eq:G-_final}
\bea\label{eq:G-_def}\nonumber\fl
G_{ab}^- = \frac{1}{|\sinh(\lambda_b+i\zeta/2)|^2}\frac{\prod_{m=1}^{M+1}\sinh(\lambda_b-\mu_m)}{\prod_{m=1,\neq b}^{M}\sinh(\lambda_b-\lambda_m)}\frac{\prod_{m=1}^{M}\sinh(\lambda_m-\lambda_b-i\zeta)}{\prod_{m=1}^{M+1} \sinh(\mu_m-\lambda_b-i\zeta)}\\\nn
\times\left(1+\frac{\sinh^N(\lambda_b+i\zeta/2)}{\sinh^N(\lambda_b-i\zeta/2)}\prod_{m=1}^{M+1}\frac{ \sinh(\mu_m-\lambda_b + i\zeta)}{ \sinh(\mu_m-\lambda_b- i\zeta)}\right)^{-1}\\
\times\left(\frac{\sinh(\lambda_b-i\zeta/2)\sinh(\lambda_b-3i\zeta/2)}{\sinh(\lambda_a-\lambda_b-i\zeta)}- c.c \right).
\eea
In this new form the expressions for the form factors (\ceq{eq:Sz_start} and \ceq{eq:S-_start}) read
\bea\label{eq:Sz_proper}\fl\nonumber
|\langle\{\mu\}_M| S_q^z |\{\lambda\}_M\rangle| = \frac{\sqrt{N}}{2} \delta_{q, P_{\mu}-P_{\lambda}}
\frac{\left|\det_M \left(\delta_{ab}+G_{ab}^z\right)\right|}
{\det_M ^{1/2}\Phi\left(\{\mu\}\right)\det_M ^{1/2}\Phi\left(\{\lambda\}\right)}\\\nonumber
\times \prod_{i,j=1}^M \left|\frac{\sinh(\mu_i-\lambda_j-i\zeta)}{\sinh^{1/2}(\mu_i-\mu_j-i\zeta)\sinh^{1/2}(\lambda_i-\lambda_j-i\zeta)}\right|\\\nonumber
\times\left|\frac{\prod_{j>i}^M\sinh(\mu_j-\mu_i)\sinh(\lambda_j-\lambda_i)}{\prod_{i,j}^M \sinh(\mu_i - \lambda_j)}\right|\\
\times\prod_{i=1}^M \left|1+\frac{\sinh^N(\lambda_i+i\zeta/2)}{\sinh^N(\lambda_i-i\zeta/2)}\frac{\prod_{m=1}^M \sinh(\mu_m-\lambda_i + i\zeta)}{\prod_{m=1}^M \sinh(\mu_m-\lambda_i- i\zeta)}\right|,\\
\label{eq:S-_proper}\fl\nonumber
|\langle\{\mu\}_{M+1}| S^- |\{\lambda\}_M\rangle| = \sqrt{N}|\sinh(i\zeta)|\delta_{q, P_{\mu}-P_{\lambda}}\frac{\left|\det_M \left(\delta_{ab}+G_{ab}^-\right)\right|}
{\det_{M+1} ^{1/2}\Phi\left(\{\mu\}\right)\det_{M} ^{1/2}\Phi\left(\{\lambda\}\right)}\\\nonumber
\times\left|\frac{\prod_{i=1}^{M+1}\prod_{j=1}^{M} \sinh(\mu_i-\lambda_j-i\zeta)}{\prod_{i,j}^{M+1}\sinh^{1/2}(\mu_i-\mu_j-i\zeta)\prod_{i,j}^{M}\sinh^{1/2}(\lambda_i-\lambda_j-i\zeta)}\right|\\\nonumber
\times\frac{\prod_{i=1}^{M}|\sinh(\lambda_i+i\zeta/2)|}{\prod_{i=1}^{M+1} |\sinh(\mu_i+i\zeta/2)|}\left|\frac{\prod_{j>i}^{M+1} \sinh(\mu_j-\mu_i)\prod_{j>i}^{M}\sinh(\lambda_j-\lambda_i)}{\prod_{i=1}^{M+1}\prod_{j=1}^{M-1}\sinh(\mu_i-\lambda_j)}\right|\\
\times\prod_{i=1}^{M}\left|1+\frac{\sinh^N(\lambda_i+i\zeta/2)}{\sinh^N(\lambda_i-i\zeta/2)}\prod_{m=1}^{M+1} \frac{\sinh(\mu_m-\lambda_i + i\zeta)}{\sinh(\mu_m-\lambda_i - i\zeta)}\right|.
\eea

\subsection{Thermodynamics}
In the continuum limit, similarly to the Lieb-Liniger case, we can define the root 
density function $\rho(\lambda)$ and backflow shift function $F$ by \cite{Korepin}
\bea\fl
\rho(\lambda) = a_1(\lambda) - \int _{-q}^q d\mu \rho(\mu) a_2(\lambda-\mu),\\\fl
F(\lambda|\mu^+, \mu^-) = - \int_{-q}^q d\mu F(\mu|\mu^+, \mu^-) a_2(\lambda-\mu) - \frac{\phi_2(\lambda-\mu^+) - \phi_2(\lambda-\mu^-)}{2\pi}.
\eea
The kernels appearing in these equations are given by (\ref{eq:theta_XXZ}) and 
\bea
a_n(\lambda) = \frac{1}{2\pi} \frac{d}{d\lambda} \phi_n(\lambda) = \frac{1}{\pi} \frac{\sin(n\zeta)}{\cosh(2\lambda)-\cos(n\zeta)}.
\eea
The rapidity shift function is related to the order $1/N$ change in rapidity positions under the addition of excitations by the definition
\bea
\mu_j = \lambda_j - \frac{F(\lambda_j)}{N\rho(\lambda_j)} + \mathcal{O}(1/N^2).
\eea 
Looking carefully at the finite size corrections one obtains in the same way as for the Bose gas (compare with section \ref{sec:TL_FS}) the following equations
\bea\fl
\rho_N(\lambda_i) = a_1(\lambda_i) - \int _{-q}^q d\mu \rho_N(\mu) a_2(\lambda_i-\mu) + \mathcal{O}(1/N^2),\\\fl
\rho_{exc,N}(\mu_i)  = \rho_N(\lambda_i) + \frac{1}{N}\left(F_N'(\lambda_j)-F_N(\lambda_j)\frac{\rho'(\lambda_j)}{\rho(\lambda_j)}\right),\\\fl\nonumber
F_N(\lambda|\mu^+, \mu^-) = - \int_{-q}^q d\mu F_N(\mu|\mu^+, \mu^-) a_2(\lambda-\mu)- \frac{\phi_2(\lambda-\mu_0^+) - \phi_2(\lambda-\mu_0^-)}{2\pi}\\\nonumber\fl
-\frac{1}{N}\left(a_2(\lambda-\mu_0^+)\left(\frac{F_N(\mu_0^+)}{\rho(\mu_0^+)}-\frac{F_N(\lambda)}{\rho(\lambda)}\right)-a_2(\lambda-\mu_0^-)\left(\frac{F_N(\mu_0^-)}{\rho(\mu_0^-)}-\frac{F_N(\lambda)}{\rho(\lambda)}\right)\right)\\\fl
-\frac{1}{2N}\int_{-q}^q d\mu \rho_N(\mu)a_2'(\lambda-\mu)\left(\frac{F_N(\lambda)}{\rho(\lambda)}-\frac{F_N(\mu)}{\rho(\mu)}\right)^2 + \frac{a_1'(\lambda)}{2N}\frac{F_N^2(\lambda)}{\rho_N^2(\lambda)}+  \mathcal{O}(1/N^2).
\eea

\subsection{Thermodynamic limit}
\subsubsection{Gaudin determinant}

The Gaudin matrix for the ground state given by the set of rapidities $\{\lambda\}$ reads
\bea\fl
\Phi_{ab} \left( \{\lambda\}_M \right) = \delta_{ab}\left(2\pi N a_1(\lambda_a) - 2\pi \sum_{k=1}^M a_2(\lambda_a-\lambda_k)\right) + 2\pi a_2(\lambda_a-\lambda_b).
\eea
Its determinant can be written in the thermodynamic limit as
\bea\label{eq:norm_lambda}\fl
\det_M \Phi_{ab} \left( \{\lambda\}_M \right) = \prod_{i=1}^M \left(2\pi N \rho_N (\lambda_i)\right)\det\left(1+\hat{a}_2\right)\times\left(1+\mathcal{O}(1/N)\right).
\eea
By analogy for an excited state given by a set of rapidities $\{\mu\}$ we get
\bea\label{eq:norm_mu}\nonumber
\det_M \Phi_{ab} \left( \{\mu\}_M \right) = \prod_{i=1}^M \left(2\pi N \rho_{exc,N} (\mu_i)\right)\det\left(1+\hat{a}_2\right)\\\fl 
=\prod_{i=1}^n\frac{\rho_{exc}(\mu_i)}{\rho(\lambda_i)}\exp\left(\int_{-q}^q d\lambda \left(F_N'(\lambda)-F_N(\lambda)\frac{\rho'(\lambda)}{\rho(\lambda)}\right)\right)\det_M \Phi_{ab} \left( \{\lambda\}_M \right).
\eea
\subsubsection{Prefactor}
Let us start by reorganising the prefactor of the $S_z$ form factor \ceq{eq:Sz_proper}. We consider first the thermodynamic limit of the following expression that resembles the Bethe equations \ceq{eq:BE_XXZ}
\bea\label{eq:M0_def}
M_0(\lambda_i) = \frac{\sinh^N(\lambda_i+i\zeta/2)}{\sinh^N(\lambda_i-i\zeta/2)}\frac{\prod_{m=1}^M \sinh(\mu_m-\lambda_i + i\zeta)}{\prod_{m=1}^M \sinh(\mu_m-\lambda_i- i\zeta)}.
\eea
By separating the excited rapidities, using the Bethe equations and taking the thermodynamic limit of the smooth part one obtains 
\bea \fl\nonumber
M_0(\lambda_i)= -\exp\left(2\pi i \int_{-q}^q d\mu F_N(\mu)a_2(\mu-\lambda_i)+\frac{2\pi i}{2}\int_{-q}^q d\mu \frac{F^2(\mu)}{N\rho(\mu)}a_2'(\mu-\lambda_i)\right)\\
\times\prod_{i=1}^n \frac{\sinh(\mu_m^+ -\lambda_i +i\zeta)}{\sinh(\mu_m^+ -\lambda_i -i\zeta)}\frac{\sinh(\mu_m^- -\lambda_i -i\zeta)}{\sinh(\mu_m^- -\lambda_i +i\zeta)}+\mathcal{O}(1/N^2).
\eea
This can be recast into a more illuminating form using the following relation
\bea\nonumber
\frac{\sinh(\mu^{\pm} -\lambda_i +i\zeta)}{\sinh(\mu^{\pm} -\lambda_i -i\zeta)} = \exp\left(i\phi_2(\lambda_i-\mu^{\pm})\right)\\ 
= \exp\left(i\phi_2(\lambda_i-\mu_0^{\pm})+i a_2(\lambda_i-\mu_0^{\pm})\frac{F_N(\mu_0^{\pm})}{N\rho_N(\mu_0^{\pm})}\right)+\mathcal{O}(1/N^2),
\eea
and together with the equation for $F(\lambda)$ we obtain
\bea\nonumber\fl
M_0(\lambda_i) = -\exp\left(-2\pi i F(\lambda_i)\right)\\\nonumber\fl
\times\exp\left(\frac{2\pi i}{N}\frac{F_N(\lambda_i)}{\rho_N(\lambda_i)}\sum_{m=1}^n\left(a_2(\lambda_i-\mu_m^+)-a_2(\lambda_i-\mu_m^-)\right)+\frac{2\pi i}{2N}\frac{F_N^2(\lambda_i)}{\rho_N^2(\lambda_i)}a_1'(\lambda_i)\right)\\\fl
\times\exp\left(\frac{2\pi i}{2N\rho_N(\lambda_i)}\int_{-q}^q d\mu a_2'(\lambda_i-\mu)\left(2F_N(\mu) F_N(\lambda_i)-F_N^2(\lambda_i)\frac{\rho_N(\mu)}{\rho_N(\lambda_i)}\right)\right).
\eea
We can make an even further simplification if we consider equations for $\rho'(\lambda)$ and $F'(\lambda)$, which read
\bea\fl
\rho'(\lambda) = a_1'(\lambda) - \int _{-q}^q d\mu \rho(\mu) a_2'(\lambda-\mu),\\\fl
F'(\lambda|\mu^+, \mu^-) = - \int_{-q}^q d\mu F(\mu|\mu^+, \mu^-) a_2'(\lambda-\mu) - a_2(\lambda-\mu^+) + a_2(\lambda-\mu^-),
\eea
and finally
\bea\label{eq:M0_final}\fl
M_0(\lambda_i) = -\exp\left[-2\pi i F(\lambda_i)\left(1+
 \frac{1}{N\rho(\lambda_i)}\left(F_N'(\lambda_i)-\frac{1}{2}F_N(\lambda_i)\frac{\rho_N'(\lambda_i)}{\rho_N(\lambda_i)}\right)\right)\right].
\eea
When applied in the prefactor one obtains
\bea \fl
\prod_{i=1}^M \left(1+\frac{\sinh^N(\lambda_i+i\zeta/2)}{\sinh^N(\lambda_i-i\zeta/2)}\frac{\prod_{m=1}^M \sinh(\mu_m-\lambda_i + i\zeta)}{\prod_{m=1}^M \sinh(\mu_m-\lambda_i- i\zeta)}\right) = \prod_{i=1}^M \left(1+M_0(\lambda_i)\right)\\\nonumber\fl
= \prod_{i=1}^M \left(2e^{i\phi_i}\sin(\pi F(\lambda_i)\right)\exp\left(\pi\int_{-q}^q d\lambda F(\lambda) \cot\left(\pi F(\lambda)\right)\left(F'(\lambda)-\frac{1}{2}F(\lambda)\frac{\rho'(\lambda)}{\rho(\lambda)}\right)\right).
\eea
In $\phi$ we gathered all purely phase factors. As we are interested only in the norm squared of the form factor, such a phase does not contribute to the final answer, and we discard it. Llet us define the following notation
\bea\label{eq:M1_def}
M_1 = \prod_{i,j=1}^M \frac{\sinh(\mu_i-\lambda_j-i\zeta)}{\sinh^{1/2}(\mu_i-\mu_j-i\zeta)\sinh^{1/2}(\lambda_i-\lambda_j-i\zeta)},\\\label{eq:M2_def}
M_2 = \prod_{j\neq i,=1}^M\frac{\sinh^{1/2}(\mu_j-\mu_i)\sinh^{1/2}(\lambda_j-\lambda_i)}{ \sinh(\mu_i - \lambda_j)},
\eea
and take the thermodynamic limit of the diagonal part of the denominator that is missing in $M_2$ as compared with the expression for the form factor, which reads
\bea\nonumber
\prod_{i=1}^M \sinh^{-1}(\mu_i-\lambda_i) = \prod_{i=1}^n \frac{\sinh(\mu_i^--\lambda_i)}{\sinh(\mu_i^+-\lambda_i)} \prod_{i=1}^M \frac{N\rho_N(\lambda_i)}{F_N(\lambda_i)}\\
= \prod_{i=1}^n \frac{F_N(\lambda_i)}{N\rho_N(\lambda_i)\sinh(\mu_i^+-\lambda_i)}\prod_{i=1}^M \frac{N\rho_N(\lambda_i)}{F_N(\lambda_i)}.
\eea

\subsection{Partial results}
\subsubsection{$S_q^z$ form factor}
Let us now summarise the results of the last section. Taking the thermodynamic limit of some of the expressions we obtain using \ceq{eq:Sz_proper}, \ceq{eq:norm_lambda}, \ceq{eq:norm_mu} and the above discussion about the prefactor
\bea\label{eq:Sz_partial}\nonumber\fl
|\langle\{\mu\}_M| S^z |\{\lambda\}_M\rangle| = \frac{\sqrt{N}}{2} \delta_{q,P_{\mu}-P_{\lambda}}\prod_{i=1}^n\left( \frac{F(\lambda_i^-)\left(\rho(\lambda_i^-)\rho_{exc}(\mu_i^+)\right)^{-1/2}}{N\sinh(\mu_i^+-\lambda_i)}\right)\\\nonumber
\times \exp\left( \pi \int_{-q}^q d\mu F(\mu)\cot(\pi F(\mu)\left(F'(\mu)-\frac{1}{2}F(\mu)\frac{\rho'(\mu)}{\rho(\mu)}\right)\right)\\\nonumber
\times \exp\left( -\frac{1}{2}\int_{-q}^q d\mu \left(F'(\mu)-F(\mu)\frac{\rho'(\lambda)}{\rho(\lambda)}\right)\right)\\
\times \frac{\left|\det_M\left(\delta_{ab}+G_{ab}^z\right)\right|}{\det(1+\hat{a}_2)}\times \left|M_1\right| \times \left(M_2 \prod_{i=1}^M \frac{\sin(\pi F(\lambda_i))}{\pi F(\lambda_i)}\right).
\eea
Calculation of the thermodynamic limit of the last line of this expression is contained in \ref{sec:App_xxzTL}. 
 
\subsubsection{$S_q^-$ form factor}
In the case of $S_q^-$ \ceq{eq:S-_proper} similar manipulations are possible and their result are shown below. Before that however we need to take a look at the shift function and labelling of rapidities. We start with the shift function. In the case of creation/annihilation of a rapidity it takes the following form (by analogy to the Bose gas, see discussion above and below \ceq{fplus}, \ceq{fminus} and \cite{PRL_08})
\bea\label{eq:Fplus}
F_+\left(\lambda\right) = F\left(\lambda; \{\mu_i^+\}_{n+1}, \{\mu_i^-\}_n\right) -Z(\lambda)/2,\\
F_-\left(\lambda\right) = F\left(\lambda; \{\mu_i^+\}_{n-1}, \{\mu_i^-\}_n\right) + Z(\lambda)/2,\label{eq:Fminus}
\eea
where $Z(\lambda)$ is a dressed charge given by the solution to the following integral equation \cite{Korepin}
\bea\label{eq:defZ}
Z(\lambda) + \int_{-q}^q d\mu a_2(\lambda-\mu) Z(\mu) = 1.
\eea
Note that the convention here is a bit misleading since acting with the operator $S_q^-$ creates an extra excitation as compared with the reference state and therefore the shift of the ground state rapidities is given by $F_{+}$. The opposite holds for the $S_q^+$. 

There is therefore always, by definition, one excited rapidity, which we will call $\mu_{n+1}^+$, where $n$ is number of additional excitations. In order to keep the notation simpler we relabel the set $\{\mu_i\}_{M+1}$ as follows. We single out the excited rapidity $\mu_{n+1}^+$ and relabel the rest as $\{\tilde{\mu}_i\}_{M}$ 
\bea
\{\mu_i\}_{M+1} = \{\tilde{\mu}_i\}_{M}\cup \mu_{n+1}^+.
\eea
Let us repeat that such a decomposition is always possible for the excited state in the $S_q^-$ form factor.

To show how it works in practice let us consider in details the following expression from the prefactor of the $S_q^-$ form factor
\bea\fl
\frac{\left(\prod_{i\neq j}^{M+1} \sinh(\mu_j-\mu_i)\prod_{i\neq j}^{M}\sinh(\lambda_j-\lambda_i)\right)^{1/2}}{\prod_{i=1}^{M+1} \prod_{j=1}^{M}\sinh(\mu_i-\lambda_j)} =  
\frac{T_{extra} \times\tilde{M}_2}{\prod_{i=1}^{M}\sinh(\tilde{\mu}_i-\lambda_i)},
\eea
where we defined a similar expression to $M_2$ \ceq{eq:M2_def} and $T_{extra}$ 
\bea\label{eq:T_extra}
\tilde{M_2} =\prod_{i\neq j}^{M} \frac{\sinh^{1/2}(\tilde{\mu}_j-\tilde{\mu}_i)\sinh^{1/2}(\lambda_j-\lambda_i)}{\sinh(\tilde{\mu}_i-\lambda_j)}. \\
T_{extra} = \prod_{i=1}^{M} \frac{\sinh{(\mu_{n+1}-\tilde{\mu}_i)}}{\sinh(\mu_{n+1}-\lambda_i)}.
\eea
The product in the denominator can be rewritten as
\bea
\prod_{i=1}^{M} \sinh^{-1}(\tilde{\mu}_i-\lambda_i) = \prod_{i=1}^{M} \frac{N\rho(\lambda_i)}{F_+(\lambda_i)}\prod_{i=1}^n \frac{F_+(\lambda_i^-)}{N\rho(\lambda_i^-)\sinh(\mu_i^+-\lambda_i^-)}
\eea
Actually, terms $M_2$ and $\tilde{M_2}$ are equivalent, the only difference being that we have to use the modified shift function $F_+(\lambda)$ in the latter.
In the same way we obtain the following equality
\bea\fl\nonumber
\left|\frac{\prod_{i=1}^{M+1}\prod_{j=1}^{M} \sinh(\mu_i-\lambda_j-i\zeta)}{\prod_{i,j}^{M+1}\sinh^{1/2}(\mu_i-\mu_j-i\zeta)\prod_{i,j}^{M}\sinh^{1/2}(\lambda_i-\lambda_j-i\zeta)}\right| = \\
=\left| \tilde{M}_1\right| \left|\frac{1}{\sinh(-i\zeta)}\right|^{1/2}\prod_{j=1}^M\left|\frac{\sinh(\mu_{n+1}-\lambda_j-i\zeta)}{\sinh(\mu_{n+1}-\mu_j-i\zeta)}\right|
\eea
Keeping in mind the relabelling of the rapidities and following the same logic as for $S_q^z$ we can obtain the following partial result for $S_q^-$ given in \ceq{eq:S-_proper}
\bea\label{eq:S-_partial}\fl\nonumber
|\langle\{\mu\}_{M+1}| S_q^- |\{\lambda\}_M\rangle| = \frac{\delta_{q,P_{\mu}-P_{\lambda}}}{\sqrt{2\pi N\rho(\mu_{n+1})}}\left|\frac{1}{\sinh^{1/2}(i\zeta)}\right|\prod_{i=1}^n\left|\frac{\sinh(\mu_{n+1}^+-\mu_i^+-i\zeta)}{\sinh(\mu_{n+1}^+-\mu_i^--i\zeta)}\right|\\\nonumber
\times\prod_{i=1}^n\frac{F_+(\lambda_i^-)\left(\rho(\lambda_i^-)\rho(\mu_i^+)\right)^{-1/2}}{N\sinh(\mu_i^+-\lambda_i^-)}\left|\exp\left(\int_{-q}^q\frac{-F_+(\lambda)d\lambda}{\tanh(\mu_{n+1}^+-\lambda-i\zeta)}\right)\right|\\\nonumber
\times\frac{\prod_{i=1}{n}\sinh(\mu_i^-+i\zeta/2)}{\prod_{i=1}^{n+1}\sinh(\mu_i^+ +i\zeta/2)}\exp\left( \int_{-q}^q \frac{F(\lambda)d\lambda}{\tanh(\lambda+i\zeta/2)}\right)\\\nn
\times\exp\left(\int_{-q}^q d\lambda \pi F_+(\lambda)\cot(\pi F_+(\lambda))\left(F_+'(\lambda)-\frac{1}{2}F_+(\lambda)\frac{\rho'(\lambda)}{\rho(\lambda)}\right)\right)\\\nonumber
\times\exp\left(-\frac{1}{2} \int_{-q}^q d\lambda\left(F_+'(\lambda)-F_+(\lambda)\frac{\rho'(\lambda)}{\rho(\lambda)}\right)\right)\times T_{extra}\\
\times\frac{\left|\det_M\left(\delta_{ab}+G_{ab}^-\right)\right|}{\det\left(1+\hat{a}_2\right)}\left|\tilde{M}_1\right|\left(\tilde{M}_2\prod_{i=1}^{M}\frac{\sinh\left(\pi F_+(\lambda_i\right))}{\pi F_+(\lambda_i)}\right).
\eea
Again the last line is the only line where we still need to take the thermodynamic limit. As mentioned before calculations of $M_{1,2}$, $\tilde{M}_{1,2}$  are similar to each other. The difference is hidden in the shift function. In the case of $S_q^z$ operator the excited state rapidities are connected to the ground state by
\bea
\mu_j = \lambda_j - \frac{F_N(\lambda_i)}{N\rho(\lambda_i)},
\eea 
while for $S_q^-$
\bea
\tilde{\mu}_j = \lambda_j - \frac{F_{+,N}(\lambda_i)}{N\rho(\lambda_i)}.
\eea
Therefore we will perform calculations only for $S_q^z$ and simply adapt the results by changing the shift function in the final formula for $S_q^-$.

\subsection{Results}
\subsubsection{$S_q^z$ form factor}
Collecting all the partial results from previous sections and \ref{sec:App_xxzTL} (\ceq{eq:Sz_partial}, \ceq{eq:M1_final}, \ceq{eq:M2_final}) we obtain a final expression for the form factor of the operator $S_q^z$
\bea\label{eq:Sz_final}\fl\nonumber
|\langle\{\mu\}_M| S_q^z |\{\lambda\}_M\rangle| = \frac{\sqrt{N}}{2} \delta_{q,P_{\mu}-P_{\lambda}} \prod_{i=1}^n \left(\frac{F(\lambda_i^-)\left(\rho(\lambda_i^-)\rho_{exc}(\mu_i^+)\right)^{-1/2}}{N\sinh(\mu_i^+-\lambda_i)}\right)\\\nonumber
\times{\prod_{i,j=1}^n}\left| \frac{\sinh(\mu_i^+-\mu_j^- -i\zeta)}{\sinh^{1/2}(\mu_i^+-\mu_j^+-i\zeta)\sinh^{1/2}(\mu_i^--\mu^-_j-i\zeta)}\right|\\\nonumber
\times\left(qN\rho(q)\right)^{-\frac{(F^2(q)+F^2(-q))}{2}}G(1+F(-q))G(1-F(q))\left(2\pi\right)^{\frac{F(q)-F(-q)}{2}}\\\nonumber
\times\exp\left(-\frac{1}{2}\int_{-q}^q d\mu d\lambda \frac{F(\lambda)F(\mu)}{\sinh^2(\lambda-\mu-i\zeta)}-\frac{1}{4}\int_{-q}^q d\lambda\, d\mu\left(\frac{F(\lambda)-F(\mu)}{\sinh(\lambda-\mu)}\right)^2\right) \\\nonumber
\times\prod_{i=1}^n \left|\frac{\exp\left(\int_{-q}^q d\lambda \frac{F(\lambda)}{\tanh(\mu_i^--\lambda-i\zeta)}\right)}
{\exp\left(\int_{-q}^q d\lambda \frac{F(\lambda)}{\tanh(\mu_i^+-\lambda-i\zeta)}\right)}\right|\\\nonumber
\times\exp\left(\frac{1}{2}P_+\int_{-1}^1 dx \frac{qF^2(qx)}{\tanh(q(x-1))}-\frac{1}{2}P_-\int_{-1}^1 dx \frac{qF^2(qx)}{\tanh(q(x+1))}\right)\\
\times \prod_{i=1}^n \left(T_{hole}^{(i)}\times T_{particle}^{(i)}\right)\times T_{cross} \frac{\left|\det\left(1+\hat{G}^z\right)\right|}{\det(1+\hat{a}_2)},
\eea
where we used the following property of the Barnes functions
\bea\nonumber
G(1+F(q))G(1-F(q))G(1+F(-q))G(1-F(-q))\times\\\nonumber\fl
\times\exp\left(\int_{-q}^q d\lambda \pi F(\lambda)F'(\lambda)\cot(\pi F(\lambda))\right)
= G^2 (1+F(-q))G^2(1-F(q))\left(2\pi \right)^{F(q)-F(-q)}.
\eea
\subsubsection*{Umklapp form factor}If the excited state is formed by $m$ consecutive umklapps, then we have a number of simplifications. First of all we do not make any mistake if we set ($i=1,\dots,m$)
\bea
\mu_i^- = -q, \\
\mu_i^+ = q.
\eea
Then going term by term we have
\bea\nonumber\fl
\prod_{i=1}^m\left(\frac{F(\lambda_i^-)\left(\rho(\lambda_i^-)\rho_{exc}(\mu_i^+)\right)^{-1/2}}{N\sinh(\mu_i^+-\lambda_i)}\right) = \left(\frac{F(-q)}{N\rho(q)\sinh(2q)}\right)^m,\\\fl
{\prod_{i,j=1}^m} \frac{\sinh(\mu_i^+-\mu_j^- -i\zeta)}{\sinh^{1/2}(\mu_i^+-\mu_j^+-i\zeta)\sinh^{1/2}(\mu_i^--\mu^-_j-i\zeta)} =\left(\frac{\sinh(2q-i\zeta)}{\sinh(-i\zeta)}\right)^{m^2},\\\fl\nonumber
\prod_{i=1}^m \frac{\exp\left(\int_{-q}^q d\lambda \frac{F(\lambda)}{\tanh(\mu_i^--\lambda-i\zeta)}\right)}
{\exp\left(\int_{-q}^q d\lambda \frac{F(\lambda)}{\tanh(\mu_i^+-\lambda-i\zeta)}\right)}\\\fl
=\exp\left(-m\int_{-q}^q d\lambda \left(\frac{F(\lambda)}{\tanh(q-\lambda-i\zeta)}+\frac{F(-\lambda)}{\tanh(q-\lambda+i\zeta)}\right)\right).
\eea
As calculated before the $T_{hole}$ term takes the following form
\bea\label{eq:Thole_Sz_umklapp}\nonumber
T_{hole} = \left(qN\rho(q)\right)^{ mF(- q)}\left(\frac{\sin\left(\pi F(q)\right)}{\pi F(q)}\right)^m\prod_{i=1}^m\frac{\Gamma\left(i - F(- q) \right)}{\Gamma(i)}\\
\times \exp\left(m P_{-}\int_{-1}^1 dx \frac{qF(qx)}{\tanh(q(x+ 1))}\right),
\eea
and $T_{particle}$ is given by
\bea\label{eq:Tparticle_Sz_umklapp}\fl
T_{particle} = \left(qN\rho(q)\right)^{m F( q)}\prod_{i=1}^m\frac{\Gamma(i - F(q))}{\Gamma(i)}
\times\exp\left(-mP_{+}\int_{-1}^{1} dx \frac{q F(qx)} {\tanh(q(x- 1))}\right).
\eea
The $T_{cross}$ term reads
\bea\label{eq:Tcross_Sz_umklapp}
T_{cross} = \left(N\rho(q)\sinh(2q)\right)^{-m^2+m}G^2(m+1).
\eea
Combining together \ceq{eq:Thole_Sz_umklapp}, \ceq{eq:Tparticle_Sz_umklapp}, \ceq{eq:Tcross_Sz_umklapp} we obtain
\bea\label{eq:Tumklapp_Sz}\fl\nonumber
T_{umklapp} = \left(qN\rho(q)\right)^{m(F(q)+F(-q))-m^2}\left(\frac{\sinh(2q)}{q}\right)^{-m^2}\left(\frac{N\rho(q)\sinh(2q)}{F(q)}\right)^m\\\nonumber
\times\exp\left(P_-\int_{-1}^1 dx \frac{mqF(qx)}{\tanh(x(x+1))}-P_+\int_{-1}^1 dx\frac{mq F(qx)}{\tanh(q(x-1))}\right)\\
\times\frac{G(m+1-F(q))G(m+1-F(-q))}{\Gamma^m(F(q))\Gamma^m(1-F(q))}.
\eea
For the umklapp excitation there is again a relation between the shift function $F(\lambda)$ and the Luttinger liquid parameter K. Function $F(\lambda)$ fulfils an analogous equation as in \ceq{fklap}.
\bea
F(\lambda) = - \int_{-q}^q d\mu \left(F(\mu)-1\right) a_2 (\lambda-\mu).
\eea 
This equation is solved by $F(\lambda) = 1-Z(\lambda)$, where $Z(\lambda)$ is a dressed charge defined in \ceq{eq:defZ}. The value of the fractional charge at the Fermi boundary is connected again with the Luttinger liquid parameter
\bea
Z(q) = \sqrt{K};
\eea
and therefore for the m-th order umklapp we obtain
\bea
F(\lambda) = -m\left(Z(\lambda)-1)\right).
\eea
Then

\bea\label{Dmresult}\fl
D_m = \frac{1}{2} \left(2\pi q\rho(q)\right)^{-2m^2 K}\left(\frac{2q}{\sinh(2q)}\right)^{2m^2}\left(\frac{\sinh^2(2q)+\sin^2\zeta}{4\sin^2\zeta}\right)^{m^2}\nonumber\\ \fl
\times \left(\frac{G^2(1+m\sqrt{K})G(1+m-m\sqrt{K})}{\Gamma^m(m-m\sqrt{K})\Gamma^m(1-m+m\sqrt{K})G(1-m+m\sqrt{K})}\right)^2\frac{\det^2\left(1+\hat{G}^z\right)}{\det^2(1+\hat{a}_2)}\nonumber\\ \fl
\times\exp\left(-\int_{-q}^q d\mu d\lambda \frac{F(\lambda)F(\mu)}{\sinh^2(\lambda-\mu-i\zeta)}-\frac{1}{2}\int_{-q}^q d\lambda\, d\mu\left(\frac{F(\lambda)-F(\mu)}{\sinh(\lambda-\mu)}\right)^2\right)\nonumber \\ \fl
\times\exp\left(P_+\int_{-1}^1 dx \frac{q(F^2(qx)-2mF(qx))}{\tanh(q(x-1))}-P_-\int_{-1}^1 dx \frac{q(F^2(qx)-2mF(qx))}{\tanh(q(x+1))}\right)\nn \\ \fl
\times\exp\left(-\int_{-q}^q d\lambda \frac{2mF(\lambda)\sinh(2(q-\lambda))}{\cosh(2(q-\lambda)-\cos(2\zeta)}\right),
\eea
where we used the following relation
\bea
\frac{1}{\tanh(a+ib)} = \frac{\sinh(2a)-i\sin(2b)}{\cosh(2a)-\cos(2b)},
\eea
and the relation between the form factor of the umklapp state and the prefactor of the correlation function (\ceq{Dmdef}, \ceq{spin_scaling})
\bea
\left|\langle m,M|S^z|M\rangle\right|^2 = \frac{D_m}{2}\left(\frac{2\pi}{N}\right)^{2m^2 K}.
\eea

\begin{figure}[htb!]
\begin{center}
\includegraphics[scale=.5, angle=-90]{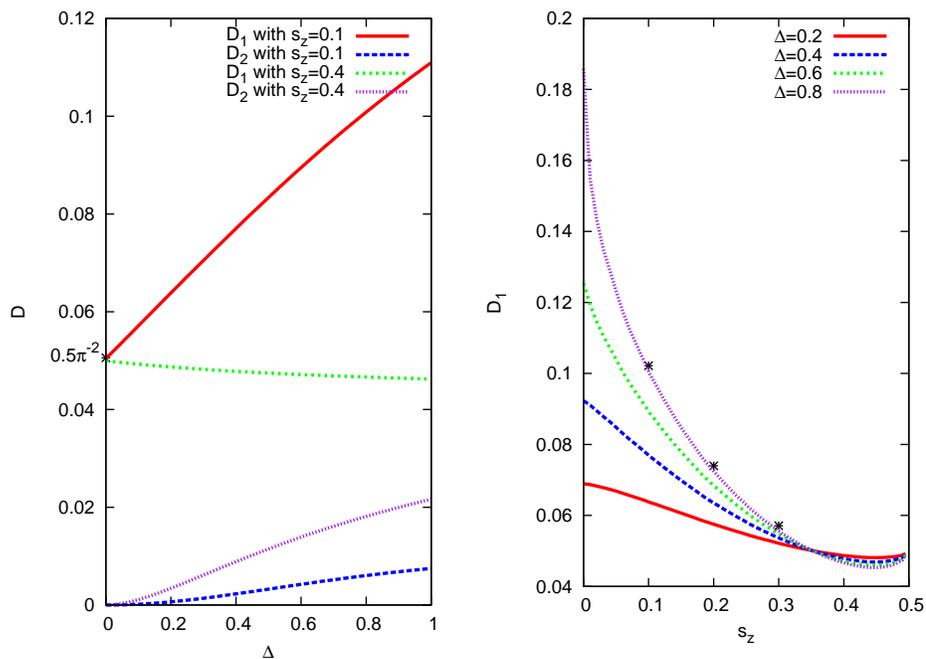}
\caption{
(Left panel) Plots of $D_1$ and $D_2$ (c.f \ceq{Dm_conclusions}) as a function of $\Delta$ for two values of $s_z$ (average magnetisation per site $s_z = 0.5-M/N$). The prefactor $D_1$ approaches the free fermions limit $\frac{1}{2\pi^2}$ as $\Delta\rightarrow 0$. (Right panel) Plots of $D_1$ as function of $s_z$ for a few values of $\Delta$. The results agree with the numerical results of \cite{XXZ_correlation_amplitude}.}
\label{D}
\end{center}
\end{figure}

\subsubsection{$S_q^-$ form factor}
Gathering results from previous sections \ceq{eq:S-_partial}, \ceq{eq:M1_final}, \ceq{eq:M2_final} we obtain
\bea\label{eq:S-_final}\fl\nonumber
|\langle\{\mu\}_{M+1}| S_q^- |\{\lambda\}_M\rangle| = \frac{\left(2\pi\right)^{\frac{F_+(q)-F_+(-q)}{2}}}{\sqrt{2\pi N\rho(\mu_{n+1})}}\delta_{q,P_{\mu}-P_{\lambda}}\left|\frac{1}{\sinh^{1/2}(i\zeta)}\right|\left(qN\rho(q)\right)^{-\frac{(F_+^2(q)+F_+^2(-q))}{2}}\\\nonumber
\times\exp\left(-\frac{1}{2}\int_{-q}^q d\mu d\lambda \left(\frac{F_+(\lambda)F_+(\mu)}{\sinh^2(\lambda-\mu-i\zeta)}+\frac{1}{2}\left(\frac{F_+(\lambda)-F_+(\mu)}{\sinh(\lambda-\mu)}\right)^2\right)\right)  \\\nonumber
\times\exp\left(\frac{1}{2}P_+\int_{-1}^1 dx \frac{qF_+^2(qx)}{\tanh(q(x-1))}-\frac{1}{2}P_-\int_{-1}^1 dx \frac{qF_+^2(qx)}{\tanh(q(x+1))}\right)\\\nonumber
\times\frac{\exp\left(\sum_{i=1}^n\int_{-q}^q d\lambda \frac{F_+(\lambda)}{\tanh(\mu_i^--\lambda-i\zeta)}\right)}{\exp\left(\sum_{i=1}^{n+1}\int_{-q}^q d\lambda \frac{F_+(\lambda)}{\tanh(\mu_i^+-\lambda-i\zeta}\right)}\frac{\det_M\left(\delta_{ab}+G_{ab}^-\right)}{\det\left(1+\hat{a}_2\right)}\\\nonumber
\times\frac{\prod_{i=1}{n}\sinh(\mu_i^-+i\zeta/2)}{\prod_{i=1}^{n+1}\sinh(\mu_i^+ +i\zeta/2)}\exp\left( \int_{-q}^q \frac{F(\lambda)d\lambda}{\tanh(\lambda+i\zeta/2)}\right)\\\nn
\times{\prod_{i,j=1}^n}\left| \frac{\sinh(\mu_i^+-\mu_j^- -i\zeta)}{\sinh^{1/2}(\mu_i^+-\mu_j^+-i\zeta)\sinh^{1/2}(\mu_i^--\mu^-_j-i\zeta)}\right|\\\nonumber
\times\prod_{i=1}^n\frac{F_+(\lambda_i^-)}{N\left(\rho(\lambda_i^-)\rho(\mu_i^+)\right)^{1/2}\sinh(\mu_i^+-\lambda_i^-)}\prod_{i=1}^n T_{hole}^{(i)} T_{particle}^{(i)}\times T_{cross}\times T_{extra}\\
\times\prod_{i=1}^n\left(\frac{\sinh(\mu_{n+1}^+-\mu_i^+-i\zeta)}{\sinh(\mu_{n+1}^+-\mu_i^--i\zeta)}\right)G(1+F_+(-q))G(1-F_+(q)).
\eea
\subsubsection*{Umklapp excitation}
Let us start with calculation of $T_{extra}$ defined in \ceq{eq:T_extra}, it gives
\bea\fl
T_{extra} = \prod_{i=1}^{M} \frac{\sinh{(\mu_{n+1}-\tilde{\mu}_i)}}{\sinh(\mu_{n+1}-\lambda_i)}
= \prod_{i=1}^M\frac{\sinh{(\mu_{n+1}-\mu_i^-)}}{\sinh\left(\mu_{n+1}-\lambda_i\right)} \prod_{i=1}^n \frac{\sinh(\mu_{n+1}-\mu_i^+)}{\sinh(\mu_{n+1}-\mu_i^-)}.
\eea
The first product resembles $T_{particle}$ and indeed for the umklapp excitation it reads
\bea\nonumber
\prod_{i=1}^M\frac{\sinh{(\mu_{n+1}-\mu_i^-)}}{\sinh\left(\mu_{n+1}-\lambda_i\right)} = \left(qN\rho(q)\right)^{ F(q)}\Gamma(1-F_+(q))\\\exp\left(-P_+\int_{-1}^{1} dx \frac{q F_+(qx)}{ \tanh(q(x- 1))}\right),
\eea
where we put the rapidity $\mu_{n+1}^+$ exactly at the right Fermi point.
The second product gives
\bea
\prod_{i=1}^m \frac{\sinh(\mu_{n+1}^+-\mu_i^+)}{\sinh(\mu_{n+1}^+-\mu_i^-)} = \frac{\Gamma(m+1)}{\left(\sinh(2q)N\rho(q)\right)^{m}}.
\eea 
$T_{extra}$ together with $T_{umklapp}$ \ceq{eq:Tumklapp_Sz} gives
\bea\fl\nonumber
\tilde{T}_{umklapp} = \left(qN\rho(q)\right)^{(m+1)F_+(q)+mF_+(-q)-m^2-m}\left(\frac{q}{\sinh(2q)}\right)^{m^2+m}\\\nonumber\fl
\times\left(N\rho(q)\sinh(2q)\right)^{m}\frac{\Gamma(m+1)G(m+1-F_+(-q))G(m+1-F_+(q))}{\Gamma^{m}(F_+(q))\Gamma^{m-1}(1-F_+(q))}\\\fl
\times\exp\left(P_{-}\int_{-1}^1 dx \frac{mq F_+(qx)}{\tanh(q(x+ 1))}-P_{+}\int_{-1}^{1} dx \frac{(m+1)q F_+(qx)}{\tanh(q(x- 1))}\right).
\eea

The exponent of $qN\rho(q)$ is given by
\bea\fl\nonumber
-\frac{1}{2}\left(F^2_+(q)+F^2_+(-q)-2(m+1)F_+(q)-2mF_+(-q)+2m^2+2m+1\right) = \\
= -m^2 K -\frac{1}{4K}.
\eea
where we used that (see \ceq{fminusfull},  \ceq{eq:Fplus})
\bea
F_+(\lambda) = mF(\lambda|\textrm{Umklapp}) + F(\lambda|q) - Z(\lambda)/2.
\eea
and
\bea
F(\pm q|q) - Z(\pm q)/2 = \frac{1}{2} \pm \frac{1}{2} \left( 1-\frac{1}{\sqrt{K}}\right).
\eea
Using the relation between umklapp form-factor and the prefactor of the correlation function (see \ceq{Emdef})
\bea
\left|\langle m,M+1|S^-|M\rangle\right|^2  =\frac{(-1)^m E_m}{2}\left(\frac{2\pi}{N}\right)^{2m^2 K+1/(2K) },
\eea
we obtain
\bea\nonumber\fl\label{Emresult}
E_m =2(-1)^m\left(2\pi\right)^{-\frac{1}{\sqrt{K}}}\frac{1}{\sin(\zeta)}\left(2\pi q\rho(q)\right)^{-2m^2 K-\frac{1}{2K}}\\\nonumber\fl
\times\left(\frac{2q}{\sinh(2q)}\right)^{2m^2-2m}\left(\frac{\sinh^2(2q)+\sin^2\zeta}{4\sin^2\zeta}\right)^{m^2-m}\\\nonumber\fl
\times\left(\Gamma(m+1)\frac{G(m-F_+(-q))G(m+1-F_+(q))G(1+F_+(-q)}{\Gamma^{m}(F_+(q))\Gamma^{m-1}(1-F_+(q))G(1-F_+(-q))}\right)^2\frac{\det^2\left(\delta_{ab}+H^-\right)}{\det^2\left(1+\hat{a}_2\right)}\\\nonumber\fl
\times\exp\left(-\int_{-q}^q d\mu d\lambda \left(\frac{F_+(\lambda)F_+(\mu)}{\sinh^2(\lambda-\mu-i\zeta)}+\frac{1}{2}\left(\frac{F_+(\lambda)-F_+(\mu)}{\sinh(\lambda-\mu)}\right)^2\right)\right)  \\\nonumber\fl
\times\exp\left(P_+\int_{-1}^1 dx \frac{q(F_+^2(qx)-2mF_+(qx))}{\tanh(q(x-1))}-P_-\int_{-1}^1 dx \frac{q(F_+^2(qx)-2(m-1)F_+(qx))}{\tanh(q(x+1))}\right)\\\fl
\times\exp\left(-\int_{-q}^q d\lambda \left(\frac{(m+1)F_+(\lambda)}{\tanh(q-\lambda-i\zeta)}+\frac{mF_+(-\lambda)}{\tanh(q-\lambda+i\zeta)}\right)+\int_{-q}^q \frac{F(\lambda)d\lambda}{\tanh(\lambda+i\zeta/2)}\right).\nn\\
\eea

\begin{figure}[htb!]
\begin{center}
\includegraphics[scale=.5, angle=-90]{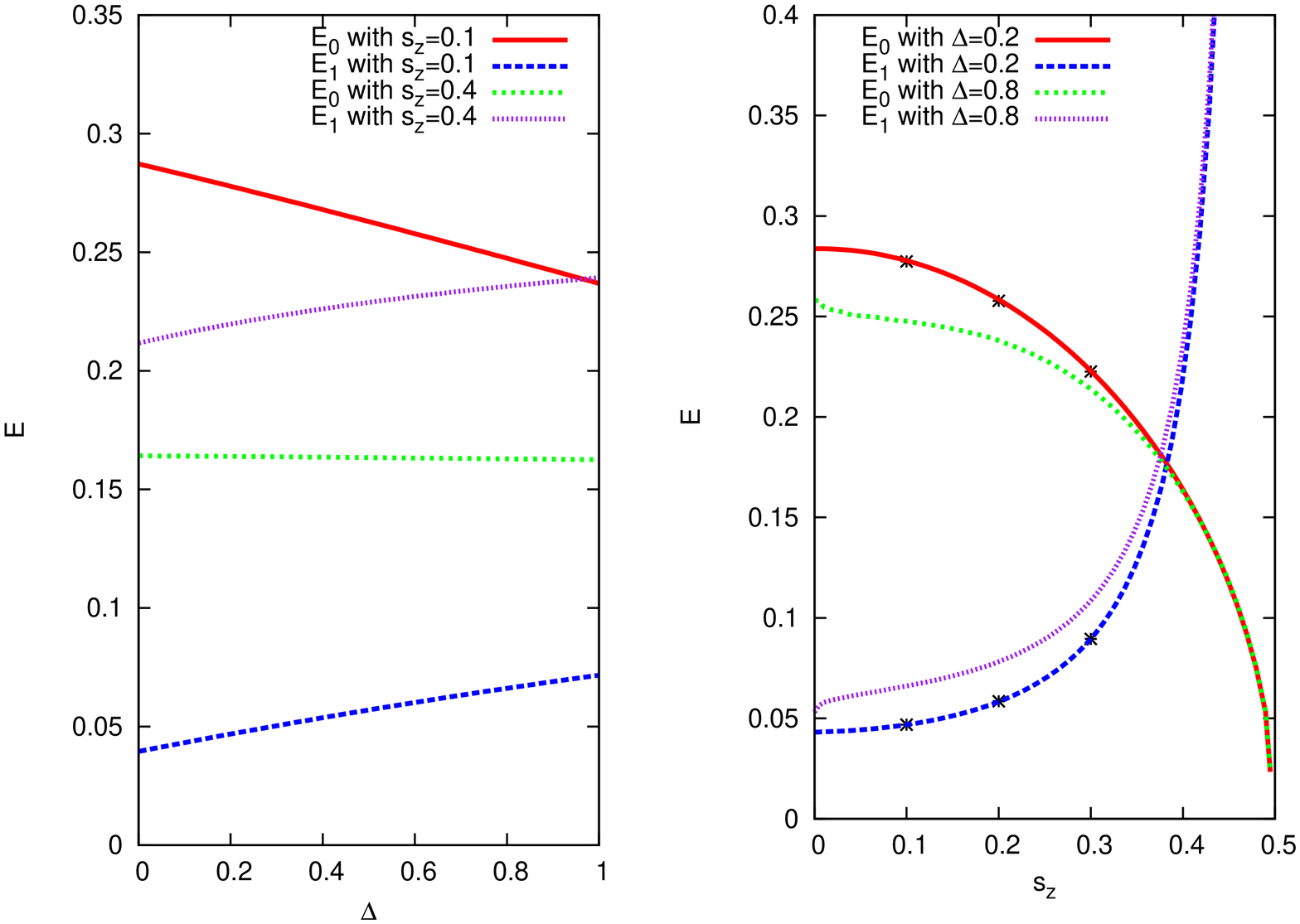}
\caption{
(Left panel) Plot of $E_0$ and $E_1$ (c.f \ceq{Em_conclusions}) as a function of $\Delta$ for different values of $s_z$. (Right panel) Plot of $E_0$ as a function of $s_z$ for different values of $\Delta$. Numerical data (crosses) from Ref.~\cite{XXZ_correlation_amplitude}.
}
\label{E}
\end{center}
\end{figure}

\section{Conclusion}
 We utilized a general method for obtaining non-universal prefactors of correlation functions from field theoretical considerations. The technique relies on the existence of a well defined relationship between lowest energy form factors of operators and the associated prefactors of their correlation functions, see e.g. Eqs.~(\ref{boson_scaling})-(\ref{spin_scaling2}). For example, in the case of equal time correlators, such a relationship can be demonstrated by an effective field theoretical description of the system  as a Luttinger liquid, when finite system size is properly accounted for. In fact such relations were already known in a few cases~\cite{Bogoliubov_J_Phys}. Moreover using the three subband model of a mobile impurity in a Luttinger liquid, we also obtained expressions for prefactors of dynamic response functions involving lowest energy form factors, see Eqs.~(\ref{DSFscaling}), (\ref{spectralscaling1}) and (\ref{spectralscaling2}). 

The universality of the field theoretical description of interacting one-dimensional quantum systems of bosons, fermions and spins, allows us to apply the relationship between the correlation prefactors and form factors to a wide variety of systems. It should be emphasized that the method is quite general, and does not rely on integrability or on any other special features of a particular system under consideration. 

In this article, we focused on three integrable models: the Calogero-Sutherland, Lieb-Liniger and XXZ models. We utilized the connection between prefactors and form factors in the case of the aforementioned systems, where explicit expressions for finite-size form factors were already available. We demonstrated how to properly take the thermodynamic limit of such form factors, which contain non-trivial power laws in system size, to obtain numerically tractable analytic expressions for non-universal prefactors of their correlation functions.

 We note here that some results for prefactors we obtained do not represent new results. On the other hand, it was our aim to demonstrate a more direct way to obtain them, with a wider range of applicability, e.g. we obtained the prefactor of the first oscillatory component of the equal-time density-density correlator for the Calogero-Sutherland model without employing the replica method or multiple integrals \cite{CS_results}, similarly our results for the correlation prefactors of the XXZ model did not require use of a master equation, or detailed asymptotic analysis of determinants~\cite{Lyon_group}. 

We summarize the results for prefactors of correlation functions in the following sub-sections. 

\subsection{Prefactors of equal-time correlators}

The equal time correlators of 1D quantum liquids can be expressed as an asymptotic series with unknown prefactors, see Eqs.~(\ref{Amdef}) - (\ref{Cmdef}), (\ref{Dmdef}) and (\ref{Emdef}). From the analysis of the finite size scaling of the field theoretical description of 1D quantum liquids one obtains a correspondence between the prefactors and form factors of the density, creation, annihilation and spin operators, see Eqs.~(\ref{boson_scaling}) - (\ref{spin_scaling2}). We present below the prefactors of the equal time density and field correlators of the bosonic Lieb-Liniger model followed by prefactors of the spin-spin correlators of the XXZ model. 

For the first prefactor $A_1$ of the oscillatory terms in the density-density correlator (see \ceq{Amdef}) of the Calogero-Sutherland model we obtained
\bea
A_1 =2 \frac{\Gamma\left(1+\frac{1}{\lambda}\right)^2}{(2\pi)^{2/\lambda}},
\eea
in agreement with the results of Ref.~\cite{CS_results} which were obtained using the replica method.

For the prefactors of the oscillatory terms in the density-density correlator (see \ceq{Amdef}) of the Lieb-Liniger model we have from we have from \ceq{density_scaling},(\ref{tgen}) and (\ref{klap})
\bea
\fl
A_{m} =& 2\gamma^2\left(\frac{q\sqrt{K}}{\rho_0}\right)^{-2m^2K}\left(\frac{4q^2+c^2}{4 c^2}\right)^{m^2}\left(\frac{G(1+m\sqrt{K})^2G(m+1-m\sqrt{K})}{\Gamma(m-m\sqrt{K})^m\Gamma(1-m+m\sqrt{K})^mG(1-m+m\sqrt{K})} \right)^2 \nn\\ \fl
& {\rm exp}\left[4mq \int_{-q}^{q}d\lambda \frac{F(\lambda) }{((\lambda + ic)^2 -q^2)}-\int_{-q}^{q}d\mu \int_{-q}^{q}d\lambda\frac{F(\lambda)F(\mu)}{(\lambda - \mu + ic)^2}\right]\nn\\ \fl
& {\rm exp}\left[2 P_{\pm} \int_{-1}^{1}dx\frac{ F^2(qx)-2mF(qx)}{x^2-1} -\frac{1}{2}\int_{-q}^{q} \int_{-q}^{q}d\lambda d\mu\left(\frac{F(\lambda)-F(\mu)}{\lambda - \mu}\right)^2\right]\frac{{\rm Det}^2(1 + \hat{G})}{{\rm Det}^2\left(1-\frac{\hat{K}}{2\pi} \right)}.
\eea
with $F(\lambda) = m(1-2\pi\rho(\lambda))$, where the quasiparticle distribution function $\rho(\lambda)$ is given by \ceq{rhoint}, $q$ is the edge of the quasimomentum distribution and satisfies $2\pi \rho(q) = K$, the Luttinger parameter. $\rho_0$ is the density and $c$ is the strength of the interaction between the bosons, with the associated dimensionless parameter $\gamma = c/\rho_0$. $G(x)$ is the Barnes G function defined in Appendix~A, and the symbol $P_{\pm}$ defined in Appendix~B is an instruction to evaluate the integral in the principal value sense where the singularity occurs at the edge of the range of integration. $\hat{G}$ is obtained from Sec.~4.5, \ceq{Gdef} while $K(\lambda) = \frac{2c}{c^2 + \lambda^2}$. The two determinants are meant to be evaluated as Fredholm determinants, see Appendix~D and Ref.~\cite{Smirnov}.

For the prefactors of the oscillatory terms in the Green's function (see \ceq{Bmdef}) of the Lieb-Liniger model we have from \ceq{boson_scaling},(\ref{tgmin}) and (\ref{klap2})
\bea 
\fl
B_{0} =&q(q\sqrt{K})^{- 1/2K}(2\pi)^{-4+2/\sqrt{K}}\frac{{\rm Det}^2(1 + \hat{H}^-)}{{\rm Det}^2\left(1-\frac{\hat{K}}{2\pi} \right)}G(1+F_-(-q))^2G(1-F_-(q))^2\Gamma(1+F_-(q))^2\nn\\ \fl
& {\rm exp}\left[-\int_{-q}^{q}d\mu \int_{-q}^{q}d\lambda\frac{F_{-}(\lambda)F_{-}(\mu)}{(\lambda - \mu + ic)^2} +2 P_{\pm} \int_{-1}^{1}dx\left(\frac{ F_{-}^2(qx)}{x^2-1} +\frac{F_{-}(qx)}{x-1}\right)\right] \nn\\ \fl
& {\rm exp} \left[ -\frac{1}{2}\int_{-q}^{q} \int_{-q}^{q}d\lambda d\mu\left(\frac{F_{-}(\lambda)-F_{-}(\mu)}{\lambda - \mu}\right)^2\right].\nn\\ 
\fl
B_{m>0} =&q\frac{(q\sqrt{K})^{-2m^2 K - 1/2K}}{2^{2m^2-2m-1}}(2\pi)^{-4+2/\sqrt{K}}\left(\frac{4q^2+c^2}{c^2}\right)^{(m-1)^2}\frac{{\rm Det}^2(1 + \hat{H}^-)}{{\rm Det}^2\left(1-\frac{\hat{K}}{2\pi} \right)}\nn\\ \fl
&\times \left(\frac{G(m-F_{-}(q))G(m+1-F_{-}(-q)))G(1+F_-(-q))}{\Gamma(F_-(-q))^{m-1}\Gamma(1-F_{-}(-q))^{m-1}G(1-F_-(-q))}\right)^2\nn\\ \fl
& {\rm exp}\left[4(m-1)q \int_{-q}^{q}d\lambda \frac{F_{-}(\lambda) }{((\lambda + ic)^2 -q^2)}-\int_{-q}^{q}d\mu \int_{-q}^{q}d\lambda\frac{F_{-}(\lambda)F_{-}(\mu)}{(\lambda - \mu + ic)^2} \right]\nn\\ \fl
& {\rm exp}\left[2 P_{\pm} \int_{-1}^{1}dx\frac{ F_{-}^2(qx)-2(m-1/2)F_{-}(qx) +x F_-(qx)}{x^2-1} -\frac{1}{2}\int_{-q}^{q} \int_{-q}^{q}d\lambda d\mu\left(\frac{F_{-}(\lambda)-F_{-}(\mu)}{\lambda - \mu}\right)^2\right].\nn\\ \fl
\eea
where $F_{-}(\lambda) =  F_m(\lambda| {\rm Umklapp}) - F(\lambda|q) - \pi\rho(\lambda)$ is the modified shift function defined in \ceq{fminus}, $K$ is the Luttinger parameter, $q$ the edge of the distribution of ground state quasimomenta, $\rho_0$ is the density and $c$ is the strength of the interaction between the bosons, with the associated dimensionless parameter $\gamma = c/\rho_0$. $G(x)$ is the Barnes G function defined in Appendix~A, and the symbol $P_{\pm}$ defined in Appendix~B is an instruction to evaluate the integral in the principal value sense where the singularity occurs at the edge of the range of integration.  $\hat{H^{-}}$ is obtained from Sec.~4.5, \ceq{Hdef}, while $K(\lambda) = \frac{2c}{c^2 + \lambda^2}$. The two determinants are meant to be evaluated as Fredholm determinants, see Appendix~D and Ref.~\cite{Smirnov}.

For the prefactors of the correlations functions of the xxz spin chain we obtained analogous expressions. The prefactor of the $S^zS^z$ correlation function (see \ceq{Dmdef}) is given by (cf. \ceq{spin_scaling}, \ceq{Dmresult})  
\bea\nonumber\fl\label{Dm_conclusions}
D_m = \frac{N}{2} \left(2\pi q\rho(q)\right)^{-2m^2 K}\left(\frac{2q}{\sinh(2q)}\right)^{2m^2}\left(\frac{\sinh^2(2q)+\sin^2\zeta}{4\sin^2\zeta}\right)^{m^2}\\\nonumber\fl
\times \left(\frac{G^2(1+m\sqrt{K})G(1+m-m\sqrt{K})}{\Gamma^m(m-m\sqrt{K})\Gamma^m(1-m+m\sqrt{K})G(1-m+m\sqrt{K})}\right)^2\frac{\det^2\left(1+\hat{G}^z\right)}{\det^2(1+\hat{a}_2)}\\\nonumber\fl
\times\exp\left(-\int_{-q}^q d\mu d\lambda \frac{F(\lambda)F(\mu)}{\sinh^2(\lambda-\mu-i\zeta)}-\frac{1}{2}\int_{-q}^q d\lambda\, d\mu\left(\frac{F(\lambda)-F(\mu)}{\sinh(\lambda-\mu)}\right)^2\right) \\\nonumber\fl
\times\exp\left(P_+\int_{-1}^1 dx \frac{q(F^2(qx)-2mF(qx))}{\tanh(q(x-1))}-P_-\int_{-1}^1 dx \frac{q(F^2(qx)-2mF(qx))}{\tanh(q(x+1))}\right)\\\fl
\times\exp\left(-\int_{-q}^q d\lambda \frac{2mF(\lambda)\sinh(2(q-\lambda))}{\cosh(2(q-\lambda)-\cos(2\zeta)}\right),
\eea
where $N$ is the number of spins, $q$ the edge of the distribution of ground state quasimomenta and $\rho(q)$ is the density at the edge of the distribution. The Luttinger parameter $K$ is given by $K=Z^2(q)$ where $Z(\lambda)$ is the dressed charge defined in \ceq{eq:defZ}. The interaction parameter $\zeta$ is connected with the anisotropy $\Delta$ by $\Delta = \cos(\zeta)$. The shift function is given by $F(\lambda) = -m(Z(\lambda)-1)$ and G stands for the Barnes function defined in Appendix~A. The symbol $P_{\pm}$ is defined in Appendix~B. The operator $\hat{G}^z$ is given by \ceq{eq:Gz_def} and $a_2(\lambda) = \frac{1}{\pi}\frac{\sin(2\zeta)}{\cosh(2\lambda)-\cos(2\zeta)}$. The two determinants are meant to be evaluated as Fredholm determinants, see Appendix~D and Ref.~\cite{Smirnov}.

For the prefactor of the $S^+S^-$ correlation function (see \ceq{Emdef}) we have (cf. \ceq{spin_scaling2}, \ceq{Emresult})
\bea\nonumber\fl\label{Em_conclusions}
E_m =2(-1)^m\left(2\pi\right)^{-\frac{1}{\sqrt{K}}}\frac{1}{\sin(\zeta)}\left(2\pi q \rho(q)\right)^{-2m^2 K-\frac{1}{2K}}\\\nonumber\fl
\times\left(\frac{2q}{\sinh(2q)}\right)^{2m^2-2m}\left(\frac{\sinh^2(2q)+\sin^2\zeta}{4\sin^2\zeta}\right)^{m^2-m}\\\nonumber\fl
\times\left(\Gamma(m+1)\frac{G(m-F_+(-q))G(m+1-F_+(q))G(1+F_+(-q)}{\Gamma^{m}(F_+(q))\Gamma^{m-1}(1-F_+(q))G(1-F_+(-q))}\right)^2\frac{\det^2\left(\delta_{ab}+H^-\right)}{\det^2\left(1+\hat{a}_2\right)}\\\nonumber\fl
\times\exp\left(-\int_{-q}^q d\mu d\lambda \left(\frac{F_+(\lambda)F_+(\mu)}{\sinh^2(\lambda-\mu-i\zeta)}+\frac{1}{2}\left(\frac{F_+(\lambda)-F_+(\mu)}{\sinh(\lambda-\mu)}\right)^2\right)\right)  \\\nonumber\fl
\times\exp\left(P_+\int_{-1}^1 dx \frac{q(F_+^2(qx)-2mF_+(qx))}{\tanh(q(x-1))}-P_-\int_{-1}^1 dx \frac{q(F_+^2(qx)-2(m-1)F_+(qx))}{\tanh(q(x+1))}\right)\\\fl
\times\exp\left(-\int_{-q}^q d\lambda \left(\frac{(m+1)F_+(\lambda)}{\tanh(q-\lambda-i\zeta)}+\frac{mF_+(-\lambda)}{\tanh(q-\lambda+i\zeta)}\right) +\int_{-q}^q \frac{F(\lambda)d\lambda}{\tanh(\lambda+i\zeta/2)} \right).\nn\\
\eea
The meaning of all the symbols in the above formula is the same as in the formula for $D_m$, with the exception that the shift function is given by $F_+(\lambda) = m F(\lambda|\textrm{Umklapp}) + F(\lambda|q) - Z(\lambda)/2$. 

The numerical evaluation of \ceq{Dm_conclusions}, \ceq{Em_conclusions} shows that only prefactors with small value of $m$ ($m=1,2$ for $D_m$ and $m=0$ for $E_m$) have significance. These prefactors are plotted in Fig. 6-9 as functions of the anisotropy parameter $\Delta$ and the filling $M/N$.   

We note that similar results for equal-time correlation prefactors were obtained in Ref.~\cite{Lyon_group}, but establishing their equivalence to ours proves non-trivial due to the appearance of multiple integrals.

By applying the techniques described in Sec.~2 above to the prefactors of singularities in dynamic response functions like the spectral function and density structure factor, we obtained a correspondence between form factors of the creation and annihilation operators and the prefactors of the response functions. We present results for the prefactors of the dynamic response functions of the Lieb-Liniger model below.

\subsection{Prefactors of the singularities of the density structure factor} 

The density structure factor $S(k,\omega)$ defined in \ceq{DSF} shows singular behavior for $\omega \approx |\varepsilon(k) = \lambda(k^2-k_F^2)|,$ with prefactor $S_{\rm CSM}(k)$ when $0<k<2k_F$. The prefactor is given by
\bea
\eqalign
\fl
S_{\rm CSM}(k) &=& \frac{2\pi}{\lambda}\Gamma\left(1+\frac{1}{\lambda}\right)\left(\frac{2k_Fk }{2k_F-k}\right)^{1-\frac{1}{\lambda}}.
\eea 

In the case of the Lieb-Liniger model, the density structure factor $S(k,\omega)$ also shows singular behavior for $\omega \approx \varepsilon_{1(2)}(k)$, i.e. near Lieb's modes \cite{PRL_08,LL}, and the behavior there is described by Eqs.~(\ref{S2def}) and (\ref{S1def}). For the prefactors $S_1(k)$ and $S_2(k)$ in those equations we obtain from Eqs.~(\ref{DSFscaling}) and (\ref{tgen}) 
\bea \fl
S_1(k) & =c^2\sqrt{K}\left(q\sqrt{K}\right)^{-(F^2(q)+F^2(-q) - 2F(q)+2)}(2\pi)^{F(q)-F(-q)}\left(\frac{G(1+F(-q))G(1-F(q))}{\Gamma(1-F(q))}\right)^2\nn\\ \fl
&{\rm exp}\left\{P_{\pm} \int_{-1}^{1}dx\frac{2 F^2(qx)}{x^2-1} + P_+ \int_{-1}^{1}dx \frac{2F(qx)}{x - 1} -\frac{1}{2}\int_{-q}^{q}d\lambda \int_{-q}^{q}d\mu\left(\frac{F(\lambda)-F(\mu)}{\lambda - \mu}\right)^2 \right\}\nn\\ \fl
& {\rm exp}\left[ \int_{-q}^{q}d\lambda \frac{2F(\lambda) (\mu^+ - q)}{(\lambda -q+ ic)(\lambda -\mu^+ + ic)}\right]\frac{(q- \mu^+)^2 +c^2}{ c^2}\left(\frac{F^2(q)} {\rho(\mu^+)(\mu^+/q-1)^2}\right)\nn\\ \fl
&{\rm exp}\left[-\int_{-q}^{q}d\mu \int_{-q}^{q}d\lambda\frac{F(\lambda)F(\mu)}{(\lambda - \mu + ic)^2}  -\int_{-q}^{q}d\lambda \frac{2F(\lambda)}{\lambda - \mu^+}\right] \frac{{\rm Det}^2(1 + \hat{G})}{{\rm Det}^2\left(1-\frac{\hat{K}}{2\pi} \right)}.
\eea
Note that the momentum dependence in the above expression is contained in the shift function $F(\lambda)=F(\lambda|\mu^+) - F(\lambda|q)$, defined in \ceq{fint} as well as the terms directly involving $\mu^+$, the quasimomentum corresponding to particle excitation (see \ceq{baex}). We may relate the physical momentum $k$ to the quasimomentum $\mu^+$ using \cite{Korepin, PRL_08}:
\bea
k = \mu^+ - \pi \rho_0 - \int_{-q}^{q}d\lambda \theta(\mu^+ - \lambda) \rho(\lambda),\eea
where, $\theta(\lambda)$ = i log $\left(\frac{ic+\lambda}{ic-\lambda}\right)$, and $q$ is the quasimomentum at the edge of the distribution; $\rho_0$ is the density and $c$ is the strength of the interaction between the bosons, with the associated dimensionless parameter $\gamma = c/\rho_0$. $G(x)$ is the Barnes G function defined in Appendix~A, and the symbol $P_{\pm}$ defined in Appendix~B is an instruction to evaluate the integral in the principal value sense where the singularity occurs at the edge of the range of integration. $\hat{G}$ is obtained from Sec.~4.5, \ceq{Gdef}, while $K(\lambda) = \frac{2c}{c^2 + \lambda^2}$. The two determinants are meant to be evaluated as Fredholm determinants, see Appendix~D and Ref.~\cite{Smirnov}.

Similarly, we get the following expression for $S_2(k)$:
\bea \fl
S_2(k) & = c^2\sqrt{K}\left(q\sqrt{K}\right)^{-(F^2(q)+F^2(-q)+2F(q)+2)}(2\pi)^{F(q)-F(-q)}\left(\frac{G(1+F(-q))G(1-F(q))\Gamma(1-F(q))}{\Gamma(1+F(\mu^-))\Gamma(1-F(\mu^-))}\right)^2\nn\\ \fl
& {\rm exp}\left[ \int_{-q}^{q}d\lambda \frac{2F(\lambda) (q-\mu^-)}{(\lambda -\mu^-+ ic)(\lambda -q + ic)}\right]\frac{(q- \mu^-)^2 +c^2}{ c^2}\left(\frac{F^2(\mu^-)} {\rho(\mu^-)(\mu^-/q -1)^2}\right)\nn\\ \fl
&{\rm exp}\left\{-\int_{-q}^{q}d\mu \int_{-q}^{q}d\lambda \frac{F(\lambda)F(\mu)}{(\lambda - \mu + ic)^2}+ P_{\pm} \int_{-1}^{1}dx\frac{2 F^2(qx)}{x^2-1} - P_+ \int_{-1}^{1}dx\frac{2F(qx)}{x - 1}\right\}\nn\\ \fl
&{\rm exp}\left[ P \int_{-1}^{1}dx \frac{2F(qx)}{x-\mu^-/q} -\frac{1}{2}\int_{-q}^{q}d\lambda \int_{-q}^{q}d\mu\left(\frac{F(\lambda)-F(\mu)}{\lambda - \mu}\right)^2\right]\times \frac{{\rm Det}^2(1 + \hat{G})}{{\rm Det}^2\left(1-\frac{\hat{K}}{2\pi} \right)}.
\eea
Here too the dependence on momentum $k$ is carried by $F(\lambda) = F(\lambda|q) - F(\lambda|\mu^-)$ and $\mu^-$, the quasimomentum of the hole excitation (see Eq.~(\ref{baex})).This time we relate the quasi momentum of the hole, $\mu^-$, to the physical momentum, $k$ using
\bea-k = \mu^-  - \pi \rho_0 - \int_{-q}^{q}d\lambda \theta(\mu^- - \lambda) \rho(\lambda).\eea

\subsection{Prefactors of the singularities of the spectral function} 
The spectral function $A(k,\omega)$ defined in \ceq{Spectral} also displays singular behavior in the vicinity of Lieb's collective modes \cite{LL}. This behavior is described by Eqs.~(\ref{overline_behavior}), (\ref{underline_behavior}) and the text in that section. There are four associated prefactors defined in Eqs.(\ref{overline_behavior}) and (\ref{underline_behavior}) for which we obtain analytic expression below.

\subsubsection{$\overline{A_{\pm}(k)}$ }

 We obtain $\overline{A_+(k)}$ from the creation operator form factor for a Bose gas with a high energy particle, and obtain $\overline{A_-(k)}$ from the annihilation operator form factor  for a Bose gas with one high energy hole. These functions give the prefactors for the spectral function singularities at $\epsilon_1(k)$ and  $-\epsilon_2(k)$ respectively. We must first relate the quasi momentum of the particle to the momentum $k$ of the excited bosonic state as follows \cite{Korepin, PRL_08}:
\bea
k = \mu^+ - \pi \rho_0 - \int_{-q}^{q}d\lambda \theta(\mu^+ - \lambda) \rho(\lambda),\eea
and for the quasi momentum of the hole
\bea
-k = \mu^-  - \pi \rho_0 - \int_{-q}^{q}d\lambda \theta(\mu^-  - \lambda) \rho(\lambda),\eea
where $\theta(\lambda)$ = i log $\left(\frac{ic+\lambda}{ic-\lambda}\right)$, and $q$ is the quasimomentum at the edge of the distribution. 

Particle-hole pairs in Eqs.~(\ref{fplus}) and (\ref{fminus}) are defined with respect to ground states of $N\pm1$ particles, so to obtain a state with a single high energy particle (hole), a particle-hole pair needs to contain a hole (particle) at $q$. Such a procedure results in the following expressions for shift functions $\overline{F_{\pm}(\lambda)}$
\bea
\overline{F_+(\lambda)} = F(\lambda|\mu^+) +\pi \rho(\lambda),\nn\\ 
\overline{F_-(\lambda)} = -F(\lambda|\mu^-) - \pi \rho(\lambda).
\eea

From \ceq{spectralscaling1}, (\ref{tgmin}) and (\ref{tgplus}) we have
\bea
\fl
\overline{A_{\pm}(k)} & = \frac{1}{\rho(\mu^-)}\left(q\sqrt{K}\right)^{-(\overline{F_{\pm}^2(q)}+\overline{F_{\pm}^2(-q)})}(2\pi)^{\overline{F_{\pm}(q)}-\overline{F_{\pm}(-q)}-2}(G(1+\overline{F_{\pm}(-q)})G(1-\overline{F_{\pm}(q)}))^2\nn\\ \fl
&{\rm exp}\left\{P_{\pm} \int_{-1}^{1}dx\frac{2 \overline{F_{\pm}^2(qx)}}{x^2-1} \mp (P) \int_{-1}^{1}dx \frac{2\overline{F_{\pm}(qx)}}{x-\mu^{\pm}/q} -\frac{1}{2}\int_{-q}^{q}d\lambda \int_{-q}^{q}d\mu\left(\frac{\overline{F_{\pm}(\lambda)}-\overline{F_{\pm}(\mu)}}{\lambda - \mu}\right)^2 \right\}\nn\\ \fl
&{\rm exp}\left[-\int_{-q}^{q}d\mu \int_{-q}^{q}d\lambda \frac{\overline{F_{\pm}(\lambda)}\ \overline{F_{\pm}(\mu)}}{(\lambda - \mu + ic)^2} \right]\times \frac{{\rm Det}^2(1 + \hat{H}^-)}{{\rm Det}^2\left(1-\frac{\hat{K}}{2\pi} \right)}\frac{((q- \mu^{\pm})^2 +c^2)}{c^2}\nn\\ \fl
& {\rm exp}\left[\pm 2 \int_{-q}^{q}d\lambda \frac{\overline{F_{\pm}(\lambda)} (\mu^{\pm} - q)}{(\lambda -\mu^{\pm}+ ic)(\lambda -q + ic)}\right].
\eea

\subsubsection{$\underline{A_{\pm}(k)}$}
 Similarly, we obtain the prefactors of the spectral function singularities, $\underline{A_+(k)}$ and $\underline{A_-(k)}$, near the Lieb modes $\epsilon_2(k)$ and $-\epsilon_1(k)$ from the form factors of the creation and annihilation operator respectively. We can obtain the prefactor $\underline{A_+(k)}$ from the creation operator form factor of a system with a high momentum hole and two particles at the right quasi-Fermi point. From the annihilation operator form factor of a system with a high momentum particle and two holes at the right quasi-Fermi point, we obtain the prefactor $\underline{A_-(k)}$. We again relate the quasimomenta of the particle and hole to the momentum $k$ using, Eqs.~(\ref{qpart}) and (\ref{qhole}) and determine $F_{\pm}(\lambda)$ using Eqs.~ (\ref{fplus}) and (\ref{fminus}). 

Particle-hole pairs in Eqs.~(\ref{fplus}) and (\ref{fminus}) are defined with respect to ground states of $N\pm1$ particles, so to obtain a state with a single high energy particle (hole), a particle-hole pair needs to contain a hole (particle) at $q$. Such a procedure results in the following expressions for shift functions  $\underline{F_{\pm}(\lambda)}$:
\bea
\underline{F_+(\lambda)} = 2F(\lambda|q) - F(\lambda|\mu^-) + \pi \rho(\lambda),\nn\\
\underline{F_-(\lambda)} = F(\lambda|\mu^+) - 2F(\lambda|q) - \pi \rho(\lambda).
\eea

 Then using Eqs.~(\ref{spectralscaling2}), (\ref{tgmin}) and (\ref{tgplus}) we get the following expressions:
\bea
\fl
\underline{A_+(k)} & =\frac{1}{\rho(\mu^-)(\mu^-/q -1)^4}\left(q\sqrt{K}\right)^{-(\underline{F_{+}^2(q)}+\underline{F_{+}^2(-q)}-4\underline{F_{+}(q)}+4)}(2\pi)^{\underline{F_{+}(q)}-\underline{F_{+}(-q)}-2} \frac{((q- \mu^-)^2 +c^2)}{c^2} \nn\\ \fl
&\times \left(\frac{G(1+\underline{F_{+}(-q)})G(1-\underline{F_{+}(q)})\Gamma(1-\underline{F_{+}(q)})\Gamma(2-\underline{F_{+}(q)})}{\Gamma(1-\underline{F_{+}(\mu^-)})\Gamma(\underline{F_{+}(\mu^-)})}\right)^2\times \frac{{\rm Det}^2(1 + \hat{H}^+)}{{\rm Det}^2\left(1-\frac{\hat{K}}{2\pi} \right)}\nn\\ \fl
&{\rm exp}\left\{P_{\pm} \int_{-1}^{1}dx\frac{2 \underline{F_{+}^2(qx)}}{x^2-1} - P_+ \int_{-1}^{1}dx\frac{4\underline{F_{+}(qx)}}{x-1}+2 \int_{-q}^{q}d\lambda \frac{\underline{F_{+}(\lambda)} (q-\mu^-)}{(\lambda -\mu^-+ ic)(\lambda -q + ic)}\right\}\nn\\ \fl
&{\rm exp}\left[ P \int_{-1}^{1}d\lambda \frac{2\underline{F_{+}(qx)}}{x-\mu^-/q} -\frac{1}{2}\int_{-q}^{q}d\lambda \int_{-q}^{q}d\mu\left(\frac{\underline{F_{+}(\lambda)}-\underline{F_{+}(\mu)}}{\lambda - \mu}\right)^2 \right]\nn\\ \fl
&{\rm exp}\left[-\int_{-q}^{q}d\mu \int_{-q}^{q}d\lambda \frac{\underline{F_{+}(\lambda)}\ \underline{F_{+}(\mu)}}{(\lambda - \mu + ic)^2} \right].
\eea\bea
\fl
\underline{A_-(k)} & = \frac{1}{\rho(\mu^+)(\mu^+/q-1)^4}\left(q\sqrt{K}\right)^{-(\underline{F_{-}^2(q)}+\underline{F_{-}^2(-q)} + 4\underline{F_{-}(q)}+4)}(2\pi)^{\underline{F_{-}(q)}-\underline{F_{-}(-q)}-2}\nn\\ \fl
&\frac{{\rm Det}^2(1 + \hat{H}^-)}{{\rm Det}^2\left(1-\frac{\hat{K}}{2\pi} \right)}\frac{((q- \mu^+)^2 +c^2)}{c^2}\left(\frac{G(1+\underline{F_{-}(-q)})G(1-\underline{F_{-}(q)})\Gamma(1+\underline{F_{-}(q)})\Gamma(2+\underline{F_-(q)})}{\Gamma(\underline{F_-(q)})\Gamma(1-\underline{F_{-}(q)})}\right)^2\nn\\ \fl
&{\rm exp}\left\{P_{\pm} \int_{-1}^{1}dx \frac{2 \underline{F_{-}^2(qx)}}{x^2-1} + P_+ \int_{-1}^{1}dx\frac{4\underline{F_{-}(qx)}}{x - 1} -\frac{1}{2}\int_{-q}^{q}d\lambda \int_{-q}^{q}d\mu\left(\frac{\underline{F_{-}(\lambda)}-\underline{F_{-}(\mu)}}{\lambda - \mu}\right)^2 \right\}\nn\\ \fl
&{\rm exp}\left[-\int_{-q}^{q}d\mu \int_{-q}^{q}d\lambda \frac{\underline{F_{-}(\lambda)}\ \underline{F_{-}(\mu)}}{(\lambda - \mu + ic)^2}  -\int_{-q}^{q}d\lambda \frac{2\underline{F_{-}(\lambda)}}{\lambda - \mu^+}\right]\nn\\ \fl
&{\rm exp}\left[2 \int_{-q}^{q}d\lambda \frac{\underline{F_{-}(\lambda)} (\mu^+ - q)}{(\lambda -q+ ic)(\lambda -\mu^+ + ic)}\right].
\eea

In the expressions above, $q$ is the quasimomentum at the edge of the distribution; $\rho_0$ is the density and $c$ is the strength of the interaction between the bosons, with the associated dimensionless parameter $\gamma = c/\rho_0$. $G(x)$ is the Barnes G function defined in Appendix~A, and the symbol $P_{\pm}$ defined in Appendix~B is an instruction to evaluate the integral in the principal value sense where the singularity occurs at the edge of the range of integration. $\hat{H}^{\pm}$ is obtained from Sec.~4.5, \ceq{Hdef}, while $K(\lambda) = \frac{2c}{c^2 + \lambda^2}$. The two determinants are meant to be evaluated as Fredholm determinants, see Appendix~D and Ref.~\cite{Smirnov}.

Calculations of the prefactors of singularities in dynamic response functions so far assumed that the field theory of mobile impurities provides an adequate description. The microscopic approach of this article allows one to also explicitly prove the existence of singularities without this assumption. If we consider the total spectral weight in a small interval of energy $\delta \omega$ in the vicinity of $\varepsilon_{1(2)}(k),$ we need to sum over states with low energy particle-hole excitations.  We may use Eq.~(\ref{mrule}), proven in Ref.~\cite{PRB} and the expansion $C(n_{r(l)}, \mu_{R(L)}) \approx n_{r(l)}^{\mu_{R(L)} - 1}/\Gamma(\mu_{R(L)})$ for $n_{r(l)} \gg 1$ to show that he total spectral weight scales with $\delta \omega$ as 
\[ 
\propto \sum_{|E-\varepsilon_{1(2)}|<\delta \omega/2} \frac{|C(n_r,\mu_R)C(n_l,\mu_L)|}{L^{\mu_R+\mu_L+1}}\propto  |\delta\omega|^{1-\mu_R-\mu_L},
\]
which proves the existence of the singularity. 

To summarize, we analytically calculated ``non-universal'' prefactors in long-distance behavior of correlation functions of the exactly solvable Lieb-Liniger model
of 1D Bose gas and the XXZ model of a 1D spin chain with anisotropic coupling. We also calculated prefactors of singularities in dynamic response functions such as the density
structure factor of the Calogero-Sutherland model of fermions with long range interactions that scale as the inverse of the square of separation, as well as for the Lieb-Liniger model and the spectral function for the Lieb-Liniger model. and proved
the existence of singularities within a continuum spectrum. Our results represent a significant step towards an
analytical calculation of the full correlators.

 \ack
The authors would like to thank L. I. Glazman and S. Lukyanov valuable comments and suggestions. This work was supported by the Texas NHARP Grant No. 01889 and the Alfred P. Sloan Foundation. This work is part of the research programme of the Foundation for Fundamental Research on Matter (FOM), which is part of the Netherlands Organisation for Scientific Research (NWO).

\newpage

\appendix
\section{$\Gamma$ products and Barnes G function}
Following are the definition of the Barnes G function and it's relation to the $\Gamma$- function as well as a few special values and asymptotic expansions:
\bea
\prod_{i = 1}^{n} \Gamma(i + a) = \frac{G(n + a + 1)}{G(1+a)}, \;\nn\\ \fl
G(1) = 1,\; \nn\\ \fl
\lim_{n\rightarrow \infty} \log(\Gamma(n)) = (n-1/2)\log(n) - n +\frac{1}{2} log(2\pi) + O\left(\frac{1}{n}\right),\; \nn\\ \fl
\lim_{n \rightarrow \infty} \log(G(n + 1)) = \frac{n^2}{2}\log(n) - \frac{3}{4}n^2 + \frac{n}{2}\log(2\pi) - \frac{1}{12}\log(n) +\zeta'(-1)+ O\left(\frac{1}{n}\right).\;
\eea

\section{Principal Value Integrals: Interior and Edge Singularities}

We define three distinct types of principal value integrals as follows,
\bea 
P \int_{a}^{b} dx \frac{f(x)}{x-c} = \lim_{\delta \rightarrow 0} \left( \int_{a}^{c-\delta} dx \frac{f(x)}{x-c} + \int_{c + \delta}^{b} dx \frac{f(x)}{x-c}\right),\nn\\ 
P_{-} \int_{a}^{b} dx \frac{f(x)}{x-a} = \lim_{\delta \rightarrow 0} \left( \int_{a+\delta}^{b} \frac{f(x)}{x-a} +f(a)\log(\delta)\right), \nn\\ 
P_{+} \int_{a}^{b} dx \frac{f(x)}{x-b} = \lim_{\delta \rightarrow 0} \left(\int_{a}^{b-\delta} \frac{f(x)}{x-b} - f(b)\log(\delta)\right),\nn\\ 
P_{\pm} \int_{a}^{b} dx \frac{f(x)(b-a)}{(x-a)(x-b)} =P_+ \int_{a}^{b} \frac{f(x)}{x-b} - P_- \int_{a}^{b} \frac{f(x)}{x-a}. 
\eea

Note that we consider all quantities being integrated to be dimensionless, and the bounds of integration to be real numbers with no physical units. All such special principal value integrals evaluated in the text are first made dimensionless by mapping the region of integration to the interval $(-1,1)$.
\section{Special Property of Barnes Function}

We will use the following property of the Barnes function,
\bea
G(1-z) = \frac{G(1+z)}{(2\pi)^z}e^{\int_0^z \pi x cot(\pi x) dx},\eea
to simplify a group of terms occurring in the final expression for the various form factors. 

Now consider the term,
\bea\
& &{\rm exp} \left\{\int_{-q}^{q} d\lambda \pi F(\lambda) F'(\lambda) cot(\pi F(\lambda)) \right\}\nn\\ \fl
&=&{\rm exp}\left\{\int_{F(-q)}^{F(q)}{\rm dx \pi x cot(\pi x)} \right\}\nn\\ 
&=&{\rm exp}\left\{\int_{0}^{F(q)}{\rm dx \pi x cot(\pi x)} - \int_{0}^{F(-q)} {\rm dx \pi x cot(\pi x)} \right\}\nn\\ 
&=&\frac{G(1-F(q))G(1+F(-q))(2\pi)^{F(q)-F(-q)}}{G(1+F(q))G(1-F(-q))}.
\eea
Thus we have
\bea
\label{lemfin}\fl
&G(1+F(q))G(1-F(q))G(1+F(-q))G(1-F(-q)){\rm exp}\left[\int_{-q}^{q} d\lambda \pi F(\lambda) F'(\lambda) cot(\pi F(\lambda)) \right ]\nn\\ \fl
&=G^2(1+F(-q))G^2(1-F(q))(2\pi)^{F(q)-F(-q)}.
\eea
\section{The Fredholm Determinant}

Given an integral equation of the form,
\bea\label{fdetstart}
\phi(x) = f(x) +\lambda \int_{a}^{b}K(x,y)\phi(y) dy,\eea
we may construct a Fredholm determinant $D(\lambda)$ from it by replacing the integral above by a finite sum. The procedure to do so is as follows and is outlined in more detail in Ref.~\cite{Smirnov}. First we discretize the interval $[a,b]$ into $n$ equal parts of length, $\delta = (b-a)/n$. We may index points in the interval using $x_j = a + j\delta$, and denote $\phi_j = \phi(x_j), f_j = f(x_j), K_{jk} = K(x_j, x_k)$ where $j,k$ run from 1 to $n$. Using these definitions, the discrete version of \ceq{fdetstart} becomes,
\bea\phi_j = f_j + \lambda  \delta\sum_{k=1}^{n} K_{jk} \phi_k, {\rm \hspace{1 cm}} j = 1,2,...,n.\eea
We note that the original integral equation in \ceq{fdetstart} gives rise to a system of $n$ equations in $n$ unknowns, $\phi_1,...,\phi_n$. We may solve the above system by constructing a determinant as per Cramer's theorem of the form, \cite{Smirnov},~\cite{Smirnov2},
\bea\label{disdet}
 D_n(\lambda) = \left| \begin{array}{cccc}
1 - \lambda K_{11}\delta & - \lambda K_{12}\delta & ...&- \lambda K_{1n}\delta  \\ 
- \lambda K_{21}\delta  & 1- \lambda K_{22}\delta&...& - \lambda K_{2n}\delta \\ 
...&...&...&...\nn\\ 
- \lambda K_{n1}\delta & - \lambda K_{n2}\delta & ...&1- \lambda K_{nn}\delta  \end{array} \right|.\eea
We wish to obtain the ``true'' Fredholm Determinant associated with the integral equation in \ceq{fdetstart} as the continuum limit of \ceq{disdet}. One way to proceed is to first expand the determinant in \ceq{disdet} as in Ref.~\cite{Smirnov2},
\bea
\fl D_n(\lambda) = 1 - \lambda\sum_{p_1=1}^n \delta K_{p_1,p_1} + ... +\frac{(-\lambda)^n}{n!} \sum_{p_1,...,p_n=1}^{n}\delta^n\left| \begin{array}{cccc}
 K_{p_1p_1} &  K_{p_1p_2} & ...&K_{p_1p_n} \\ 
 K_{p_2p_1} &  K_{p_2p_2} & ...&K_{p_2p_n} \\ 
...&...&...&...\\
 K_{p_np_1} &  K_{p_np_2} & ...&K_{p_np_n}  \end{array} \right|,\nn\eea
 and take $n\rightarrow \infty$. This allows us to replace the finite sums by integrals. Thus we may define the Fredholm determinant as the entire function $D(\lambda)$, where
\bea\label{fred}
D(\lambda) = 1 + \sum_{n=1}^{\infty} \frac{(-\lambda)^n}{n!} d_n,\eea
with
\bea
d_n = \int_{a}^{b}dx_1 \int_{a}^{b} dx_2 ... \int_{a}^{b}dx_n \left| \begin{array}{cccc}
 K(x_1,x_1) &  K(x_1,x_2) & ...&K(x_1,x_n) \\
 K(x_2,x_1) &  K(x_2,x_2) & ...&K(x_2,x_n) \\
...&...&...&...\\
 K(x_n,x_1) &  K(x_n,x_2) & ...&K(x_n,x_n)  \end{array} \right|.\nn
 \eea
 The above derivation also illustrates an elementary numerical method for calculating Fredholm determinants associated with various integral operators. 

\section{Representation of the determinants for the xxz spin chain}\label{sec:App_xxzDet}
Expressions for the form factors of operators $\hat{S}^z$ and $\hat{S}^{\pm}$ involve determinants of some matrices. In general the thermodynamic limit of the determinant of an arbitrary matrix is not a well-defined object as it may diverge when the size of matrix goes to infinity. The purpose of this calculations is to find different representations for the matrices and therefore for the determinants as well. In the final representation we should have only determinants of Fredholm-type. The matrices we are interested in are as follows
\bea\fl
H_{ab}^z(\{\mu\}_{i=1}^M,\{\lambda\}_{i=1}^M) = \frac{ \mathcal{Y}_a(\lambda_b, \{\mu\})}{\sinh(\mu_a-\lambda_b)} -2P_{ab}(\{\mu\},\{\lambda\}), \,\, a,b = 1,\dots, M,\\\fl
H_{ab}^-(\{\mu\}_{i=1}^M,\{\lambda\}_{i=1}^{M-1}) = \frac{ \mathcal{Y}_a(\lambda_b, \{\mu\})}{\sinh(\mu_a-\lambda_b)}, \,\,\, a = 1,\dots, M,\,\, b=1,\dots,M-1,\\\fl
H_{aM}^-(\{\mu\}_{i=1}^M,\{\lambda\}_{i=1}^{M-1}) = \frac{1}{\sinh^2(\mu_a)+\sin^2(\zeta/2)} \,\,\, a = 1,\dots, M.
\eea
In the first matrix sets $\{\mu\}$ and $\{\lambda\}$ consist of $M$ elements, in the second matrix set $\{\lambda\}$ is smaller and consists of $M-1$ elements. Matrix $P_{ab}$ is given by
\bea
P_{ab}(\{\mu\},\{\lambda\}) = \frac{\prod_{m=1}^M \sinh(\lambda_m-\lambda_b-i\zeta)}{\sinh^2\mu_a+\sin^2 \zeta/2},
\eea
Function $\mathcal{Y}_a(x, \{\alpha\})$ is given by
\bea\fl
\mathcal{Y}_a(x, \{\alpha\}) = \prod_{j\neq a}^M \sinh(\alpha_j-x-i\zeta)-\frac{\sinh^N(x+i\zeta/2)}{\sinh^N(x-i\zeta/2)}\prod_{j\neq a}^M \sinh(\alpha_j-x+i\zeta).
\eea
The size of the products above is determined by the size of a set $\{\alpha\}$. Whenever set $\{\alpha_i\}_{i=1}^M$ fulfills the Bethe equations then
\bea
\mathcal{Y}_i(\alpha_i, \{\alpha\}) = 0,\,\,\forall i = 1,\dots, M .
\eea
The two matrices $H_{ab}^z$, $H_{ab}^-$ can be related to each other. Let us consider a matrix given by the following elements
\bea
H_{ab}\left(\{\alpha_i\}_{i=1}^M,\{\beta_i\}_{i=1}^M\right) = \frac{ \mathcal{Y}_a(\beta_b, \{\alpha\})}{\sinh(\alpha_a-\beta_b)}, \,\,\, a,b = 1,\dots,M.
\eea
Then in order to obtain $H_{ab}^z$ we simply have to subtract matrix $P_{ab}$ and let $\{\alpha\}_{i=1}^M=\{\mu\}_{i=1}^M$ and $\{\beta\}_{i=1}^M=\{\lambda\}_{i=1}^M$. In order to obtain $H_{ab}^-$ we take $\{\alpha\}_{i=1}^M=\{\mu\}_{i=1}^M$ and $\{\beta\}_{i=1}^{M-1}=\{\lambda\}_{i=1}^{M-1}$, as well as we set $\beta_M = -i\zeta/2$. Then the determinants are related by the following relation
\bea
\det_M H_{ab}^- \left(\{\mu\}_{i=1}^M,\{\lambda\}_{i=1}^{M-1}\right) =\nn\\
= \left(\prod_{i=1}^M \sinh(\mu_j-i\zeta/2)\right)^{-1} \det_M H_{ab}\left(\{\mu\}_{i=1}^M,\{\lambda\}_{i=1}^{M-1}, -i\zeta/2\right).
\eea

All these show that we can consider first the general case of matrix $H_{ab}$ and specialise only later. The main idea is as follows. We want to extract the diverging part and calculate its determinant separately.
\bea\label{eq:def_ApB}
\det_M H_{ab} = \det_M\left(C_{ab}\right)\det_M\left(A_a\delta_{ab}+B_{ab}\right),\\
C_{ab} = \frac{1}{\sinh(\mu_a-\lambda_b)}.
\eea
Then the determinants on the r.h.s can be calculated even in the thermodynamic limit. The first determinant is Cauchy-like, the second one can be represented as a Fredholm determinant. 
 
Let us consider all the quantities as functions of arbitrary real numbers $\{\alpha_i\}^M$ and $\{\beta_i\}_{i=1}^M$. Only later we will connect them with rapidities. In order to find $A_a$ and $B_{ab}$ we simply use the inverse of $C$ to find that
\bea
\sum_{i=1}^M C_{ai}^{-1}H_{ib} = A_a\delta_{ab}+B_{ab}.
\eea
Such a sum can be computed by introducing an auxiliary integral of a complex function. This function and contour of integration should fulfill two criteria. Firstly the integral should be equal to 0, secondly the function should have poles that generate the sum. If this is the case then the sum is simply equal to the sum of all other residues taken with a minus sign. For the integral to be equal to 0 it is enough for the function to be $i\pi$ periodic. As it will become clear later the function which would generate the above sum is actually odd under shift by $i\pi$ and therefore we introduce by hand an odd function $f(z)$ which ensures that the integral is indeed equal to 0. 
Therefore we rewrite \ceq{eq:def_ApB} as 
\bea\label{eq:def_ApB_correct}
\det_M H_{ab} = \prod_{m=1}^M{f^{-1}(\alpha_i)}\det_M\left(C_{ab}\right)\det_M\left(A_a\delta_{ab}+B_{ab}\right).
\eea
and the vector $A_a$ and the matrix $B_{ab}$ are now given by
\bea
\sum_{i=1}^M C_{ai}^{-1}H_{ib}f(\alpha_i) = A_a\delta_{ab}+B_{ab}.
\eea
Function $f(z)$ should fulfill one more assumption, namely it should have no poles in the region of integration.
 
\subsubsection*{Inverse and the determinant of $C_{ab}$}
Motivated by the fact that $C_{ab}$ is similar to a Cauchy matrix, we simply write its inverse as
\bea\fl
C_{ab}^{-1} = \frac{1}{\sinh(\beta_a-\alpha_b)}\frac{\prod_{m=1}^M\sinh(\beta_a-\alpha_m)}{\prod_{m=1,\neq a}^M\sinh(\beta_a-\beta_m)}
\frac{\prod_{m=1}^M\sinh(\alpha_b-\beta_m)}{\prod_{m=1,\neq b}^M\sinh(\alpha_b-\alpha_m)}.
\eea
One can prove that it is indeed the inverse directly from the definition. We have to show that
\bea
\sum_{i=1}^M C_{ai}C_{ib}^{-1} = \delta_{ab},
\eea
which after substituting the expressions for matrix elements gives the following equality
\bea\nn
\sum_{i=1}^M \frac{-1}{\sinh(\beta_i-\alpha_a)\sinh(\beta_i-\alpha_b)}\frac{\prod_{m=1}^M \sinh(\beta_i-\alpha_m)}{\prod_{m=1,\neq i}^M \sinh(\beta_i-\beta_m)}=\\=\delta_{ab} \frac{\prod_{m=1,\neq b}^M \sinh(\alpha_b-\alpha_m)}{\prod_{m=1}^M\sinh(\alpha_b-\beta_m)}.
\eea
In order to perform the summation we consider an auxiliary integral defined as
\bea
\frac{1}{2\pi i} \oint_{\gamma} dz \frac{-1}{\sinh(z-\alpha_a)\sinh(z-\alpha_b)}\frac{\prod_{m=1}^M \sinh(z-\alpha_m)}{\prod_{m=1}^M \sinh(z-\beta_m)}  = 0.
\eea 
The contour of integration $\gamma$ is chosen in such a way that the integration gives 0. On the other hand the integral is equal to the sum of residues. One type of residues, these at $z\rightarrow \beta_i$ gives the sum
\bea\fl
Res_{z\rightarrow \alpha_i} = \frac{-1}{\sinh(\beta_i-\alpha_a)\sinh(\beta_i-\alpha_b)}\frac{\prod_{m=1}^M \sinh(\beta_i-\alpha_m)}{\prod_{m=1,\neq i}^M \sinh(\beta_i-\beta_m)}, \;\;\; i=1,\dots\,M.
\eea
There is one more residue when $a=b$ then
\bea
Res_{z\rightarrow \alpha_a} = -\delta_{ab}\frac{\prod_{m=1,\neq a}^M\sinh(\alpha_a-\alpha_m)}{\prod_{m=1}^M\sinh(\alpha_a-\beta_m)}.
\eea
The sum of residues must be 0 therefore 
\bea
\sum_{i=1}^M Res_{z\rightarrow \alpha_i} = - Res_{z\rightarrow \alpha_a}.
\eea
This shows that $C_{ab}^{-1}$ is indeed the inverse of $C_{ab}$.

We need also the determinant of $C_{ab}$. One can write
\bea\fl\nonumber
\det_M C_{ab}^{-1} = \prod_{j=1}^M \frac{\prod_{m=1}^M\sinh(\beta_j-\alpha_m)\sinh(\alpha_j-\beta_m)}{\prod_{m=1,\neq j}^M\sinh(\beta_j-\beta_m)\sinh(\alpha_j-\alpha_m)}
\det_M \left(\frac{1}{\sinh(\beta_j-\alpha_k)}\right)\\\nonumber
=\left(\frac{\prod_{i,j}^M \sinh(\alpha_i - \beta_j)}{\prod_{j>i}^M\sinh(\alpha_j-\alpha_i)\sinh(\beta_j-\beta_i)}\right)^2\det_M\left(\frac{1}{\sinh(\alpha_j-\beta_k)}\right)\\
=\left(\frac{\prod_{i,j}^M \sinh(\alpha_i - \beta_j)}{\prod_{j>i}^M\sinh(\alpha_j-\alpha_i)\sinh(\beta_j-\beta_i)}\right)^2 \det_M C_{ab}.
\eea
Therefore
\bea
\det_M C_{ab} = \frac{\prod_{j>i}^M\sinh(\alpha_j-\alpha_i)\sinh(\beta_j-\beta_i)}{\prod_{i,j}^M \sinh(\alpha_i - \beta_j)},
\eea
where we used that
\bea
\det_M C_{ab}^{-1} = \left(\det_M C_{ab}\right)^{-1}.
\eea

\subsubsection*{Calculation of $A_a$ and $B_{ab}$}
Having the inverse of $C_{ab}$ it is rather straightforward to calculate $A_a$ and $B_{ab}$. We follow the same logic as for the inverse of a Cauchy matrix and after similar calculations we obtain that
\bea\fl
A_a(\{\alpha\}, \{\beta\}) = f(\beta_b)\frac{\prod_{m=1}^M \sinh(\alpha_m-\beta_a- i\zeta)}{\sinh(- i\zeta)}\\
-f(\beta_b)\frac{\sinh^N(\beta_a+i\zeta/2)}{\sinh^N(\beta_a-i\zeta/2)}\frac{\prod_{m=1}^M \sinh(\alpha_m-\beta_a + i\zeta)}{\sinh(+ i\zeta)},\nonumber
\eea
and for $B_{ab}$
\bea\fl
B_{ab}(\{\alpha\}, \{\beta\}) = f(\beta_b+i\zeta)\frac{\prod_{m=1}^M\sinh(\beta_b-\alpha_m)}{\prod_{m=1,\neq b}^M\sinh(\beta_b-\beta_m)}\frac{\prod_{m=1,\neq b}^M \sinh(\beta_m-\beta_b- i\zeta)}{\sinh(\beta_a-\beta_b- i \zeta)}\nonumber\\\fl
-f(\beta_b-i\zeta)\frac{\prod_{m=1}^M\sinh(\beta_b-\alpha_m)}{\prod_{m=1,\neq b}^M\sinh(\beta_b-\beta_m)}\frac{\sinh^N(\beta_b+i\zeta/2)}{\sinh^N(\beta_b-i\zeta/2)}\frac{\prod_{m=1,\neq b}^M \sinh(\beta_m-\beta_b+ i\zeta)}{\sinh(\beta_a-\beta_b+ i \zeta)}.\nonumber\\
\eea
In the formula above, we have changed the indices in the prefactor of matrix $B_{ab}$ such that it is only $b$-dependent. Such an operation does not change the determinant and simplifies calculations.

We have solved the problem of representing the generalised matrix in terms of Cauchy-like and Fredholm-like matrices. Now we specialise to the two cases of our interest, we start with $H_{ab}^z$.

\subsubsection*{Results for $H_{ab}^z$}
In order to calculate $H_{ab}^z$ we have to include the extra matrix $P_{ab}$, this is done easily because all the necessary operations are linear so they don't change $A_a$ nor $B_{ab}$ and therefore
\bea\fl
\prod_{m=1}^M \frac{1}{f(\alpha_m)}\det_M \left(H_{ab}f(\alpha_a)- 2P_{ab}f(\alpha_a)\right) = \det_M C_{ab} \det_M \left(A_a\delta_{ab}+B_{ab}+D_{ab}\right),
\eea
where 
\bea
D_{ab} = -2\sum_{i=1} C_{ai}^{-1} P_{ib} f(\alpha_i).
\eea
Again we introduce an auxiliary integral and follow the same spirit as before to obtain
\bea\fl\nonumber
D_{ab}(\{\mu\}, \{\lambda\}) = 2 \frac{\prod_{m=1}^M\sinh(\lambda_a-\mu_m)}{\prod_{m=1,\neq a} \sinh(\lambda_a-\lambda_m)}\prod_{m=1}^M \sinh(\lambda_m-\lambda_b-i\zeta)\\\fl
\times \left(f(-i\zeta/2)\frac{\prod_{m=1,\neq a}\sinh(\lambda_m+i\zeta/2)}{\prod_{m=1}\sinh(\mu_m+i\zeta/2)}-f(i\zeta/2)\frac{\prod_{m=1,\neq a}\sinh(\lambda_m-i\zeta/2)}{\prod_{m=1}\sinh(\mu_m-i\zeta/2)}\right).
\eea
where we have already substituted the sets of rapidities $\{\mu_{i=1}^M\}$ and $\{\lambda_{i=1}^M\}$ describing Bethe states for the arbitrary sets of real numbers $\{\alpha_{i=1}^M\}$ and $\{\beta_{i=1}^M\}$.  
If we do the same with expressions for $A_a$ and $B_{ab}$ we obtain
\bea\nonumber
A_a(\{\mu\}, \{\lambda\}) = f(\lambda_b)\frac{\prod_{m=1}^M \sinh(\mu_m-\lambda_a- i\zeta)}{\sinh(- i\zeta)}\\
\times\left(1+\frac{\sinh^N(\lambda_a+i\zeta/2)}{\sinh^N(\lambda_a-i\zeta/2)}\frac{\prod_{m=1}^M \sinh(\mu_m-\lambda_a + i\zeta)}{\prod_{m=1}^M \sinh(\mu_m-\lambda_a- i\zeta)}\right),\\
B_{ab}(\{\mu\}, \{\lambda\}) = \frac{\prod_{m=1}^M\sinh(\lambda_b-\mu_m)}{\prod_{m=1,\neq b}^M\sinh(\lambda_b-\lambda_m)}\prod_{m=1,\neq b}^M \sinh(\lambda_m-\lambda_b- i\zeta)\nonumber\\
\times\left(\frac{f(\lambda_b+i\zeta)}{\sinh(\lambda_a-\lambda_b- i \zeta)} -\frac{f(\lambda_b-i\zeta)}{\sinh(\lambda_a-\lambda_b+ i \zeta)}\right).
\eea
As the last step we transform the determinant
\bea\nonumber\fl
\prod_{m=1}^M\frac{1}{f(\mu_i)}\det_M \left(H_{ab}f(\mu_a)- 2P_{ab}f(\mu_a)\right) = \prod_{m=1}^M\frac{1}{f(\mu_i)}\det_M C_{ab} \det_M \left(A_a\delta_{ab}+B_{ab}+D_{ab}\right)\\ = \prod_{i=1}^M \frac{A_i}{f(\mu_i)}  \det_M C_{ab} \det_M \left(\delta_{ab}+G_{ab}^z\right),
\eea
and upon choosing $f(z) = \sinh(z+i\zeta/2)$ we obtain
\bea\label{eq:Gz_final}\nonumber\fl
G_{ab}^z = -\frac{\sinh(i\zeta)}{\sinh(\lambda_b+i\zeta/2)} \frac{\prod_{m=1}^M\sinh(\lambda_b-\mu_m)}{\prod_{m=1,\neq b}^M\sinh(\lambda_b-\lambda_m)} \prod_{m=1}^M \frac{\sinh(\lambda_m-\lambda_b-i\zeta)}{\sinh(\mu_m-\lambda_b- i\zeta)} \\
\times\left(1+\frac{\sinh^N(\lambda_b+i\zeta/2)}{\sinh^N(\lambda_b-i\zeta/2)}\prod_{m=1}^M\frac{ \sinh(\mu_m-\lambda_b + i\zeta)}{\sinh(\mu_m-\lambda_b- i\zeta)}\right)^{-1}\nonumber\\
\times\left(\frac{\sinh(\lambda_b+3/2i\zeta)}{\sinh(\lambda_a-\lambda_b-i\zeta)}-\frac{\sinh(\lambda_b-i\zeta/2)}{\sinh(\lambda_a-\lambda_b+i\zeta)}-\right. \nn\\
\left. -2\prod_{m=1}^M\frac{\sinh(\lambda_m-i\zeta/2)}{\sinh(\mu_m-i\zeta/2)}\frac{\sinh(i\zeta)}{\sinh(\lambda_a-i\zeta/2)}\right).
\eea
Finally we can write
\bea\label{eq:detHz_final}\fl\nonumber
\det_M H_{ab}^z(\{\mu\},\{\lambda\}) = (-1)^M \sinh^{-M}(i\zeta) \;\prod_{i,j}^M \sinh(\mu_i-\lambda_j-i\zeta)\det_M(\delta_{ab}+G_{ab}^z)\\\nonumber
\times\prod_{i=1}^M \left(1+\frac{\sinh^N(\lambda_i+i\zeta/2)}{\sinh^N(\lambda_i-i\zeta/2)}\frac{\prod_{m=1}^M \sinh(\mu_m-\lambda_i + i\zeta)}{\prod_{m=1}^M \sinh(\mu_m-\lambda_i- i\zeta)}\right)\\
\times\prod_{m=1}^M\frac{\sinh(\lambda_m+i\zeta/2)}{\sinh(\mu_m+i\zeta/2)}\frac{\prod_{j>i}^M\sinh(\mu_j-\mu_i)\sinh(\lambda_j-\lambda_i)}{\prod_{i,j}^M \sinh(\mu_i - \lambda_j)}.
\eea

\subsubsection*{Results for $H_{ab}^-$}
Following the prescription described in the beginning  we set $\alpha_i = \mu_i$, and $\beta_i = \lambda_i$ where sets $\{\mu_{i=1}^N\}$ and $\{\lambda_{i=1}^{M-1}\}$ solve Bethe equations and $\lambda_{M} = -i\zeta/2$. Then we obtain ($a,b = 1,\dots,M-1$)
\bea\fl
A_a(\{\mu\}, \{\lambda\}) &= \frac{f(\lambda_a)}{\sinh(- i\zeta)}\prod_{m=1}^M \sinh(\mu_m-\lambda_a- i\zeta)\nn\\
&\times\left(1+\frac{\sinh^N(\lambda_a+i\zeta/2)}{\sinh^N(\lambda_a-i\zeta/2)}\prod_{m=1}^M\frac{ \sinh(\mu_m-\lambda_a + i\zeta)}{\sinh(\mu_m-\lambda_a- i\zeta)}\right),\\\fl
A_M(\{\mu\}, \{\lambda\}) &= \frac{f(-i\zeta/2)}{\sinh(- i\zeta)}\prod_{m=1}^M \sinh(\mu_m - i\zeta/2),\\\fl
B_{ab}(\{\mu\}, \{\lambda\}) &= \frac{1}{\sinh(\lambda_b+i\zeta/2)}\frac{\prod_{m=1}^{M}\sinh(\lambda_b-\mu_m)}{\prod_{m=1,\neq b}^{M}\sinh(\lambda_b-\lambda_m)}\prod_{m=1,\neq b}^{M-1} \sinh(\lambda_m-\lambda_b- i\zeta)\nn\\
&\times\left(\frac{\sinh(\lambda_b-i\zeta/2)f(\lambda_b-i\zeta)}{\sinh(\lambda_a-\lambda_b+ i \zeta)}-\frac{\sinh(\lambda_b+3i\zeta/2)f(\lambda_b+i\zeta)}{\sinh(\lambda_a-\lambda_b- i \zeta)}\right),\\\fl
B_{Mb}(\{\mu\}, \{\lambda\}) &= \left(f(\lambda_b+i\zeta)-f(\lambda_b-i\zeta)\right)\frac{\prod_{m=1}^{M}\sinh(\lambda_b-\mu_m)}{\prod_{m=1,\neq b}^{M}\sinh(\lambda_b-\lambda_m)}\nn\\
&\times\prod_{m=1,\neq b}^{M-1} \sinh(\lambda_m-\lambda_b- i\zeta),\\\fl
B_{aM}(\{\mu\}, \{\lambda\}) &= -\frac{f(i\zeta/2)}{\sinh(\lambda_a-i\zeta/2)}\prod_{m=1}^{M-1}\frac{\sinh(\lambda_m-i\zeta/2)}{\sinh(\lambda_m+i\zeta/2)}\prod_{m=1}^M \sinh(\mu_m+i\zeta/2),\\\fl
B_{MM}(\{\mu\}, \{\lambda\}) &= -\frac{f(i\zeta/2)}{\sinh(-i\zeta)}\prod_{m=1}^{M-1}\frac{\sinh(\lambda_m-i\zeta/2)}{\sinh(\lambda_m+i\zeta/2)}\prod_{m=1}^M \sinh(\mu_m+i\zeta/2).
\eea
By setting $f(z) = \sinh(z-i\zeta/2)$ we obtain
\bea
\det_M(\delta_{ab}A_a + B_{ab}) = \prod_{i=1}^{M}A_i\det_{M-1}\left(\delta_{ab}+G_{ab}^-\right),
\eea
where the $(M-1)\times(M-1)$ matrix $G_{ab}^-$ is given by the following elements
\bea\nn\label{eq:G-_final}\fl
G_{ab}^- = \frac{1}{|\sinh(\lambda_b+i\zeta/2)|^2}\frac{\prod_{m=1}^M\sinh(\lambda_b-\mu_m)}{\prod_{m=1,\neq b}^{M-1}\sinh(\lambda_b-\lambda_m)}\frac{\prod_{m=1}^{M-1}\sinh(\lambda_m-\lambda_b-i\zeta)}{\prod_{m=1}^M \sinh(\mu_m-\lambda_b-i\zeta)}\\
\times\left(1+\frac{\sinh^N(\lambda_b+i\zeta/2)}{\sinh^N(\lambda_b-i\zeta/2)}\prod_{m=1}^M\frac{ \sinh(\mu_m-\lambda_b + i\zeta)}{ \sinh(\mu_m-\lambda_b- i\zeta)}\right)^{-1}\nn\\
\times\left(\frac{\sinh(\lambda_b-i\zeta/2)\sinh(\lambda_b-3i\zeta/2)}{\sinh(\lambda_a-\lambda_b-i\zeta)}- c.c \right).
\eea

The final expression reads
\bea\label{eq:detH-_final}\nonumber\fl
\det_M H_{ab}^- = \frac{1}{\sinh^{M-1}(i\zeta)}\prod_{i=1}^M \prod_{j=1}^{M-1}\sinh(\mu_i-\lambda_j-i\zeta)\det_{M-1}\left(\delta_{ab}+G_{ab}^-\right)\\\fl\nn
 \times\prod_{i=1}^{M-1}\left(1+\frac{\sinh^N(\lambda_i+i\zeta/2)}{\sinh^N(\lambda_i-i\zeta/2)}\prod_{m=1}^M\frac{ \sinh(\mu_m-\lambda_i + i\zeta)}{\sinh(\mu_m-\lambda_i- i\zeta)} \right)\\\fl
\times \frac{\prod_{i=1}^{M-1}|\sinh(\lambda_i+i\zeta/2)|^2}{\prod_{i=1}^M |\sinh(\mu_i+i\zeta/2)|^2}\frac{\prod_{j>i}^M \sinh(\mu_j-\mu_i)\prod_{j>i}^{M-1}\sinh(\lambda_j-\lambda_i)}{\prod_{i=1}^M\prod_{j=1}^{M-1}\sinh(\mu_i-\lambda_j)}.
\eea

\section{Thermodynamics limits for the xxz spin chain}\label{sec:App_xxzTL}
In this Appendix we consider the thermodynamic limit of various expressions that are part of the form factors of the operators $S_q^z$ and $S_q^-$ of the xxz spin chain.
\subsection{Calculation of $M_1$}
Let us repeat one more time the formula for $M_1$
\bea
M_1 = \prod_{i,j=1}^M \frac{\sinh(\mu_i-\lambda_j-i\zeta)}{\sinh^{1/2}(\mu_i-\mu_j-i\zeta)\sinh^{1/2}(\lambda_i-\lambda_j-i\zeta)}.
\eea
The crucial point in the calculation of double products like $M_1$ and $M_2$ is to divide them into 3 different parts as follows
\bea
p_1  = {\prod_{i,j=1}^M}{''} \frac{\sinh(\mu_i-\lambda_j-i\zeta)}{\sinh^{1/2}(\mu_i-\mu_j-i\zeta)\sinh^{1/2}(\lambda_i-\lambda_j-i\zeta)}\\
p_2 = \prod_{i=1}^n{\prod_{m=1}^M}{''}\frac{\sinh(\mu_i^--\mu_m-i\zeta)}{\sinh(\mu_i^+-\mu_m-i\zeta)}\frac{\sinh(\mu_i^+-\lambda_m-i\zeta)}{\sinh(\mu_i^--\lambda_m-i\zeta)}\\
p_3 = {\prod_{i,j=1}^n} \frac{\sinh(\mu_i^+-\mu_j^- -i\zeta)}{\sinh^{1/2}(\mu_i^+-\mu_j^+-i\zeta)\sinh^{1/2}(\mu_i^--\mu^-_j-i\zeta)}.
\eea
In terms of these $M_1$ is simply
\bea
M_1 = p_1 p_2 p_3.
\eea
In $p_1$ we can take the thermodynamic limit of both products, in $p_2$ only of the inner product, and $p_3$ is a discrete part which depends on the details of the excited state. We start with $p_1$, which gives
\bea\fl
p_1 = {\prod_{i,j=1}^M}{''} \frac{\sinh(\mu_i-\lambda_j-i\zeta)}{\sinh^{1/2}(\mu_i-\mu_j-i\zeta)\sinh^{1/2}(\lambda_i-\lambda_j-i\zeta)}\\\nonumber\fl
= \prod_{i,j =1}^M \frac{\sinh\left(\lambda_i-\lambda_j-i\zeta-\frac{F_N(\lambda_i)}{N\rho_N(\lambda_i)}\right)}{\sinh^{1/2}\left(\lambda_i-\lambda_j-i\zeta-\frac{F_N(\lambda_i)}{N\rho_N(\lambda_i)}+\frac{F_N(\lambda_j)}{N\rho_N(\lambda_j)}\right)\sinh^{1/2}(\lambda_i-\lambda_j-i\zeta)}\\\nonumber\fl
=\prod_{i,j=1}^M \frac{\cosh\left(\frac{F_N(\lambda_i)}{N\rho_N(\lambda_i)}\right)}{\cosh^{1/2}\left(\frac{F_N(\lambda_i)}{N\rho_N(\lambda_i)}-\frac{F_N(\lambda_j)}{N\rho_N(\lambda_j)}\right)}
\frac{1-\coth(\lambda_i-\lambda_j-i\zeta)\tanh\left(\frac{F_N(\lambda_i)}{N\rho_N(\lambda_i)}\right)}{\left(1-\coth(\lambda_i-\lambda_j-i\zeta)\tanh\left(\frac{F_N(\lambda_i)}{N\rho(\lambda_i)}-\frac{F_N(\lambda_i)}{N\rho(\lambda_i)}\right)\right)^{1/2}}.
\eea
In the following calculations we will again skip some expression that contribute only to the phase of the form factor. The first product gives
\bea\fl
\prod_{i,j=1}^M \frac{\cosh\left(\frac{F_N(\lambda_i)}{N\rho_N(\lambda_i)}\right)}{\cosh^{1/2}\left(\frac{F_N(\lambda_i)}{N\rho_N(\lambda_i)}-\frac{F_N(\lambda_j)}{N\rho_N(\lambda_j)}\right)} =\exp\left(\frac{1}{2} \int _{-q}^q d\lambda d\mu F(\lambda) F(\mu)\right)+ \mathcal{O}(1/N).
\eea
The second product gives
\bea
\prod_{i,j=1}^M\frac{1-\coth(\lambda_i-\lambda_j-i\zeta)\tanh\left(\frac{F_N(\lambda_i)}{N\rho_N(\lambda_i)}\right)}{\left(1-\coth(\lambda_i-\lambda_j-i\zeta)\tanh\left(\frac{F_N(\lambda_i)}{N\rho(\lambda_i)}-\frac{F_N(\lambda_i)}{N\rho(\lambda_i)}\right)\right)^{1/2}}\\\nonumber 
=\exp\left[-\frac{1}{2}\sum_{ij}\frac{F_N(\lambda_i)}{N\rho_N(\lambda_i)}\left(\coth(\lambda_i-\lambda_j-i\zeta)- c.c \right) \right]\\\nonumber
\times\exp\left[ -\frac{1}{4}\sum_{i,j} \left(\frac{F_N^2(\lambda_i)}{N^2\rho_N^2(\lambda_i)}\left(\coth(\lambda_i-\lambda_j-i\zeta)-c.c.\right)\right)\right]\\\nonumber
\times \exp\left(-\frac{1}{2}\sum_{i,j} \frac{F_N(\lambda_i)F_N(\lambda_j)}{N^2\rho_N(\lambda_i)\rho_N(\lambda_j)}\coth^2(\lambda_i-\lambda_j-i\zeta)\right)+\mathcal{O}(1/N)\\\nonumber
= \exp\left(-\frac{1}{2}\int_{-q}^q d\mu d\lambda F_N(\lambda)F_N(\mu)\left( \coth^2(\lambda-\mu-i\zeta)\right)+i\phi\right)+\mathcal{O}(1/N),
\eea
where in $\phi$ we gathered all the purely phase factors. They do not contribute to the final answer and can be dropped. Note that despite factor $i$ the integral is a real number. Altogether we obtain for $p_1$ the following answer
\bea
\left|p_1\right| = \exp\left(-\frac{1}{2}\int_{-q}^q d\mu d\lambda \frac{F_N(\lambda)F_N(\mu)}{\tanh^2(\lambda-\mu-i\zeta)}\right) + \mathcal{O}(1/N).
\eea
We move on now to $p_2$
\bea
p_2 = \prod_{i=1}^n{\prod_{m=1}^M}''\frac{\sinh(\mu_i^--\mu_m-i\zeta)}{\sinh(\mu_i^+-\mu_m-i\zeta)}\frac{\sinh(\mu_i^+-\lambda_m-i\zeta)}{\sinh(\mu_i^--\lambda_m-i\zeta)}\\\nonumber
= \prod_{i=1}^n \prod_{m=1}^M\frac{\sinh\left(\mu_i^--\lambda_m-i\zeta+\frac{F_N(\lambda_m)}{N\rho_N(\lambda_m)}\right)\sinh(\mu_i^+-\lambda_m-i\zeta)}{\sinh(\mu_i^--\lambda_m-i\zeta)\sinh\left(\mu_i^+-\lambda_m-i\zeta+\frac{F_N(\lambda_m)}{N\rho_N(\lambda_m)}\right)}\\\nonumber
= \prod_{i=1}^n \prod_{m=1}^M \frac{1+\coth(\mu_i^--\lambda_m-i\zeta)\sinh\left(\frac{F_N(\lambda_m)}{N\rho_N(\lambda_m)}\right)}{1+\coth(\mu_i^+-\lambda_m-i\zeta)\sinh\left(\frac{F_N(\lambda_m)}{N\rho_N(\lambda_m)}\right)}+\mathcal{O}(1/N)\\\nonumber
= \prod_{i=1}^n \exp\left(\int_{-q}^q d\lambda F(\lambda)\left(\coth(\mu_i^--\lambda-i\zeta)-\coth(\mu_i^+-\lambda-i\zeta)\right)\right).
\eea
Which gives
\bea
p_2= \prod_{i=1}^n \frac{\exp\left(\int_{-q}^q d\lambda \frac{F(\lambda)}{\tanh(\mu_i^--\lambda-i\zeta)}\right)}
{\exp\left(\int_{-q}^q d\lambda \frac{F(\lambda)}{\tanh(\mu_i^+-\lambda-i\zeta)}\right)}.
\eea
The final answer for $M_1$ is then given by the following expression
\bea\label{eq:M1_final}
\left|M_1\right| = \exp\left(-\frac{1}{2}\int_{-q}^q d\mu d\lambda \frac{F_N(\lambda)F_N(\mu)}{\tanh^2(\lambda-\mu-i\zeta)}\right)  \\\nonumber
\times\left|\prod_{i=1}^n \frac{\exp\left(\int_{-q}^q d\lambda \frac{F(\lambda)}{\tanh(\mu_i^--\lambda-i\zeta)}\right)}
{\exp\left(\int_{-q}^q d\lambda \frac{F(\lambda)}{\tanh(\mu_i^+-\lambda-i\zeta)}\right)}\right|\\\nonumber
\times{\prod_{i,j=1}^n}\left| \frac{\sinh(\mu_i^+-\mu_j^- -i\zeta)}{\sinh^{1/2}(\mu_i^+-\mu_j^+-i\zeta)\sinh^{1/2}(\mu_i^--\mu^-_j-i\zeta)}\right|+\mathcal{O}(1/N).
\eea
A similar formula holds for $\tilde{M}_1$ with $F_+(\lambda)$ replacing $F(\lambda)$.

\subsection{Calculation of $M_2$}
Let us recall once again the expression we are going to consider in this section
\bea
M_2 = \prod_{j\neq i}\frac{\sinh^{1/2}(\mu_j-\mu_i)\sinh^{1/2}(\lambda_j-\lambda_i)}{ \sinh(\mu_i - \lambda_j)}.
\eea 
The double product can be written as a product of 3 different expressions
\bea
p_1 = {\prod_{j\neq i}^M}{''} \frac{\sinh^{1/2}(\mu_j-\mu_i)\sinh^{1/2}(\lambda_j-\lambda_i)}{\sinh(\mu_j-\lambda_i)},\\
p_2 = \prod_{h=1}^n {\prod_{i=1}^M}{''} \frac{\sinh(\lambda_i-\mu_h^-)}{\sinh(\mu_i-\mu_h^-)}\frac{\sinh(\mu_i-\mu_h^+)}{\sinh(\lambda_i-\mu_h^+)},\\
p_3 = \prod_{i\neq j}^n \frac{\sinh^{1/2}(\mu_j^+-\mu_j^+)\sinh^{1/2}(\mu_j^--\mu_j^-)}{\sinh(\mu_j^+-\mu_j^-)},
\eea
which we obtained from rewriting the expression under the double product as
\bea\nonumber
\prod_{j\neq i}^M \sinh(\mu_j-\mu_i) = {\prod_{j\neq i}^M}{''} \sinh(\mu_j-\mu_i)\prod_{i=1}^n {\prod_{j=1,\neq i}^M}{''} \frac{\sinh^2(\mu_j-\mu_i^+)}{\sinh^2(\mu_j-\mu_i^-)}\\\nonumber
\times\prod_{i\neq j}^n \frac{\sinh(\mu_j^+-\mu_j^+)\sinh(\mu_j^--\mu_j^-)}{\sinh^2(\mu_j^+-\mu_j^-)},\\\nonumber
\prod_{j\neq i}^M \sinh(\mu_i-\lambda_j) = {\prod_{j\neq i}^M}{''}\sinh(\mu_i-\lambda_j)\prod_{i=1}^n \prod_{j=1,\neq i}^M {''} \frac{\sinh(\mu_i^+-\lambda_j)}{\sinh(\mu_i^--\lambda_j)}
\eea
Let's focus on the first expression. We start with
\bea\fl\nonumber
p_1 = \prod_{j\neq i}^M \frac{\sinh^{1/2}(\mu_j^--\mu_i^-)\sinh^{1/2}(\lambda_j-\lambda_i)}{\sinh(\mu_j^--\lambda_i)}\\\nonumber\fl
=\prod_{j\neq i}^M \frac{\cosh^{1/2}\left(\frac{F(\lambda_j)}{N\rho(\lambda_j)}-\frac{F(\lambda_i)}{N\rho(\lambda_i)}\right)}{\cosh(\frac{F(\lambda_j)}{N\rho(\lambda_j)})}
\frac{\left(1-\coth(\lambda_j-\lambda_i)\tanh\left(\frac{F(\lambda_j)}{N\rho(\lambda_j)}-\frac{F(\lambda_i)}{N\rho(\lambda_i)}\right)\right)^{1/2}}{\left(1-\coth(\lambda_j-\lambda_i)\tanh\left(\frac{F(\lambda_j)}{N\rho(\lambda_j)}\right)\right)}\\\fl
=\exp\left[-\frac{1}{2}\left(\int_{-q}^q d\lambda F(\lambda)\right)^2\right]
\prod_{j\neq i}^M\frac{\left(1-\coth(\lambda_j-\lambda_i)\tanh\left(\frac{F(\lambda_j)}{N\rho(\lambda_j)}-\frac{F(\lambda_i)}{N\rho(\lambda_i)}\right)\right)^{1/2}}{\left(1-\coth(\lambda_j-\lambda_i)\tanh\left(\frac{F(\lambda_j)}{N\rho(\lambda_j)}\right)\right)},\nn\\
\eea
where we used that
\bea\fl
\prod_{j\neq i}^M \frac{\cosh^{1/2}\left(\frac{F(\lambda_j)}{N\rho(\lambda_j)}-\frac{F(\lambda_i)}{N\rho(\lambda_i)}\right)}{\cosh(\frac{F(\lambda_j)}{N\rho(\lambda_j)})} = \exp\left[-\frac{1}{2}\left(\int_{-q}^q d\lambda F(\lambda)\right)^2\right] + \mathcal{O}(1/N).
\eea
Calculation of the second product is more difficult. Let us use the following notation
\bea
T = \prod_{j\neq i}^M\frac{\left(1-\coth(\lambda_j-\lambda_i)\tanh\left(\frac{F(\lambda_j)}{N\rho(\lambda_j)}-\frac{F(\lambda_i)}{N\rho(\lambda_i)}\right)\right)^{1/2}}{\left(1-\coth(\lambda_j-\lambda_i)\tanh\left(\frac{F(\lambda_j)}{N\rho(\lambda_j)}\right)\right)},
\eea
and consider different regions. We follow the same logic as in the case of the Bose gas. The names of the regions are also the same. The details of the computation are outlined in section \ref{sec:M_2}.

\subsubsection{Region I}
In the first region, where we can expand the product logarithmically, we obtain
\bea\nonumber\fl
\log T_{I} = \frac{1}{2}\sum_{j\neq i \in I} \coth^2(\lambda_j-\lambda_i)\frac{F(\lambda_j)F(\lambda_i)}{N^2\rho(\lambda_i)\rho(\lambda_j)}\\\fl
=\frac{1}{2}\left(\int_{-q+\nu^*(-q)}^{q} d\lambda \int_{-q}^{\lambda-\nu^*(\lambda)} d\mu + \int_{-q}^{q-\nu^*(q)} d\lambda \int_{\lambda+\nu^*(\lambda)}^{q} d\mu\right) \coth^2(\lambda_j-\lambda_i)F(\lambda)F(\mu).\nonumber\\\fl
=\frac{1}{2}\int_{-q+\nu^*(-q)}^q d\lambda\, F^2(\lambda )\int_{-q}^{\lambda-\nu^*(\lambda)} d\mu\coth^2(\lambda-\mu)\fl\nonumber\\
+\frac{1}{2}\int_{-q}^{q-\nu^*(q)} d\lambda\, F^2(\lambda )\int_{\lambda+\nu^*(\lambda)}^{q} d\mu\coth^2(\lambda-\mu)\fl\nonumber\\\fl
-\frac{1}{4}\int_{-q+\nu^*(-q)}^q d\lambda\, \int_{-q}^{\lambda-\nu^*(\lambda)} d\mu\coth^2(\lambda-\mu)\left(F(\lambda)-F(\mu)\right)^2\fl\nonumber\\\fl
-\frac{1}{4}\int_{-q}^{q-\nu^*(q)} d\lambda\, \int_{\lambda+\nu^*(\lambda)}^{q} d\mu\coth^2(\lambda-\mu)\left(F(\lambda)-F(\mu)\right)^2.
\eea
The last two integrals are actually cut-off independent, one can extend limits of integration without introducing any error nor divergence, therefore
\bea
\log T_{I} = \frac{1}{2}\int_{-q+\nu^*(-q)}^q d\lambda\, F^2(\lambda )\int_{-q}^{\lambda-\nu^*(\lambda)} d\mu\coth^2(\lambda-\mu)\nonumber\\
+\frac{1}{2}\int_{-q}^{q-\nu^*(q)} d\lambda\, F^2(\lambda )\int_{\lambda+\nu^*(\lambda)}^{q} d\mu\coth^2(\lambda-\mu)\nonumber\\
-\frac{1}{4}\int_{-q}^q d\lambda\, \int_{-q}^{q} d\mu\coth^2(\lambda-\mu)\left(F(\lambda)-F(\mu)\right)^2.
\eea 
The first two integrals are cut-off dependent and should simplify once we combine answers from different regions. Now let us move to  region II.

\subsubsection{Region II}
In this region we can make the following approximation
\bea
\lambda_j-\lambda_k = \frac{1}{N\rho(\lambda_j)}(j-k)-\frac{C_j^b}{N^2\rho(\lambda_j)} (j-k)^2 + \mathcal{O}(1/N^3),
\eea
where
\bea
C_j^b = -\frac{\rho'(\lambda_j)}{2\rho(\lambda_j)}.
\eea
First one observes that
\bea
\frac{F(\lambda_j)}{\rho(\lambda_j)}-\frac{F(\lambda_i)}{\rho(\lambda_i)} = \frac{d}{d\lambda}\left(\frac{F(\lambda_j)}{\rho(\lambda_j)}\right)|_{\lambda=\lambda_j}(\lambda_j-\lambda_i) \equiv C_j^a (\lambda_j-\lambda_i),
\eea
where we have defined $C_j^a$. We only need to consider terms up to $\mathcal{O}(1/N)$ because they are the only ones contributing in the thermodynamic limit. Then we have
\bea\fl\nonumber
T_{II} = \prod_{i\neq j \in II} \frac{\tanh^{1/2}(\lambda_j-\lambda_i)\left(\tanh(\lambda_j-\lambda_i)-\tanh\left(\frac{F(\lambda_j)}{N\rho(\lambda_j)}-\frac{F(\lambda_i)}{N\rho(\lambda_i)}\right)\right)^{1/2}}{\left(\tanh(\lambda_j-\lambda_i)-\tanh\left(\frac{F(\lambda_j)}{N\rho(\lambda_j)}\right)\right)} \\\fl
= \left(\prod_{i\neq j \in II} \frac{\tanh(\lambda_j-\lambda_i)}{\tanh(\lambda_j-\lambda_i)-\tanh\left(\frac{F(\lambda_j)}{N\rho(\lambda_j)}\right)}
\frac{\tanh(\lambda_j-\lambda_i)-\tanh\left(\frac{F(\lambda_j)}{N\rho(\lambda_j)}-\frac{F(\lambda_i)}{N\rho(\lambda_i)}\right)}{\tanh(\lambda_j-\lambda_i)+\tanh\left(\frac{F(\lambda_i)}{N\rho(\lambda_i)}\right)}\right)^{1/2}.\nn\\
\eea
For the first product we obtain
\bea\fl
\frac{\tanh(\lambda_j-\lambda_i)}{\tanh(\lambda_j-\lambda_i)-\tanh\left(\frac{F(\lambda_j)}{N\rho(\lambda_j)}\right)}
=\frac{j-i}{j-i-\frac{F(\lambda_j)}{1-\frac{(j-i)C_j^b}{N}}}+ \mathcal{O}(1/N^2).
\eea
For the second
\bea\fl
\frac{\tanh(\lambda_j-\lambda_i)-\tanh\left(\frac{F(\lambda_j)}{N\rho(\lambda_j)}-\frac{F(\lambda_i)}{N\rho(\lambda_i)}\right)}{\tanh(\lambda_j-\lambda_i)+\tanh\left(\frac{F(\lambda_i)}{N\rho(\lambda_i)}\right)}
=\frac{j-i}{j-i+\frac{F(\lambda_j)}{1-\frac{(j-i)C_j^b-C_j^a}{N}}}+ \mathcal{O}(1/N^2)
\eea
Therefore
\bea\fl
T_{II} = \prod_{i\neq j \in II} \frac{j-i}{\left(j-i-\frac{F(\lambda_j)}{1-\frac{(j-i)C_j^b}{N}}\right)^{1/2}\left(j-i-\frac{F(\lambda_j)}{1-\frac{(j-i)C_j^b-C_j^a}{N}}\right)^{1/2}}.
\eea
We obtain the same expression as for the Bose gas (see \ceq{t2a}, therefore as a final answer in this region we have (\ceq{t2region})
\bea\fl\nonumber
T_{II} = \prod_{j=1}^M \left(\frac{\pi F(\lambda_j)}{\sin(\pi F(\lambda_j)}\right) \exp\left(\frac{1}{2}\int_{-q}^q d\lambda \left(1-\pi F(\lambda)\cot(\pi F(\lambda))\right)\left(F'(\lambda)-\frac{F(\lambda)\rho'(\lambda)}{\rho(\lambda)}\right)\right)\\\nonumber
\times\exp\left(-\frac{N}{n^*}\int_{-q}^{q-\nu^*(q)} d\lambda\rho(\lambda) F^2(\lambda)- \frac{N}{n^*}\int_{-q+\nu^*(-q)}^{q} d\lambda\rho(\lambda) F^2(\lambda)\right)\\
\times\left(\frac{1}{\Gamma(1-F(q))\Gamma(1+F(q))\Gamma(1+F(-q))\Gamma(1-F(-q))}\right)^{n^*/2}.
\eea

\subsubsection{Region III}
In region III we again obtain the same answer as for the Bose gas (\ceq{t3a}, \ceq{t3b}), hence
\bea\fl\nn
T_{III} = \left[G(1+F(q))G(1-F(q))G(1+F(-q))G(1-F(-q))\right]^{1/2}\\\fl
\times (n^*)^{\frac{-F^2(q)-F^2(-q)}{2}}\left[\Gamma(1-F(q))\Gamma(1+F(q))\Gamma(1+F(-q))\Gamma(1-F(-q))\right]^{n^*/2}.
\eea

\subsubsection{Cutoff dependent part}
In the end, let us collect all the cut-off dependent parts of $T_{I}$, $T_{II}$ and $T_{III}$, altogether they give
\bea\fl\nonumber
log(T_{cutoff}) = -\frac{1}{2}(F^2(q)+F^2(-q))\log(n^*)\\\nonumber
+\frac{1}{2}\int_{-q+\nu^*(-q)}^q d\lambda\, F^2(\lambda )\left(\int_{-q}^{\lambda-\nu^*(\lambda)} d\mu\coth^2(\lambda-\mu)-\frac{1}{\nu^*(\lambda)}\right)\nonumber\\
+\frac{1}{2}\int_{-q}^{q-\nu^*(q)} d\lambda\, F^2(\lambda )\left(\int_{\lambda+\nu^*(\lambda)}^{q} d\mu\coth^2(\lambda-\mu)-\frac{1}{\nu^*(\lambda)}\right).
\eea
Let us start with one of the inner integrals
\bea\fl
\int_{-q}^{\lambda-\nu^*(\lambda)} d\mu\coth^2(\lambda-\mu)-\frac{1}{\nu^*(\lambda)} = \lambda-\frac{2}{3}\nu^*(\lambda)+q-\coth(\lambda+q)+\mathcal{O}\left((\nu^*(\lambda))^3\right)\nn\\
\eea
and then
\bea\nonumber
\int_{-q+\nu^*(-q)}^q d\lambda\, F^2(\lambda )\left(\int_{-q}^{\lambda-\nu^*(\lambda)} d\mu\coth^2(\lambda-\mu)-\frac{1}{\nu^*(\lambda)}\right)\\\nonumber
=-\int_{-q+\nu^*(-q)}^q d\lambda F^2(\lambda)\coth(\lambda+q) + \int_{-q}^q d\lambda F^2(\lambda) (q+\lambda) +\mathcal{O}(\nu^*(\lambda))\\\nonumber
=-P_-\int_{-1}^1 dx \frac{qF^2(qx)}{\tanh(q(x+1))} +F^2(-q)\log(\nu^*(-q))\\
+\int_{-q}^q d\lambda F^2(\lambda) (q+\lambda) +\mathcal{O}(\nu^*(\lambda)).
\eea
Similar calculation for the second integral gives
\bea\nonumber
\int_{-q}^{q-\nu^*(q)} d\lambda\, F^2(\lambda )\left(\int_{\lambda+\nu^*(\lambda)}^{q} d\mu\coth^2(\lambda-\mu)-\frac{1}{\nu^*(\lambda)}\right)\\\nonumber
=\int_{-q}^{q-\nu^*(q)} d\lambda\, F^2(\lambda ) \coth(\lambda-q)+\int_{-q}^q F^2(\lambda)(q-\lambda) + \mathcal{O}(\nu^*(\lambda))\\\nonumber
=P_+\int_{-1}^1 dx \frac{qF^2(qx)}{\tanh(q(x-1))} +F^2(q)\log(\nu^*(q))\\
+\int_{-q}^q d\lambda F^2(\lambda) (q-\lambda) +\mathcal{O}(\nu^*(\lambda)).
\eea
Finally for the cut-off dependent part we get ($\rho(q)=\rho(-q)$)
\bea\nonumber
log(T_{cutoff}) = -\frac{1}{2}(F^2(q)+F^2(-q))\log(qN\rho(q))+q\int_{-q}^q d\lambda F^2(\lambda)\\
+ \frac{1}{2}P_+\int_{-1}^1 dx \frac{qF^2(qx)}{\tanh(q(x-1))}-\frac{1}{2}P_-\int_{-1}^1 dx \frac{qF^2(qx)}{\tanh(q(x+1))}.
\eea
Gathering together results from all the regions we obtain
\bea\fl\nonumber
p_1 = \left(qN\rho(q)\right)^{-\frac{(F^2(q)+F^2(-q))}{2}}\left(G(1+F(q))G(1-F(q))G(1+F(-q))G(1-F(-q))\right)^{1/2}\\\nonumber\fl
\times\prod_{j=1}^M \left(\frac{\pi F(\lambda_j)}{\sin(\pi F(\lambda_j)}\right)\exp\left(-\frac{1}{4}\int_{-q}^q d\lambda\, d\mu\left(\frac{F(\lambda)-F(\mu)}{\sinh(\lambda-\mu)}\right)^2\right)\\\nonumber\fl  
\times\exp\left(\frac{1}{2}\int_{-q}^q d\lambda \left(1-\pi F(\lambda)\cot(\pi F(\lambda))\right)\left(F'(\lambda)-\frac{F(\lambda)\rho'(\lambda)}{\rho(\lambda)}\right)\right)\\\fl
\times\exp\left(\frac{1}{2}P_+\int_{-1}^1 dx \frac{qF^2(qx)}{\tanh(q(x-1))}-\frac{1}{2}P_-\int_{-1}^1 dx \frac{qF^2(qx)}{\tanh(q(x+1))}\right),
\eea
where we used that
\bea\fl
q\int_{-q}^q d\lambda F^2(\lambda) = \frac{1}{4}\int_{-q}^q d\lambda d\mu \left(F(\lambda)-F(\mu)\right)^2 +\frac{1}{2}\left(\int_{-q}^q d\lambda F(\lambda)\right)^2
\eea

\subsubsection{Calculation of $p_2$}
The calculation of $p_2$ is divided into 2 steps. First we will calculate the hole terms, thereafter the particles terms. We start with
\bea\nonumber
T_{hole} = \prod_{h=1}^n \prod_{i=1,\neq h}^M \frac{\sinh(\lambda_i-\mu_h^-)}{\sinh(\mu_i^--\mu_h^-)}\\
=\prod_{h=1}^n \prod_{i=1,\neq h}^M \left(1-\coth(\lambda_i-\mu_h)\frac{F(\lambda_i)}{N\rho(\lambda_i)}\right)^{-1}\times\left(1+\mathcal{O}(1/N)\right).
\eea 
There are again two qualitatively different situations, first when holes are near the Fermi edges, second when they are located deep inside the Fermi sea. Without making any mistake we can consider a single hole and then simply generalise to many holes by multiplying single contributions. Divide now the product into two parts, first when the hole rapidity is well separated from other rapidities, and second when the hole rapidity is close to the other rapidity. We obtain
\bea
T_{hole}^I = \exp\left(\sum_{|j-h|> n^*}\frac{\coth(\lambda_j-\lambda_h)F(\lambda_i)}{N\rho(\lambda_i)}\right)+\mathcal{O}(1/N),\\
T_{hole}^{II} = \prod_{|j-h|<n^*} \frac{i-h+F(\lambda_h)}{i-h}+\mathcal{O}(1/N),\\
T_{hole} = T_{hole}^I \times T_{hole}^{II}.
\eea

{\bf Case I : Hole is near the quasi-Fermi point}

We obtain a similar expression as for the Bose gas (\ceq{hfer})
\bea\nonumber
T_{hole}^{h,edge} = \left(qN\rho(q)\right)^{\mp F(\pm q)}\frac{\sin\left(\pi F(q)\right)}{\pi F(q)}\frac{\Gamma\left(\mp q_h^{\pm} \pm F(\pm q) \right)}{\Gamma(\mp q_h^{\pm})}\\
\times\exp\left(P_{\pm}\int_{-1}^1 dx \frac{qF(qx)}{\tanh(q(x\mp 1))}\right).
\eea
Number $q_h^{\pm}$ is a quantum number that parametrizes the position of the hole and is defined in the same way as for the Bose gas.

{\bf Case II : Hole is deep inside the distribution}

The same situation here (\ceq{tholedeep})
\bea
T_{hole}^{h,deep} = \frac{\sin\left(\pi F(\mu_h^-)\right)}{\pi F(\mu_h^-)}\exp\left(P\int_{-1}^1 dx \frac{q F(qx)} {\tanh\left(q\left(x-\frac{\mu_h^-}{q}\right)\right)}\right).
\eea

{\bf Particle term}

The computation of the particle part goes along the same lines and gives (compare with \ceq{tpart1}, \ceq{tpart2})
\bea\nonumber
T_{particle}^{h, near} = \left(qN\rho(q)\right)^{\pm F(\pm q)}\frac{\Gamma(\pm p_h^{\pm}+1\mp F(\pm q))}{\Gamma(\pm p_h^{\pm}+1)}\\
\times\exp\left(-P_{\pm}\int_{-1}^{1} dx \frac{q F(qx)} {\tanh(q(x\mp 1))}\right),\\
T_{particle}^{h, far} = \exp\left(-\int _{-1}^1 dx \frac{qF(qx)} {\tanh\left(q\left(x-\frac{\mu_h^+}{q}\right)\right)}\right).
\eea 
Here again $p_h^{\pm}$ parametrizes the position of a particle in the same way as for the Bose gas.
\subsubsection{Final answer}
Let us summarise now the calculation of $M_2$. In order to follow the same notation as in the case of the Bose gas we redefine term $p_3$ as $T_{cross}$ (see \ceq{tcross})
\bea
T_{cross} = \prod_{i\neq j}^n \frac{\sinh^{1/2}(\mu_j^+-\mu_j^+)\sinh^{1/2}(\mu_j^--\mu_j^-)}{\sinh(\mu_j^+-\mu_j^-)}.
\eea
The final answer for $M_2$ is
\bea\label{eq:M2_final}\fl\nonumber
M_2 = \left(qN\rho(q)\right)^{-\frac{(F^2(q)+F^2(-q))}{2}}\left(G(1+F(q))G(1-F(q))G(1+F(-q))G(1-F(-q))\right)^{1/2}\\\nonumber
\times\prod_{j=1}^M \left(\frac{\pi F(\lambda_j)}{\sin(\pi F(\lambda_j)}\right)\exp\left(-\frac{1}{4}\int_{-q}^q d\lambda\, d\mu\left(\frac{F(\lambda)-F(\mu)}{\sinh(\lambda-\mu)}\right)^2\right)\\\nonumber
\times\exp\left(\frac{1}{2}\int_{-q}^q d\lambda \left(1-\pi F(\lambda)\cot(\pi F(\lambda))\right)\left(F'(\lambda)-\frac{F(\lambda)\rho'(\lambda)}{\rho(\lambda)}\right)\right)\\\nonumber
\times\exp\left(\frac{1}{2}P_+\int_{-1}^1 dx \frac{qF^2(qx)}{\tanh(q(x-1))}-\frac{1}{2}P_-\int_{-1}^1 dx \frac{qF^2(qx)}{\tanh(q(x+1))}\right)\\
\times \prod_{i=1}^n \left(T_{hole}^{(i)}\times T_{particle}^{(i)}\right)\times T_{cross}.
\eea 

For the form factor of $S^-$ operator calculation of $\tilde{M}_2$ follows the same lines and the final answer is the same up to a change in shift function, $F_+(\lambda)$ instead of $F(\lambda)$.

\subsection{Fredholm determinants }
The last crucial step that we need to perform is take the thermodynamic limit of matrix elements  $G_{ab}^z$ and write its determinant in form of a Fredholm determinant. The expression for $G_{ab}^z$ reads
\bea\fl
G_{ab}^z = \frac{\sinh(i\zeta)}{\sinh(\lambda_b+i\zeta/2)} \frac{\prod_{m=1}^M\sinh(\lambda_b-\mu_m)}{\prod_{m=1,\neq b}^M\sinh(\lambda_b-\lambda_m)} \prod_{m=1}^M \frac{\sinh(\lambda_m-\lambda_b-i\zeta)}{\sinh(\mu_m-\lambda_b- i\zeta)} \\\fl
\times\left(1+\frac{\sinh^N(\lambda_a+i\zeta/2)}{\sinh^N(\lambda_b-i\zeta/2)}\frac{\prod_{m=1}^M \sinh(\mu_m-\lambda_b + i\zeta)}{\prod_{m=1}^M \sinh(\mu_m-\lambda_b- i\zeta)}\right)^{-1}\nonumber\\\fl\nn
\times\left(\frac{\sinh(\lambda_b+3i\zeta/2)}{\sinh(\lambda_a-\lambda_b-i\zeta)}-\frac{\sinh(\lambda_b-i\zeta/2)}{\sinh(\lambda_a-\lambda_b+i\zeta)}-\prod_{m=1}^M \frac{\sinh(\lambda_m-i\zeta/2)}{\sinh(\mu_m-i\zeta/2)}\frac{2\sinh(i\zeta)}{\sinh(\lambda_a-i\zeta/2)}\right)
\eea
We start by considering the first product
\bea\fl\nonumber
\frac{\prod_{m=1}^M\sinh(\lambda_a-\mu_m)}{\prod_{m=1,\neq a}^M\sinh(\lambda_a-\lambda_m)} = \sinh(\lambda_a-\mu_a)\prod_{m\neq a}\frac{\sinh(\lambda_a-\mu_m)}{\sinh(\lambda_a-\lambda_m)}\\\fl
=  \sinh(\lambda_a-\mu_a)\prod_{i=1}^n \frac{\sinh(\lambda_a-\mu_i^+)}{\sinh(\lambda_a-\mu_i^-)}
\prod_{m\neq a}\frac{\sinh\left(\lambda_a-\lambda_m+\frac{F_N(\lambda_m)}{N\rho_N(\lambda_m)}\right)}{\sinh(\lambda_a-\lambda_m)}
\eea
In order to calculate the thermodynamic limit of the last expression we need to be careful and split calculations into two regions. Let's define the cutoff distance $n^*$ and associate to it two rapidities $\lambda_L$ and $\lambda_R$ corresponding to rapidities with indices $a-n^*$ and $a+n^*$ respectively. Then we have
\bea\fl\nonumber
\prod_{m\neq a}\frac{\sinh\left(\lambda_a-\lambda_m+\frac{F_N(\lambda_m)}{N\rho_N(\lambda_m)}\right)}{\sinh(\lambda_a-\lambda_m)} = \prod_{m\neq a}\left(1+\coth(\lambda_a-\lambda_m)\frac{F_N(\lambda_m)}{N\rho(\lambda_m)}\right) + \mathcal{O}(1/N)\\\nonumber\fl
= \prod_{|m-a|>n^*}\left(1+\coth(\lambda_a-\lambda_m)\frac{F_N(\lambda_m)}{N\rho(\lambda_m)}\right)\prod_{|m-a|<n^*}\left(1+\coth(\lambda_a-\lambda_m)\frac{F_N(\lambda_m)}{N\rho(\lambda_m)}\right)\\\nonumber\fl
=\exp\left(\int_{-q}^{\lambda_R} d\lambda F(\lambda)\coth(\lambda_a-\lambda) + \int_{\lambda_L}^{q} d\lambda F(\lambda)\coth(\lambda_a-\lambda)\right)\\\nonumber\fl
\times\prod_{|m-a|<n^*}\left(\frac{a-m+F_N(\lambda_m)}{a-m}\right)+\mathcal{O}(1/N)\\\nonumber\fl
=\exp\left(P\int_{-q}^{q} d\lambda F(\lambda)\coth(\lambda_a-\lambda)\right)\frac{\Gamma(1+n^*+F_N(\lambda_a))\Gamma(1+n^*-F_N(\lambda_a))}{\Gamma^2(1+n^*)\Gamma(1+F_N(\lambda_a))\Gamma(1-F_N(\lambda_a))}\\\fl
=\exp\left(P\int_{-q}^{q} d\lambda F(\lambda)\coth(\lambda_a-\lambda)\right)\frac{\sin(\pi F(\lambda_a))}{\pi F(\lambda_a)}+\mathcal{O}(1/N).
\eea
Therefore 
\bea\fl\nonumber
\frac{\prod_{m=1}^M\sinh(\lambda_a-\mu_m)}{\prod_{m=1,\neq a}^M\sinh(\lambda_a-\lambda_m)} = \frac{\sin(\pi F(\lambda_a))}{N\rho(\lambda_a)}\exp\left(P\int_{-q}^{q} d\lambda F(\lambda)\coth(\lambda_a-\lambda)\right)\\
\times\prod_{i=1}^n \frac{\sinh(\lambda_a-\mu_i^+)}{\sinh(\lambda_a-\mu_i^-)}.
\eea
Now we move on to the next product. Here the thermodynamic limit is straightforward and gives
\bea\nonumber
\prod_{m=1}^M \frac{\sinh(\lambda_m-\lambda_a-i\zeta)}{\sinh(\mu_m-\lambda_a- i\zeta)} = \prod_{i=1}^n \frac{\sinh(\mu_i^- -\lambda_a-i\zeta)}{\sinh(\mu_i^+ - \lambda_a - i\zeta)}\\ 
\times\exp\left(\int_{-q}^q d\lambda F(\lambda) \coth(\lambda-\lambda_a -i\zeta)\right)+\mathcal{O}(1/N).
\eea
The next expression can be written using $M_0$ defined earlier \ceq{eq:M0_def}, and reads
\bea\fl\nonumber
\left(1+M_0(\lambda_a)\right)^{-1} = \left(1+\frac{\sinh^N(\lambda_a+i\zeta/2)}{\sinh^N(\lambda_a-i\zeta/2)}\frac{\prod_{m=1}^M \sinh(\mu_m-\lambda_a + i\zeta)}{\prod_{m=1}^M \sinh(\mu_m-\lambda_a- i\zeta)}\right)^{-1}\\
=\frac{\exp\left(i\pi F(\lambda_a)\right)}{2\sin(\pi F(\lambda_a))} + \mathcal{O}(1/N).
\eea
Finally, the last product gives
\bea\nn
\prod_{m=1}^M \frac{\sinh(\lambda_m-i\zeta/2)}{\sinh(\mu_m+i\zeta/2)} =\\= \prod_{i=1}^n \frac{\sinh(\mu_i^- +i\zeta/2)}{\sinh(\mu_i^+ +i\zeta/2)}\exp\left( \int _{-q}^q  \frac{F(\lambda)d\lambda}{\tanh(\lambda+i\zeta/2)}\right) + \mathcal{O}(1/N).
\eea
Therefore, the matrix elements $G_{ab}^z$ take form
\bea\fl\nn
G_{ab}^z = \frac{1}{2\pi N\rho(\lambda_a)}\frac{\sinh(i\zeta)}{\sinh(\lambda_b+i\zeta/2)}\prod_{i=1}^n \frac{\sinh(\mu_i^- -\lambda_a-i\zeta)}{\sinh(\mu_i^+ - \lambda_a - i\zeta)} \frac{\sinh(\lambda_a-\mu_i^+)}{\sinh(\lambda_a-\mu_i^-)}\\\nonumber\fl
\times\exp \left(P\int_{-q}^{q} d\lambda \frac{F(\lambda)}{\tanh(\lambda_a-\lambda)} +\int_{-q}^q d\lambda \frac{F(\lambda)}{ \tanh(\lambda-\lambda_a -i\zeta)}+ i\pi F(\lambda_a)\right)\\\nonumber\fl
\times\left(\prod_{i=1}^n \frac{\sinh(\mu_i^- -i\zeta/2)}{\sinh(\mu_i^+ -i\zeta/2)}\exp\left( \int _{-q}^q  \frac{F(\lambda)d\lambda}{\tanh(\lambda-i\zeta/2)}\right)\frac{-2\sinh(i\zeta)}{\sinh(\lambda_a-i\zeta/2)}+ \right. \\\fl \left. + \frac{\sinh(\lambda_b+3i\zeta/2)}{\sinh(\lambda_a-\lambda_b-i\zeta)}-\frac{\sinh(\lambda_b-i\zeta/2)}{\sinh(\lambda_a-\lambda_b+i\zeta)}\right).
\eea
Now we can define an operator $\hat{G}$ that acts on $(-q,q)$ and is given by
\bea\fl\nn\label{eq:Gz_def}
\hat{G}^z(\mu, \nu) = a(\nu)\left( \prod_{i=1}^n \frac{\sinh(\mu_i^- -i\zeta/2)}{\sinh(\mu_i^+ -i\zeta/2)}\exp\left( \int _{-q}^q  \frac{F(\lambda)d\lambda}{\tanh(\lambda-i\zeta/2)}\right)\frac{-2\sinh(i\zeta)}{\sinh(\mu-i\zeta/2)}+ \right. \\ \left. + \frac{\sinh(\nu+3i\zeta/2)}{\sinh(\mu-\nu-i\zeta)}-\frac{\sinh(\nu-i\zeta/2)}{\sinh(\mu-\nu+i\zeta)}\right),
\eea
where
\bea\fl\nonumber
a(\nu) = \frac{1}{2\pi}\frac{\sinh(i\zeta)}{\sinh(\nu+i\zeta/2)}\prod_{i=1}^n \frac{\sinh(\mu_i^- -\nu-i\zeta)}{\sinh(\mu_i^+ - \nu - i\zeta)} \frac{\sinh(\nu-\mu_i^+)}{\sinh(\nu-\mu_i^-)}\\\fl
\times\exp \left(P\int_{-q}^{q} d\lambda \frac{F(\lambda)}{\tanh(\nu-\lambda)} +\int_{-q}^q d\lambda \frac{F(\lambda)}{ \tanh(\lambda-\nu -i\zeta)}+ i\pi F(\nu)\right).
\eea
Finally
\bea
\det_M (\delta_{ab}+G_{ab}^z) = \det\left(1+\hat{G}^z\right) + \mathcal{O}(1/N),
\eea
where it is understood that the determinant on the r.h.s is a Fredholm determinant.

\subsubsection{Determinant in $S_q^-$ form factor}
The matrix appearing in the determinant in the $S_q^-$ form factor is given by \ceq{eq:G-_def}
\bea\nonumber\fl
G_{ab}^- = \frac{1}{|\sinh(\lambda_b+i\zeta/2)|^2}\frac{\prod_{m=1}^{M+1}\sinh(\lambda_b-\mu_m)}{\prod_{m=1,\neq b}^{M}\sinh(\lambda_b-\lambda_m)}\frac{\prod_{m=1}^{M} \sinh(\lambda_m-\lambda_b- i\zeta)}{\prod_{m=1}^{M+1}\sinh(\mu_m-\lambda_b-i\zeta)}\\\nonumber
\times\left(1+\frac{\sinh^N(\lambda_b+i\zeta/2)}{\sinh^N(\lambda_b-i\zeta/2)}\prod_{m=1}^M \frac{\sinh(\mu_m-\lambda_a + i\zeta)}{\sinh(\mu_m-\lambda_a - i\zeta)}\right)^{-1}\\
\times\left(\frac{\sinh(\lambda_b-i\zeta/2)\sinh(\lambda_b-3i\zeta/2)}{\sinh(\lambda_a-\lambda_b- i \zeta)} - c.c.\right).
\eea
The thermodynamic limit here is similar to the case studied above. The only difference is the extra rapidity $\mu_{n+1}^+$ appearing now and then. In the thermodynamic limit we have
\bea
\det_M\left(\delta_{ab}+G_{ab}^-\right) = \det\left(1+H^-(\mu,\nu)\right) + \mathcal{O}(1/N).
\eea
The operator $H^-(\mu,\nu)$ acts again on $(-q,q)$ and is given by
\bea\label{eq:H-_def}
H^-(\mu,\nu)  = b^-(\nu)\left(\frac{\sinh(\nu-i\zeta/2)\sinh(\nu-3i\zeta/2)}{\sinh(\mu-\nu- i \zeta)}- c.c\right),
\eea 
where 
\bea\fl\nonumber
b^-(\nu) = \frac{1}{2\pi |\sinh(\nu+i\zeta/2)|^2}\prod_{i=1}^n \frac{\sinh(\mu_i^--\nu-i\zeta)}{\sinh(\mu_i^--\nu)}\prod_{i=1}^{n+1}\frac{\sinh(\mu_i^+-\nu)}{\sinh(\mu_i^+-\nu-i\zeta)}\\\fl
\times \exp\left(\int_{-q}^q d\lambda \frac{F_+(\lambda)}{\tanh(\lambda-\nu-i\zeta)}-P\int_{-q}^q d\lambda \frac{F_+(\lambda)}{\tanh(\lambda-\nu)}+i\pi F_+(\lambda)\right).
\eea

\end{document}